\documentclass[preprint2]{exoplanet} 

\usepackage{times}
\usepackage{amsmath}
\usepackage{amssymb}
\usepackage{amsthm}

\def\m{\rm \,m}

\def\K{\rm \,K}

\def\W{\rm \,W}

\def\km{\rm \, km}

\def\tt{\bf }

\begin{document}

\title{\textbf{\LARGE Atmospheric Circulation of Terrestrial Exoplanets}}

\author {\textbf{\large Adam P. Showman}}
\affil{\small\em University of Arizona}

\author {\textbf{\large Robin D. Wordsworth}}
\affil{\small\em University of Chicago}

\author {\textbf{\large Timothy M. Merlis}}
\affil{\small\em Princeton University}

\author{\textbf{\large Yohai Kaspi}}
\affil{\small\em Weizmann Institute of Science}

\begin{abstract}
\begin{list}{ } {\rightmargin 1in}
\baselineskip = 11pt
\parindent=1pc {\small The investigation of planets around other stars
  began with the study of gas giants, but is now extending to the
  discovery and characterization of super-Earths and terrestrial
  planets. Motivated by this observational tide, we survey the basic
  dynamical principles governing the atmospheric circulation of
  terrestrial exoplanets, and discuss the interaction of their
  circulation with the hydrological cycle and global-scale climate
  feedbacks.  Terrestrial exoplanets occupy a wide range of physical
  and dynamical conditions, only a small fraction of which have yet
  been explored in detail. Our approach is to lay out the fundamental
  dynamical principles governing the atmospheric circulation on
  terrestrial planets---broadly defined---and show how they can provide a
  foundation for understanding the atmospheric behavior of these
  worlds. We first survey basic atmospheric dynamics, including the
  role of geostrophy, baroclinic instabilities, and jets in the
  strongly rotating regime (the ``extratropics'') and the role of the
  Hadley circulation, wave adjustment of the thermal structure, and
  the tendency toward equatorial superrotation in the slowly rotating
  regime (the ``tropics''). We then survey key elements of the
  hydrological cycle, including the factors that control
  precipitation, humidity, and cloudiness. Next, we summarize key
  mechanisms by which the circulation affects the global-mean climate,
  and hence planetary habitability. In particular, we discuss the
  runaway greenhouse, transitions to snowball states, atmospheric
  collapse, and the links between atmospheric circulation and CO$_2$
  weathering rates.  We finish by summarizing the key questions and
  challenges for this emerging field in the future.
 \\~\\~\\~}
\end{list}
\end{abstract}

\section{INTRODUCTION}
\label{Intro}

The study of planets around other stars is an exploding field.  To
date, numerous exoplanets have been discovered, spanning a wide range
of masses, incident stellar fluxes, orbital periods, and orbital
eccentricities.  A variety of observing methods have allowed
observational characterization of the atmospheres of these exoplanets,
opening a new field in comparative climatology \citep[for
  introductions, see][]{deming-seager-2009, seager-deming-2010}.
Because of their relative ease of observability, this effort to date
has emphasized transiting giant planets with orbital semi-major axes
of $\sim$0.1 AU or less. The combination of radial velocity and
transit data together allow estimates of the planetary mass, radius,
density, and surface gravity.  Wavelength-dependent observations of
the transit and secondary eclipse, when the planet passes in front of
and behind its star, respectively, as well as the full-orbit light
curves, allow the atmospheric composition, vertical temperature
structure, and global temperature maps to be derived---at least over
certain ranges of pressure.  These observations provide evidence for a
vigorous atmospheric circulation on these worlds \citep[e.g.][]
{knutson-etal-2007b}.

Despite a major emphasis to date on extrasolar giant planets (EGPs),
super Earths and terrestrial planets are increasingly becoming
accessible to discovery and characterization.  Over 50 super Earths
have been confirmed from groundbased and spacebased planet searches,
including planets that are Earth sized \citep{fressin-etal-2012} and
planets that orbit within the classical habitable zones (HZ) of their
stars \citep{borucki-etal-2012, borucki-etal-2013}. Hundreds of
additional super-Earth candidates have been found by the NASA {\it
  Kepler} mission \citep{borucki-etal-2011}.  For observationally
favorable systems, such as super Earths orbiting M dwarfs
\citep{charbonneau-etal-2009}, atmospheric characterization has
already begun \citep[e.g.][]{bean-etal-2010, desert-etal-2011,
  berta-etal-2012}, placing constraints on the atmospheric composition
of these objects.  Methods that are used today for EGPs to obtain
dayside infrared spectra, map the day-night temperature pattern, and
constrain cloudiness and albedo will be extended to the
characterization of smaller planets in coming years.  This
observational vanguard will continue over the next decade with
attempts to determine the composition, structure, climate, and
habitability of these worlds.  Prominent next-generation observational
platforms include NASA's James Webb Space Telescope (JWST) and Transiting Exoplanet
Survey Satellite (TESS), as well as a wide range of upcoming groundbased
instruments.

This observational tide provides motivation for
understanding the circulation regimes of extrasolar terrestrial planets.  
Fundamental motivations are threefold.  First, we wish to 
understand current and future spectra, lightcurves, and other observations,
and understand the role of dynamics in affecting these observables.
Second, the circulation---and climate generally---play a key role in
shaping the habitability of terrestrial exoplanets, a subject of crucial
importance in understanding our place in the Cosmos.  Third, the wide
range of conditions experienced by exoplanets allows an opportunity to
extend our theoretical understanding of atmospheric circulation and climate
to a much wider range of conditions than encountered in the solar system.
Many of the dynamical mechanisms controlling the atmospheric circulation
of terrestrial exoplanets will bear similarity to those operating
on Earth, Mars, Venus, and Titan---but with different details and
spanning a much wider continuum of behaviors.  Understanding this
richness is one of the main benefits to be gained from the
study of terrestrial exoplanets.

Here we survey the basic dynamical principles governing the
atmospheric circulation of terrestrial exoplanets, discuss the
interaction of their circulation with global-scale climate feedbacks,
and review the specific circulation models of terrestrial exoplanets
that have been published to date.  Much of our goal is to provide a
resource summarizing basic dynamical processes, as distilled from not
only the exoplanet literature but also the solar system and
terrestrial literature, that will prove useful for understanding these
fascinating atmospheres. Our intended audience includes graduate students
and researchers in the fields of astronomy, planetary science, and
climate---without backgrounds in atmospheric dynamics---who wish to learn about, and
perhaps enter, the field.  This review builds on the previous survey
of the atmospheric circulation on exoplanets by
\citet{showman-etal-2010} and complements other review articles in the
literature describing the dynamics of solar-system atmospheres
\citep[e.g.,][as well as the other chapters in this
  volume]{gierasch-etal-1997, leovy-2001, read-lewis-2004,
  ingersoll-etal-2004, vasavada-showman-2005, schneider-2006,
  flasar-etal-2009, delgenio-etal-2009, read-2011}.

Section~\ref{overview} provides an overview of dynamical fundamentals
of terrestrial planet atmospheres, including both the rapidly and
slowly rotating regimes, with a particular emphasis on aspects
relevant to terrestrial exoplanets.  Section~\ref{hydrological}
reviews basic aspects of the hydrological cycle relevant to
understanding exoplanets with moist atmospheres.
Section~\ref{climate} addresses the interaction between atmospheric
dynamics and global-scale climate, and Section~\ref{conclusion}
summarizes outstanding questions.

\begin{figure*}
 \epsscale{1.6}
\plotone{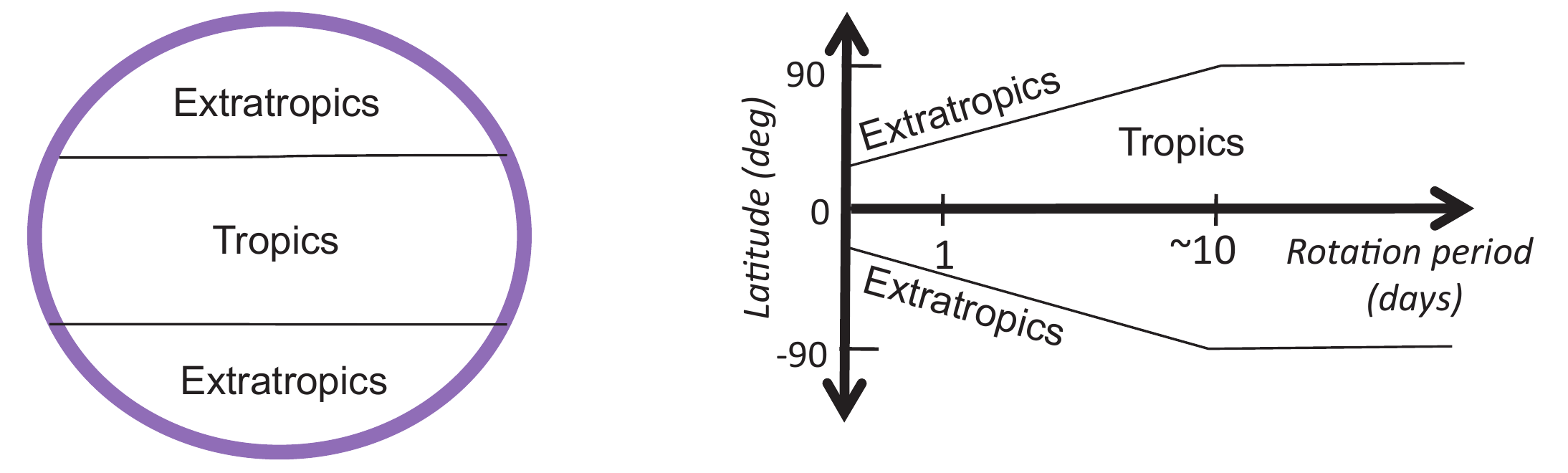}
 \caption{Schematic illustration of the regimes of extratropics 
(defined as $Ro \ll 1$) and tropics (defined as $Ro\gtrsim 1$). 
For an Earth- or Mars-like planet, the boundary between the
regimes occurs at $\sim$20--$30^{\circ}$ latitude ({\it left panel});
however, the transition occurs at higher latitudes when the rotation
period is longer, and terrestrial planets with rotation periods longer than
$\sim$10 Earth days may represent ``all tropics'' worlds
({\it right panel}).}
\label{tropics-extratropics-schematic}
 \end{figure*}

\section{OVERVIEW OF DYNAMICAL FUNDAMENTALS}
\label{overview}

Key to understanding atmospheric circulation is understanding the
extent to which rotation dominates the dynamics.  This can be
quantified by the Rossby number, $Ro=U/fL$, defined as the ratio
between advection forces and Coriolis forces in the horizontal
momentum equation.  Here, $U$ is a characteristic horizontal wind
speed, $f=2\Omega\sin\phi$ is the Coriolis parameter, $L$ is a
characteristic horizontal scale of the flow ($\sim$$10^3\,$km or more
for the global-scale flows considered here), $\Omega$ is the 
planetary rotation rate ($2\pi$ over the rotation period), and
$\phi$ is the latitude.  

In nearly inviscid atmospheres, the dynamical regime differs greatly
depending on whether the Rossby number is small or large.  When
$Ro\ll1$, the dynamics are rotationally dominated.  Coriolis forces
approximately balance pressure-gradient forces in the horizontal
momentum equation.  This so-called geostrophic balance supports the
emergence of large horizontal temperature gradients; as a result, the
atmosphere is generally unstable to a type of instability known as
baroclinic instability.  These instabilities generate eddies that
dominate much of the dynamics, controlling the equator-to-pole heat
fluxes, temperature contrasts, meridional mixing rates, vertical
stratification, and the formation of zonal jets.\footnote{The terms
zonal and meridional
refer to the east-west and north-south directions, respectively; thus,
the zonal wind is eastward wind, meridional wind is northward wind,
and meridional mixing rates refer to mixing rates in the north-south
direction.  A zonal average is an average in longitude.}
\footnote{Except for regions close to the surface, atmospheres are 
generally stably stratified, meaning they are stable to dry convection:
air parcels that are displaced upward or downward will return
to their original location rather than convecting.}  On the other hand,
when $Ro\gtrsim 1$, rotation plays a modest role; the dynamics are
inherently ageostrophic, horizontal temperature contrasts tend to be
small, and baroclinic instability is less important or negligible.
The temperature structure is regulated by a large-scale overturning
circulation that transports air latitudinally---that is, the Hadley
circulation---as well as by adjustment of the thermal structure due to
atmospheric waves.  These two regimes differ sufficiently that they
are best treated separately.

It is useful to define terms for these regimes.
In Earth's atmosphere, the regime of $Ro\gtrsim 1$ approximately
coincides with the tropics, occurring equatorward of
$\sim$20--$30^{\circ}$ latitude, whereas the regime of $Ro\ll 1$ approximately
coincides with the extratropics, occuring poleward of $\sim$$30^{\circ}$
latitude.  Broadening our scope to other planets, we {\it define}
the ``tropics'' and ``extratropics'' as the dynamical regimes---regardless
of temperature---where
large-scale circulations exhibit $Ro\gtrsim1$ and $Ro\ll 1$, respectively.


Figure~\ref{tropics-extratropics-schematic} illustrates how the extent
of the tropics and extratropics depend on rotation rate for a typical
terrestrial planet.  The sphere on the left depicts an Earth- or
Mars-like world where the boundary between the regimes occurs at
$\sim$20--$30^{\circ}$ latitude.  At longer rotation periods, the
tropics occupy a greater fraction of the planet, and idealized general
circulation model (GCM) experiments\footnote{A GCM solves the global
  three-dimensional (3D) fluid-dynamics equations relevant to a
  rotating atmosphere, coupled to calculations of the atmospheric
  radiative-transfer everywhere over the full 3D grid (necessary for
  determining the radiative heating/cooling rate, which affects the
  dynamics), and parameterizations of various physical processes
  including frictional drag against the surface, sub-grid-scale
  turbulence, and (if relevant) clouds.  ``Idealized'' GCMs refer to
  GCMs where these components are simplified, e.g., adopting a gray
  radiative-transfer scheme rather than solving the full non-gray
  radiative transfer.} show that, for Earth-like planetary radii,
gravities, and incident stellar fluxes, planets exhibit $Ro\gtrsim 1$
everywhere when the rotation period exceeds $\sim$10 Earth days
\citep[e.g.,][]{delgenio-suozzo-1987}
(Figure~\ref{tropics-extratropics-schematic}, right panel).
Dynamically, such slowly-rotating planets are essentially ``all
tropics'' worlds.\footnote{The term was first coined by
  \citet{mitchell-etal-2006} in reference to Titan.}  Venus and Titan
are examples in our own solar system, exhibiting near-global Hadley
cells, minimal equator-pole temperature differences, little role for
baroclinic instabilities, and a zonal jet structure that differs
significantly from those on Earth and Mars.  Terrestrial exoplanets
characterizable by transit techniques will preferentially be close to
their stars and tidally locked, implying slow rotation rates; many of
these exoplanets should likewise be ``all tropics'' worlds.

\begin{figure*}
 \epsscale{1.5}
\plotone{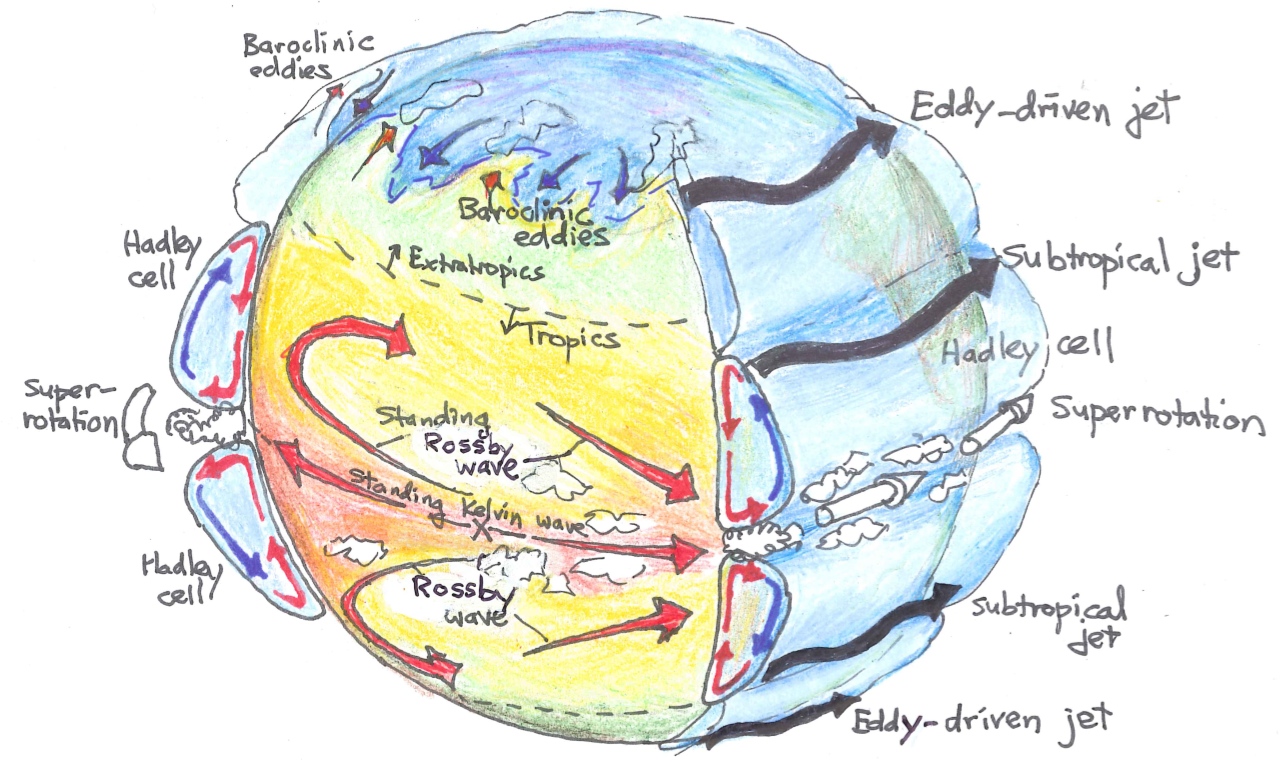}
 \caption{Schematic illustration of dynamical processes occurring
on a generic terrestrial exoplanet.  These include baroclinic
eddies, Rossby waves, and eddy-driven jet streams in the extratropics,
and Hadley circulations, large-scale Kelvin and Rossby waves,
and (in some cases) equatorial superrotation in the tropics.  The ``X'' at the equator
marks the substellar point, which will be fixed in longitude on
synchronously rotating planets.  Cloud formation, while complex, will
likely be preferred in regions of mean ascent, including the rising
branch of the Hadley circulation, within baroclinic eddies, and---on
synchronously rotating planets---in regions of ascent on the dayside.}
\label{atmospheric-processes-schematic}
 \end{figure*}

Figure~\ref{atmospheric-processes-schematic} previews several of the
key dynamical processes occurring at large scales on a generic
terrestrial exoplanet, which we survey in more detail in the
subsections that follow.  In the extratropics, the baroclinic eddies
that dominate the meridional heat transport (Section~\ref{baroclinic})
generate meridionally propagating Rossby waves
(Section~\ref{rossby}), which leads to a
convergence of momentum into the instability latitude, generating an
eddy-driven jet stream (Section~\ref{jet-formation}). Multiple zones
of baroclinic instability, and multiple eddy-driven jets, can emerge
in each hemisphere if the planet is sufficiently large or the
planetary rotation is sufficiently fast.  In the tropics, the Hadley
circulation (Section~\ref{hadley}) dominates the meridional heat
transport; in idealized form, it transports air upward near the
equator and poleward in the upper troposphere, with a return flow to
the equator along the surface.  Due to the relative weakness of
rotational effects in the tropics, atmospheric waves can propagate
unimpeded in longitude, and adjustment of the thermal structure by
these waves tends to keep horizontal temperature gradients weak in the
tropics (Section~\ref{wave-adjustment}).  Many exoplanets will rotate
synchronously and therefore exhibit permanent day- and nightsides; the
resulting, spatially locked day-night heating patterns will generate
large-scale, standing equatorial Rossby and Kelvin waves, which in
many cases will lead to equatorial superrotation, that is, an eastward
flowing jet at the equator (Section~\ref{superrotation}).  Significant
communication between the tropics and extratropics can occur, among
other mechanisms, via meridionally propagating Rossby waves that
propagate from one region to the other.

We review the extratropical and tropical regimes, along with the
key processes shown in Figure~\ref{atmospheric-processes-schematic},
in this section.

\subsection{Extratropical regime}

\subsubsection{Force balances and geostrophy}

The extratropical regime corresponds to $Ro\ll 1$.   For
typical terrestrial-planet wind speeds of $\sim$$10\rm\,m\,s^{-1}$
and Earth-like planet sizes, planets will have extratropical
zones for rotation periods of a few (Earth) days or shorter.
When $Ro\ll1$ and friction is weak, the Coriolis force and
pressure-gradient force will approximately balance in the horizontal momentum
equation; the resulting balance, called geostrophic balance, is
given by \citep[e.g.,][pp.~85-88]{vallis-2006}
\begin{equation}
fu = -\left({\partial\Phi\over\partial y}\right)_p \qquad
fv = \left({\partial\Phi\over\partial x}\right)_p 
\end{equation}
where $\Phi$ is the gravitational potential, $x$ and $y$ are eastward
and northward distance, $u$ and $v$ are zonal and meridional wind
speed, $f$ is the Coriolis parameter, and the derivatives are
evaluated at constant pressure.  The implication is that winds tend to
flow along, rather than across, isobars.  In our solar system, the
atmospheres of Earth, Mars, and all four giant planets exhibit
geostrophic balance away from the equator, and the
same will be true for a wide range of terrestrial exoplanets.  See
\citet{pedlosky-1987}, \citet{holton-2004}, or \citet{vallis-2006} for
introductions to the dynamics of rapidly rotating atmospheres in the
geostrophic regime.

 In a geostrophic flow, there exists a tight link between winds and
 temperatures.  Differentiating the geostrophic equations in pressure
 (which here acts as a vertical coordinate) and invoking hydrostatic
 balance and the ideal-gas law, we obtain the thermal-wind equations
 \citep[][pp.~89-90]{vallis-2006}
\begin{equation}
f{\partial u\over \partial \ln p} = {\partial(RT)\over\partial y}
\qquad\qquad f{\partial v\over\partial \ln p}=-{\partial (RT)\over\partial x}
\label{thermal-wind}
\end{equation}
where $T$ is temperature and $R$ is the specific gas constant.  The
equation implies that, for the geostrophic component of the flow,
meridional temperature gradients accompany vertical shear (that is,
vertical variation) of the zonal
wind, and zonal temperature gradients accompany vertical shear (i.e.,
vertical variation) of the
meridional wind.  Because the surface wind is generally weak compared
to that at the tropopause, one can thus obtain an estimate of the wind
speed in the upper troposphere---given the equator-pole temperature
gradient---by integrating (\ref{thermal-wind}) vertically.  

To order of magnitude, for example, Equation~(\ref{thermal-wind})
implies a zonal wind $\Delta u\sim R \Delta T_{\rm eq-pole}\Delta\ln
p/(fa)$, where $\Delta T_{\rm eq-pole}$ is the temperature difference
between the equator and pole, $\Delta \ln p$ is the number of scale
heights over which this temperature difference extends, and $a$ is the
planetary radius.  Inserting Earth parameters ($\Delta T_{\rm
  eq-pole}\approx 20\rm\,K$, $a\approx 6000\rm\,km$,
$R=287\rm\,J\,kg^{-1}\,K^{-1}$, $f\approx 10^{-4}\rm\,s^{-1}$, and
$\Delta\ln p = 1$), we obtain $\Delta u \approx 10\rm\,m\,s^{-1}$,
which is indeed a characteristic value of the zonal wind in Earth's
upper troposphere.

What are the implications of thermal-wind balance for a planet's
global circulation pattern?  Given the thermal structure expected on
typical, low-obliquity, rapidly rotating exoplanets, with a warm
equator and cool poles, Equation~(\ref{thermal-wind}) makes several
useful statements:
\begin{itemize}
\item It implies that the zonal wind increases
(i.e., becomes more eastward) with altitude.  Assuming the surface
winds are weak, this explains the predominantly eastward nature of
the tropospheric winds on Earth and Mars, especially in midlatitudes,
and suggests an analogous pattern on rapidly rotating exoplanets.
\item Because temperature gradients---at least on Earth and
  Mars---peak in midlatitudes, thermal-wind balance helps explain why
  the zonal wind shear---hence the upper-tropospheric zonal winds
  themselves---are greater at midlatitudes than at the equator or
  poles.  This is the latitude of the jet streams, and thermal-wind
  balance therefore describes how the winds increase with height in
  the jet streams.  Because the tropospheric meridional temperature gradient is
  greater in the winter hemisphere than the summer hemisphere,
  thermal-wind balance also implies that the upper-tropospheric,
  mid-latitude winds should be faster in the winter hemisphere,
  as in indeed occurs on Earth \citep{peixoto-oort-1992} 
  and Mars \citep{smith-2008}.
\item If the mean zonal temperature gradients are small compared to
  mean meridional temperature gradients, which tends to be true when
  the rotation is fast (and in particular when the solar day is
  shorter than the atmospheric radiative time constant), thermal wind
  implies that the time-mean zonal winds are stronger than the
  time-mean meridional winds.  This provides a partial explanation for
  the zonal (east-west) banding of the wind structure on Earth and
  Mars and suggests that a similarly banded wind pattern will occur on
  rapidly rotating exoplanets.\footnote{In the presence of topography
or land-ocean contrasts, some local regions may exhibit meridional
winds that, even in a time average, are not small compared to zonal winds.
In such a case, $u\gg v$ only in the zonal mean.}
\end{itemize}

The fact that Coriolis forces
can balance pressure gradients in a geostrophic flow implies that such
flows can sustain larger horizontal pressure and temperature contrasts
than might otherwise exist.  To order of magnitude, hydrostatic
balance of the dynamical pressure and density fluctuations $\delta p$
and $\delta\rho$ \citep[see] [pp.~41-42]{holton-2004} implies that
these fluctuations satisfy
\begin{equation}
\delta p\approx \delta\rho g D
\end{equation}
where $D$ is the vertical scale of the circulation.  In the horizontal
momentum equation, the pressure-gradient force is
approximately $\delta\rho \,g D/(\rho L)$, where $L$ is a
characteristic horizontal length scale of the flow.   Noting
that $\delta\rho/\rho \sim \delta\theta_h/\theta$, where 
$\delta\theta_h$ is the characteristic horizontal potential temperature
difference and $\theta$ is the characteristic potential 
temperature,\footnote{Potential temperature is defined as
the temperature an air parcel would have if transported adiabatically
to a reference pressure $p_0$ (often taken to be 1 bar). When
the ratio of gas constant to specific heat $R/c_p$ is constant,
as is often approximately true in atmospheres, then it is defined
by $\theta = T(p_0/p)^{R/c_p}$.  It is conserved following adiabatic
reversible processes and is thus a measure of atmospheric entropy.}
geostrophy implies \citep[cf.][]{charney-1963}
\begin{equation} 
{\delta\theta_h\over\theta} \sim {f U L\over gD} \sim {Fr \over Ro},
\qquad\qquad Ro \lesssim 1
\label{charney-extratropics}
\end{equation} 
where $Fr\equiv U^2/gD$ is a
dimensionless quantity known as a Froude number.  For a typical,
rapidly rotating terrestrial planet where $U\approx 10\rm\,m\,s^{-1}$,
$D\approx10\rm\,km$, $g\approx10\rm\,m\,s^{-2}$,
$f\approx10^{-4}\rm\,s^{-1}$ (implying rotation periods of an
Earth day), the fractional temperature contrasts over distances
comparable to the planetary radius approach $\sim$0.1, implying
temperature contrasts of $\sim$$20\rm\,K$.  This is similar to the
actual equator-to-pole temperature contrast in Earth's troposphere.
In contrast, these values greatly exceed typical horizontal
temperature contrasts in the tropical regime of $Ro\sim 1$ (see
Eq.~\ref{charney-tropics}).

An important horizontal length scale in atmospheric dynamics is
the Rossby deformation radius, defined in the extratropics as
\begin{equation}
L_D = {ND\over f},
\end{equation}
where $N$ is the Brunt-Vaisala frequency, which is a measure of vertical
stratification.  Because the deformation
radius is a natural length scale that emerges from the interaction of
gravity (buoyancy) and rotation in stably stratified atmospheres, many
phenomena, including geostrophic adjustment, baroclinic instabilities,
and the interaction of convection with the environment, produce
atmospheric structures with horizontal sizes comparable to the
deformation radius. As a result, the circulations in atmospheres (and
oceans) often exhibit predominant length scales not too different from
the deformation radius.

\subsubsection{Baroclinic instabilities and their effect on thermal structure}
\label{baroclinic}

\begin{figure*}
 \epsscale{1.4}
\plotone{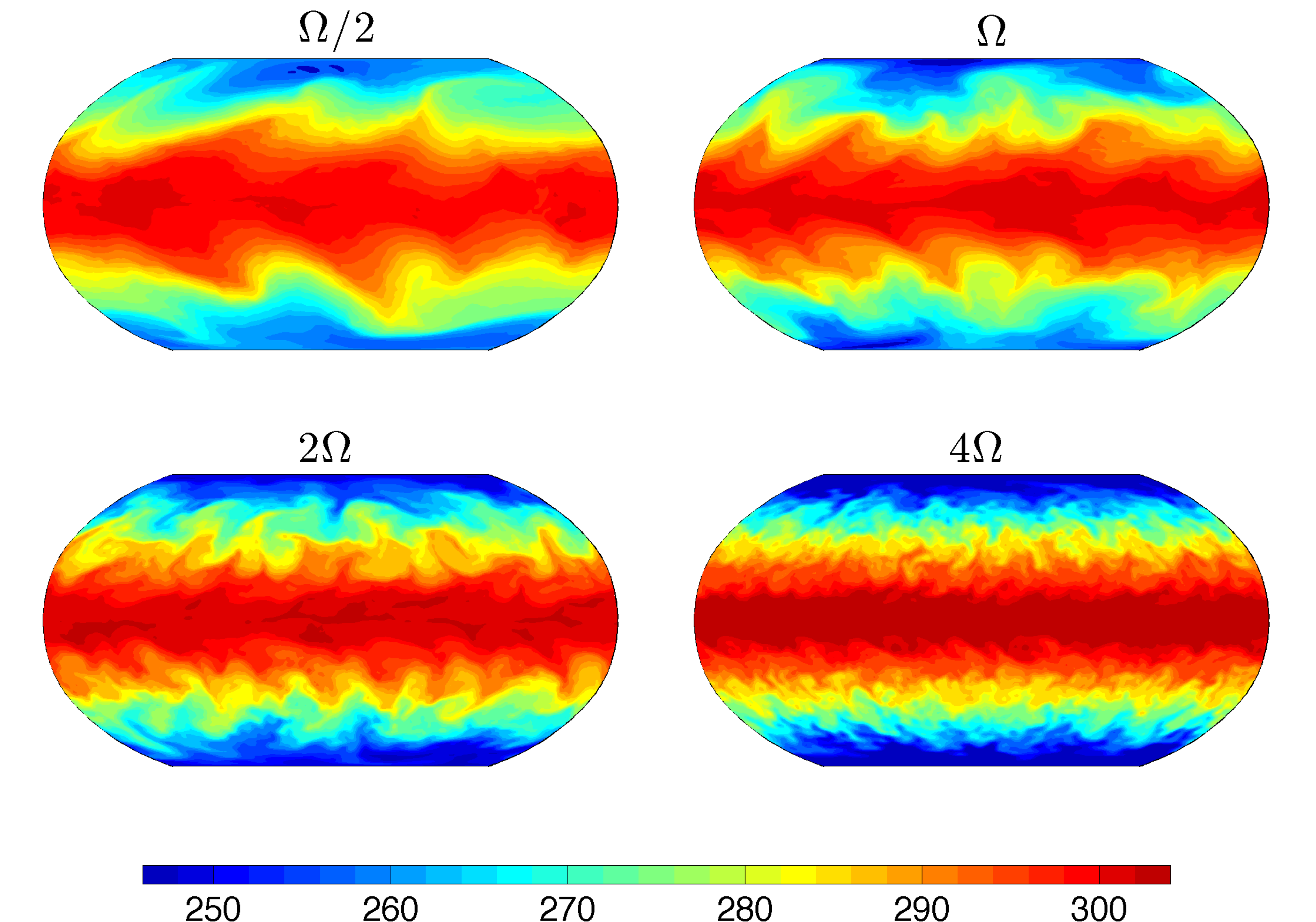}
 \caption{Surface temperature (colorscale, in K) from GCM
   experiments in \citet{kaspi-showman-2012}, illustrating the
   dependence of temperature and jet structure on rotation rate.
   Experiments are performed using the Flexible Modeling System (FMS)
   model analogous to those in \citet{frierson-etal-2006} and
   \citet{kaspi-schneider-2011}; radiative transfer is represented by
   a two-stream, grey scheme with no diurnal cycle (i.e., the incident
   stellar flux depends on latitude but not longitude).  A
   hydrological cycle is included with a slab ocean.  Planetary
   radius, gravity, atmospheric mass, incident stellar flux, and
   atmospheric thermodynamic properties are the same as on Earth;
   models are performed with rotation rates from half (upper left) to
   four times that of Earth (lower right).  Baroclinic instabilities
   dominate the dynamics in mid- and high-latitudes, leading to
   baroclinic eddies whose length scales decrease with increasing
   planetary rotation rate.}
\label{earthlike-circulation}
 \end{figure*}

Planets experience meridional gradients in the net radiative heating
rate.  At low obliquities, this gradient corresponds to net heating at
the equator and net cooling at the poles, leading to meridional
temperature contrasts with, generally, a hot equator and cold poles.
Even if individual air columns are convectively stable (i.e., if the
potential temperature increases with height in the troposphere),
potential energy can be extracted from the system if the cold polar
air moves downward and equatorward and if the warm equatorial air
moves upward and poleward.  The question is then whether dynamical
mechanisms actually exist to extract this energy and thereby transport
heat from the equator to the poles.  In the tropics, the dominant
meridional heat-transport mechanism is a thermally direct Hadley
circulation (see Figure~\ref{atmospheric-processes-schematic} and
Section~\ref{hadley}), but such circulations tend to be suppressed by
planetary rotation in the extratropics.  At $Ro\ll 1$, the meridional
temperature gradients are associated with an upward-increasing zonal
wind in thermal-wind balance (Equation~\ref{thermal-wind}).  It is
useful to think about the limit where longitudinal gradients of
heating and temperature are negligible---in which case the
(geostrophic) wind is purely zonal---and where the meridional wind
is zero, as occurs when these zonal winds are perfectly balanced.  In
such a hypothetical solution, the temperature at each latitude would
be in local radiative equilibrium, and no meridional heat transport
would occur.

It turns out that this hypothetical steady state is not dynamically
stable: small perturbations on this steady solution grow over time,
producing eddies that extract potential energy from the horizontal
temperature contrast (transporting warm low-latitude air upward and
poleward, transporting cool high-latitude air downward and
equatorward, thereby lowering the center of mass, flattening
isentropes,\footnote{Isentropes are surfaces of constant entropy, which
are equivalent to surfaces of constant potential temperature.  In a
stably stratified atmosphere where entropy increases with altitude,
isentropes will bow downward (upward) in regions that, as measured
on an isobar, are hot (cold).} and
reducing the meridional temperature gradient).  This is baroclinic
instability, so named because it depends on the baroclinicity of the
flow---i.e., on the fact that surfaces of constant density incline
significantly with respect to surfaces of constant pressure.  The
instabilities are inherently three-dimensional and manifest locally as
tongues of cold and warm air penetrating equatorward and poleward in
the extratropics.  Figure~\ref{atmospheric-processes-schematic}
illustrates such eddies schematically and
Figure~\ref{earthlike-circulation} provides examples from GCM
experiments under Earth-like conditions.  The fastest-growing modes
have zonal wavelengths comparable to the deformation radius and growth
rates proportional to $(f/N)\partial u/\partial z$, where $\partial
u/\partial z$ is the vertical shear of the zonal wind that exists in
thermal wind balance with the meridional temperature gradient.  For
Earth-like conditions, these imply length scales of $\sim$$4000\,$km
and growth timescales of $\sim$3--5 days for the dominant modes
\citep[for reviews, see, e.g.,][Chapter 6]{pierrehumbert-swanson-1995,
  vallis-2006}.

As expected from baroclinic instability theory, the dominant length
scale of the baroclinic, heat-transporting eddies in the extratropics
scales inversely with the planetary rotation rate.  This is
illustrated in Figure~\ref{earthlike-circulation}, which shows
instantaneous snapshots of the temperature in idealized Earth-like GCM
experiments where the rotation rate is varied between half and four
times that of Earth \citep{kaspi-showman-2012}.  The smaller eddies
in the rapidly rotating models are less efficient at transporting
energy meridionally, leading to a greater equator-to-pole temperature
difference in those cases.  A similar dependence has been found
by other authors as well \citep[e.g.][]{schneider-walker-2006,
kaspi-schneider-2011}.

\begin{figure*}
 \epsscale{1.}
\plotone{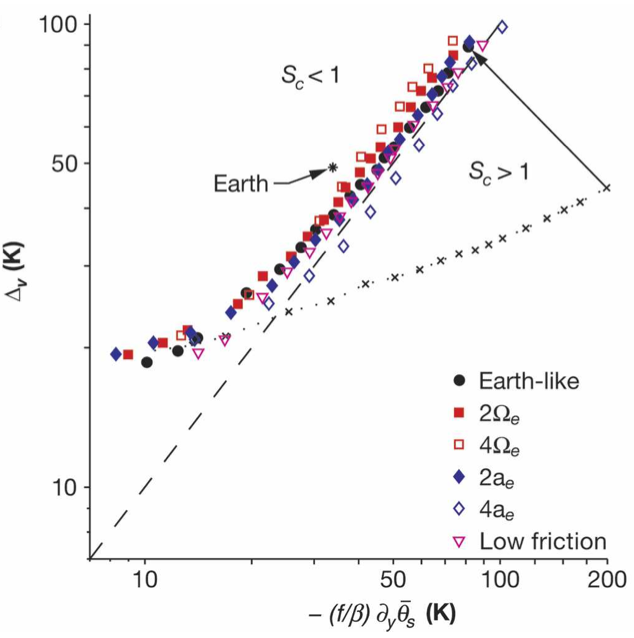}
 \caption{A measure of the near-surface meridional potential
   temperature difference across the extratropics (abscissa) and
   potential temperature difference in the extratropics taken
   vertically across the troposphere (ordinate) in models of planets
   forced by equator-to-pole heating gradients from
   \citet{schneider-walker-2006}. When the equator-to-pole forcing is
   sufficiently great, the extratropical dynamics are dominated by
   baroclinic eddies, which adjust the thermal structure to a state
   where horizontal and vertical potential temperature differences in
   the extratropics are comparable.  Different symbols denote models
   with differing rotation rates, planetary radii, and/or
   boundary-layer friction, coded in the legend.  Crosses depict the
   radiative-equilibrium states for the Earth-like models.  The dashed
   line denotes a slope of 1.}
\label{supercriticality}
 \end{figure*}

In the extratropics of Earth and Mars, baroclinic instabilities play a
key role in controlling the thermal structure, including the
equator-to-pole temperature gradient and the vertical stratification
(i.e., the Brunt-Vaisala frequency);
this is also likely to be true in the extratropics of terrestrial
exoplanets.  In particular, GCM experiments suggest that the
extratropics---when dominated by baroclinic instability---adjust to a
dynamic equilibrium in which meridional temperature gradients and
tropospheric stratifications scale together (atmospheres with larger
tropospheric stratification exhibit larger meridional temperature
gradients and vice versa).  Figure~\ref{supercriticality} illustrates
this phenomenon from a sequence of idealized GCM experiments from
\citet{schneider-walker-2006} for terrestrial planets forced by
equator-to-pole heating gradients.  The abscissa shows the meridional
temperature difference across the baroclinic zone and the ordinate
shows the vertical stratification through the troposphere; each symbol
represents the equilibrated state of a particular model including the
effects of dynamics.  Although the radiative-equilibrium thermal
structures fill a significant fraction of the parameter space, the
baroclinic eddy entropy fluxes adjust the thermal structure to a state
where the vertical stratification and meridional temperature
difference are comparable, corresponding to a line with a slope of one
in Figure~\ref{supercriticality}. The extent to which this
relationship holds in general and parameter regimes where it may break
down are under investigation \citep{zurita-gotor-vallis-2009,
  jansen-ferrari-2012}.

Most work on this problem has so far investigated planets forced only
by equator-pole heating gradients
\citep[e.g.,][]{schneider-walker-2006, kaspi-showman-2012}, but the
zonal (day-night) heating gradients could significantly influence the
way that baroclinic eddies regulate the thermal structure,
particularly on synchronously rotating planets.
\citet{edson-etal-2011} compared GCM simulations for a control,
Earth-like simulation forced only by an equator-pole heating gradient
and an otherwise identical synchronously rotating planet with
day-night thermal forcing.  They found that the (zonal-mean)
meridional isentrope slopes were gentler in the model with day-night
forcing than in the control model. This differing behavior presumably
results from the fact that the synchronously rotating model exhibits
meridional heat fluxes that are primarily confined to the dayside
(rather than occurring at all longitudes as in the control model) and
from the fact that the day-night forcing generates planetary-scale
standing waves (see Section~\ref{superrotation}) in the synchronous
model that are absent in the control model.  Both traits can
influence the (zonal-mean) meridional heat transport and therefore the
isentrope slopes and zonal-mean temperature differences between the
equator and poles.  Additional work quantifying the behavior for
tidally locked planets over a wider parameter regime would be highly
beneficial.

Over the past several decades, many authors have attempted to
elucidate theoretically how the meridional heat fluxes due to
baroclinic eddies depend on the background meridional temperature
gradient, tropospheric stratification, planetary rotation rate, and
other parameters.  Almost all of this work has emphasized planets in
an Earth-like regime, forced by equator-pole heating gradients (e.g.,
\citet{green-1970}, \citet{stone-1972}, \citet{larichev-held-1995},
\citet{held-larichev-1996}, \citet{pavan-held-1996},
\citet{haine-marshall-1998}, \citet{barry-etal-2002},
\citet{thompson-young-2006},\citet{thompson-young-2007},
\citet{schneider-walker-2008}, \citet{zurita-gotor-vallis-2009}; for
reviews, see \citet{held-1999}, and \citet{showman-etal-2010}).  If a
theory for this dependence could be developed, it would constitute a
major step toward a predictive theory for the dependence of the
equator-to-pole temperature difference on planetary parameters.
However, this is a challenging problem, and no broad consensus has yet
emerged.

\begin{figure*}
 \epsscale{1.5}
\plotone{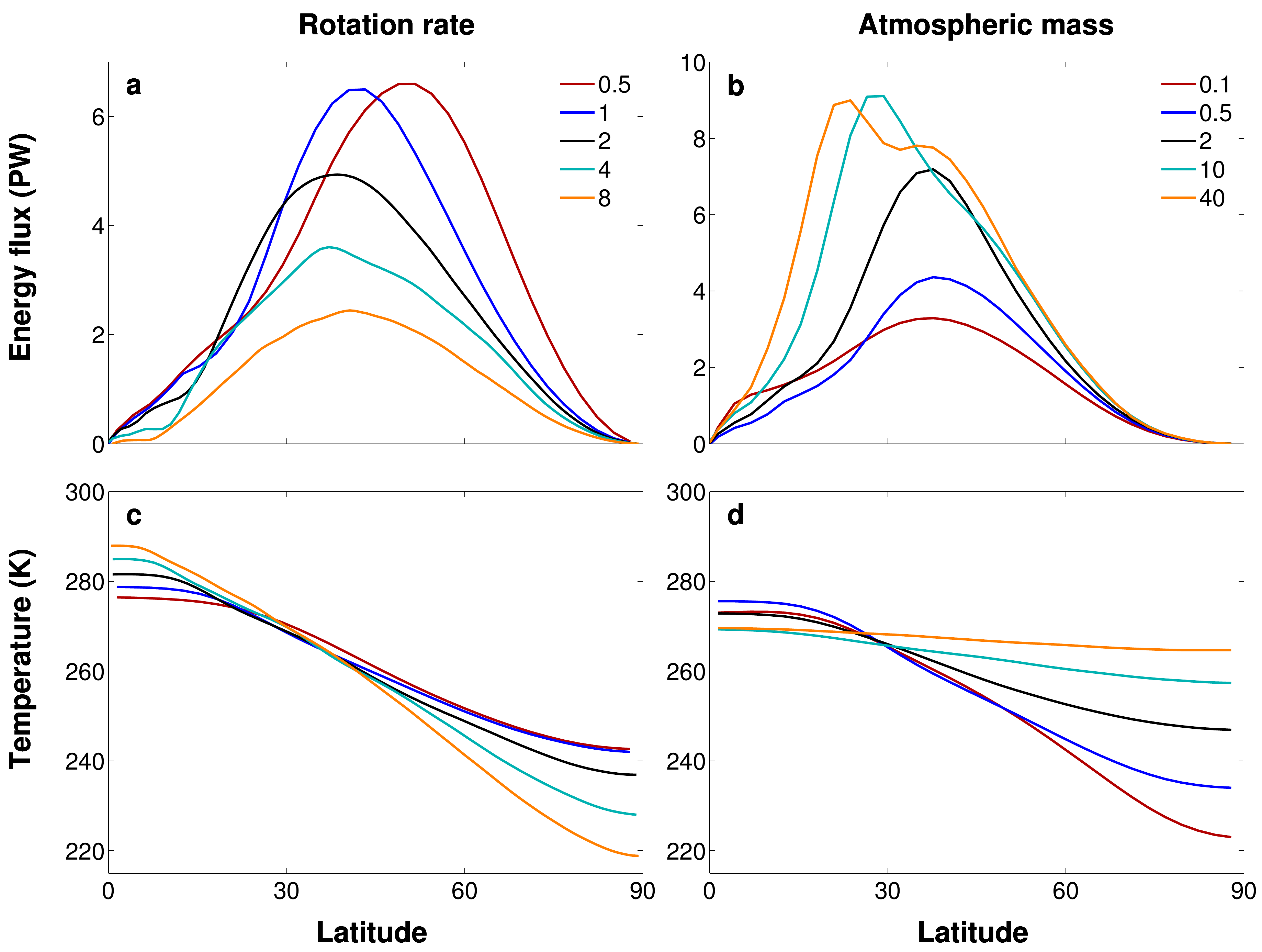}
 \caption{Latitude dependence of the vertically and zonally integrated
   meridional energy flux (a and b) and vertically and zonally
   averaged temperature (c and d) in GCM experiments from
   \citet{kaspi-showman-2012} for models varying rotation rate and
   atmospheric mass.  The energy flux shown in (a) and (b) is the flux
   of moist static energy, defined as $c_pT + gz + Lq$, where $c_p$ is
   specific heat at constant pressure, $g$ is gravity, $z$ is height,
   $L$ is latent heat of vaporization of water, and $q$ is the water
   vapor abundance.  The left column (a and c) explores sensitivity to
   rotation rate; models are shown with rotation rates ranging from
   half to eight times that of Earth.  In these experiments, the
   atmospheric mass is held constant at the mass of Earth's
   atmosphere.  Right column (b and d) explores the sensitivity to
   atmospheric mass; models are shown with atmospheric masses from 0.1
   to 40 times the mass of Earth's atmosphere.  In these experiments,
   the rotation rate is set to that of Earth.  The equator-to-pole
   temperature difference is smaller, and the meridional energy flux
   is larger, when the planetary rotation rate is slower, and/or when
   the atmospheric mass is larger.  Other model parameters, including
   incident stellar flux, optical depth of the atmosphere in the
   visible and infrared, planetary radius, and gravity, are Earth-like
   and are held fixed in all models.}
\label{deltaT-vary-rotation}
 \end{figure*}

Nevertheless, recent GCM studies demonstrate how the equator-to-pole
temperature differences and other aspects of the dynamics depend on
planetary radius, gravity, rotation rate, atmospheric mass, and
incident stellar flux \citep{kaspi-showman-2012}.  These authors
performed idealized GCM experiments of planets forced by
equator-to-pole heating gradients, including a hydrological cycle and
representing the radiative transfer using a two-stream, grey approach.
\citet{kaspi-showman-2012} found that the equator-to-pole temperature
difference decreases with increasing atmospheric mass
and increases with increasing rotation rate, planetary radius (at constant interior
density) or planetary density (at constant interior mass).
Figure~\ref{deltaT-vary-rotation} shows the zonal-mean surface
temperature and meridional eddy energy fluxes versus latitude for the
cases with differing rotation rates (left) and atmospheric masses
(right).  Meridional eddy energy fluxes are weaker at faster rotation
rates (Figure~\ref{deltaT-vary-rotation}a), presumably because the
smaller eddy length scales (Figure~\ref{earthlike-circulation}) lead
to less efficient energy transport.  This helps to explain the fact
that larger equator-to-pole temperature differences occur in the
faster rotating simulations, as is evident in
Figure~\ref{deltaT-vary-rotation}c and
Figure~\ref{earthlike-circulation}.  Likewise, massive atmospheres
exhibit a greater thermal storage capacity than less massive
atmospheres, allowing a greater meridional energy flux at a given
baroclinic eddy amplitude.  Everything else being equal, planets with greater
atmospheric mass therefore exhibit smaller equator-to-pole temperature
differences than planets with lesser atmospheric mass
(Figure~\ref{deltaT-vary-rotation}d).  Additional
detailed work is warranted to clarify the physical processes
controlling these trends and to seek a predictive theory for the
dependences.

\subsubsection{Rossby waves}
\label{rossby}

Much of the structure of the large-scale atmospheric circulation can
be understood in terms of the interaction of Rossby waves with the
background flow; they are the most important wave type for the
large-scale circulation. In this section we survey their linear dynamics,
and follow in subsequent sections with discussions of how they
help to shape the structure of the extratropical circulation via
nonlinear interactions.

Rossby waves are best understood through conservation of potential
vorticity (PV), which for a shallow fluid layer of
thickness $h$ can be written as
\begin{equation}
q = {\zeta + f\over h}
\label{pv}
\end{equation}
where $\zeta = \mathbf{k}\cdot \nabla \times \mathbf{v}$ is the
relative vorticity, $\mathbf{k}$ is the vertical (upward) unit vector,
and ${\bf v}$ is the horizontal velocity vector.  As written, this is
the conserved form of PV for the shallow-water equations
\citep[e.g.][]{pedlosky-1987, vallis-2006, showman-etal-2010}, but the
same form holds in the three-dimensional primitive equations if the
relative vorticity is evaluated at constant potential temperature and
$h$ is appropriately defined in terms of the gradient of pressure with
respect to potential temperature \citep[][p.~187-188]{vallis-2006}.
To illuminate the dynamics, we consider the simplest system that
supports Rossby waves, namely a one-layer barotropic model governing
two-dimensional, non-divergent flow.  For this case, $h$ can be
considered constant, leading to conservation of absolute vorticity,
$\zeta + f$, following the flow.  This can be written, adopting Cartesian
geometry, as
\begin{equation}
{\partial\zeta\over\partial t} + {\bf v}\cdot\nabla\zeta + v\beta = F
\label{barotropic-vorticity}
\end{equation}
where $v$ is the meridional velocity, $\beta\equiv df/dy$
is the gradient of the Coriolis parameter $f$ with northward distance 
$y$, and $F$ represents any sources or sinks of potential vorticity.  For
a brief review of equation sets used in atmospheric dynamics,
see \citet{showman-etal-2010}; more detailed treatments can be
found in \citet{vallis-2006}, \citet{mcwilliams-2006}, or \citet{holton-2004}.

To investigate how the $\beta$ effect allows wave propagation,
we consider the version of Eq.~(\ref{barotropic-vorticity}) linearized
about a state of no zonal-mean flow, which
amounts to dropping the term involving advection of $\zeta$.
 Given the assumption that the flow is horizontally nondivergent,
we can define a streamfunction, $\psi$, such
that $u=-{\partial \psi\over\partial y}$ and $v={\partial\psi\over
\partial x}$.  This definition implies that $\zeta=\nabla^2\psi$, allowing
the linearized equation to be written
\begin{equation}
{\partial\nabla^2\psi\over\partial t} + \beta {\partial\psi
\over\partial x}=0
\label{linear-barotropic}
\end{equation}
This is useful because it is now an equation in only one variable,
$\psi$, which can easily be solved to determine wave behavior.
Assuming that $\beta$ is constant (an approximation known as the ``beta
plane''), and adopting plane-wave solutions $\psi=\psi_0\exp[i(kx + ly -\omega t)]$, we
obtain a dispersion relation\footnote{For introductions to wave equations
and dispersion relations, see for example \citet{vallis-2006}, \citet{holton-2004},
\citet{pedlosky-1987}, or \citet{pedlosky-2003}.}
\begin{equation}
\omega = -{\beta k\over k^2 + l^2}
\label{dispersion}
\end{equation}
where $k$ and $l$ are the zonal and meridional wavenumbers (just
$2\pi$ over the longitudinal and latitudinal wavelengths,
respectively) and $\omega$ is the oscillation frequency.  The resulting
waves are called Rossby waves.  They are large-scale (wavelengths
commonly $\sim$$10^3\km$ or more), low-frequency (periods of order a
day or longer), and---away from the equator on a rapidly rotating
planet---are in approximate geostrophic balance.  As demonstrated by
Equation~(\ref{dispersion}), they exhibit westward phase speeds.

The wave-induced zonal and meridional velocities are defined,
respectively, as $u'=u'_0 \exp[i(kx + ly -\omega t)]$ and $v'=v'_0
\exp[i(kx + ly - \omega t)]$, where $u'_0$ and $v'_0$ are complex
amplitudes.  The above solution implies that
\begin{eqnarray}
u'_0 = -i l \psi_0 \qquad\qquad
v'_0 = i k \psi_0.
\label{u-v-rossby}
\end{eqnarray}
The velocities represented by these relations are parallel to lines of
constant phase and therefore perpendicular to the direction of phase
propagation.  Equation~({\ref{u-v-rossby}) implies that the zonal and
  meridional wave velocities are correlated; this provides a mechanism
  by which Rossby wave generation, propagation, and dissipation can
  transport momentum and thereby modify the mean flow, a point we
  return to in Section~\ref{jet-formation}.

Physically, the restoring force for Rossby waves is the $\beta$
effect, namely, the variation with latitude of the planetary
vorticity.  Because the Coriolis parameter varies with latitude, PV
conservation requires that any change in latitude of fluid parcels
will cause a change in the relative vorticity.  The flow associated
with these relative vorticity perturbations leads to oscillations
in the position of fluid parcels and thereby allows wave propagation.
The meridional velocities
deform the phase surfaces in a manner that leads to the westward
phase velocities captured in Equation~(\ref{dispersion})(see
discussion in \citet{vallis-2006} or \citet{holton-2004}).

\subsubsection{Jet formation I: Basic mechanisms}
\label{jet-formation}

A wide range of numerical experiments and observations show that zonal
jets tend to emerge spontaneously on rapidly rotating planets.  On
many planets, such zonal jets dominate the circulation, and thus
understanding them is crucial to understanding the circulation as a whole.
The dynamics controlling jets is intimately connected to the dynamics
controlling the meridional temperature gradients, vertical stratification,
and other aspects of the dynamics.  Sufficiently strong jets exhibit 
sharp gradients of the meridional PV gradient, and as such they can
act as barriers to the meridional mixing of heat, moisture, and
chemical tracers \citep[e.g.,][]{beron-vera-etal-2008}, significantly
affecting the meridional structure of the atmosphere.

While there exist many mechanisms of jet formation, among the most
important is the interaction of Rossby waves with the mean flow.  This
mechanism plays a key role in causing the extratropical jets on Earth,
Mars, and perhaps Jupiter and Saturn; the mechanism is similarly
expected to play an important role in the atmospheres of terrestrial
exoplanets.

\begin{figure*}
 \epsscale{1.}
\plotone{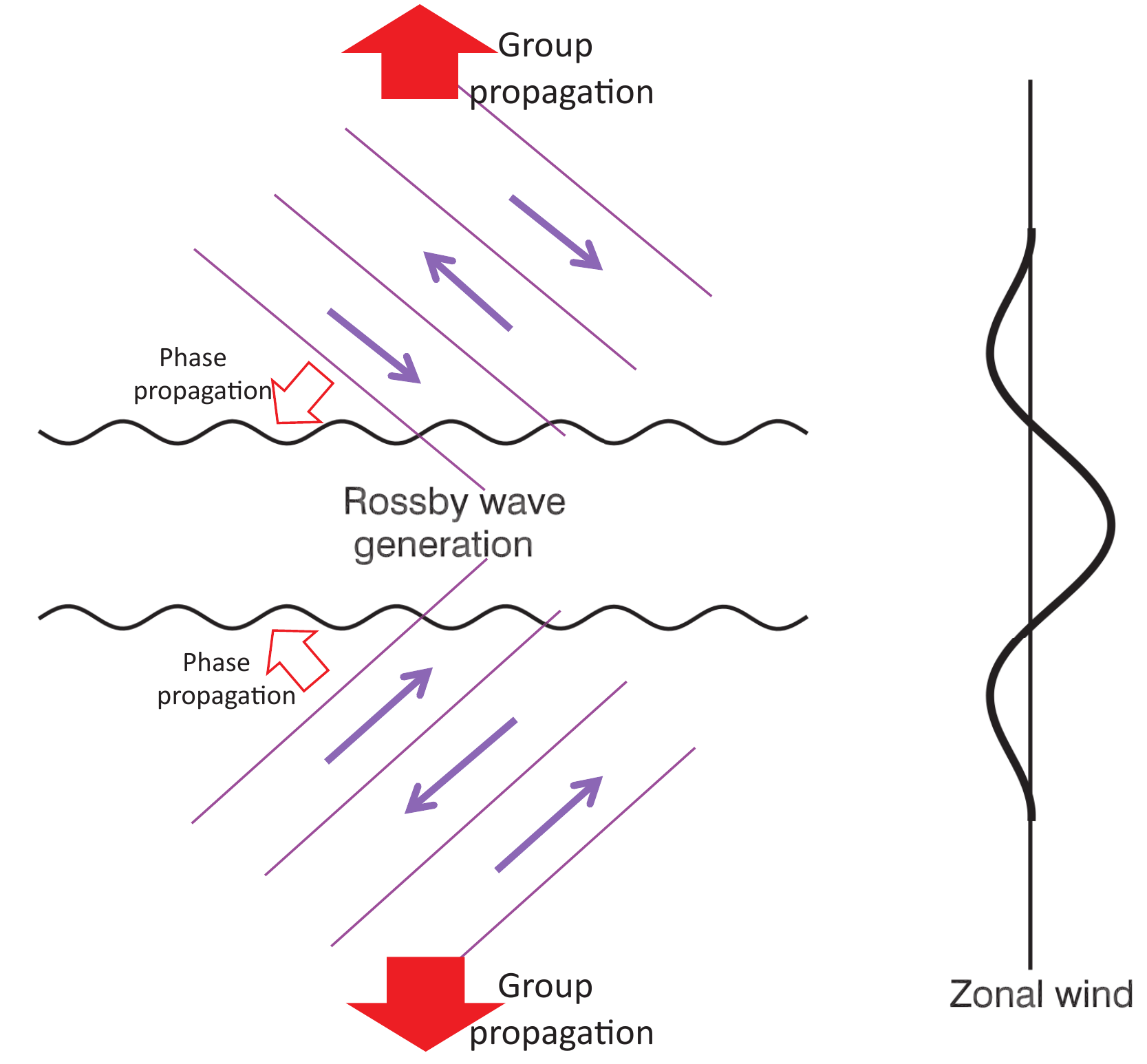}
 \caption{Schematic illustration of how Rossby wave generation can
   lead to the formation of zonal jets.  Imagine that some
   process---such as baroclinic instability---generates Rossby waves
   at a specific latitude, and that the Rossby waves then propagate
   north and south from their latitude of generation.  Rossby waves
   with northward group propagation exhibit eddy velocities tilting
   northwest-to-southeast, whereas those with southward group
   propagation exhibit eddy velocities tilting southwest-to-northeast.
   These correlations imply that the waves transport eastward eddy
   momentum into the latitude region where Rossby waves are generated,
   resulting in formation of an eddy-driven jet.}
\label{rossby-wave-schematic}
 \end{figure*}

As \citet{thompson-1971}, \citet{held-1975}, and many subsequent
authors have emphasized, a key property of meridionally propagating
Rossby waves is that they
induce a meridional flux of prograde (eastward) eddy angular momentum
into their generation latitude.  To illustrate, we again consider the
solutions to Eq.~(\ref{linear-barotropic}).  
The latitudinal transport of (relative) eastward eddy momentum per mass is
$\overline{u'v'}$, where $u'$ and $v'$ are the deviation of the zonal
and meridional winds from their zonal mean, and the overbar denotes a
zonal average.  Using Eq.~(\ref{u-v-rossby}) shows that
\begin{equation}
\overline{u'v'}=-{1\over 2}\hat\psi^2 kl.
\label{u'v'}
\end{equation}
Now, the dispersion relation (\ref{dispersion}) implies that the
meridional group velocity is $\partial\omega/\partial l=2\beta k l/
(k^2+l^2)^2$.  Since the group velocity must point away from the
region where the Rossby waves are generated, we must have
$kl>0$ north of this wave source and $kl<0$ south of the wave source.  Combining
this information with Eq.~(\ref{u'v'}) shows that $\overline{u'v'}<0$
north of the source and $\overline{u'v'}>0$ south of the source.
Thus, the Rossby waves flux eastward momentum into the latitude 
of the wave source.  

This process is illustrated in Fig.~\ref{rossby-wave-schematic}.
Suppose some process generates Rossby wave packets at a specific
latitude, which propagate north and south from the latitude of
generation.  The northward propagating wave packet exhibits eddy
velocities tilting northwest-southeast, while the southward
propagating packet exhibits eddy velocities tilting
southwest-northeast.  The resulting eddy velocities visusally resemble
an eastward-pointing chevron pattern centered at the latitude of wave
generation.  The correlation between these velocities leads to
non-zero Reynolds stresses (that is, a non-zero $\overline{u'v'}$)
and a flux of angular momentum into the
wave generation latitude.  An example of this schematic chevron
pattern in an actual GCM experiment is shown in Figure~\ref{frierson}.

\begin{figure*}
 \epsscale{1.6}
\plotone{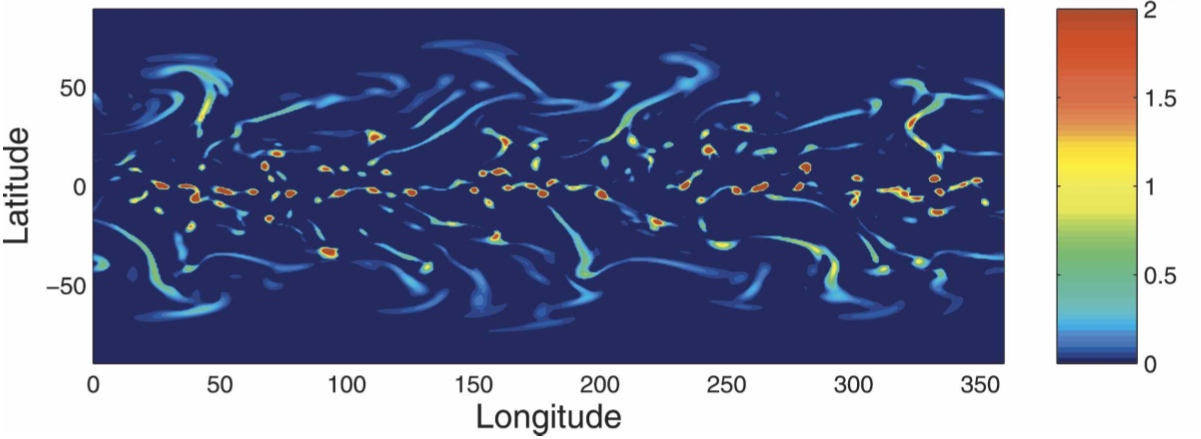}
 \caption{Instantaneous precipitation (units
   $10^{-3}\rm\,kg\,m^{-2}\,s^{-1}$) in an idealized Earth GCM by
   \citet{frierson-etal-2006}, illustrating the generation of phase
   tilts by Rossby waves as depicted schematically in
   Figure~\ref{rossby-wave-schematic}.  Baroclinic instabilities in
   midlatitudes ($\sim$40--50$^{\circ}$) generate Rossby waves that
   propagate meridionally.  On the equatorward side of the
   baroclinically unstable zone (latitudes of $\sim$20--50$^{\circ}$),
   the waves propagate equatorward, leading to characteristic
   precipitation patterns tilting southwest-northeast in the northern
   hemisphere and northwest-southeast in the southern hemisphere.  In
   contrast, the phase tilts are reversed (though less well organized)
   poleward of $\sim$50--60$^{\circ}$ latitude, indicative of poleward
   Rossby wave propagation. Equatorward of $20^{\circ}$ latitude,
   tropical convection dominates the precipitation pattern.
%
}
\label{frierson}
 \end{figure*}

The above reasoning is for free (unforced) waves but can be
extended to an atmosphere forced by vorticity sources/sinks
and damped by frictional drag \citep[see reviews in][]{held-2000,
vallis-2006, showman-polvani-2011}.  
In the nondivergent, barotropic system, the 
zonal-mean zonal momentum equation is given by, adopting a Cartesian
coordinate system for simplicity,
\begin{equation}
{\partial\overline{u}\over\partial t}=-{\partial(\overline{u'v'})\over
\partial y} - {\overline{u}\over\tau_{\rm drag}}
\label{eulerian-mean-barotropic}
\end{equation}
where we have decomposed the winds into their zonal means
(given by overbars) and the deviations therefrom (given
by primes), such that $u=\overline{u}+u'$, $v=\overline{v}+v'$,
$\zeta=\overline{\zeta}+\zeta'$, etc.
Here, $y$ is northward distance and we have parameterized drag
by a linear (Rayleigh) friction that relaxes the winds toward zero
over a specified drag time constant $\tau_{\rm drag}$.
To link the momentum budget to the vorticity sources and 
sinks, we first note that, for the horizontally nondivergent
system, the definition of vorticity implies that $\overline{v'\zeta'}=
-\partial(\overline{u'v'})/\partial y$.  Second, we multiply
the linearized version of Eq.~(\ref{barotropic-vorticity}) by
$\zeta'$ and zonally average, yielding an equation for the
so-called pseudomomentum
\begin{equation}
{\partial{\cal A}\over \partial t} + \overline{v'\zeta'}
= {\overline{\zeta'F'}\over 2(\beta - {\partial^2\overline{u}\over
\partial y^2})}
\label{pseudomomentum}
\end{equation}
where ${\cal A}=(\beta - \partial^2\overline{u}/\partial y^2)^{-1}
\overline{\zeta'^2}/2$ is the pseudomomentum, which characterizes the 
amplitude of the eddies in a statistical (zonal-mean) sense.  Here,
$F'$ is the eddy component of the vorticity source/sink defined in
Equation~(\ref{barotropic-vorticity}). 
We then combine Eqs.~(\ref{eulerian-mean-barotropic}) and
(\ref{pseudomomentum}). If the zonal-mean flow equilibrates to
a statistical steady state (i.e., if
$\partial\overline{u}/\partial t$ and $\partial {\cal A}/\partial t$
are zero), this yields a relationship between the zonal-mean zonal
wind, $\overline{u}$, and the eddy generation of vorticity:
\begin{equation}
{\overline{u}\over\tau_{\rm drag}}={\overline{\zeta'F'}\over
2(\beta - {\partial^2\overline{u}\over\partial y^2})}.
\label{u-barotropic}
\end{equation}
(Note that this equilibration does not require the {\it eddies themselves}
to be steady, but rather simply that their zonally averaged mean amplitude,
characterized by ${\cal A}$, is steady.) 

What are the implications of this equation?  Consider a region
away from wave sources where the eddies are dissipated.  Dissipation
acts to reduce the eddy amplitudes, implying that $F'$ and
$\zeta'$ exhibit opposite signs 
(see Eq.~\ref{barotropic-vorticity}).\footnote{For example, 
for the specific case of linear drag, represented in the zonal and
meridional momentum equations as $-u/\tau_{\rm drag}$ and $-v/\tau_{\rm drag}$,
respectively, then in the absence of forcing $F=-\zeta/\tau_{\rm drag}$.}
Therefore, regions of net wave dissipation will exhibit
$\overline{\zeta'F'}<0$ and the zonal-mean zonal wind will be
westward.  On the other hand, in region where eddies
are generated, the wave sources act to increase the wave
amplitudes,
implying that $F'$ and $\zeta'$ exhibit the same signs.  In such a
region, $\overline{\zeta'F'}>0$ and the zonal-mean zonal wind will be
eastward.  The two regions are linked because waves propagate (in the
sense of their group velocity) from their latitude of generation to
their latitude of dissipation; this allows a statistical steady state
in both the zonal-mean eddy amplitudes and the zonal-mean zonal wind
to be achieved despite the tendency of damping (forcing) to locally
decrease (increase) the eddy amplitudes.  In summary, we thus recover
the result that eastward net flow occurs in regions of net wave
generation while westward net flow occurs in regions of net wave
damping.

In the extratropics, baroclinic instabilities constitute a primary
mechanism for generating Rossby waves in the free troposphere, which
propagate from the latitude of instability to surrounding latitudes
(see Figure~\ref{frierson}).  To the extent that the extratropics of
exoplanets are baroclinically unstable, this mechanism thus implies
the emergence of the extratropical eastward jets; indeed, the
midlatitude jets in the tropospheres of Earth and Mars result from
this process.  Such ``eddy-driven'' jets are illustrated schematically
in Figure~\ref{atmospheric-processes-schematic}.\footnote{In the
  terrestrial literature, these are referred to as ``eddy-driven''
  jets to distinguish them from the subtropical jets that occur at the
  poleward edge of the Hadley cell (see \S\ref{tropics}).}  To be more
precise, the eddy-momentum flux convegence should be thought of as
driving the surface wind \citep[e.g.,][]{held-2000}.  The only torque
that can balance the vertically integrated eddy-momentum flux
convergence---and the associated eastward acceleration---is frictional
drag at the surface.  Thus, in latitudes where baroclinic instability
or other processes leads to strong Rossby wave generation, and the
radiation of those waves to other latitudes, eastward zonal-mean winds
emerge at the surface.  The zonal wind in the upper troposphere is
then set by the sum of the surface wind and the vertically integrated
wind shear, which is in thermal-wind balance with the meridional
temperature gradients.

\begin{figure*}
 \epsscale{1.}
\plotone{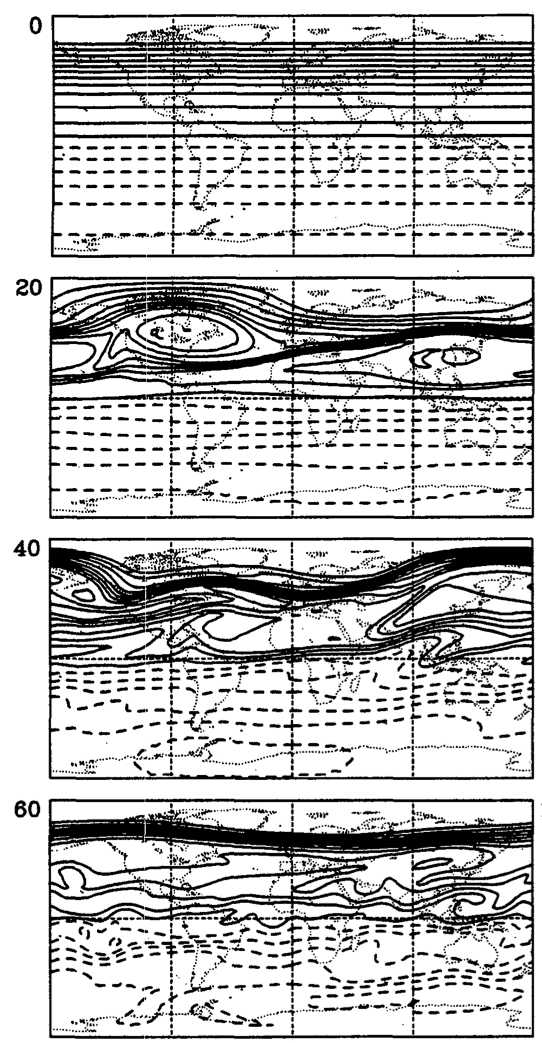}
 \caption{Demonstration of how breaking Rossby waves can shape jet
   dynamics.  Plots show time evolution of the potential vorticity (in
   contours, with positive solid, negative dashed, and a contour
   interval of $0.25\times10^{-8}\rm\,m^{-1}\,s^{-1}$) in a
   shallow-water calculation of Earth's stratospheric polar vortex by
   \citet{polvani-etal-1995}.  The model is initialized from a
   zonally-symmetric initial condition with broadly distributed PV
   organized into a polar vortex centered over the north pole (top
   panel).  A topographic perturbation is then introduced to generate
   a Rossby wave, which manifests as undulations in the PV contours
   (second panel).  The wave amplitude becomes large enough for the
   undulations to curl up, leading to wave breaking (second and third
   panels).  The edge of the vortex (corresponding to the region of
   tighly spaced PV contours) is resistant to such wave breaking, but
   the regions of weaker meridional PV gradient on either side are
   more susceptible.  This breaking mixes and homogenizes the PV in
   those regions, thereby lessening the meridional gradient still
   further (this manifests as a widening in the latitudinal separation
   of the PV contours, especially from the equator to
   $\sim$$60^{\circ}$N latitude, in the third and fourth panels).  In
   contrast, stripping of material from the vortex edge by this mixing
   process causes a {\it sharpening} of the PV jump associated with
   the vortex edge (visible as a tightening of the contour spacing at
   $\sim$70--$80^{\circ}$N latitude in the fourth panel). This sharp
   PV jump is associated with a narrow, fast eastward jet---the polar
   jet---at the edge of the polar vortex. }
\label{surf-zone}
 \end{figure*}

\begin{figure*}
 \epsscale{0.6}
\plotone{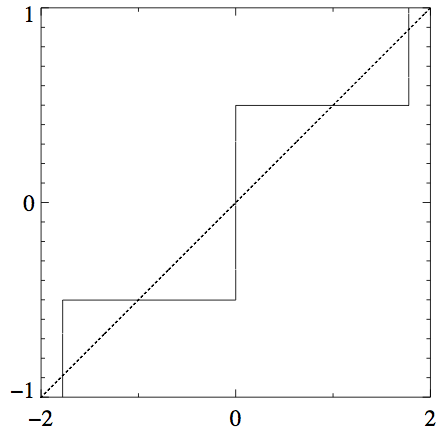}
\put(-100.,-10.){\normalsize Potential vorticity}
\plotone{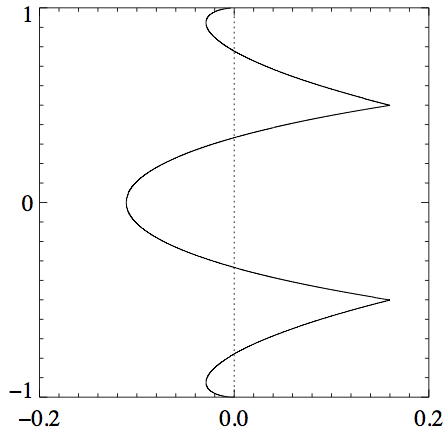}
\put(-90.,-10.){\normalsize Zonal wind}
\put(-290.,50){\normalsize \rotatebox{90}{sin(Latitude)}}
 \caption{Relationship between PV (left) and zonal winds (right)
   versus the sine of latitude for a flow consisting of a PV
   ``staircase'' with constant-PV strips separated by regions of sharp
   PV gradients.  In a motionless fluid on a spherical planet, the
   zonal wind is zero, and the PV increases smoothly with sine of
   latitude (dotted lines).  But if PV is homogenized into strips, as
   shown on the left, then the implied zonal wind structure (demanded
   by the relationship between PV and winds) is as shown on the right
   (solid curves).  Thus, homogenization of PV into strips on rapidly
   rotating planets implies the emergence of zonal jets. This is for
   the specific case of a 2D, horizontally non-divergent flow governed
   by Equation~(\ref{barotropic-vorticity}), but a similar
   relationship holds even in more realistic models: regions of weak
   PV gradients correspond to regions of broad westward flow, while PV
   discontinuities correspond to sharp eastward jets.  For the
   specific case of a zonally symmetric flow governed by
   Equation~(\ref{barotropic-vorticity}), in Cartesian geometry with
   constant $\beta$, it is straightforward to show that the zonal-wind
   profile consists of parabolas connected end-to-end.  For analytic
   solutions in more general cases, see \citet{marcus-lee-1998},
   \citet{dunkerton-scott-2008}, and \citet{wood-mcintyre-2010}.  From
   \citet{scott-2010}.  }
\label{analytic}
 \end{figure*}

In the above theory, the only source of explicit damping was a linear
drag, but in the real extratropical atmosphere, wave breaking will
play a key role and can help explain many aspects of extratropical
jets.  Indeed, much of jet dynamics can be understood in terms of
spatially inhomogeneous mixing of potential vorticity (PV, see
Equation~\ref{pv}) caused by this wave breaking
\citep{dritschel-mcintyre-2008}.  Rossby waves manifest as meridional
undulations in PV contours, and Rossby wave breaking implies that the
PV contours become so deformed that they curl up and overturn in the
longitude-latitude plane (see Figure~\ref{surf-zone} for an example).  
Such overturning generally requires large
wave amplitudes; in particular, the local, wave-induced perturbation
to the meridional PV gradient must become comparable to the background
(zonal-mean) gradient, so that, locally, the meridional PV gradient
changes sign.  This criterion implies that Rossby wave breaking occurs
more easily (i.e., at smaller wave amplitude) in regions of weak PV
gradient than in regions of strong PV gradient.  Now, the relationship
between PV and winds implies that
eastward zonal jets comprise regions where the
(zonal-mean) meridional PV gradient is large, whereas westward zonal
jets comprise regions where the meridional PV gradient is small 
(Figure~\ref{analytic}).  Therefore, vertically and meridionally
propagating Rossby waves are more likely to break between (or on the
flanks of) eastward jets, where the PV gradient is small, than at the
cores of eastward jets, where the PV gradient is large.  This wave
breaking causes irreversible mixing and homogenization of PV.  The mixing is
thus spatially inhomogeneous---mixing preferentially homogenizes the
PV in the regions where its gradient is already weak, and sharpens the
gradients in between.  As emphasized by
\citet{dritschel-mcintyre-2008}, this is a positive feedback: by
modifying the background PV gradient, such mixing {\it promotes}
continued mixing at westward jets but {\it inhibits} future mixing at
eastward jets.
Because the PV jumps are associated with eastward jets, this process
tends to sharpen eastward jets and leads to broad ``surf zones'' of
westward flow in between.  This is precisely the behavior evident in
Figure~\ref{surf-zone}.  Because of this positive feedback, one
generally expects that when an initially homogeneous system is
stirred, robust zonal jets should spontaneously emerge---even if the
stirring itself is not spatially organized
\citep[e.g.,][]{dritschel-mcintyre-2008, dritschel-scott-2011,
  scott-dritschel-2012, scott-tissier-2012}.  Nevertheless, strong
radiative forcing and/or frictional damping represent sources and
sinks of PV that can prevent a pure PV staircase (and the associated
zonal jets) from being achieved.  The factors that determine the
equilibrium PV distribution in such forced/damped cases is an area of
ongoing research.

\subsubsection{Jet formation II: Rossby wave interactions with turbulence}

Additional insights on the formation of zonal jets emerge from a
consideration of atmospheric turbulence.
In the $Ro\ll 1$ regime, the interaction of large-scale atmospheric
turbulence with planetary rotation---and in particular with the
$\beta$ effect---generally leads to the formation of a zonally banded
appearance and the existence of zonal jets (for important
examples, see \citet{rhines-1975}, \citet{williams-1978},
\citet{maltrud-vallis-1993}, \citet{cho-polvani-1996a},
\citet{huang-robinson-1998}, and \citet{sukoriansky-etal-2007};
reviews can be found in \citet{vasavada-showman-2005},
\citet{showman-etal-2010}, and \citet[][chapter 9]{vallis-2006}).  By
allowing Rossby waves, the $\beta$ effect introduces a fundamental
anisotropy between the zonal and meridional direction that often leads
to jets.

Consider the magnitudes of the terms in the vorticity equation
(\ref{barotropic-vorticity}).  Being a curl of the wind field, the
relative vorticity has characteristic magnitude ${U\over L}$.
Therefore, the advection term ${\bf v}\cdot\nabla\zeta$ has
characteristic magnitude ${U^2\over L^2}$, where $L$ is some
horizontal lengthscale of interest.  The $\beta$ term has
characteristic magnitude $\beta U$.  For a given wind amplitude, the
advection term dominates at small scales (i.e., as $L\to0$).  At these
scales, the $\beta$ term is unimportant, and the equation therefore
describes two-dimensional turbulence.  Such turbulence will be
isotropic, because (at scales too small for $\beta$ to be important)
the equation contains no terms that would distinguish the east-west
from the north-south directions.  On the other hand, at large
scales, the $\beta$ term will dominate over the nonlinear advection
term, and this implies that Rossby waves dominate the
dynamics.  The transition between these regimes occurs at the Rhines
scale,
\begin{equation}
L_R =  \left({U\over\beta}\right)^{1/2}.
\label{rhines}
\end{equation}

The Rhines scale is traditionally interpreted as giving the transition
scale between the regimes of turbulence (at small scales) and Rossby
waves (at large scales).  Generally, two-dimensional and
quasi-two-dimensional fluids exhibit an upscale energy transfer from
small scales to large scales \citep[e.g.,][Chapter 8]{vallis-2006}.
Therefore, if turbulent energy is injected into the fluid at small
scales (e.g., through convection, baroclinic instabilities, or other
processes), turbulent interactions can drive the energy toward larger
scales where it can be affected by $\beta$.  At scales close to the
Rhines scale, the $\beta$ effect forces the turbulence to become
anisotropic: since the term $v\beta$ involves only the meridional
speed but not the zonal speed, the dynamics becomes different in the
east-west and north-south directions.  

\begin{figure*}
 \epsscale{1.6}
\plotone{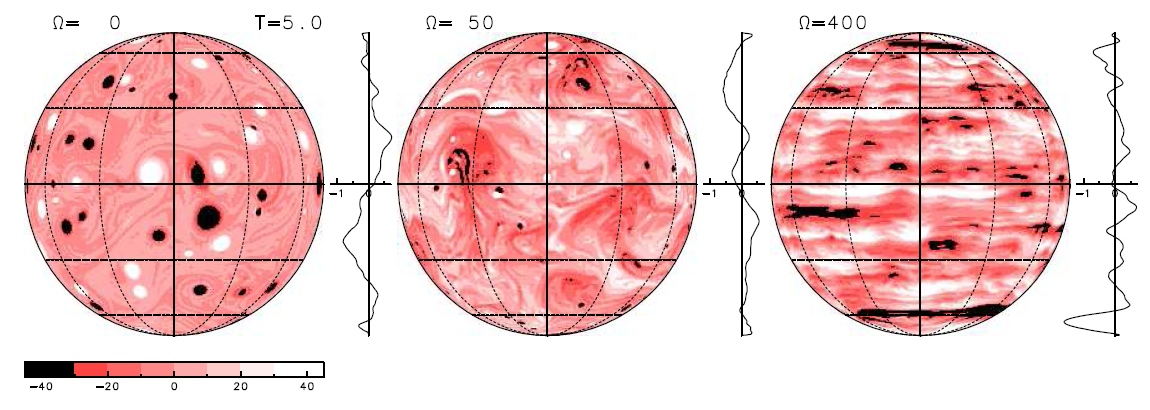}
 \caption{Solutions of the 2D non-divergent barotropic vorticity
   equation (\ref{barotropic-vorticity}) illustrating the effect of
   planetary rotation on large-scale atmospheric turbulence.  Three
   simulations are shown, initialized from identical initial
   conditions containing turbulence at small length scales.  Only the
   rotation rate differs between the three models; from left to right,
   the rotation rates are zero, intermediate, and rapid.  There is no
   forcing or large-scale damping (save for a numerical viscosity
   required for numerical stability) so that the total flow energy is
   nearly constant in all three cases.  In the non-rotating case,
   energy cascades to larger length scales as vortices merge, but the
   flow remains isotropic.  In the rapidly rotating case, the flow is
   highly anisotropic and zonal banding develops.  From
   \citet{hayashi-etal-2000}; see also \citet{yoden-etal-1999},
   \citet{ishioka-etal-1999}, and \citet{hayashi-etal-2007}.  }
\label{hayashi}
 \end{figure*}

This anisotropy causes the development of zonal banding in planetary
atmospheres. As shown by \citet{vallis-maltrud-1993} and other
authors, the transition between turbulence and Rossby waves is
anisotropic, occurring at different length scales for different
wavevector orientations.  (Equation~\ref{rhines} gives the
characteristic value ignoring this geometric effect.)  Essentially,
non-linear interactions are better able to transfer turbulent energy
upscale for structures that are zonally elongated than for structures
that are isotropic or meridionally elongated.  Preferential
development of zonally elongated structures at large scales, often
consisting of zonal jets, results.  
In many cases, the turbulent interactions do not involve
the gradual transport of energy to incrementally ever-larger
lengthscales (a transport that would be ``local'' in spectral space)
but rather tend to involve spectrally {\it non}-local interactions wherein
small-scale turbulence directly pumps the zonal jets \citep[e.g.,][]
{nozawa-yoden-1997a, huang-robinson-1998, sukoriansky-etal-2007,
read-etal-2007, wordsworth-etal-2008}.

These processes are illustrated in Figure~\ref{hayashi}, which
shows the results of three solutions of the 2D, non-divergent 
vorticity equation (\ref{barotropic-vorticity}) starting from
an initial condition containing numerous small-scale, close-packed, isotropic
vortices.  The model on the left is non-rotating, the model in
the middle has intermediate rotation rate, and the model on the
right is rapidly rotating.  In the non-rotating case (left panel), the 2D
turbulence involves vortex mergers that drive
turbulent energy from  small scales toward
larger scales.  The flow remains isotropic and no jets form.
In the rapidly rotating case, however, the turbulence interacts
with Rossby waves at scales comparable to $L_R$, leading to
robust zonal banding (right panel).

A wide variety of studies have shown that zonal jets generated in this
way exhibit characteristic meridional widths comparable to the Rhines
scale [for reviews, see \citet{vasavada-showman-2005}, \citet[Chapter
    9]{vallis-2006}, and \citet{delgenio-etal-2009}].  Nevertheless,
there has been significant debate about the relationship of the Rhines
scale to other important lengthscales, such as Rossby deformation
radius.  The characteristic lengthscale for baroclinic instabilities
is the deformation radius; therefore, in a flow driven by baroclinic
instabilities, turbulent energy is injected at scales close to $L_D$.
In principle, it is possible that $L_R$ greatly exceeds $L_D$, in
which case nonlinear interactions would transfer the energy upscale
from the deformation radius to the scale of the jets
themselves. However, atmospheric GCMs under Earth-like conditions have
generally found that the circulation adjusts to a state where the
deformation radius is comparable to the Rhines scale
\citep[e.g.,][]{schneider-walker-2006, schneider-walker-2008}.  In
this case, baroclinic instabilties directly inject turbulent
energy into the flow at scales comparable to the
meridional width of the zonal jets.  The injected energy interacts
nonlinearly with the mean flow to generate jets, but significant
upscale transfer of the energy is not necessarily involved.  Still,
situations exist (including in Earth's ocean) where
the Rhines scale and deformation radius differ substantially and
significant upscale energy transfer occurs between the two scales
\citep[e.g.][]{jansen-ferrari-2012}.  

\begin{figure*}
 \epsscale{1.2}
\plotone{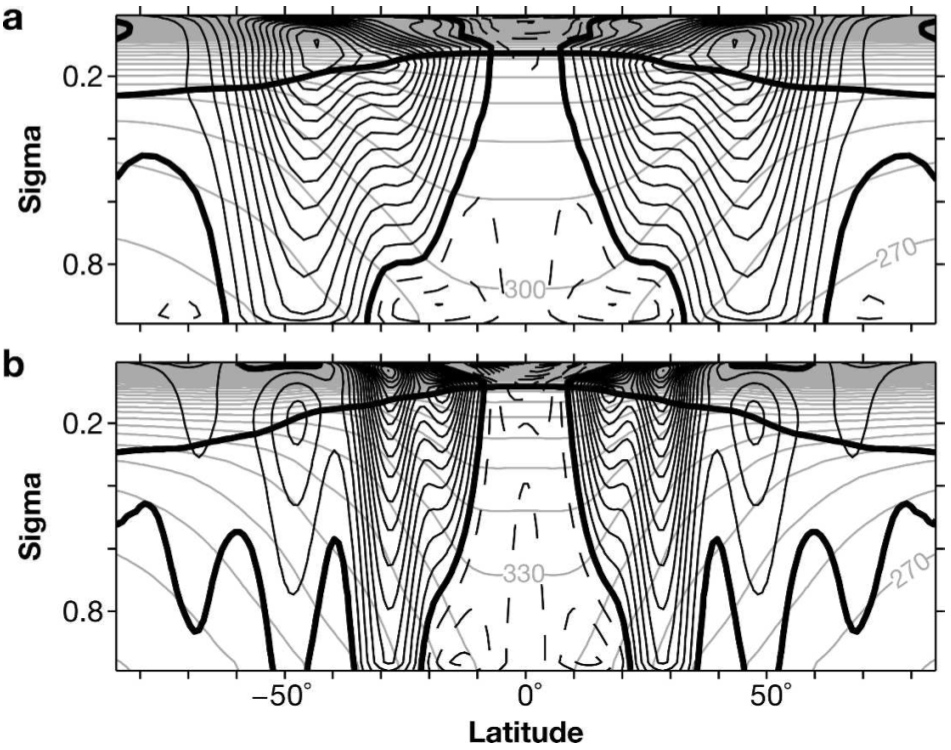}
 \caption{Idealized GCM simulations of Earth-like planets exhibiting
jet formation.  These models are forced by equator-pole heating
gradients.  Black contours show zonal-mean zonal wind (positive solid,
negative dashed, zero contour is thick; contour interval is 
$2.5\rm\,m\,s^{-1}$).  Grey contours show zonal-mean potential temperature
(K, contour interval is $10\rm\,K$).  Heavy line near top denotes the
tropopause. Vertical coordinate is ratio
of pressure to surface pressure.  The top panel shows an Earth-like
case, while the bottom panel shows a case at 4 times the Earth
rotation rate.   The Earth-like case exhibits one eddy-driven jet
in each hemisphere.  In the rapidly rotating case, this jet is confined
much closer to the equator and additional, eddy-driven jets have
emerged in each hemisphere. From \citet{schneider-walker-2006}.}
\label{schneider-walker}
 \end{figure*}

These arguments suggest that, for extratropical jets driven by
baroclinic instability, the relative sizes of the deformation radius
and of the extratropics itself determines the number of jets.  When a
planet is small, like Earth or Mars, only one strip of baroclinic
instabilities---and one eddy-driven jet---can fit into the baroclinic
zone (as illustrated schematically in
Figure~\ref{atmospheric-processes-schematic}).  When the planet is
relatively large relative to the eddy size, however, the baroclinic
zone breaks up into multiple bands of baroclinic instability, and
hence multiple jet streams.  Figure~\ref{schneider-walker} shows an
example from \citet{schneider-walker-2006} showing an Earth-like case
(top) and a case at four times the Earth rotation rate (bottom).  The
Earth case exhibits only one, midlatitude eddy-driven jet in each
hemisphere.  In the rapidly rotating case, the deformation radius is
four times smaller, and the baroclinic zone breaks up into three jets
in each hemisphere.  This process coincides with a steepening of the
isentropes (thin grey contours), which is associated with a greater
equator-to-pole temperature difference, consistent with those shown in
Figures~\ref{earthlike-circulation} and \ref{deltaT-vary-rotation}.

\subsection{Tropical Regime}
\label{tropics}

The tropics, defined here as regimes of $Ro\gtrsim 1$, are inherently
ageostrophic, and their dynamics differ significantly from those
of the extratropics.  Unlike the case of geostrophic flow, horizontal
temperature gradients in the $Ro\gtrsim1$ regime tend to be 
modest, which affects the dynamics in myriad ways.  

The tendency toward weak temperature gradients can be motivated by
considering arguments analogous to those leading up to
Equation~(\ref{charney-extratropics}).  In the absence of a
significant Coriolis force, large-scale horizontal pressure-gradient
forces tend to be balanced by advection, represented to order of
magnitude as $U^2/L$.  The fractional horizontal potential temperature
difference is then \citep{charney-1963}
\begin{equation}
{\delta\theta_h\over \theta} \sim {U^2\over g D} \sim Fr \qquad\qquad Ro\gtrsim1
\label{charney-tropics}
\end{equation}
Comparison with Equation~(\ref{charney-extratropics}) immediately
shows that, in the rapidly rotating regime characterized by $Ro\ll1$,
the lateral temperature contrasts are a factor of $Ro^{-1}$ bigger
than they are in the slowly rotating regime of $Ro\gtrsim 1$.  These
trends are evident in the GCM experiments in
Figure~\ref{earthlike-circulation} and \ref{deltaT-vary-rotation}c.
For a given wind speed, slowly rotating planets tend to be more
horizontally isothermal than rapidly rotating planets.  In the case of
a typical terrestrial planet where $U\sim 10\rm\,m\,s^{-1}$, $D\sim
10\rm\,km$, and $g\approx 10\rm\,m\,s^{-2}$, we have $Fr\sim 10^{-3}$.
One might thus expect that, in the tropics of Earth, and globally on
Venus, Titan, and slowly rotating exoplanets, the lateral temperature
contrasts are $\sim$$1\rm\,K$.  Note, however, that because of the
quadratic dependence of $\delta \theta_h$ on wind speed, large
temperature differences could occur if the winds are sufficiently
fast.  Of course, additional arguments would be needed to obtain a
self-consistent prediction of both $\delta \theta_h$ {\it and} $U$ on
any given planet.

We here discuss several of the major dynamical
mechanisms relevant for the tropical regime.

\subsubsection{Hadley circulation and subtropical jets}
\label{hadley}

Planets generally exhibit meridional gradients of the mean incident
stellar flux, with, at low obliquity, highly irradiated, warm conditions at
low latitudes and poorly irradiated, cooler conditions at high
latitudes.  The Hadley circulation represents the tropical response to
this insolation gradient and is the primary mechanism for meridional
heat transport in the tropical atmosphere.  Stripped to its essence, the Hadley
circulation in an atmosphere forced primarily by an equator-pole
heating gradient can be idealized as an essentially two-dimensional
circulation in the latitude-height plane: hot air rises near the
equator, moves poleward aloft, descends at higher latitudes, and
returns to the equator along the surface (see
Figure~\ref{atmospheric-processes-schematic} for a schematic).  All
the terrestrial planets with thick atmospheres in the solar
system---Earth, Venus, Mars, and Titan---exhibit Hadley cells, and
Hadley circulations will likewise play an important role on terrestrial
exoplanets.

The Hadley circulations exert a significant effect on the mean
planetary climate.   Because of its meridional energy transport, the
meridional temperature gradient across the Hadley circulation tends to
be weak.  Moreover, on planets with hydrological cycles
(Section~\ref{hydrological}), the ascending branch tends to be a
location of cloudiness and high rainfall, whereas the descending
branches tend to be drier and more cloud-free.  In this way, the
Hadley circulation exerts control over regional climate.  On Earth,
most tropical rainforests occur near the equator, at the latitude of
the ascending branch, while many of the world's major deserts (the
Sahara, the American southwest, Australia, and South Africa) occur
near the latitudes of the descending branches.  Because the
latitudinal and vertical distribution of cloudiness and humidity can
significantly influence the planetary albedo and greenhouse effect,
the mean planetary surface temperature---as well as the distribution
of temperature, cloudiness, and humidity across the planet---depend
significantly on the structure of the Hadley circulation.  Moreover,
by helping to control the equator-pole distribution of temperature,
humidity, and clouds, the Hadley circulation will influence the
conditions under which terrestrial planets can experience global-scale
climate feedbacks, including transitions to globally glaciated
``snowball'' states, atmospheric collapse, and runaway greenhouses
(Section~\ref{climate}).

The structure of the Hadley cell is strongly controlled by planetary
rotation.  Frictional drag acts most strongly in the surface branch of
the circulation, and the upper branch, being decoupled from the
surface, is generally less affected by frictional drag.  A useful
limit to consider is one where the upper branch is frictionless, such
that individual air parcels ascending at the equator conserve their
angular momentum per unit mass about the planetary rotation axis,
given by $m=(\Omega a \cos\phi + u)a\cos\phi$, as they move poleward
\citep{held-hou-1980}.  If the ascending branch is at the equator and
exhibits zero zonal wind, then the zonal wind in the upper
(poleward-flowing) branch of such an angular-momentum conserving
circulation is
\begin{equation}
u = \Omega a {\sin^2\phi\over \cos\phi}
\label{uM}
\end{equation}
where $a$ is the planetary radius and $\phi$ is latitude.  Thus, in
the upper branch of the Hadley cell, the zonal wind increases rapidly
with latitude, the more so the faster the planet rotates or the larger 
the planetary size.  Under
modern Earth conditions, this equation implies that the zonal wind is
$134\rm\,m\,s^{-1}$ at $30^{\circ}$ latitude, reaches
$1000\rm\,m\,s^{-1}$ at a latitude of $67^{\circ}$, and becomes
infinite at the poles.  This is of course impossible, and implies that
planetary rotation, if sufficiently strong, confines the Hadley
circulation to low latitudes.  Real Hadley cells do
not conserve angular momentum, and exhibit zonal winds increasing
more slowly with latitude than expressed by Eq.~(\ref{uM}); nevertheless,
rotation generally confines Hadley circulations to low latitudes
even in this case.

\begin{figure*}
 \epsscale{0.9}
\plotone{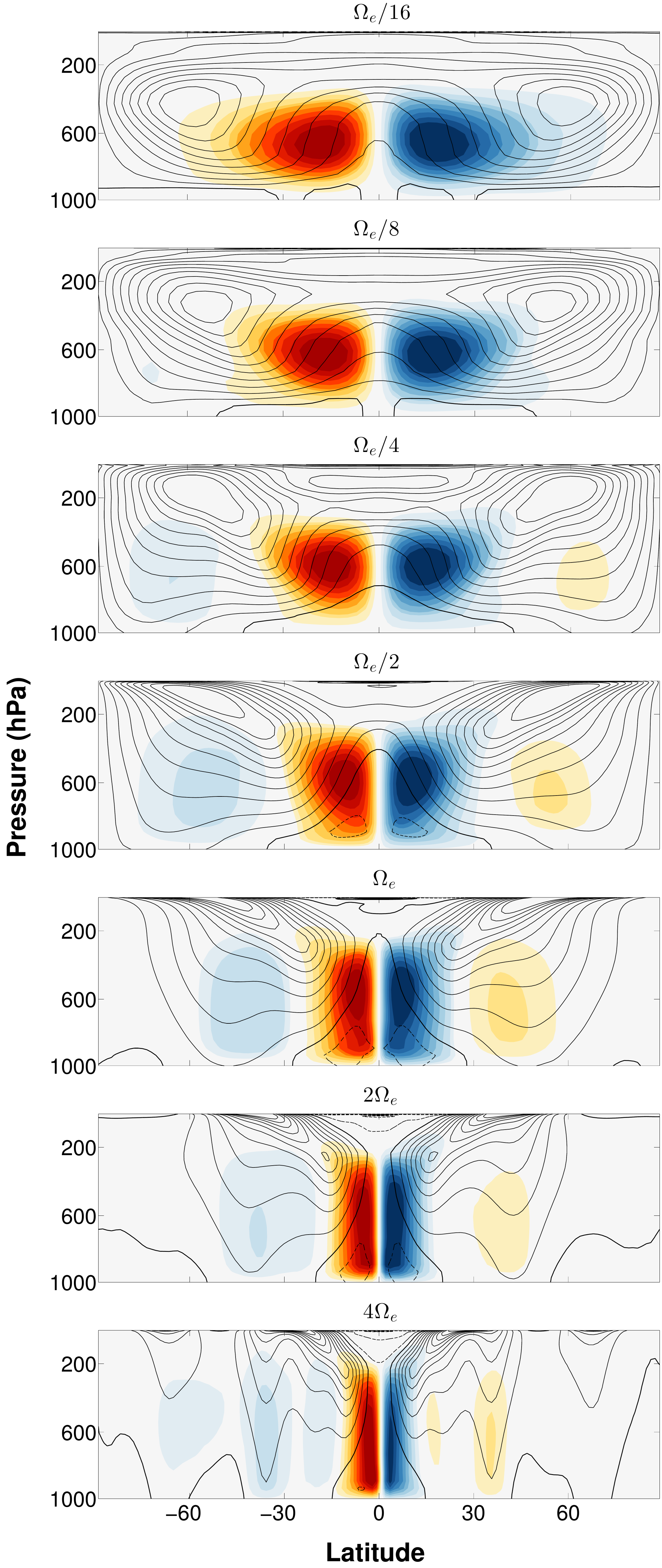}
\epsscale{0.7677}
\plotone{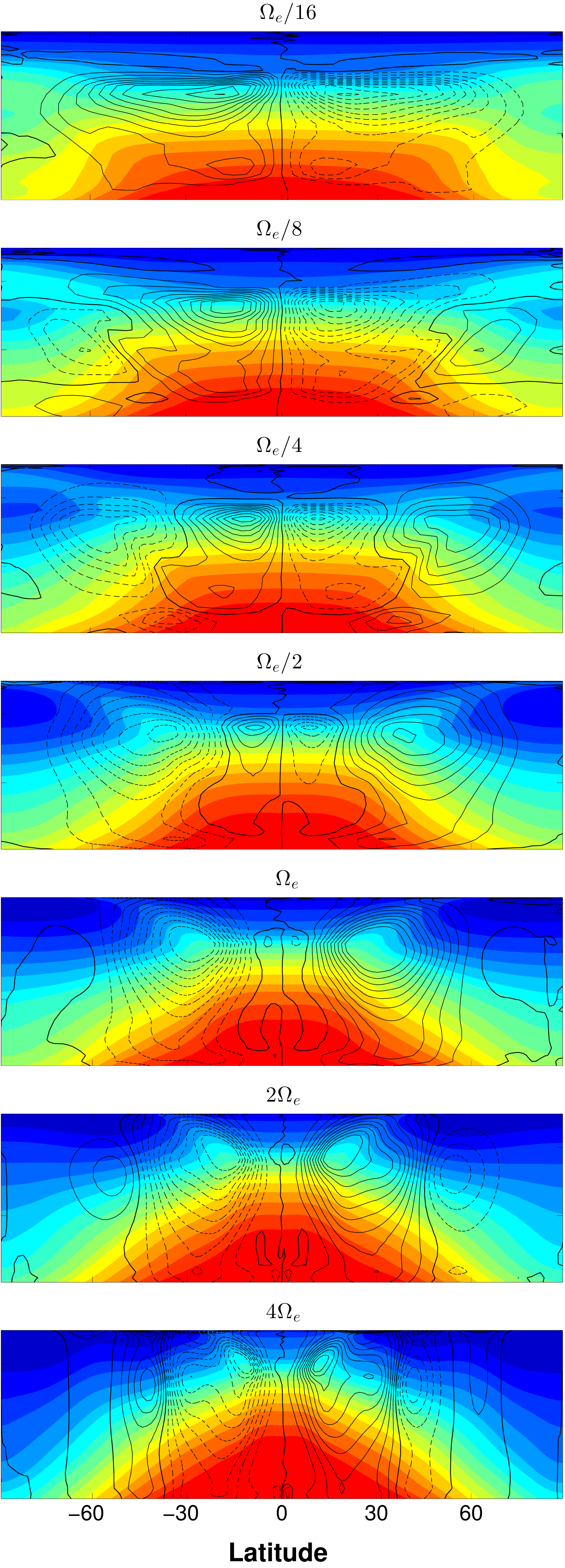}
 \caption{Zonal-mean circulation for a sequence of idealized GCM
   experiments of terrestrial planets from \citet{kaspi-showman-2012},
   showing the dependence of the Hadley circulation on planetary
   rotation rate.  The models are driven by an imposed equator-pole
   insolation pattern with no seasonal cycle. The figure shows seven
   experiments with differing planetary rotation rates, ranging from
   1/16$^{\rm th}$ to four times that of Earth from top to bottom,
   respectively.  {\it Left column:} Thin black contours show
   zonal-mean zonal wind; the contour interval is $5\rm\,m\,s^{-1}$,
   and the zero-wind contour is shown in a thick black contour.
   Orange/blue colorscale depicts the mean-meridional streamfunction,
   with blue denoting clockwise circulation and orange denoting
   counterclockwise circulation.  {\it Right column:} colorscale shows
   zonal-mean temperature.  Contours show zonal-mean meridional
   eddy-momentum flux, $\overline{u'v'}\cos\phi$.  Solid and dashed
   curves denote positive and negative values, respectively (implying
   northward and southward transport of eastward eddy momentum,
   respectively).  At slow rotation rates, the Hadley cells are nearly
   global, the subtropical jets reside at high latitude, and the
   equator-pole temperature difference is small.  The low-latitude meridional
   momentum flux is equatorward, leading to equatorial superrotation
   (eastward winds at the equator) in the upper troposphere.  At faster rotation
   rates, the Hadley cells and subtropical jets contract toward the
   equator, an extratropical zone, with eddy-driven jets, develops at
   high latitudes, and the equator-pole temperature difference is
   large.  The low-latitude meridional momentum flux is poleward,
   resulting from the absorption of equatorward-propagating Rossby waves
   coming from the extratropics.}
\label{hadley-cell}
 \end{figure*}

Over the past 30 years, many GCM studies have been performed to
investigate how the Hadley circulation depends on planetary rotation
rate and other parameters \citep{hunt-1979, williams-holloway-1982,
  williams-1988a, williams-1988b, delgenio-suozzo-1987,
  navarra-boccaletti-2002, walker-schneider-2005,
  walker-schneider-2006, kaspi-showman-2012}.
Figure~\ref{hadley-cell} illustrates an example from
\citet{kaspi-showman-2012}, showing GCM experiments for planets with
no seasonal cycle forced by equator-pole heating gradients.  Models
were performed for rotation rates ranging from 1/16$^{\rm th}$ to 4
times that of Earth (top to bottom panels in Figure~\ref{hadley-cell},
respectively).  Other parameters in these experiments, including solar
flux, planetary radius and gravity, and atmospheric mass are all
Earth-like.  In general, the Hadley circulation consists of a cell in
each hemisphere extending from the equator toward higher latitudes
(dark red and blue regions in Figure~\ref{hadley-cell}).  As the air
in the upper branch moves poleward, it accelerates eastward by the
Coriolis force, leading to the so-called subtropical jets whose
amplitudes peak at the poleward edge of the Hadley cell (such
subtropical jets are shown schematically in
Figure~\ref{atmospheric-processes-schematic}).  At sufficiently low
rotation rates (top two panels of Figure~\ref{hadley-cell}), the Hadley
circulation is nearly global, extending from equator to pole in both
hemispheres, with the subtropical jets peaking at latitudes of
$\sim$$60^{\circ}$.  Such models constitute ``all tropics'' worlds
lacking any high-latitude extratropical baroclinic zone.  At faster
rotation rates (bottom five panels), the Hadley circulation---and the
associated subtropical jets---become confined closer to the equator,
and the higher latitudes develop an extratropical zone with baroclinic
instabilities.  The emergence of mid- and high-latitude eddy-driven
jets in the extratropics, with eastward surface wind, can be seen in
these rapidly rotating cases (cf Section~\ref{jet-formation}).  As
shown in Figure~\ref{hadley-cell}, the tropospheric temperatures are
relatively constant with latitude across the Hadley cell and begin
decreasing poleward of the subtropical jets near its poleward edge.
The resulting equator-to-pole temperature differences are small at
slow rotation and increase with increasing rotation rate
(Figure~\ref{hadley-cell}).

Seasonality exerts a strong effect on the Hadley circulation.  When a
planet's obliquity is non-zero, the substellar latitude oscillates
between hemispheres, crossing the equator during equinox and reaching
a peak excursion from the equator at solstice.  The rising branch
of the Hadley circulation tends to follow the latitude of maximum
sunlight and therefore oscillates between hemispheres as 
well.\footnote{When the atmospheric heat capacity is large, the
response will generally be phase lagged with respect to the stellar
heating pattern.}  Near
equinox, the rising branch lies close to the equator, with Hadley
cells of approximately equal strength in each hemisphere (cf,
Figure~\ref{hadley-cell}).  Near solstice, the rising branch lies in
the summer hemisphere.  As before, two cells exist, with air in one
cell (the ``winter'' cell) flowing across the equator and descending
in the winter hemisphere, and air in the other cell (the ``summer''
cell) flowing poleward toward the summer pole.  Generally, for
obliquities relevant to Earth, Mars, and Titan, the (cross-equatorial)
winter cell is much stronger than the summer cell \citep[see
  e.g.,][pp.~158-160, for the Earth case]{peixoto-oort-1992}.  The
zonal wind structure at solstice and equinox also differ significantly.
The zonal wind in the ascending branch is generally weak, and, the greater
its latitude, the lower its angular momentum per unit mass. If the
ascending branch lies at latitude $\phi_0$ and exhibits zero zonal
wind, and if the upper branch of the Hadley circulation conserves
angular momentum, the zonal wind in the upper branch is
\begin{equation}
u = \Omega a {\cos^2\phi_0 - \cos^2\phi\over \cos\phi}.
\label{uM-season}
\end{equation}
Notice that this equation implies strong westward (negative) wind
speeds near the equator if $\phi_0$ is displaced off the equator---a
result of the fact that, in the winter cell, air is moving away from
the rotation axis as it approaches the equator.  Just such a
phenomenon of strong westward equatorial wind under solsticial
conditions can be seen in both idealized, axisymmetric models of the
Hadley cell \citep[e.g.,][]{lindzen-hou-1988, caballero-etal-2008,
  mitchell-etal-2009} and full GCM simulations performed under
conditions of high obliquity \citep{williams-pollard-2003}.
Nevertheless, eddies can exert a considerable effect on the Hadley
cell, which will cause deviations from Eq.~(\ref{uM-season}).

The Hadley circulation can exhibit a variety of possible behaviors
depending on the extent to which the circulation in the upper branch
is angular-momentum conserving.  The different regimes can be
illuminated by considering the zonal-mean zonal wind equation, written
here for the 3D primitive equations in pressure coordinates on the sphere
\citep[cf][Section 6]{held-2000}
\begin{equation}
{\partial \overline{u}\over\partial t} = (f+\overline{\zeta})
\overline{v} - \overline{\omega}{\partial\overline{u}\over
\partial p} - {1\over a\cos^2\phi}{\partial(\cos^2\phi \overline{u'v'})
\over\partial\phi} - {\partial(\overline{u'\omega'})\over\partial p}
\label{eulerian-mom}
\end{equation}
where $\omega = dp/dt$ is the vertical velocity in pressure
coordinates, $d/dt$ is the advective derivative, and as before the
overbars and primes denote zonal averages and deviations therefrom.
In a statistical steady state, the lefthand side of
Equation~(\ref{eulerian-mom}) is zero.  Denoting the eddy terms (i.e.,
the sum of the last two terms on the righthand side) by $-S$, we obtain
\begin{equation}
(f+\overline{\zeta})\overline{v} = \overline{\omega}
{\partial\overline{u}\over\partial p} + S.
\label{hadley-zonal-balance1}
\end{equation}
For Earth, it turns out that the vertical advection by the mean flow
(first term on righthand side) does not play a crucial role, at least
for a discussion of the qualitative behavior, so that the zonal
momentum balance for the upper branch of the Hadley circulation can be
written \citep[e.g.,][]{held-2000, schneider-2006,
  walker-schneider-2006}
\begin{equation}
(f+\overline{\zeta})\overline{v} =  f(1 - Ro_H)\overline{v}\approx S
\label{hadley-zonal-balance2}
\end{equation}
where the Rossby number associated with the Hadley circulation is defined
as $Ro_H = -\overline{\zeta}/f$.

The Hadley circulation exhibits distinct behavior depending on whether
the Rossby number, $Ro_H$, is large or small.  Essentially, $Ro_H$ is
a non-dimensional measure of the effect of eddies on the Hadley cell
\citep[e.g.,][] {held-2000, schneider-2006, walker-schneider-2006,
  schneider-bordoni-2008}. The meridional width, amplitude,
temperature structure, and seasonal cycle of the Hadley circulation
depend on the relative roles of thermal forcing (in the form of
meridional heating gradients) and mechanical forcing (in the form of
$S$).  The Hadley circulation will respond differently to a given
change in thermal forcing depending on whether eddy forcing is
negligible or important.

 Theories currently exist for the $Ro_H\to 1$
and $Ro_H\to 0$ limits, although not yet for the more complex
intermediate case which seems to be representative of Earth and
perhaps planets generally.

When eddy-induced accelerations are negligible, then $S = 0$, and for
non-zero circulations the absolute vorticity must then be zero within
the upper branch, i.e., $f+\overline{\zeta} = 0$, or, in other words,
$Ro_H \to 1$.  The definitions of relative vorticity and angular
momentum imply that $f+\overline{\zeta} = (a^2\cos\phi)^{-1}\partial
\overline{m}/\partial\phi$, from which it follows that a circulation
with zero absolute vorticity exhibits angular momentum per mass that
is constant with latitude.  This is the angular-momentum conserving
limit mentioned previously and, for atmospheres experiencing a heating
maximum at the equator, would lead to a zonal-wind profile obeying
(\ref{uM}) in the upper branch.  Several authors have explored
theories of such circulations, both for annual-mean conditions and
including a seasonal cycle \citep[e.g.,][]{held-hou-1980,
  lindzen-hou-1988, fang-tung-1996, polvani-sobel-2002,
  caballero-etal-2008, adam-paldor-2009, adam-paldor-2010}.  Given the
angular-momentum conserving wind in the upper branch
(Equation~\ref{uM}), and assuming the near-surface wind is weak---a
result of surface friction---the mean vertical shear of the zonal wind
between the surface and upper troposphere is therefore known. The
tropospheric meridional temperature gradient can then be obtained from
the thermal-wind balance (\ref{thermal-wind}), or from generalizations
of it that include curvature terms on the sphere
\citep[cf][]{held-hou-1980}.  When the Hadley cell is confined to low
latitudes, this leads to a quartic dependence of the temperature on
latitude; for example, in the \citet{held-hou-1980} model, the
potential temperature at a mid-tropospheric level is
\begin{equation}
\theta = \theta_{\rm equator} - {\Omega^2\theta_0\over 2ga^2H} y^4
\label{temp-hh}
\end{equation}
where $H$ is the vertical thickness of the Hadley cell, $y$ is northward
distance from the equator, and $\theta_{\rm equator}$ is the 
potential temperature at the equator.   This dependence implies
that the temperature varies little with latitude across most of
the Hadley cell but plummets rapidly near the poleward edges of
the cell.  In this $Ro_H\to 1$ regime, the strength of the 
Hadley cell follows from the thermodynamic energy equation,
and in particular from the radiative heating gradients (with 
heating near the equator and cooling in the subtropics).  The Hadley
cell also has finite meridional extent that is determined by the
latitude at which the integrated cooling away from the equator
balances the integrated heating near the equator.
The Hadley circulation in this limit is thermally driven.

On the other hand, eddy-momentum accelerations $S$ are often important
and can play a defining role in shaping the Hadley cell properties. In
the limit $Ro_H\ll 1$, the strength of the Hadley circulation is
determined not by the thermal forcing (at least directly) but rather
by the eddy-momentum flux divergences: because the absolute vorticity
at $Ro_H\ll 1$ is approximately $f$, the meridional velocity in the
Hadley cell is given, via Equation~(\ref{hadley-zonal-balance2}), by
$\overline{v}\approx S/f$, at least under conditions where the
vertical momentum advection of the mean flow can be neglected.
Generally, in the Earth and Mars-like context, the primary eddy
effects on the Hadley cell result from the equatorward propagation of
Rossby waves generated in the baroclinincally unstable zone in the
midlatitudes (Section~\ref{jet-formation}).  The zonal phase
velocities of these waves, while generally westward relative to the
peak speeds of the eddy-driven jet, are eastward relative to the
ground.  These Rossby waves propagate into the subtropics (causing
a poleward flux of eddy angular momentum visible in the bottom
three panels of Figure~\ref{hadley-cell}), where they
reach critical levels on the flanks of the subtropical jets
\citep{randel-held-1991}.  The resulting wave absorption generally
causes a westward wave-induced acceleration of the zonal-mean zonal
flow, which removes angular momentum from the poleward-flowing air in
the upper branch of the Hadley circulation.  Therefore, the angular
momentum decreases poleward with distance away from the equator in the
upper branch of the Hadley cell.  As a result, although the zonal wind
still increases with latitude away from the equator, it does so more
gradually than predicted by Equation~(\ref{uM}); for example, at
$20^{\circ}$ latitude, the zonal-mean zonal wind speed in Earth's
upper troposphere is only $\sim$$20\rm\,m\,s^{-1}$, significantly
weaker than the $\sim$$60\rm\,m\,s^{-1}$ predicted by
Equation~(\ref{uM}).  In turn, the weaker vertical shear of the zonal
wind implies a weaker meridional temperature gradient through thermal
wind (Equation~\ref{thermal-wind}), leading to a meridional
temperature profile than can remain flatter over a wider range of
latitudes than predicted by Equation~(\ref{temp-hh}) (see
\citet{farrell-1990} for examples of this phenomenon in the context of
a simple, axisymmetric model).

Real planets probably lie between these two extremes.  On Earth,
observations and models indicate that eddy effects on the Hadley cell
are particularly important during spring and fall equinox and in the
summer hemisphere cell, whereas the winter hemisphere
(cross-equatorial) Hadley cell is closer to the angular momentum
conserving limit \citep[e.g.,][]{kim-lee-2001b, walker-schneider-2005,
  walker-schneider-2006, schneider-bordoni-2008,
  bordoni-schneider-2008, bordoni-schneider-2010}.  In particular, the
Rossby numbers $Ro_H$ vary from $\lesssim 0.3$--0.4 in the equinoctal
and summer cells to $\gtrsim 0.7$ in the winter cell.  Physically, the
mechanisms for the transition between these regimes involve
differences in an eddy-mean-flow feedback between equinoctal and
solsticial conditions \citep{schneider-bordoni-2008,
  bordoni-schneider-2008}. During the equinox, the ascending branch of
the Hadley cell lies near the equator, and the upper branch transports
air poleward---i.e., toward the rotation axis---into both hemispheres.
Angular momentum conservation therefore implies that the upper branch
of the Hadley cell exhibits eastward zonal flow (Eq.~\ref{uM}), allowing the
existence of critical layers and the resultant absorption of the
equatorward-propagating Rossby waves from midlatitudes.  Eddies
therefore play a crucial role ($S$ is large and $Ro_H$ is small).
During solstice, the ascending branch of the Hadley cell is displaced
off the equator into the summer hemisphere.  Air flows poleward in the
upper branch of the summer cell, again leading to eastward zonal flow,
the existence of critical layers, and significant eddy influences.  On
the other hand, air in the winter-hemisphere cell rises in the
subtropics of the summer hemisphere and flows across the equator to
the winter hemisphere, where it descends in the subtropics.  Because
this equatorward motion moves the air away from the rotation axis,
angular momentum conservation leads to a broad region of {\it
  westward} zonal winds across much of the winter cell
\citep[Eq.~\ref{uM-season};][]{lindzen-hou-1988}.  Because the
equatorward-propagating midlatitude eddies exhibit eastward zonal
phase speeds (relative to the ground), the winter-hemisphere cell
therefore lacks critical layers and is largely transparent to these
waves. (Indeed, most of them have already reached critical levels and
been absorbed before even reaching the latitude of the winter cell).
The result is a much smaller net absorption of eddies, and therefore a
winter hemisphere Hadley cell whose upper branch is much closer to the
angular-momentum conserving limit.


Other feedbacks between Hadley cells and eddies are possible,
particularly on tidally locked exoplanets where the day-night heating
pattern dominates.  In particular, tidally locked exoplanets exhibit a
strong day-night heating pattern that induces standing,
planetary-scale tropical waves, which can drive an eastward
(superrotating) jet at the equator (Section~\ref{superrotation}).
Such superrotation corresponds to a local maximum at the equator of
angular momentum per mass about the planetary rotation axis; by
definition, this means that the meridional circulation in a Hadley
cell, if any, must cross contours of constant angular momentum.  This
can only occur in the presence of eddy accelerations that alter the
angular momentum of the air in the upper branch as it flows
meridionally.  Therefore, Hadley cells in the presence of
superrotation jets tend to be far from the angular-momentum conserving
limit.  The term $\overline{\omega} {\partial\overline{u}/\partial p}$
in Equation~(\ref{hadley-zonal-balance1}) may be crucial on such
planets, unlike the typical case on Earth.

\citet{shell-held-2004} explored the interaction between superrotation
and the Hadley circulation in a zonally symmetric, 1-1/2 layer
shallow-water model where the effect of the tropical eddies was
parameterized as a specified eastward acceleration at the equator.
They found four classes of behavior depending on the strength of the
eddy forcing.  For sufficiently weak zonal eddy acceleration, the
model's Hadley circulation approached the angular-momentum conserving
limit.  For eastward eddy accelerations exceeding a critical value,
however, two stable equilibria emerged for a given eddy acceleration.
In one equilibrium, the Hadley cell is strong, and the massive
advection from below of air with minimal zonal wind inhibits the
ability of the eddy forcing to generate a strongly superrotating jet
at the equator.  As a result, angular momentum varies only modestly
with latitude near the equator, promoting the ability of the Hadley
cell to remain strong.  In the other equilibrium, the Hadley cell and
the associated vertical momentum advection are weak, so the eddy
acceleration is therefore able to produce a fast superrotating jet.
The corresponding thermal profile is close to radiative equilibrium,
maintaining the Hadley cell in its weak configuration.
\citet{shell-held-2004} showed that when the eddy forcing exceeds a
second critical threshold, only the latter solution exists,
corresponding to a weak Hadley cell and a strongly superrotating jet.
For eddy forcing exceeding a third critical threshold, the Hadley
circulation collapses completely, replaced by an eddy-driven
meridional circulation with {\it sinking} motion at the equator and
{\it ascending} motion off the equator.  It will be interesting to use
more sophisticated models to determine the extent to which these
various regimes apply to terrestrial exoplanets, and, if so, the
conditions for transitions between them.

\subsubsection{Wave adjustment}
\label{wave-adjustment}

As discussed earlier in this section, the slowly rotating regime generally
exhibits small horizontal temperature contrasts, which results from the
existence of dynamical mechanisms that efficiently regulate the
thermal structure.  The Hadley circulation, discussed previously,
is one such mechanism, and is particularly important in the latitudinal direction at large
scales.   Another key mechanism is adjustment of
the thermal structure by gravity waves.

How does such wave adjustment work?  
When moist convection, radiative heating
gradients, or other processes generate horizontal temperature
variations, the resulting pressure gradients lead to the radiation of
gravity waves.  The horizontal convergence/divergence induced by these
waves adjusts air columns up or down in such a way as to flatten the
isentropes and therefore erase the horizontal temperature constrasts.
This process, which is essentially the non-rotating endpoint of
geostrophic adjustment, plays a key role in minimizing the horizontal
temperature contrasts in the tropics; the resulting dynamical regime
is often called the ``weak temperature gradient'' or WTG regime
\citep{sobel-etal-2001, sobel-2002}.

\begin{figure*}
 \epsscale{1.0}
\plotone{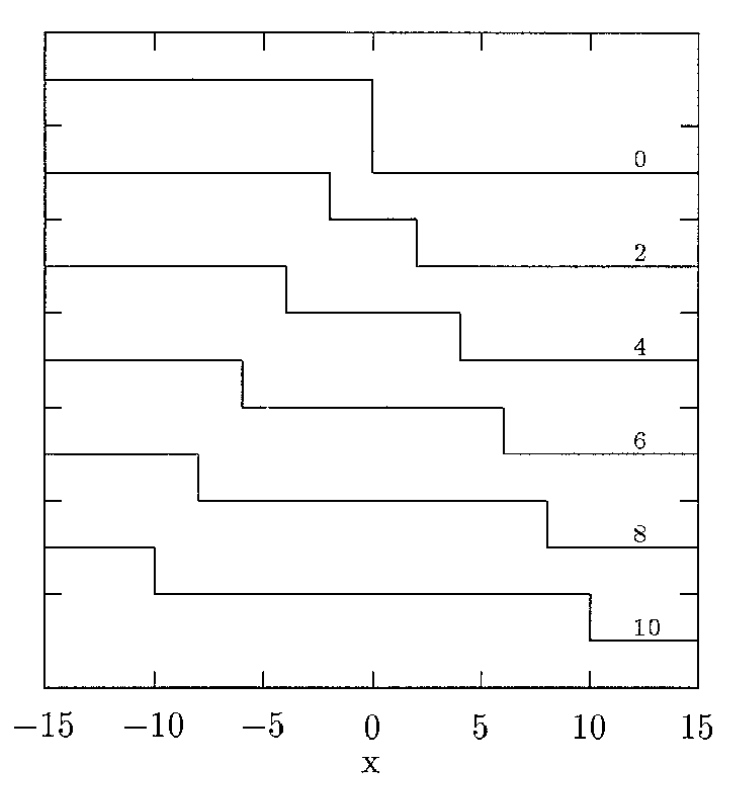}
 \caption{\small An analytic solution of the wave adjustment process
in a one-dimensional, nonrotating shallow-water fluid.   Each curve shows
the elevation of the water surface (with an arbitrary vertical offset for clarity)
versus horizontal distance at
a particular time.  Numbers labeling each curve give time.  The initial 
condition is shown at the top, and subsequent states are shown underneath.
Note how all the topography is ``carried away'' by the waves.
From \citet{kuo-polvani-1997}.}
\label{kuo-polvani}
 \end{figure*}

This adjustment process is most simply visualized in the context of a
one-layer fluid.  Imagine a shallow, nonrotating layer of water in one
spatial dimension, whose initial surface elevation is a step function.
This step causes a sharp horizontal pressure gradient force, and this
will lead to radiation of gravity waves (manifested here as surface
water waves) in either direction.  Figure~\ref{kuo-polvani} shows an
analytical solution of this process from \citet{kuo-polvani-1997}.
The horizontal convergence/divergence induced by the waves changes the
fluid thickness.
Assuming the waves can radiate to infinity (or break at some distant
location), the final state is a flat layer.  When the fractional
height variation is small, this adjustment process causes only a small
horizontal displacement of fluid parcels; the adjustment is done by
waves, not long-distance horizontal advection.


This wave-adjustment mechanism also acts to regulate the thermal
structure in three-dimensional, continuously stratified atmospheres
\citep[e.g.,][]{bretherton-smolarkiewicz-1989, sobel-2002}.  In a
stratified atmosphere, horizontal temperature differences are
associated with topographic variations of isentropes, which play a
role directly analogous to the topography of the water surface in the
example described above.  Gravity waves induce horizontal
convergence/divergence, which changes the vertical thickness of the
fluid columns, thereby pushing the isentropes up or down.
Assuming that planetary rotation is sufficiently weak, and that the
waves can radiate to infinity, the final state is one with flat
isentropes.   Note that, if the initial fractional temperature contrast
is small, this adjustment causes only a small lateral motion of fluid
parcels.  

\begin{figure*}
\epsscale{0.8}\plotone{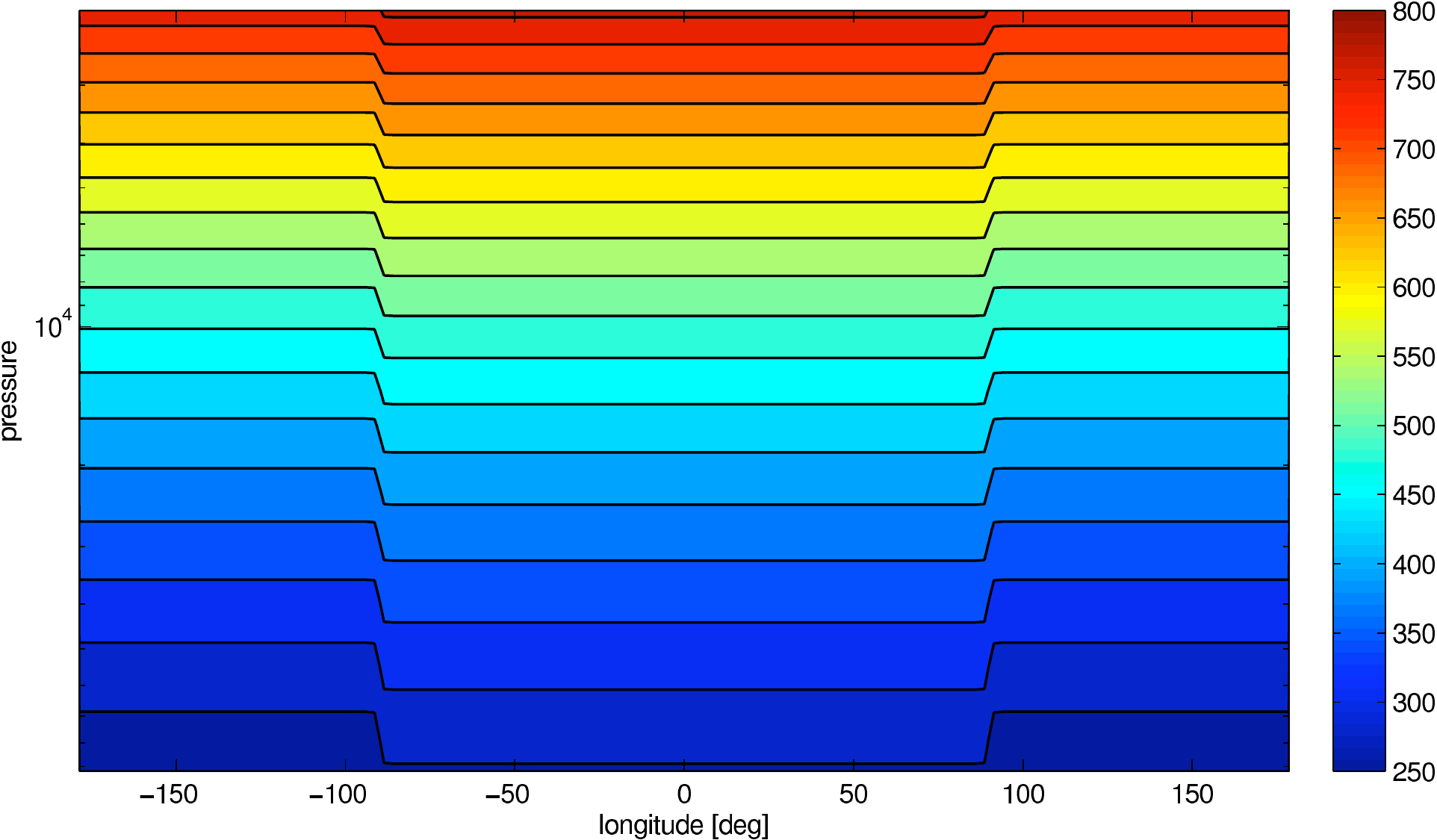}
\epsscale{0.6}\plotone{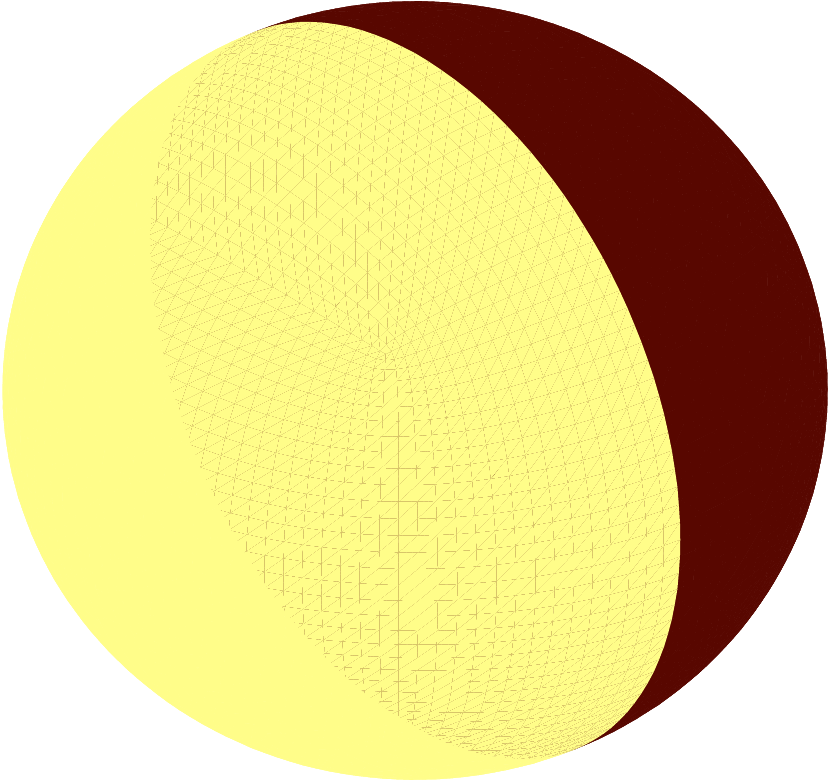}\\
\epsscale{0.8}\plotone{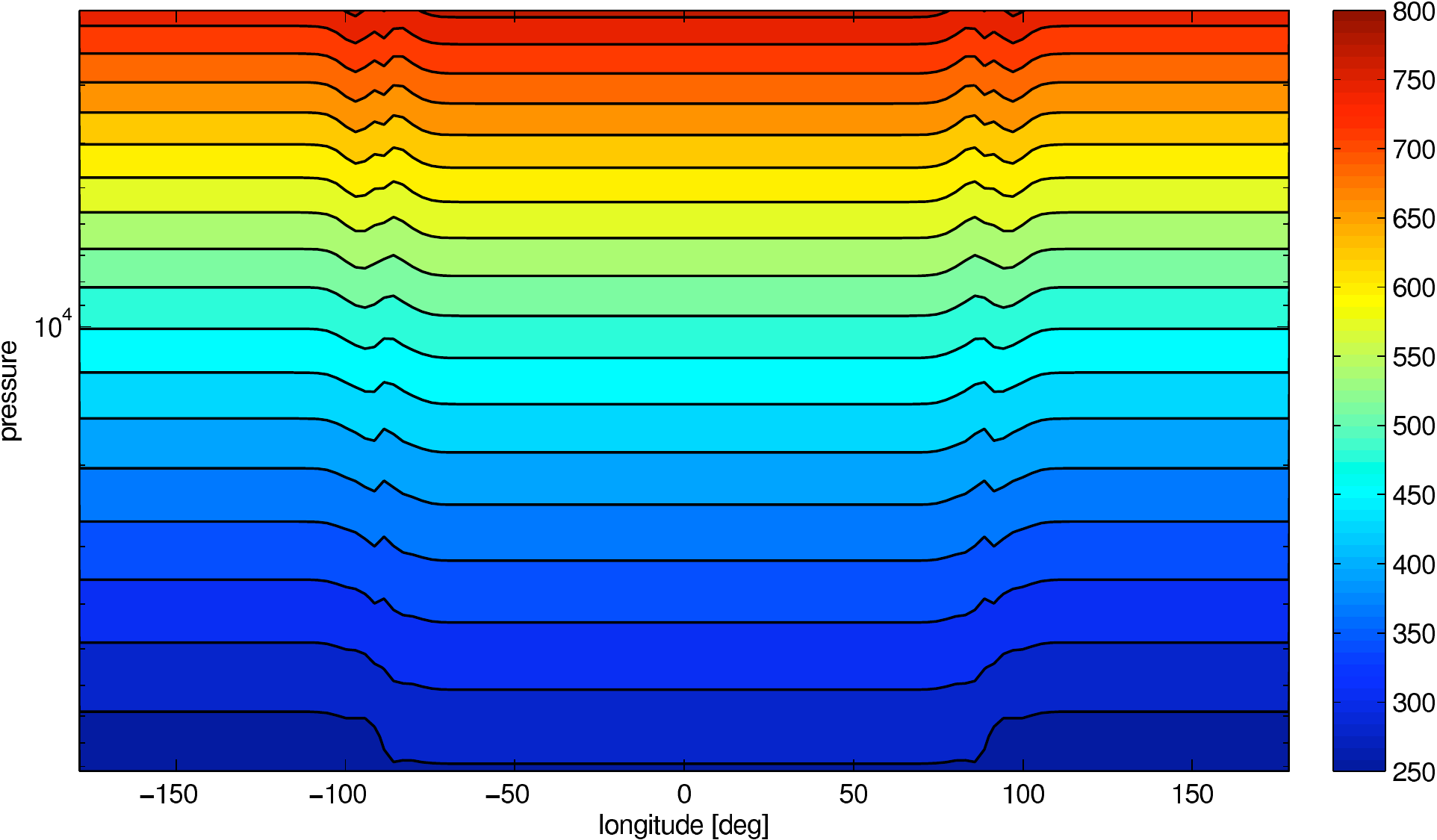}
\epsscale{0.6}\plotone{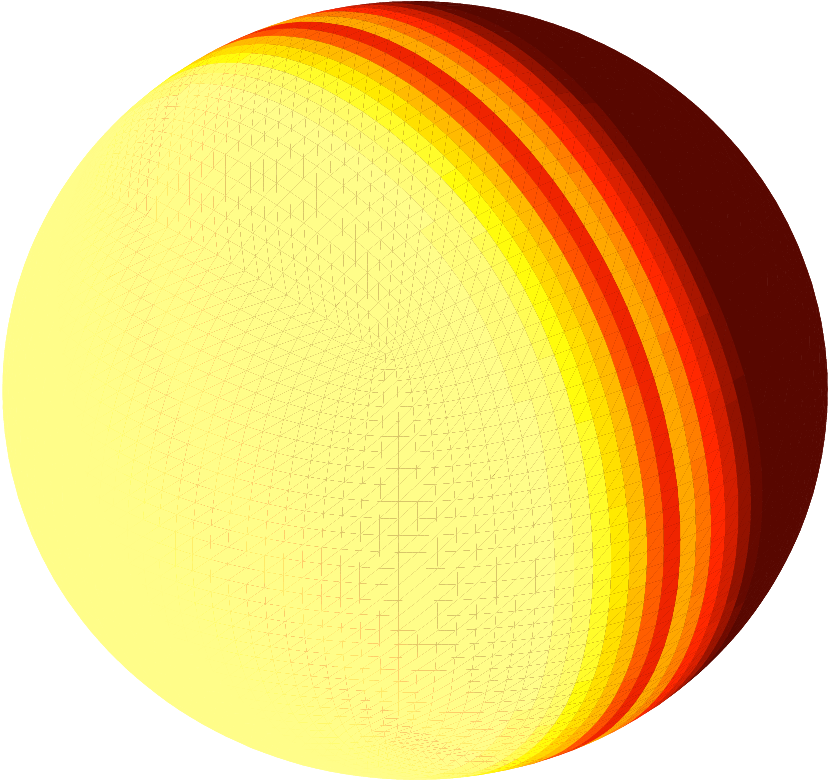}\\
\epsscale{0.8}\plotone{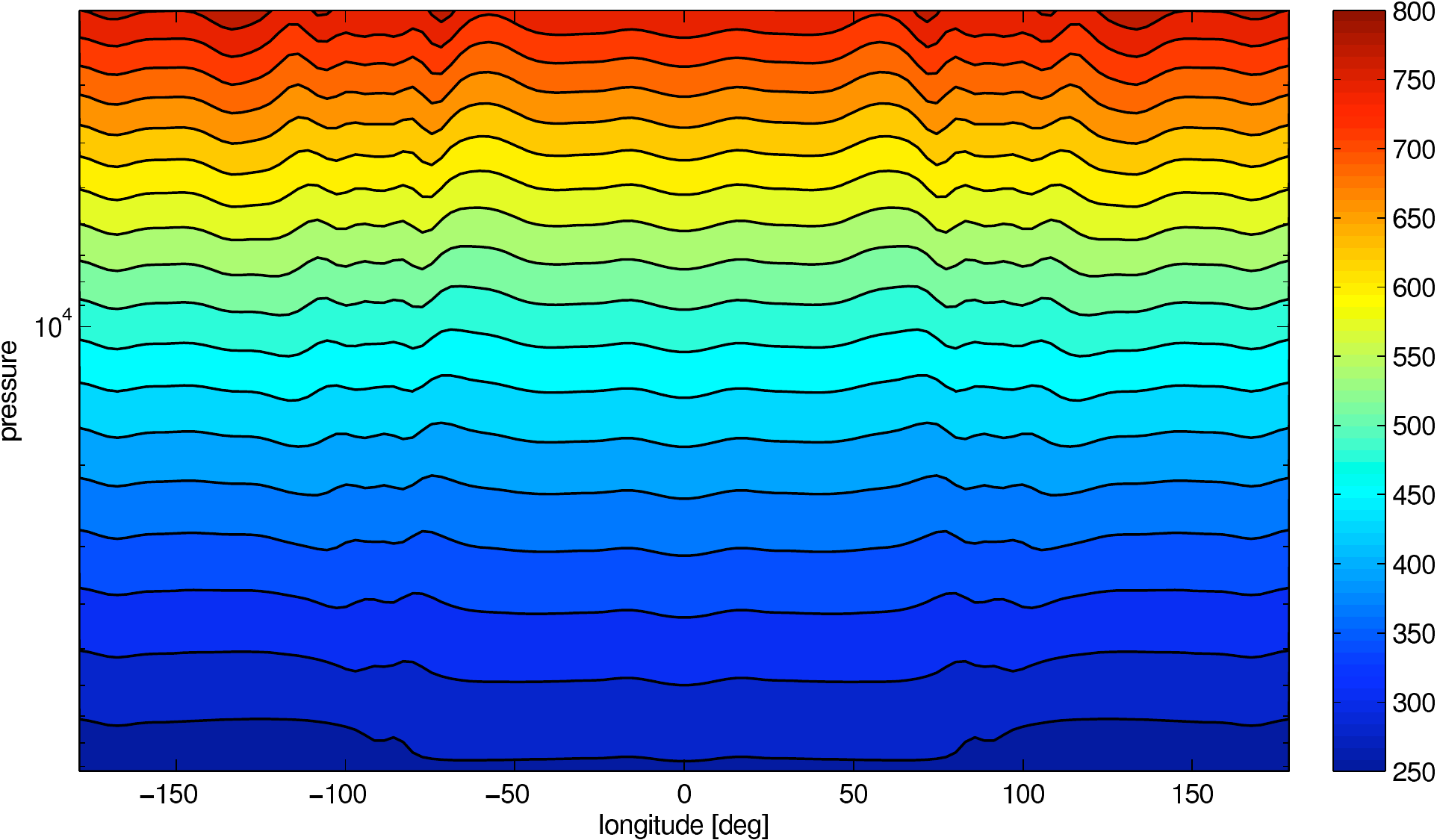}
\epsscale{0.6}\plotone{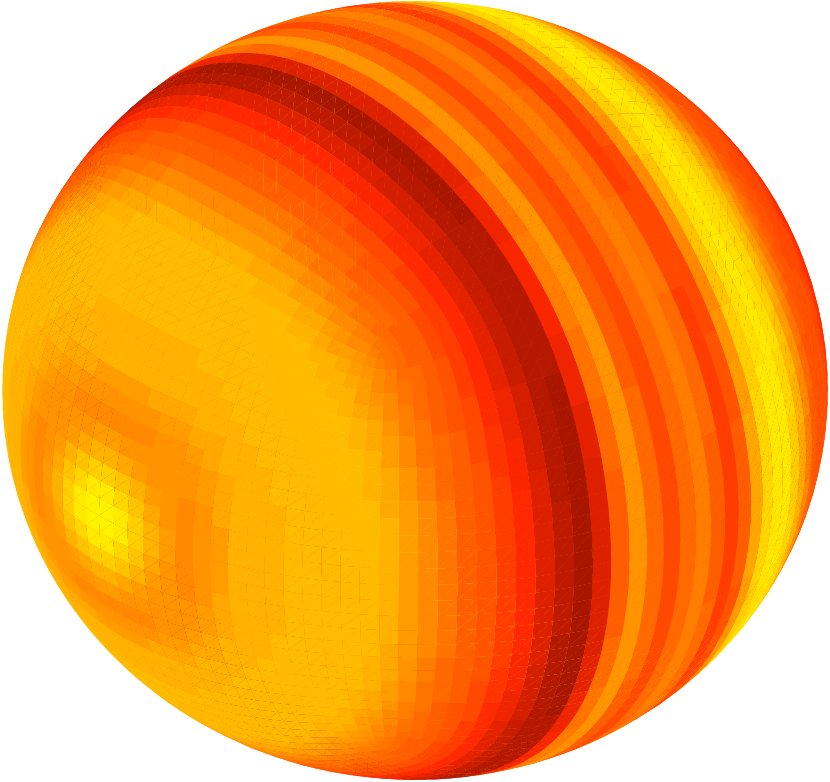}\\
\epsscale{0.8}\plotone{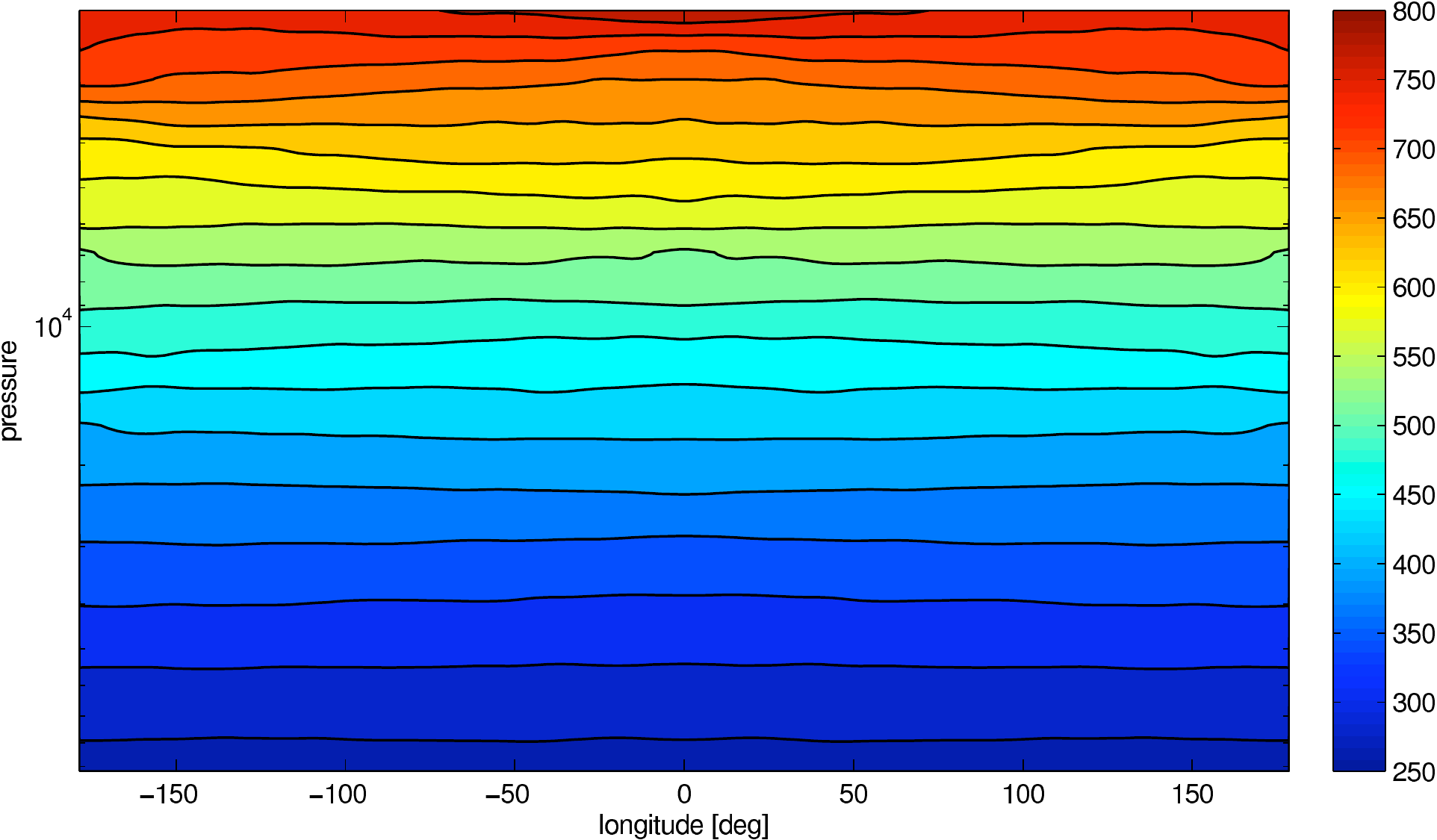}
\epsscale{0.6}\plotone{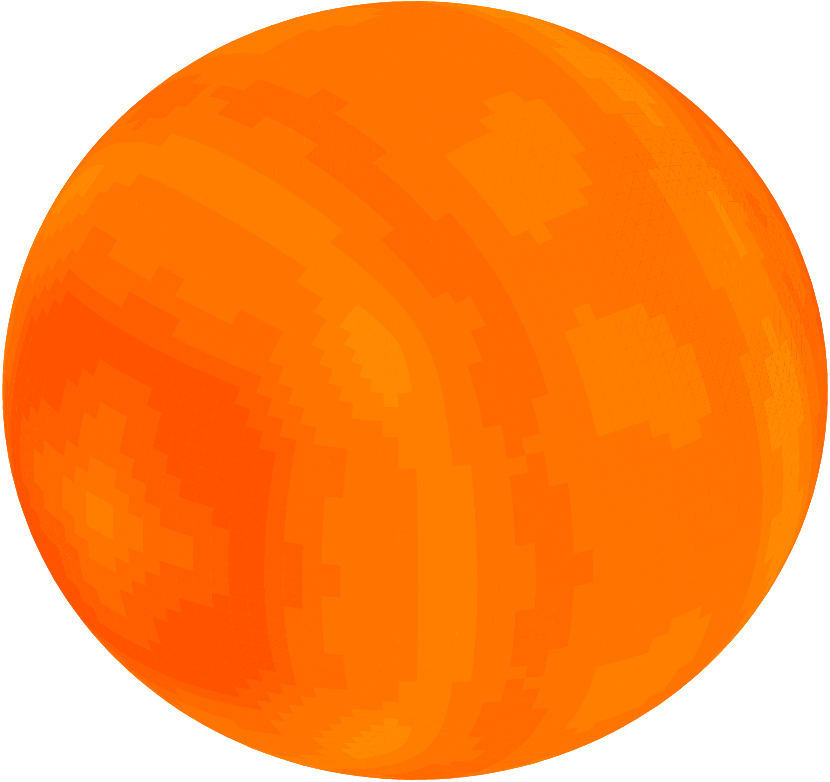}

 \caption{Numerical solution of wave adjustment on a spherical,
   non-rotating terrestrial planet with the radius and gravity of
   Earth.  We solved the global, three-dimensional primitive
   equations, in pressure coordinates, using the MITgcm.  Half of the
   planet (the ``nightside'') was initialized with a constant
   (isothermal) temperature of $T_{\rm night}=250\rm\,K$,
   corresponding to a potential temperature profile $\theta_{\rm
     night}=T_{\rm night}(p_0/p)^{\kappa}$, where $\kappa=R/c_p = 2/7$
   and $p_0=1\rm\,bar$ is a reference pressure.  The other half of the
   planet (the ``dayside'') was initialized with a potential temperature
   profile $\theta_{\rm night}(p)+\Delta\theta$, where
   $\Delta\theta=20\rm\,K$ is a constant. Domain extends from
   approximately 1 bar at the bottom to 0.001 bar at the top;
   equations were solved on a cubed-sphere grid with horizontal
   resolution of C32 ($32\times32$ cells per cube face, corresponding
   to an approximate resolution of $2.8^{\circ}$) and 40 levels in the
   vertical, evenly spaced in log-$p$.  The model includes a sponge at
   pressures less than 0.01 bar to absorb upward-propagating waves.  This is
   an initial value problem; there is no radiative heating/cooling so that 
   the flow is adiabatic. {\bf Left:} Potential temperature
(colorscale and contours) at the equator versus longitude and pressure;
  {\bf Right:} Temperature at a pressure of 0.2 bar over the globe
  at times of 0 (showing the initial condition), $0.5\times10^4\rm\,s$,
$3\times10^4\rm\,s$, and the final long-term state once the waves
have propagated into the upper atmosphere.  Air parcels move by only
a small fraction of a planetary radius during the adjustment process,
but the final state nevertheless corresponds to nearly flat isentropes
with small horizontal temperature variations on isobars.}
\label{wave-adjust-3d}
 \end{figure*}

Figure~\ref{wave-adjust-3d} demonstrates this process explicitly in
a three-dimensional, continously stratified atmosphere.
There we show a solution of the global, three-dimensional primitive
equations for an initial-value problem in which half of the planet
(the ``dayside'') is initialized to have a potential temperature that
is $20\K$ hotter than the other half (the ``nightside'').  The
heating/cooling is zero so that the flow is adiabatic and isentropes
are material surfaces.  The top row shows the initial condition, and
subsequent rows show the state at later times during the evolution.
At early times ($0.5\times10^4\rm\,s$, second row), waves begin to
radiate away from the day-night boundary; they propagate across most
of the planet by $\sim$$2\times10^4\rm\,s$ (third row).  They also propagate
upward where they are damped in the upper atmosphere, leaving the
long-term state well-adjusted (fourth row).  Note that, in the
final state, the isentropes are approximately flat and the day-night
temperature differences are greatly reduced over their initial
values.


The timescales for this wave-adjustment process can differ
significantly from the relevant advection and mixing timescales.
Because the wave-adjustment mechanism requires the propagation of
gravity waves, the timescale to adjust the temperatures over some
distance $L$ is essentially the gravity wave propagation time over
distance $L$, i.e., $\tau_{\rm adjust}\sim L/NH$, where $N$ is the
Brunt-Vaisala frequency and $H$ is a scale height.  For example,
\citet{bretherton-smolarkiewicz-1989} show how gravity waves adjust
the thermal structure in the environment surrounding tropical cumulus
convection; they demonstrate that the process operates on a timescale
much shorter than the lateral mixing timescale.  Moreover, if the wave
propagation speeds differ significantly from the wind speeds, then
$\tau_{\rm adjust}$ can differ significantly from the horizontal
advection time, $\tau_{\rm adv}\sim L/U$.  In the example shown
in Fig.~\ref{wave-adjust-3d}, for example, the wave speeds are
$c\sim NH\sim 200\rm\,m\,s^{-1}$ (using $N\approx0.02\rm\,s^{-1}$
and $H\approx10\rm\,km$), but the peak tropospheric wind speeds in
this simulation are
almost ten times smaller.   This implies that the timescale
for wave propagation is nearly ten times shorter than the timescale
for air to advect over a given distance.

Understanding the conditions under which the WTG regime breaks down is
critical for understanding the atmospheric stability, climate, and
habitability of exoplanets.  Synchronously rotating planets exhibit
permanent day and night sides, and for such atmospheres to remain
stable against collapse, they must remain sufficiently warm on the
nightside \citep{joshi-etal-1997}.  Crudely, one might expect that
dynamics fails to erase the day-night temperature contrast when the
radiative timescale becomes shorter than the relevant dynamical
timescale.  Most previous exoplanet literature has assumed that the
relevant comparison is between the radiative and advective timescales
\citep[e.g.,][]{showman-etal-2010, cowan-agol-2011}.  However, when
the wave-adjustment time is short, a comparison between the radiative
and wave-adjustment timescales may be more appropriate.  For typical
terrestrial-planet parameters ($N\approx 10^{-2}\rm \,s^{-1}$,
$H=10\rm\,km$ and taking $L$ to be a typical terrestrial-planet radius
of $6000\rm\,km$) yields $\tau_{\rm adjust}\sim 10^5\,$sec.  This
would suggest that on planets with radiative time constants $\lesssim
10^5\,$sec, the waves are damped, the WTG regime breaks down, and
large day-night temperature differences may occur.  Earth, Venus, and
Titan are safely out of danger, but Mars is transitional, and any
exoplanet whose atmosphere is particularly thin and/or hot is also at
risk.  Nevertheless, subtleties exist.  For example, there exists a
wide range of wave-adjustment timescales associated with waves of
differing wavelengths and phase speeds; moreover, the timescale for waves to propagate
vertically out of the troposphere is not generally equivalent to the
timescale for them to propagate horizontally across a hemisphere.
Although the wave-adjustment process is fundamentally a linear one,
nonlinearities may become important at high amplitude as the WTG
regime breaks down and the fractional day-night temperature difference
becomes large. The horizontal advection time may play a key role
in this case.  Further work is warranted on the precise conditions for
WTG breakdown and the extent to which they can be packaged as
timescale comparisons.

For simplicity, we have so far framed the discussion around
non-rotating planets; to what extent does the mechanism carry over to
rotating planets?  Because horizontal Coriolis forces are zero at the
equator, the picture has broad relevance for tropical meteorology even
on rapidly rotating planets.  More specifically, planetary rotation
tends to trap tropical wave modes into an equatorial waveguide, whose
meridional width is approximately the equatorial Rossby deformation
radius, $L_{\rm eq}=(c/\beta)^{1/2}$, where $c$ is a typical gravity
wave speed and $\beta=df/dy$ is the derivative of the Coriolis
parameter with northward distance; this tends to yield a
characteristic waveguide width of order $(NH/\beta)^{1/2}$ in a
continuously stratified atmosphere (see \citet{matsuno-1966},
\citet[][pp.~394-400, 429-432]{holton-2004}, or
\citet[][pp.~200-208]{andrews-etal-1987} for a discussion of
equatorial wave trapping).  Typical values are $L_{\rm
  eq}\sim$$10^3\rm\,km$ for Earth and Mars.  These equatorially
trapped modes, including the Kelvin wave, Rossby waves, and mixed
Rossby-gravity waves---as well as smaller-scale gravity waves
triggered by convection and other processes---can adjust the thermal
state in a manner analogous to that described here.  Because
large-scale, equatorially trapped waves can propagate in longitude and
height but not latitude, the adjustment process will occur more
efficiently in the zonal than the meridional direction, and it will
tend to be confined to within an equatorial deformation radius of the
equator.  Indeed, the mechanism described here helps to explain why
the Earth's tropospheric tropical temperatures are nearly zonally
uniform, and moreover shows how moist convection can regulate the
thermal structure over wide areas of the tropics despite its sporadic
occurrence \citep{bretherton-smolarkiewicz-1989}.  On slowly rotating
planets---including Venus, Titan, and tidally locked super Earths,
where the deformation radius is comparable to or greater than the
planetary radius---this wave-adjustment processes will not be confined
to low latitudes but will act globally to mute horizontal temperature
contrasts.  In large measure, this mechanism is responsible for the
small horizontal temperature contrasts observed in the tropospheres of
Venus and Titan. 

 Note that the WTG regime (when it occurs) applies
best in the free troposphere, where friction is weak and waves are
free to propagate; in the frictional boundary layer near a planet's
surface, the scaling will not generally hold and there may exist
substantial horizontal temperature gradients. For example,
\citet{joshi-etal-1997} and \citet{merlis-schneider-2010} present
terrestrial exoplanet simulations exhibiting weak
temperature gradients in the free troposphere (day-night contrasts
$\lesssim 3\rm\,K$) but larger day-night temperature contrasts at the
surface (reaching $\sim$$30$--$50\rm\,K$).

\subsubsection{Equatorial superrotation}
\label{superrotation}

Recent theoretical work suggests that many tidally locked terrestrial
exoplanets will exhibit a fast eastward, or {\it superrotating}, jet
stream at the equator.  More specifically, superrotation is defined as
atmospheric flow whose angular momentum (per unit mass) about the
planet's rotation axis exceeds that of the planetary surface at the
equator.\footnote{According to this definition, most eastward jets at
  mid- to high latitudes are {\it not} superrotating.}
In our solar system, the tropical atmospheres
of Venus, Titan, Jupiter, and Saturn all superrotate.  Even localized
layers within Earth's equatorial stratosphere exhibit superrotation,
part of the so-called ``Quasi-Biennial Oscillation''
\citep{andrews-etal-1987}.  In contrast, Uranus and Neptune, as well
as the tropospheres of Earth and Mars, exhibit mean westward
equatorial flow (subrotation).  Interestingly, three-dimensional
circulation models of synchronously rotating exoplanets, which are
subject to a steady, day-night heating pattern, have consistently
showed the emergence of such superrotation---both for terrestrial
planets \citep{joshi-etal-1997, merlis-schneider-2010, heng-vogt-2011,
edson-etal-2011, wordsworth-etal-2011} and hot Jupiters
\citep{showman-guillot-2002, cooper-showman-2005, showman-etal-2008a,
showman-etal-2009, showman-etal-2013, menou-rauscher-2009, rauscher-menou-2010,
heng-etal-2011, heng-etal-2011b, perna-etal-2010, perna-etal-2012}.
Figure~\ref{superrotate} shows examples from several recent studies.

\begin{figure*}
 \epsscale{1.0}
\begin{minipage}[c]{0.5\textwidth}
\center{\Large Synchronously locked}
\epsscale{0.99}\plotone{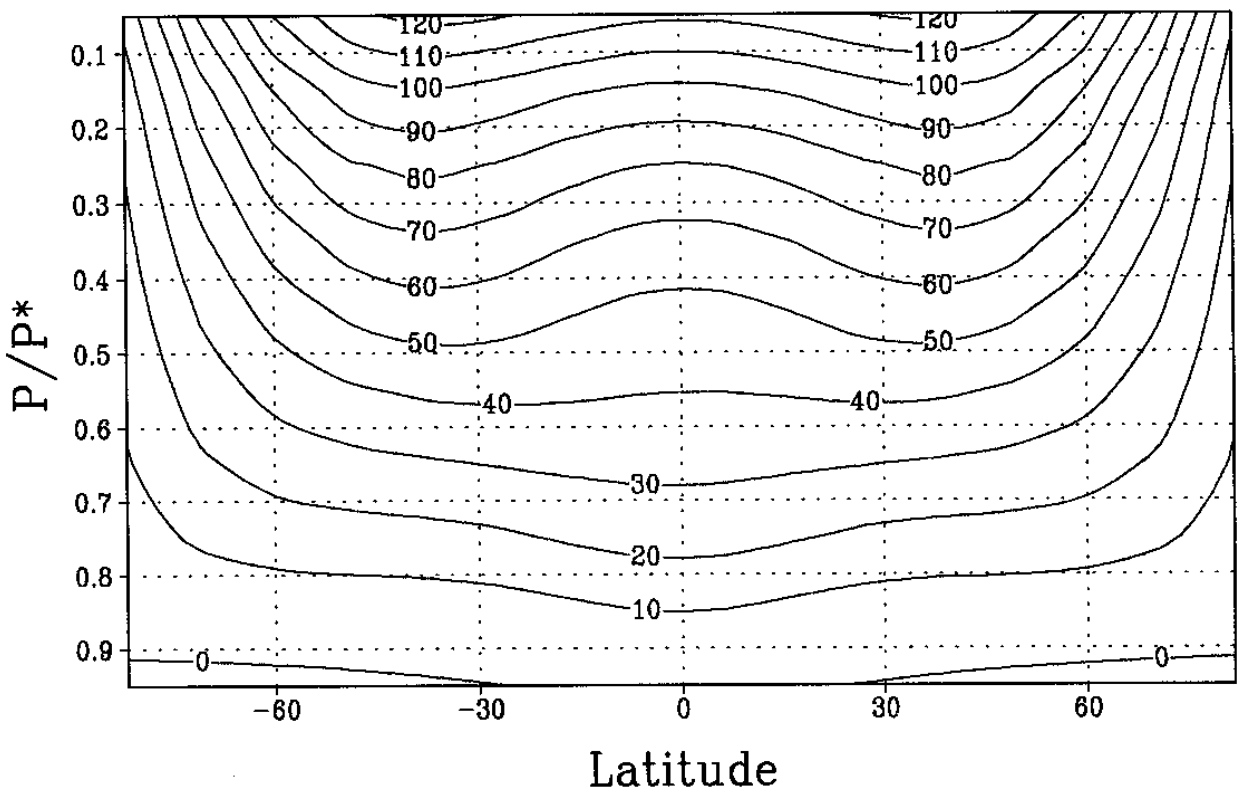}
\epsscale{0.98}\plotone{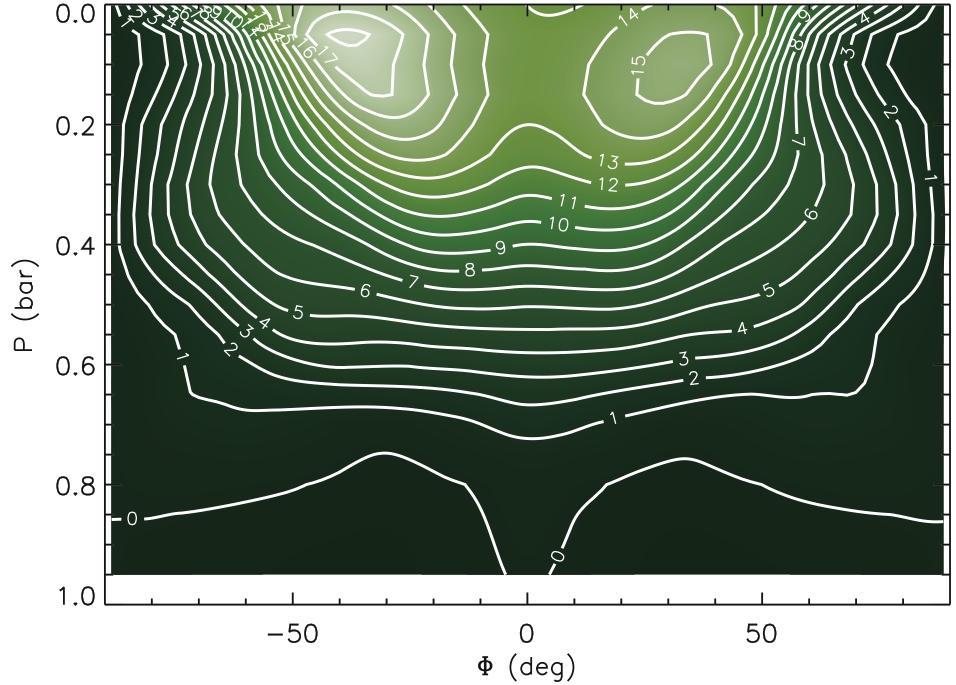}
\epsscale{0.88}\plotone{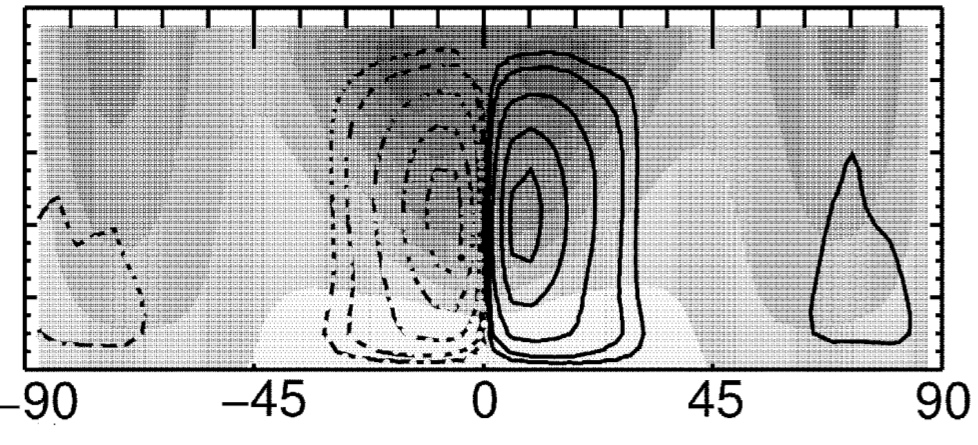}
\end{minipage}
\begin{minipage}[c]{0.5\textwidth}
\center{\Large Axisymmetric forcing}
\epsscale{1.0}\plotone{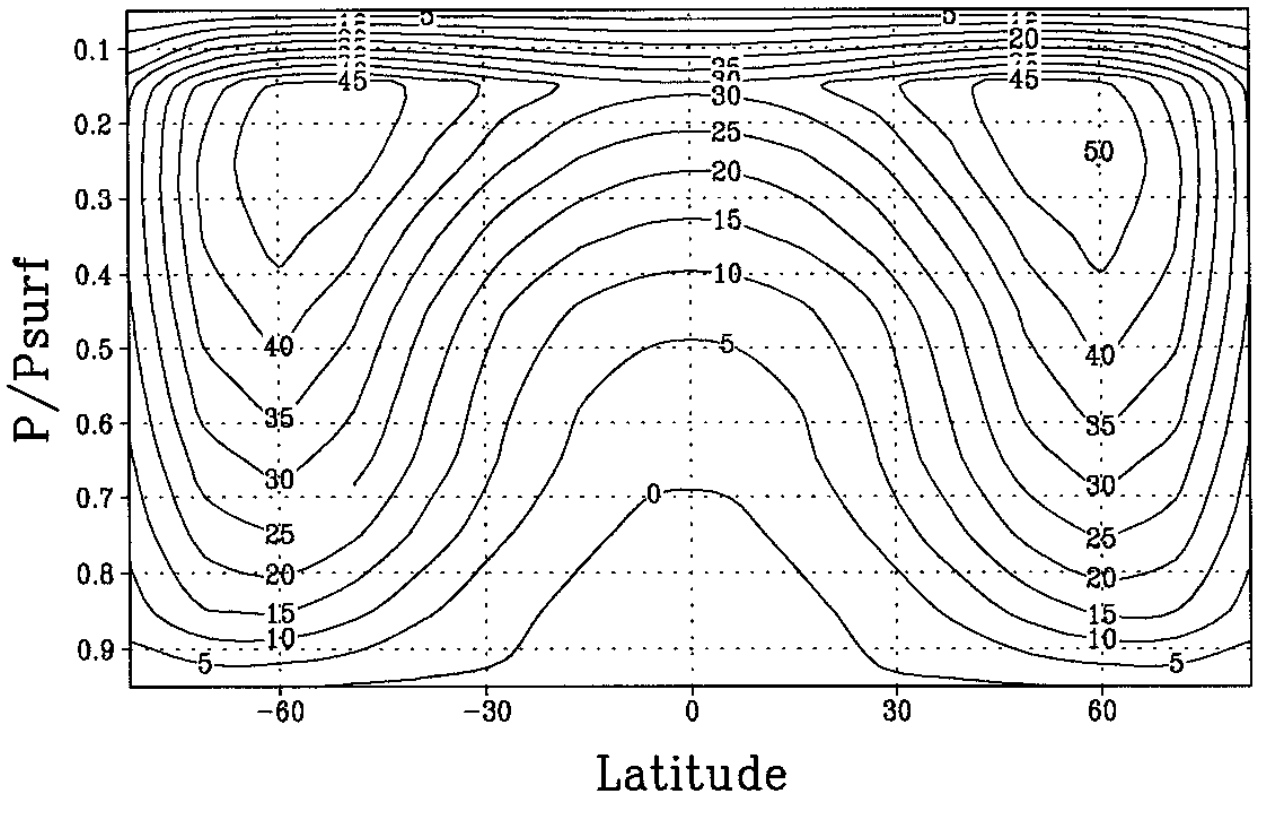}
\epsscale{1.0}\plotone{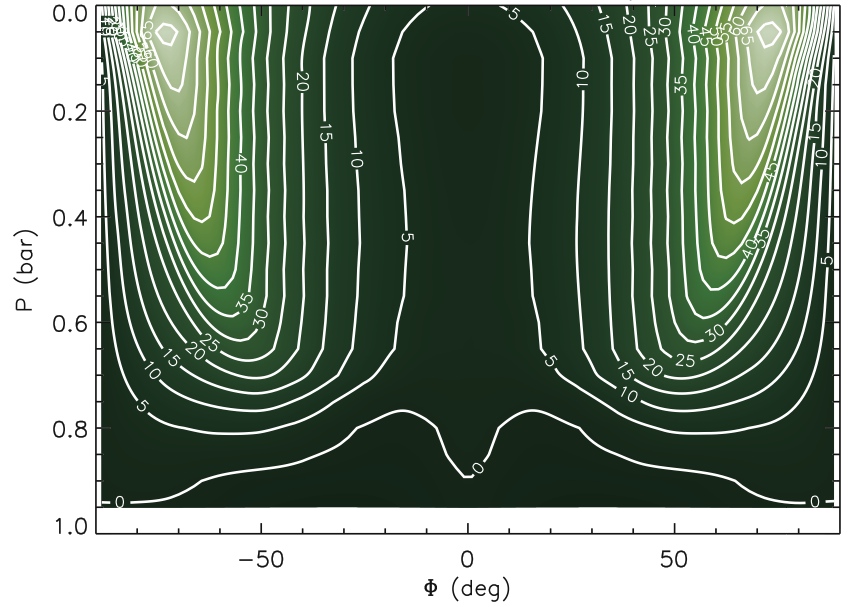}
\epsscale{0.85}\plotone{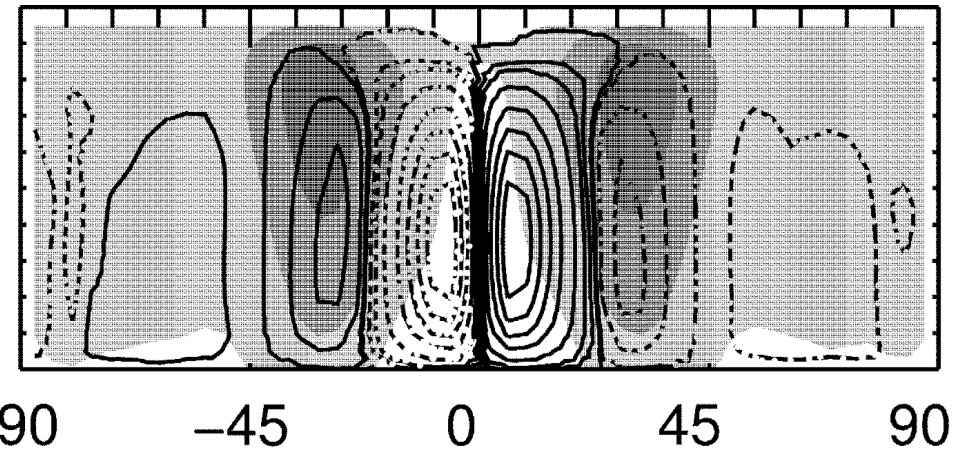}
\end{minipage}
 \caption{Zonal-mean zonal winds (contours) versus latitude (abscissa)
and pressure (ordinate) from recent studies illustrating the development
of equatorial superrotation in models of synchronously rotating
terrestrial exoplanets.  In each case, the left column shows a synchronously
rotating model with a steady, day-night heating pattern, and the right
model shows an otherwise similar control experiment with axisymmetric
heating (i.e., no day-night pattern).  {\it Top:} From \citet{joshi-etal-1997};
rotation period is 16 Earth days.  Contours give zonal-mean zonal wind
in $\rm m\,s^{-1}$.  {\it Middle:} \citet{heng-vogt-2011}; 
rotation period is 37 days (left) and 20 days (right).  Contours give
zonal-mean zonal wind in $\rm m\,s^{-1}$.  {\it Bottom:} 
\citet{edson-etal-2011}; rotation period is 1 Earth day.  Greyscale
gives zonal-mean zonal wind; eastward is shaded, with peak values
(in dark shades) reaching $\sim$$30\rm\,m\,s^{-1}$.  Contours give
mean-meridional streamfunction.  In all the models, the synchronously
rotating variants exhibit a strong, broad eastward jet centered at the equator, whereas
the axisymmetrically forced variants exhibit weaker eastward or westward flow
at the equator, with eastward jets peaking in the mid-to-high latitudes.}
\label{superrotate}
 \end{figure*}

Equatorial superrotation is interesting for several reasons.  First,
it influences the atmospheric thermal structure and thus plays an
important role in shaping observables.  When radiative and advective
timescales are similar, the superrotation can cause an eastward
displacement of the thermal field that influences infrared light
curves and spectra \citep[e.g.,][]{showman-guillot-2002}.  An eastward
displacement of the hottest regions from the substellar point has been
observed in light curves of the hot Jupiter HD 189733b
\citep{knutson-etal-2007b} and it may also be detectable for
synchronously rotating super Earths with next-generation observatories
\citep{selsis-etal-2011}.  The thermal structure of the leading and
trailing terminators may also differ.  Second, superrotation is
dynamically interesting; understanding the mechanisms that drive
superrotation in the exoplanet context may inform our understanding of
superrotation within the solar system (and vice versa).  The equator
is the region farthest from the planet's rotation axis; therefore, a
superrotating jet corresponds to a local maximum of angular momentum
per mass about the planet's rotation axis.  Maintaining a
superrotating equatorial jet against friction or other processes
therefore requires angular momentum to be transported up-gradient from
regions where it is low (outside the jet) to regions where it is high
(inside the jet).  \citet{hide-1969} showed that the necessary
up-gradient angular-momentum transport must be accomplished by waves
or eddies.

In models of synchronously rotating exoplanets, the defining feature
that allows emergence of strong superrotation is the steady day-night
heating contrast.  Models that include strong dayside heating and
nightside cooling---fixed in longitude due to the synchronous
rotation---generally exhibit a broad, fast eastward jet centered at
the equator.  In contrast, otherwise similar models with axisymmetric
forcing (i.e., an equator-to-pole heating gradient with no diurnal
cycle) exhibit only weak eastward or westward winds at the equator,
often accompanied by fast eastward jets in the mid-to-high latitudes
(Figure~\ref{superrotate}).  

Although the Earth's troposphere is not
superrotating, it does exhibit tropical zonal heating anomalies due to
longitudinal variations in the surface type (land versus ocean),
sea-surface temperature, and prevalence of cumulus convection near the
equator \citep{schumacher-etal-2004, kraucunas-hartmann-2005,
  norton-2006}.  Qualitatively, these tropical heating/cooling
anomalies resemble the day-night heating contrast on a synchronously
rotating exoplanet, albeit at higher zonal wavenumber and lower
amplitude.  Starting in the 1990s, several authors in the terrestrial
literature demonstrated using GCMs that, if sufficiently
strong, these types of tropical heating anomalies can drive equatorial
superrotation \citep{suarez-duffy-1992, saravanan-1993,
  hoskins-etal-1999, kraucunas-hartmann-2005, norton-2006}.  The
qualitative similarity in the forcing and the response suggests that
the mechanism for superrotation is the same in both the terrestrial
and the exoplanet models.

What is the mechanism for the equatorial superrotation occurring in
these models?  Motivated by arguments analogous to those in
Section~\ref{jet-formation}, \citet{held-1999b} and
\citet{hoskins-etal-1999} suggested heuristically that the
superrotation results from the poleward propagation of Rossby waves
generated at low latitudes by the tropical zonal heating anomalies;
subsequent authors have likewise invoked this conceptual framework in
qualitative discussions of the topic
\citep[e.g.][]{tziperman-farrell-2009, edson-etal-2011,
  arnold-etal-2012}.  At its essence, the hypothesis is attractive,
since it is simple, based on the natural relationship of the
meridional propagation directions of Rossby waves to the resulting
eddy velocity phase tilts, and seemingly links extratropical and
tropical dynamics.  Nevertheless, the idea remains qualitative, and
its relevance to the type of equatorial superrotation seen in 
exoplanet GCMs has not been demonstrated.

Moreover, challenges exist.  First, unlike in the extratropics,
large-scale baroclinic equatorial wave modes in the tropics are
equatorially trapped, confined to a wave guide whose meridional width
is approximately the equatorial Rossby deformation radius.  Such
waves---including the Kelvin waves, equatorial Rossby waves, and the
mixed Rossby-gravity wave---can propagate in longitude and height but
not latitude.  
Unlike the case of the barotropic Rossby waves discussed in
Section~\ref{rossby}, simple analytic solutions for such waves
\citep[see, e.g.,][]{matsuno-1966, holton-2004, andrews-etal-1987}
exhibit no meridional momentum flux.  Second, in the context of slowly
rotating exoplanets, these waves exhibit meridional scales typically
extending from the equator to the pole
\citep[e.g.,][]{mitchell-vallis-2010}, leaving little room for
meridional propagation.  Given these issues, it is unclear that the
paradigm of waves propagating from one latitude to another applies.

It can be shown explicitly that the barotropic theory for
the interaction of meridionally propagating Rossby waves with the mean
flow---which explains the emergence of eddy-driven jets in the
extratropics (Section~\ref{jet-formation})---fails to explain the
equatorial superrotation emerging from exoplanet GCMs
\citep{showman-polvani-2011}. Many GCMs of synchronously rotating
exoplanets exhibit circulation patterns---including equatorial
superrotation---that, to zeroth order, are mirror symmetric about the
equator \citep[e.g.,][]{showman-guillot-2002, cooper-showman-2005,
  showman-etal-2008a, showman-etal-2009, showman-etal-2013,
  heng-etal-2011, heng-etal-2011b, showman-polvani-2010,
  showman-polvani-2011, rauscher-menou-2010, rauscher-menou-2012,
  perna-etal-2010, perna-etal-2012}.  For a flow with such symmetry,
the relative vorticity is antisymmetric about, and zero at, the
equator---at all longitudes, not just in the zonal mean. Under
these conditions, the forcing $\overline{\zeta'F'}$ equals zero
in Equations~(\ref{pseudomomentum}) and (\ref{u-barotropic}).
Equation~(\ref{u-barotropic}) therefore predicts that the zonal-mean
zonal wind is zero at the equator---inconsistent with the equatorial
superrotation emerging in the GCMs.  

\begin{figure*}
 \epsscale{1.0}
\plotone{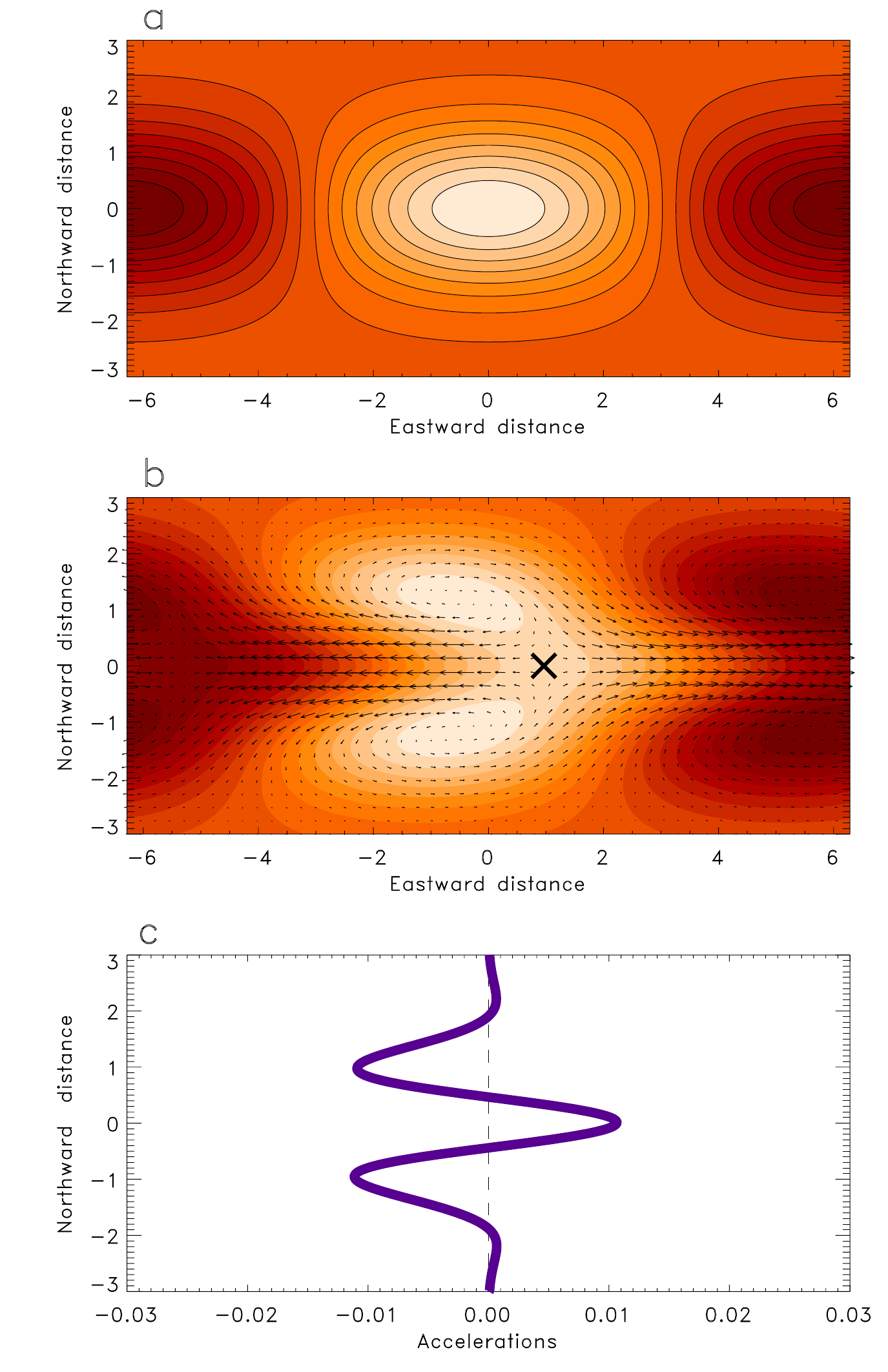}
 \caption{Analytical solutions of the shallow-water equations from
   \citet{showman-polvani-2011} showing how day-night thermal forcing
   on a synchronously rotating exoplanet can induce equatorial
   superrotation.  (a) Spatial structure of the radiative-equilibrium
   height field versus longitude (abscissa) and latitude (ordinate).
   (b) Height field (colors) and eddy velocities (arrows) for steady,
   analytic solutions forced by radiative relaxation and linear drag,
   performed on the equatorial $\beta$ plane and analogous to the
   solutions of \citet{matsuno-1966} and \citet{gill-1980}.  The
   equatorial behavior is dominated by a standing, equatorially
   trapped Kelvin wave and the mid-to-high latitude behavior is
   dominated by an equatorially trapped Rossby wave.  The
   superposition of these two wave modes leads to eddy velocities
   tilting northwest-southeast (southwest-northeast) in the northern
   (southern) hemisphere, as necessary to converge eddy momentum onto
   the equator and generate equatorial superrotation.  (c) The zonal
   accelerations of the zonal-mean wind implied by the linear
   solution.  The acceleration is eastward at the equator, implying
   that equatorial superrotation will emerge.  Note that the
   equatorward momentum flux that induces superrotation results from
   differential {\it zonal}, rather than {\it meridional}, propagation
   of the wave modes; indeed, these modes are equatorially trapped and
   exhibit no meridional propagation at all.}
\label{showman-polvani}
 \end{figure*}

\citet{showman-polvani-2011} offered an alternate theory to overcome
these obstacles and show how the equatorial superrotation can emerge
from the day-night thermal forcing on synchronously rotating
exoplanets.  They presented an analytic theory using the 1-1/2 layer
shallow-water equations, representing the flow in the upper
troposphere; they demonstrated the theory with a range of linear and
non-linear shallow-water calculations and full, 3D GCM experiments.
In their theory, the day-night forcing generates standing,
planetary-scale Rossby and Kelvin waves, analogous to those described
in analytic solutions by \citet{matsuno-1966} and \citet{gill-1980}.
See Figure~\ref{atmospheric-processes-schematic} for a schematic
illustration of these waves in the context of the overall circulation
of a synchronously rotating exoplanet.
In \citet{showman-polvani-2011}'s model, these baroclinic wave modes
are equatorially trapped and, in contrast to the theory described in
Section~\ref{jet-formation}, exhibit no meridional propagation.  The
Kelvin waves, which straddle the equator, exhibit eastward (group)
propagation, while the Rossby modes, which lie on the poleward flanks
of the Kelvin wave, exhibit westward (group) propagation.  Relative to
radiative-equilibrium solutions, this latitudinally varying zonal
propagation causes an eastward displacement of the thermal and
pressure fields at the equator, and a westward displacement of them at
higher latitudes (see Figure~\ref{showman-polvani} for an analytic
solution demonstrating this behavior).  In turn, these displacements
naturally generate an eddy velocity pattern with velocities tilting
northwest-southeast in the northern hemisphere and southwest-northeast
in the southern hemisphere, which induces an equatorward flux of eddy
momentum (Figure~\ref{showman-polvani}) and equatorial superrotation.

It is crucial to note that, in the \citet{showman-polvani-2011}
theory, the superrotation results not from wave tilts associated with
{\it meridional} wave propagation (which cannot occur for these
equatorially trapped modes), but rather from the differential {\it
  zonal} propagation of these Kelvin and Rossby modes.  Indeed, in
full, three-dimensional GCMs exhibiting equatorial superrotation in
response to tropical heating anomalies, the tropical eddy response
bears striking resemblance to these analytical, ``Gill-type''
solutions \citep{kraucunas-hartmann-2005, norton-2006,
  caballero-huber-2010, showman-polvani-2011, arnold-etal-2012},
providing further evidence that superrotation in these 3D models
results from this mechanism.  Interestingly, though,
\citet{showman-polvani-2011} and \citet{showman-etal-2013} showed
that, even when thermal damping is sufficiently strong to inhibit
significant zonal propagation of these wave modes, the multi-way force
balance between pressure-gradient forces, Coriolis forces, advection,
and (if present) frictional drag can in some cases lead to
prograde-equatorward and retrograde-poleward tilts in the eddy
velocities, allowing an equatorward momentum convergence and
equatorial superrotation.  At face value, this latter mechanism does
not seem to require appeal to wave propagation at all.  See
\citet{showman-polvani-2011} and \citet{showman-etal-2013} for further
discussion.

Nevertheless, real atmospheres contain a wide variety of wave modes,
and there remains a need to better test and integrate these various
mechanisms.  Linear studies show that tropical convection on Earth
generates two broad classes of wave mode---baroclinic, equatorially
trapped waves that propagate in longitude and height but remained
confined near the equator, and barotropic Rossby waves that propagate
in longitude and latitude but exhibit no vertical propagation
\citep[e.g.,][]{hoskins-karoly-1981, salby-garcia-1987,
  garcia-salby-1987}.  
By focusing on the 1-1/2 layer shallow-water model,
\citet{showman-polvani-2011}'s study emphasized the equatorially
trapped component and did not include a barotropic (meridionally
propagating) mode.  It would therefore be worth revisiting this issue
with a multi-layer model that includes both classes of wave mode;
such a model would allow a better test of the \citet{held-1999b}
hypothesis and allow a determination of the relative importance
of the two mechanisms under various scenarios for day-night
heating on exoplanets.

Several wave-mean-flow feedbacks exist that may influence the strength
and properties of equatorial superrotation on exoplanets. 
First are
the possible feedbacks with the Hadley circulation 
\citep{shell-held-2004} described in Section~\ref{hadley}.  These
feedbacks have yet to be explored in an exoplanet context.  

Second, on planets rotating sufficiently rapidly to exhibit
baroclinincally active extratopical zones, there is a possible
feedback involving the effect of midlatitude eddies on the equatorial
flow.  In the typical Earth-like (non-superrotating) regime,
equatorward-propagating Rossby waves generated by midlatitude
baroclinic instabilities can be absorbed on the equatorward flanks of
the subtropical jets (Section~\ref{hadley}), causing a westward
acceleration that helps inhibit a transition to superrotation---even
in the presence of tropical forcing like that shown in
Figure~\ref{showman-polvani}.  On the other hand, if such tropical
forcing becomes strong enough to overcome this westward torque, a
transition to superrotation nevertheless becomes possible.  Once this
occurs, the equatorward-propagating Rossby waves no longer encounter
critical levels in the tropics, and the tropics become transparent to
these waves, eliminating any westward acceleration associated with
their absorption.  This feedback suggests that, near the transition,
modest increases in the tropical wave source can cause sudden and
massive changes in the equatorial jet speed.  Two-level Earth-like
GCMs with imposed tropical eddy forcing confirm this general picture
and show that, near the transition, hysteresis can occur
\citep{suarez-duffy-1992, saravanan-1993, held-1999b}.  Nevertheless,
otherwise similar GCM experiments with a more continuous vertical
structure suggest that the feedback in practice is not strong
\citep{kraucunas-hartmann-2005, arnold-etal-2012}, hinting that the
2-level models may not properly capture all the relevant dynamics.
Additional work is needed to determine the efficacy of this feedback
in the exoplanet context.

Third, \citet{arnold-etal-2012} pointed out a  feedback
between the tropical waves (like those in
Figure~\ref{showman-polvani}) and the superrotation.  In the presence
of stationary, zonally asymmetric tropical heating/cooling anomalies,
the linear response comprises steady, planetary-scale Rossby waves (cf
Figure~\ref{showman-polvani}), whose thermal extrema are phase shifted
to the west of the substellar longitude due to the westward (group)
propagation of the waves.  In the presence of modest superrotation,
the phase shifts are smaller, and the wave amplitudes are larger.  A
resonance occurs when the (eastward) speed of the superrotation equals
the (westward) propagation speed of the Rossby waves, leading to very
large Rossby wave amplitude for a given magnitude of thermal forcing.
\citet{arnold-etal-2012} show that this feedback leads to an increased
convergence of eddy momentum onto the equator as the superrotation
develops, accelerating the superrotation still further.  This positive
feedback drives the atmosphere toward the resonance; beyond the
resonance, the feedback becomes negative.  \citet{arnold-etal-2012}
demonstrated these dynamics in linear shallow-water calculations and
full GCM experiments.  Nevertheless, a puzzling aspect of the
interpretation is that, for the equatorially trapped wave modes
considered by \citet{showman-polvani-2011} and
\citet{arnold-etal-2012}, the ability of the waves to cause
equatorward momentum convergence depends not just on the Rossby waves
but on {\it both} equatorially trapped Rossby and Kelvin waves, whose
superposition---given the appropriate zonal phase offset of the two
modes---leads to the necessary phase tilts and equatorward momentum
fluxes to cause superrotation.  The resonance described above would
not apply to the Kelvin wave component, since its (group) propagation
is eastward.  Additional theoretical work may clarify the issue.

\citet{edson-etal-2011} performed GCM simulations of synchronously
rotating but otherwise Earth-like exoplanets over a broad range of
rotation rates, and they found the existence of a sharp transition in
the speed of superrotation as a function of rotation rate.  The system
exhibited hysteresis, meaning the transition from weak to strong
superrotation occurred at a different rotation rate than the
transition from strong to weak superrotation.  The specific mechanisms
remain unclear; nevertheless, the model behavior suggests that
positive feedbacks, possibly analogous to those described above, help
to control the superrotation in their models.  The speed of
superrotation likewise differed at differing rotation rates in the GCM
simulations of \citet{merlis-schneider-2010}, but the transition was a
more gradual function of rotation rate, and these authors did not
suggest the existence of any hysteresis.

Finally, we note that slowly rotating planets exhibit a natural
tendency to develop equatorial superrotation even in the absence of
day-night or other imposed eddy forcing.  This has been demonstrated
in a variety of GCM studies where the imposed thermal forcing does not
include a diurnal cycle (i.e., where the radiative equilibrium
temperature depends on latitude and pressure but not longitude)
\citep[e.g.,][]{delgenio-etal-1993, delgenio-zhou-1996,
  yamamoto-takahashi-2003, herrnstein-dowling-2007,
  richardson-etal-2007, lee-etal-2007, hollingsworth-etal-2007,
  mitchell-vallis-2010, parrish-etal-2011, lebonnois-etal-2012}.
Superrotation of this type can be seen in the axisymmetrically forced,
slowly rotating models from \citet{kaspi-showman-2012} in
Figure~\ref{hadley-cell} (top four panels) and \citet{joshi-etal-1997}
and \citet{heng-vogt-2011} in Figure~\ref{superrotate}.  Although the
zonal wind at the equator corresponds to a local minimum with respect
to latitude, it is eastward, and the equatorial upper troposphere
still comprises a local maximum of angular momentum per mass.  Thus,
even this configuration (unlike a case where the zonal-mean zonal wind
at the equator is zero or westward) must be maintained by eddy
transport of angular momentum to the equator.  Just such eddy-momentum
fluxes can be seen in Figure~\ref{hadley-cell}: in the slowly
rotating cases exhibiting superrotation, eddy momentum converges
onto the equator (opposite to the sign of the low-latitude 
eddy-momentum fluxes in the faster-rotating 
cases lacking superrotation). Although the details
are subtle, several authors have suggested that the superrotation in
this class of model results from equatorward angular momentum
transport by a barotropic instability of the subtropical jets
\citep[e.g.][]{delgenio-etal-1993, delgenio-zhou-1996,
  mitchell-vallis-2010}.  This mechanism is relevant to the
superrotation on Venus and Titan (although the thermal tides, which
represent a day-night forcing that can trigger global-scale wave
modes, are relvant to those planets as well).  See the chapter by
Lebonnois et al. for more details.

\subsection{Effect of day-night forcing}

Many observationally accessible exoplanets will be
sufficiently close to their stars to be tidally despun to a
synchronous rotation state.  For such exoplanets, the slowly rotating,
tropical dynamical regime will often go hand-in-hand with a strong
day-night thermal forcing arising from the fact that they have
permanent daysides and nightsides.  A major question is how the 
atmospheric circulation depends on the strength of this day-night forcing.
Will the circulation exhibit an essentially day-night flow pattern?
Or, instead, will the circulation respond primarily to the {\it zonal-mean}
radiative heating/cooling, developing a zonally banded flow in response
to the planetary rotation?  What is the amplitude of the day-night
temperature difference?  Is there a danger of the atmosphere freezing
out on the night side? 


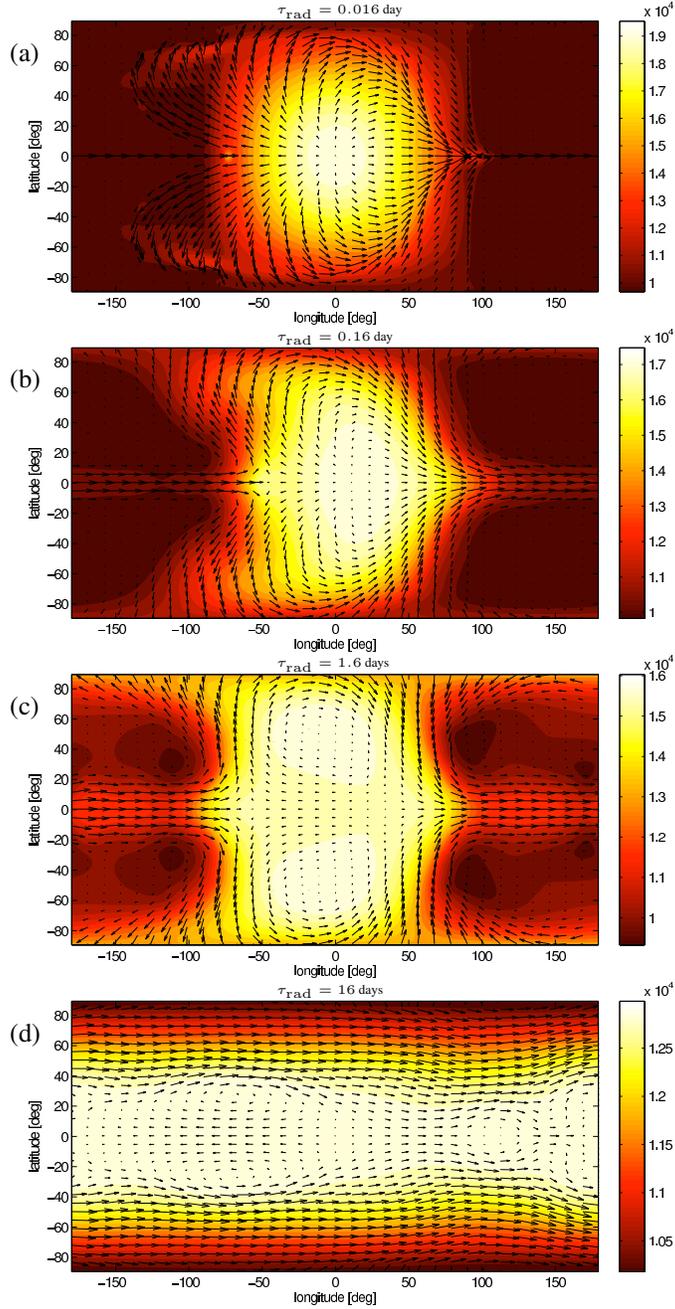
\begin{figure*}
\begin{minipage}[c]{1.0\textwidth}
\centering

\vspace{-1in}
\includegraphics[scale=0.47, angle=0]{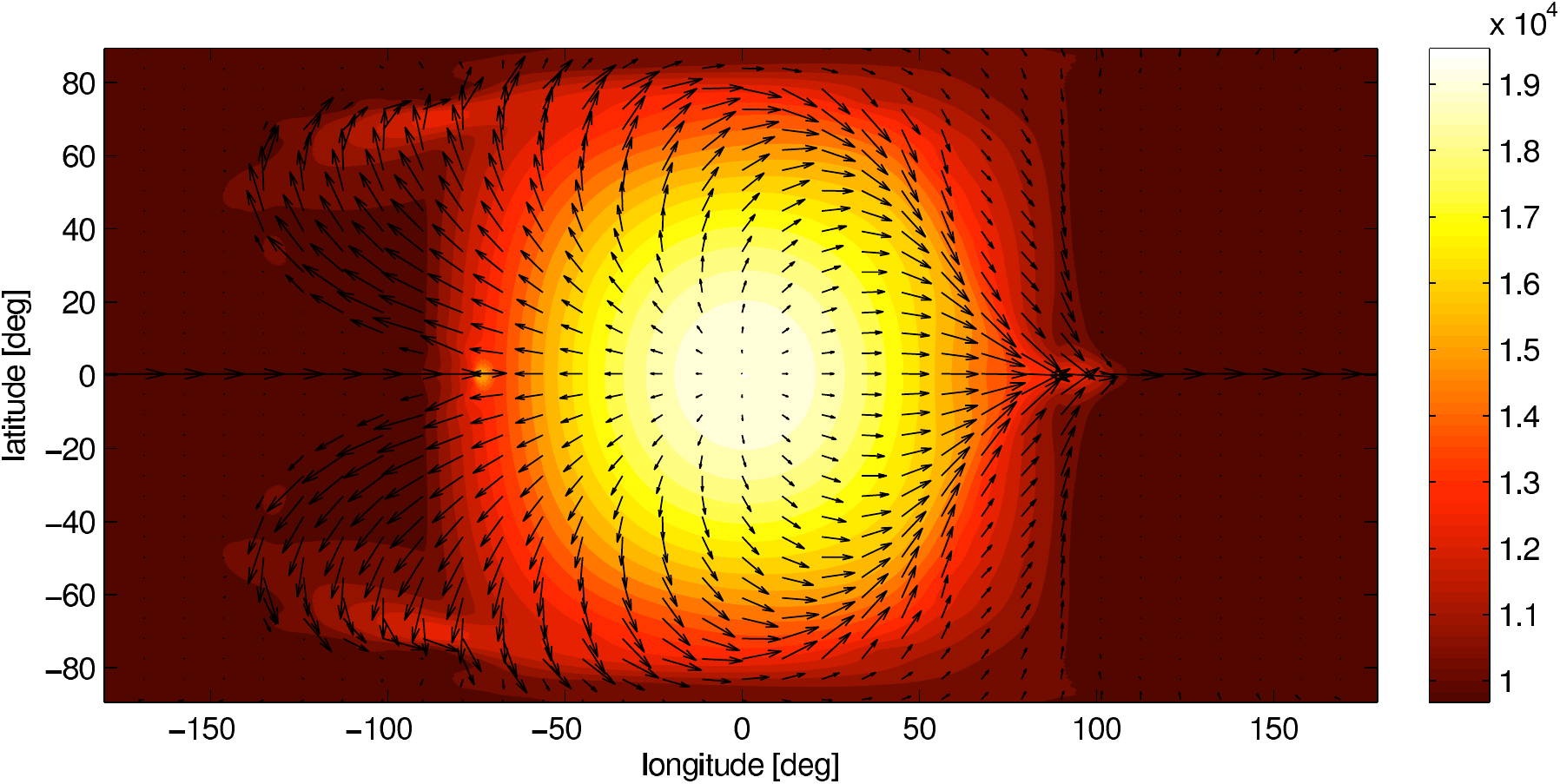}
\put(-251.,100.){\normalsize (a)}
\put(-150.,118.){\tiny $\tau_{\rm rad}=0.016$ day}

\includegraphics[scale=0.47, angle=0]{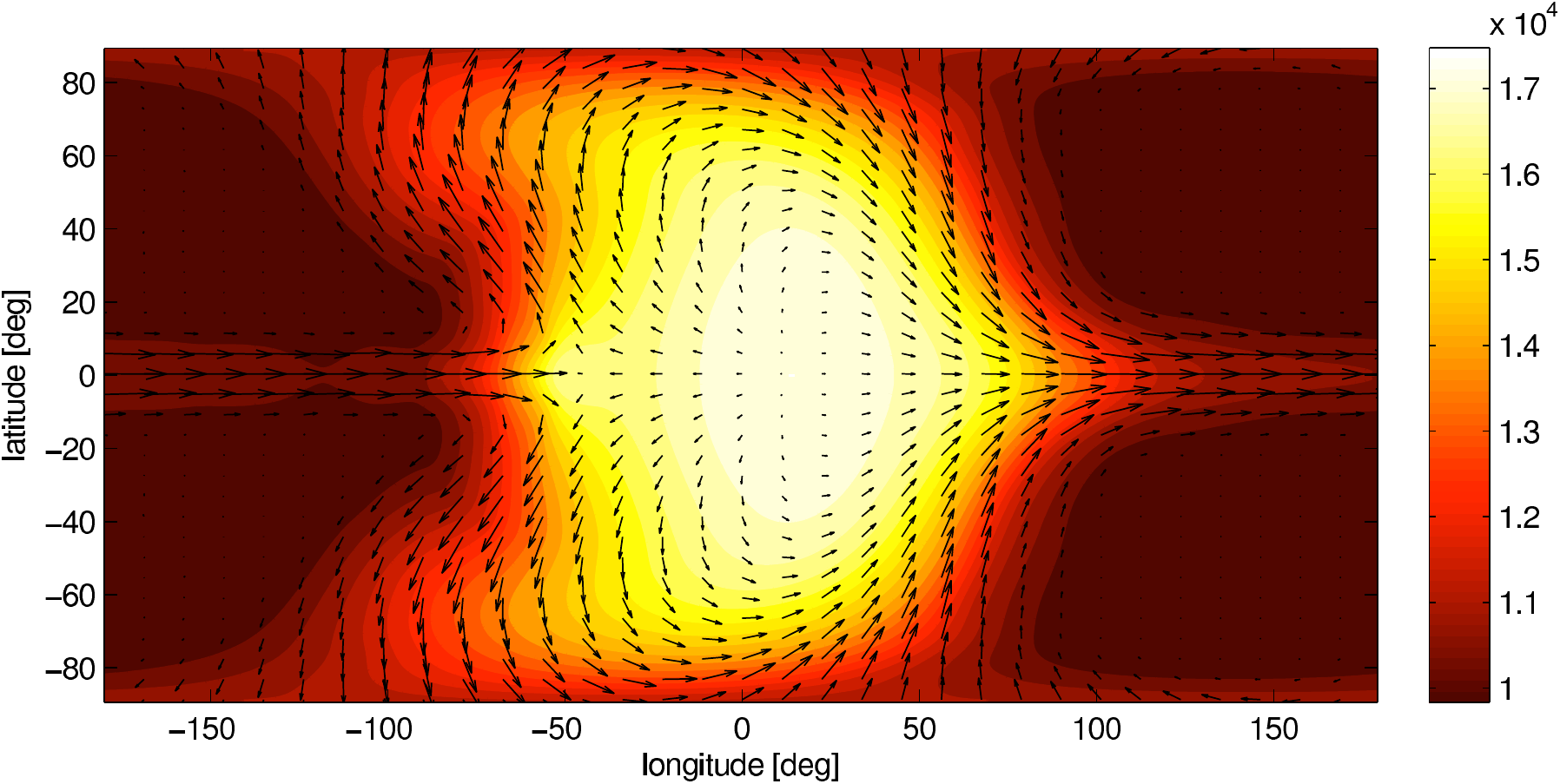}
\put(-251.,100.){\normalsize (b)}
\put(-150.,118.){\tiny $\tau_{\rm rad}=0.16$ day}


\includegraphics[scale=0.47, angle=0]{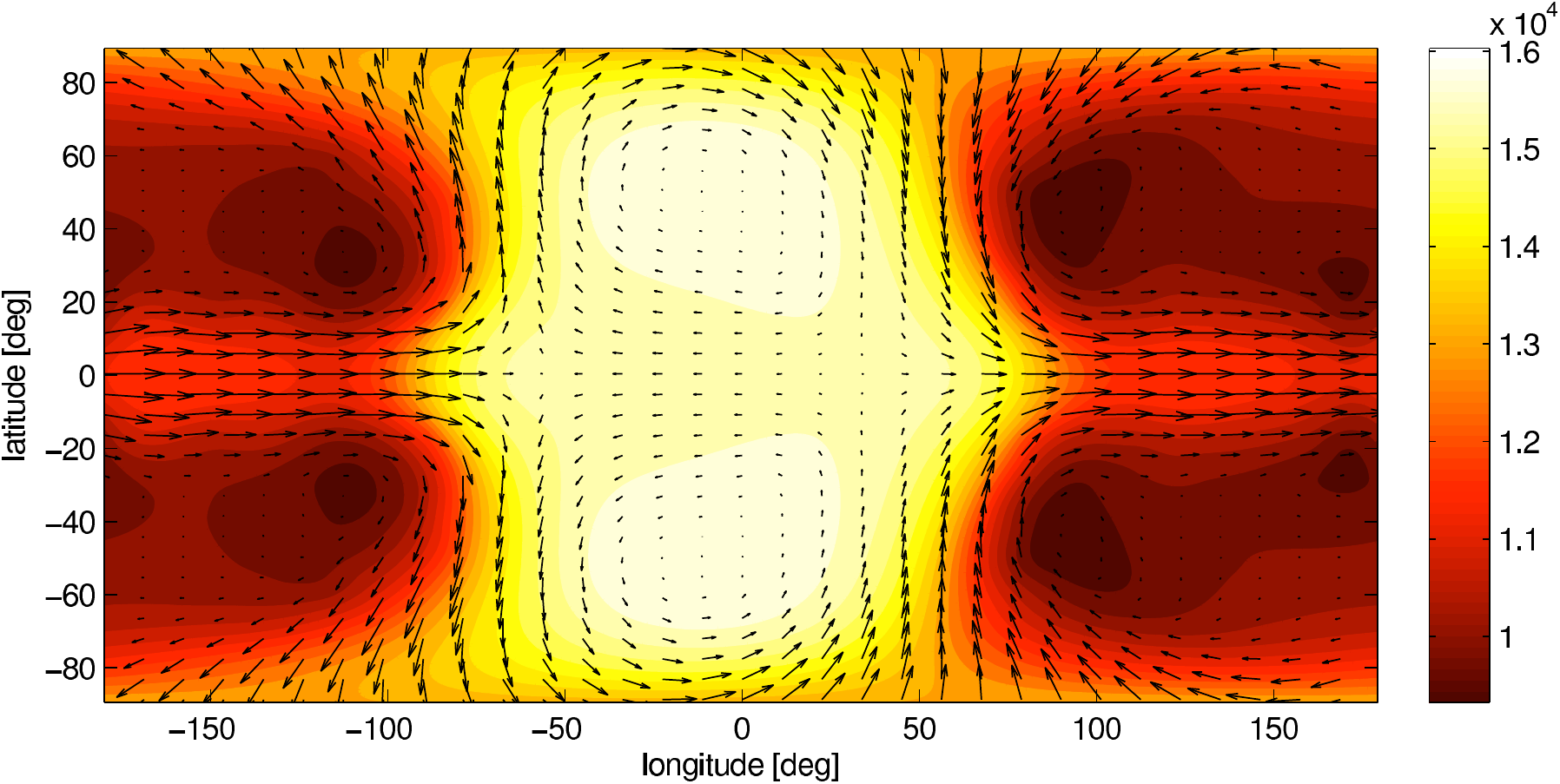}
\put(-251.,100.){\normalsize (c)}
\put(-150.,118.){\tiny $\tau_{\rm rad}=1.6$ days}


\includegraphics[scale=0.47, angle=0]{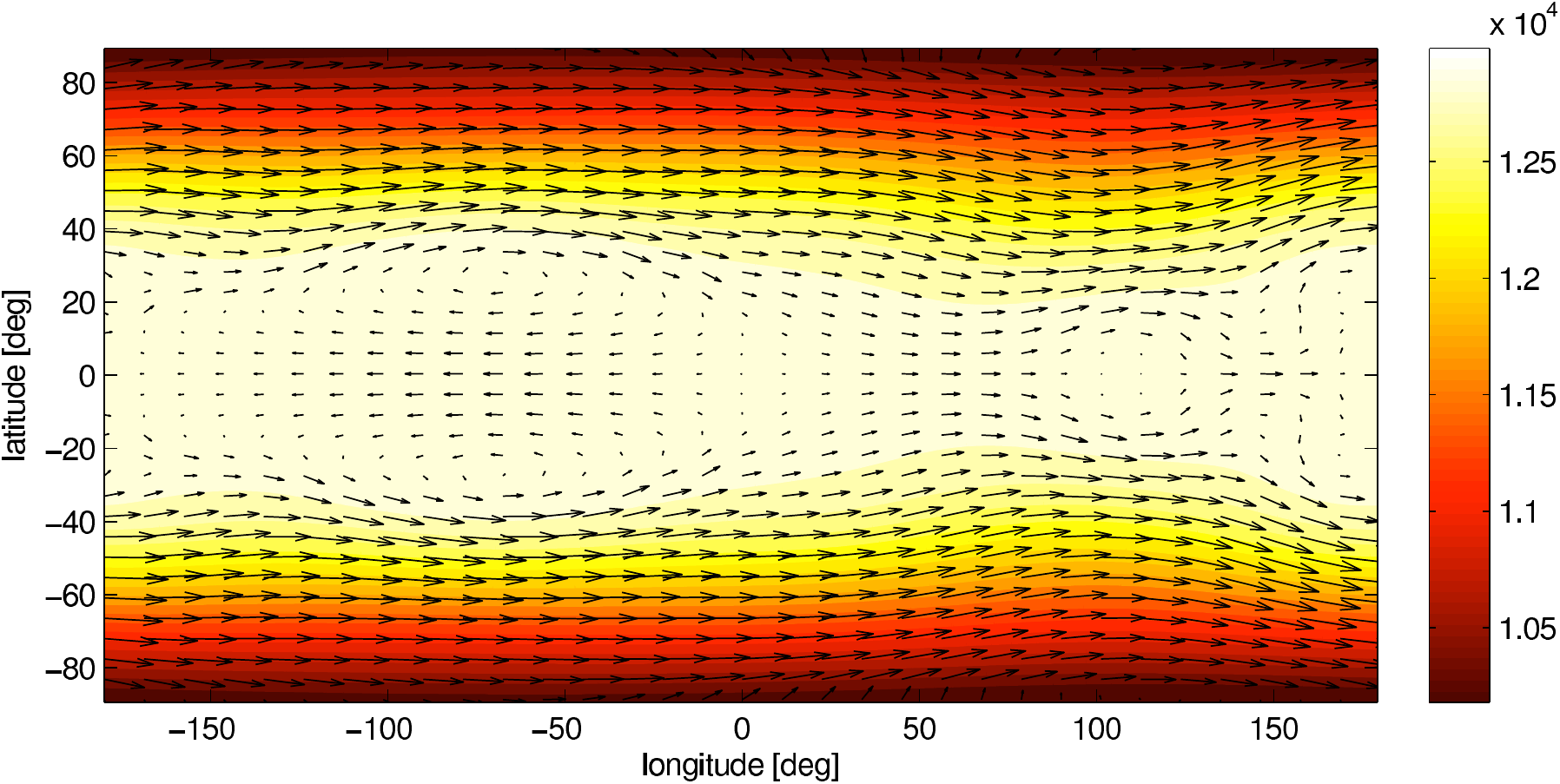}
\put(-251.,100.){\normalsize (d)}
\put(-150.,118.){\tiny $\tau_{\rm rad}=16$ days}

\caption{Layer thickness $gh$ (orange scale, units $\rm m^2\,s^{-2}$)
  and winds (arrows) for the equilibrated (steady-state) solutions to
  the shallow-water equations in full spherical geometry for
  sychronously rotating terrestrial exoplanets driven by a day-night
  thermal forcing.  Specifically, to drive the flow, a mass
  source/sink term $(h_{\rm eq}-h)/\tau_{\rm rad}$ is included in the
  continuity equation, where $h_{\rm eq}$ is a specified
  radiative-equilibrium layer thickness, $h$ is the actual layer
  thickness (depending on longitude $\lambda$, latitude $\phi$, and
  time), and $\tau_{\rm rad}$ is a specified radiative time constant.
  Here $h_{\rm eq}$ equals a constant, $H$, on the nightside, and
  equals $H+\cos\lambda\cos\phi$ on the dayside.  The substellar point
  is at longitude, latitude ($0^{\circ},0^{\circ}$) \citep[for further
    details about the model, see][]{showman-etal-2013}.  Here, the
  planet radius is that of Earth, rotation period is 3.5 Earth days,
  and $gH=10^4\rm\,m^2\,s^{-2}$.  These values imply an equatorial
  deformation radius, $(\sqrt{gH}/\beta)^{1/2}$, which is 60\% of the
  planetary radius.  The four models are identical in all ways except
  the radiative time constant, which is varied from $0.016$ to $16$
  Earth days from top to bottom, respectively.  When the radiative
  time constant is short, the flow exhibits a predominantly day-night
  circulation pattern with a large day-night contrast close to
  radiative equilibrium (a).  Intermediate radiative time constants (b
  and c) still exhibit considerable day-night contrasts, yet allow the
  emergence of significant zonal flows.  Long radiative time constants
  (d) lead to a zonally banded flow pattern with little variation of
  thermal structure in longitude---despite the day-night forcing.}
\label{stswm}
\end{minipage}
\end{figure*}

To pedagogically illustrate the dynamics involved, we present in
Figure~\ref{stswm} four numerical solutions of the shallow-water
equations, representing the upper tropospheric flow on a sychronously
rotating, terrestrial exoplanet.  The shallow-water layer represents
the mass above a given isentrope in the mid-troposphere.  Since
atmospheres are stably stratified, the entropy increases with height,
and thus radiative heating (which increases the entropy of air)
transports mass upward across isentropes, while radiative cooling
(which decreases the entropy of air) transports mass downward across
isentropes.  Thus, day-night heating in the shallow-water system is
parameterized as a mass source/sink that adds mass on the dayside
(representing its transport into the upper-tropospheric shallow-water
layer from the lower troposphere) and removes it on the nightside
(representing its transport from the shallow-water layer into the
lower troposphere).  Here, we represent this process as a Newtonian
relaxation with a characteristic radiative timescale, $\tau_{\rm
  rad}$, that is a specified free parameter.  The models are identical
to those in \citet{showman-etal-2013} except that the planetary
parameters are chosen to be appropriate to a terrestrial exoplanet.

Figure~\ref{stswm} demonstrates that the amplitude of the day-night
forcing exerts a major effect on the resulting circulation patterns.
The four simulations are identical except for the imposed radiative
time constant, which ranges from short in the top panels to long in
the bottom panels.  When the radiative time constant is extremely
short (Figure~\ref{stswm}a), the circulation consists primarily of
day-night flow in the upper troposphere, and the thermal structure is
close to radiative equilibrium, with a hot dayside and a cold
nightside.  The radiative damping is so strong that any global-scale
waves of the sort shown in Figure~\ref{showman-polvani} are damped
out.  This simulation represents a complete breakdown of the WTG
regime described in Section~\ref{wave-adjustment}.  Intermediate
radiative time constants (Figure~\ref{stswm}b and c) lead to large
day-night contrasts coexisting with significant dynamical structure,
including planetary scale waves that drive a superrotating equatorial
jet.  When the radiative time constant is long (Figure~\ref{stswm}d),
the circulation exhibits a zonally banded pattern, with little
variation of the thermal structure in longitude---despite the
day-night nature of the imposed thermal forcing.  In this limit, the
circulation is firmly in the WTG regime.  Interestingly, the day-night
forcing is sufficiently weak that superrotation does not occur in this
case.


In a planetary atmosphere, the radiative time constant should increase
with increasing temperature or decreasing atmospheric mass.
Figure~\ref{stswm} therefore suggests that synchronously rotating
planets that are particularly hot, or have thin atmospheres, may
exhibit day-to-night flow patterns with large day-night temperature
differences, while planets that are cooler, or have thicker
atmospheres, may exhibit banded flow patterns with smaller day-night
temperature differences \citep[e.g.,][] {showman-etal-2013}.  In the
context of hot Jupiters, these arguments suggest that the most
strongly irradiated planets should exhibit less efficient day-night
heat redistribution (with temperatures closer to radiative
equilibrium) than less-irradiated planets, a trend that already
appears to be emerging in secondary-eclipse observations
\citep{cowan-agol-2011}.  

In the context of terrestrial planets, the trend in Figure~\ref{stswm}
could likewise describe a transition in circulation regime occurring
as a function of incident stellar flux, but it might also describe a
transition as a function of atmospheric mass {\it for a given stellar
  flux}.  Because of its thin atmosphere, Mars, example, has a
radiative time constant considerably shorter than that on Earth, which
helps to explain the much larger day-night temperature differences on
Mars compared to Earth.  Even more extreme is Jupiter's moon Io, due
to its particularly tenuous atmosphere (surface pressure
$\sim$nanobar).  SO$_2$ is liberated by sublimation on the dayside but
collapses into surface frost on the nightside, and the resulting
day-night pressure differences are thought to drive supersonic flows
from day to night \citep{ingersoll-etal-1985,
  ingersoll-1989}.\footnote{Neither Mars nor Io is synchronously
  rotating with respect to the Sun, but one can still obtain large
  day-night temperature differences in such a case if the radiative
  time constant is not long compared to the solar day.}  In the
exoplanet context, close-in rocky planets subject to extreme stellar
irradiation may lie in a similar regime, with tenuous atmospheres
maintained by sublimating silicate rock on the dayside, fast day-night
airflows, and condensation of this atmospheric material on the
nightside \citep{leger-etal-2011, castan-menou-2011}.  Prominent
examples include CoRoT-7b, Kepler-10b, and 55 Cnc-e.  Even terrestrial
exoplanets with temperatures similar to those of Earth or Mars could
experience sufficiently large day-night temperature differences to
cause freeze-out of the atmosphere on the nightside if the atmosphere
is particularly thin \citep{joshi-etal-1997}, an issue we return to in
Section~\ref{climate}.  

\section{HYDROLOGICAL CYCLE}
\label{hydrological}
\newcommand  {\about} {\mathop{\sim}\!}

The hydrological cycle---broadly defined, where the condensate may be
water or other chemical species---plays a fundamental role in
determining the surface climate of a planet. It directly influences
the surface climate by setting the spatial and temporal distributions
of precipitation and evaporation and indirectly influences the surface
temperature by affecting the horizontal energy transport, vertical
convective fluxes, atmospheric lapse rate, and planetary albedo; these
are key aspects of planetary habitability.  The hydrological cycle is
also fundamental in determining the radiative balance of the planet
through the humidity and cloud distributions.  The planetary radiative
balance both affects the surface climate and is what is remotely
observed.

Here, we survey the role of the hydrological cycle in affecting the
atmospheric circulation and climate of terrestrial planets.  We begin
with a brief discussion of the role of the ocean
(Section~\ref{oceans}), as this is the ultimate source of atmospheric
moisture and can influence atmospheric circulation in a variety of
ways.  We next discuss the thermodynamics of phase change 
(Section~\ref{thermo}), which has significant implications for
how a hydrological cycle affects the circulation.   Global 
precipitation is discussed next (Section~\ref{global-precip}),
with an emphasis on energetic constraints
that are independent of details of the atmospheric circulation.
The following section (\ref{regional-precip}) highlights the
important role of the atmospheric circulation for regional
precipitation, in contrast to global precipitation.
We then follow with a survey of how the atmospheric circulation helps
to control the distribution of humidity (Section~\ref{humidity})
and clouds (Section~\ref{clouds}).  Lastly, we summarize how
the hydrological cycle in turn {\it affects} the structure of the
atmospheric circulation (Section~\ref{moist-circulation}).


\subsection{Oceans}
\label{oceans}

The presence of an ocean affects planetary climate and atmospheric
circulations in several key ways.  We begin the discussion of the
hydrological cycle with the ocean in recognition that it is the source
of water vapor in an atmosphere, and the
influences of the hydrological cycle on climate (presented in what follows)
may be modulated by differences in water availability on other
planets.  Oceanography
is, of course, a vast field of research, and we will only briefly
discuss aspects that may be of immediate interest in the
exoplanet context.  For overviews of physical oceanography
and the ocean's effect on climate in the terrestrial context, see for example
\citet{peixoto-oort-1992}, \citet{siedler-etal-2001}, \citet{pedlosky-2004}, 
\citet{vallis-2006, vallis-2011}, \citet{marshall-plumb-2007},
\citet{olbers-etal-2012}, and \citet{williams-follows-2011},
among others.

Oceans on exoplanets will span a wide range.  At one extreme are
planets lacking surface condensate reservoirs (Venus) or with liquid
surface reservoirs small enough to form only disconnected lakes
(Titan).  In such a case, the surface condensate should not contribute
significantly to the meridional or day-night heat transport, but could
still significantly impact the climate via energy exchanges and
evaporation into the atmosphere.  At the other extreme are super
Earths whose densities are sufficiently low to require significant
fractions of volatile materials in their interiors.  Prominent
examples include GJ 1214b \citep{charbonneau-etal-2009,
  rogers-seager-2010b, nettelmann-etal-2011} and several planets in
the Kepler-11 system \citep{lissauer-etal-2011, lopez-etal-2012};
many of the hundreds of super Earths discovered by Kepler
also likely fall into this category \citep[e.g.,][]{rogers-etal-2011}.
Many such planets will contain water-rich fluid envelopes thousands of
km thick.  If their interior temperature profiles are sufficiently hot
(in particular, if, at the pressure of the critical point, the
atmospheric temperature exceeds the temperature of the critical
point\footnote{For a pure water system, the critical point pressure
  and temperature are 221 bars and $647\K$ respectively, but they
  depend on composition for a system containing other components such
  as hydrogen \citep[e.g.,][]{wiktorowicz-ingersoll-2007}.}), then
these planets will exhibit a continuous transition from a
supercritical fluid in the interior to a gas in the atmosphere, with
no phase boundary; on the other hand, if their interior temperature
profiles are sufficiently cold, then their water-rich envelopes will
be capped by an oceanic interface, overlain by an atmosphere
\citep{leger-etal-2004, wiktorowicz-ingersoll-2007}.  Earth lies at
intermediate point on this continuum, with an ocean that is a small
fraction of the planet's mass and radius, yet is continuously
interconnected and thick enough to cover most of the surface area.

Our understanding of ocean dynamics and ocean-atmosphere interactions
stem primarily from Earth's ocean, so to provide context we first
briefly overview the ocean structure.  The Earth's oceans have a mean
thickness of $3.7\rm\,km$ and a total mass $\sim$260 times that of the
atmosphere.  At the surface, the ocean is warm at the equator
($\sim$$27^{\circ}$C in an annual and longitudinal average) and cold at
the poles ($\sim$$0^{\circ}$C).  The deep ocean temperature is
relatively uniform at a temperature of $\sim$1--2$^{\circ}$C, thoughout
the world---even in the tropics.  The transition from warmer surface
water to cooler deep water is called the thermocline and occurs at
$\sim$$0.5\rm\,km$ depth, depending on latitude.  Because low-density,
warmer water generally lies atop high-density, cooler water, the ocean
is stably stratified and does not convect except at a few localized
regions near the poles.  However, turbulence caused by wind and waves
homogenizes the top $\sim$10--$100\rm\,m$ (depending on weather
conditions and latitude), leading to profiles of density, salinity,
and composition that vary little across this so-called ``mixed
layer.''  Because of its efficient communication with the atmosphere,
the thickness of the mixed layer exerts a strong effect on climate
(e.g., on the extent to which the ocean modulates seasonal cycles).
Nevertheless, the deeper ocean also interacts with the atmosphere on a
wide range of timescales up to thousands of years.

The Earth's oceans affect the climate in numerous ways.  For
discussion, it is useful to decompose the role of the ocean into
time-dependent (e.g., modification of heat capacity) and
time-independent (e.g., time-mean energy transports) categories.

Consider the time-dependent energy budget of the ocean mixed layer:
\begin{equation}
\alpha {\partial T_{\rm surf}\over\partial t} = R_{\mathrm{sfc}} - LE - SH
 - \nabla \cdot F_o,
\label{eqn-surface_energy_budget}
\end{equation}
where $\alpha$ is the heat capacity per unit area of the mixed layer
(approximately equal to $\rho c_p h$, where $\rho$ is the density,
$c_p$ is the specific heat, and $h$ is the thickness of the mixed
layer), net surface radiative flux $R_{\mathrm{sfc}}$, latent enthalpy flux
$LE$, sensible enthalpy flux $SH$, and ocean energy flux divergence
$\nabla \cdot F_o$, including vertical advection into the base of the
mixed layer.  Earth's mixed layer depth has geographic and seasonal
structure because it is connected both to the interior ocean's thermal
structure and currents from below and is forced by the atmosphere and
radiation from above.

From this budget, a clear time-dependent effect of the ocean is that
it acts as a thermal surface reservoir, which reduces the seasonal and
diurnal cycle amplitude compared to possible land-covered planets
\citep[$\alpha$ for the ocean is substantially larger than
for land;][]{hartmann-1994, pierrehumbert-2010}.  In addition to the
direct modification of the heat capacity, the presence of an ocean
leads to weaker temperature fluctuations through evaporation, which is
a more sensitive function of temperature than surface longwave
radiation or sensible surface fluxes
\citep[e.g.,][]{pierrehumbert-2010}.  Oceans further influence a
planet's thermal evolution on long timescales through the storage of
heat in the interior ocean (i.e., the vertical component of the ocean
energy flux divergence in Eq.~\ref{eqn-surface_energy_budget}) and
through their role as a chemical reservoir
\citep{sarmiento-gruber-2006}; the ocean storage and release of
carbon is important for Earth's glacial cycles, for example.

In addition to the effect of the ocean in determining a planet's
time-dependent evolution, oceans modify a planet's time-mean climate.
Evaporation of water from the ocean is the essential moisture source
to the atmosphere and to land regions with net precipitation.  Oceans
affect the radiation balance of a planet directly through their albedo
and indirectly through providing the moisture source for the atmospheric
hydrological cycle.  Finally, ocean energy transport is important in
determining the surface temperature.

Many authors have examined the sensitivity of ocean energy transport
to changes in surface wind stress and surface buoyancy gradients,
often decomposing the energy transport into components associated with
deep meridional overturning circulations and shallow, wind-driven
circulations \citep[cf.][]{ferrari-ferreira-2011}.  Here, we note that
ocean energy transport can affect surface climate differently than
atmospheric energy transport.  For example,
\citet{enderton-marshall-09} presented simulations with different
simplified ocean basin geometries; the ocean energy transport, surface
temperature, and sea ice of the simulated equilibrium climates
differed, but the total energy transport was relatively unchanged
(i.e., the changes in atmospheric energy transport largely offset
those of the ocean\footnote{Changes in atmospheric and oceanic energy
  transport need not compensate in general
  \citep[e.g.,][]{vallis-farneti-2009}}).  An implication of these
results is that while observations of a planet's top-of-atmosphere
radiation allow the total (ocean and atmosphere) energy transport to
be determined, this alone is not sufficient to constrain the surface
temperature gradient.

\subsection{Thermodynamics of phase change}
\label{thermo}

Before describing the ways the hydrological cycle can interact with
the atmospheric dynamics (which we do in subsequent subsections), 
we first summarize the pure thermodynamics of a system
exhibiting phase changes.  Most atmospheres in the Solar System
have contituents that can condense: water on Earth, CO$_2$ on Mars,
methane on Titan, N$_2$ on Triton and Pluto, and several species,
including H$_2$O, NH$_3$, and H$_2$S, on Jupiter, Saturn, Uranus,
and Neptune.  Because atmospheric air parcels change temperature
and pressure over time (due both to atmospheric motion and to
day-night or seasonal temperature swings), air parcels make
large excursions across the phase diagram, and often strike one
or more phase boundaries---leading to condensation.  This condensation
can exert major influence over the circulation.

The starting point for understanding these phase changes is the
Clausius-Clapeyron equation, which relates changes in the saturation
vapor pressure of a condensable vapor, $e_s$, to the latent heat of
vaporization $L$, temperature $T$, density of vapor $\rho_{\rm vap}$,
and density of condensate $\rho_{\rm cond}$
\begin{equation}
{d e_s\over dT} = {1\over T} {L \over \rho_{\rm vap}^{-1} - \rho_{\rm cond}^{-1}}.
\label{eqn-CC0}
\end{equation}
For condensate-vapor interactions far from the critical point,
the condensate density greatly exceeds the vapor density and
can be ignored in Eq.~(\ref{eqn-CC0}).  Adopting the ideal-gas
law, the equation can then be expressed
\begin{equation}
\frac{1}{e_s} \frac{d e_s}{d T} = \frac{L}{R_v T^2},
\label{eqn-CC}
\end{equation}
where $R_v$ is the gas constant of the condensable vapor.

The saturation vapor pressure is the pressure of vapor if the air were
in equilibrium with a saturated surface; it increases with temperature
and is a purely thermodynamic quantity. In contrast, the vapor
pressure of an atmosphere is not solely a thermodynamic quantity, as
atmospheres are generally subsaturated. Therefore, the vapor pressure
depends on the relative humidity $\mathcal{H} = e/e_s$, which in turn
depends on the atmospheric circulation (discussed in what follows).
Condensation occurs when air becomes supersaturated, although there
are also microphysical constraints such as the availability of
condensation nuclei.  A consequence of \eqref{eqn-CC} is that
atmospheric humidity will increase with temperature, if relative
humidity is unchanged.

The vertical structure of water vapor or other condensible
species can be quantified by comparing it to that of pressure.
The pressure scale height $H_p$ for hydrostatic atmospheres is 
given by
\begin{equation}
\frac{1}{H_p} = \frac{\partial \ln p}{\partial z} =
\frac{g}{R T}.
\label{eqn-p_scale}
\end{equation}
For Earth, $g \about 10 \, \mathrm{m \, s^{-2}}$, 
$T \about 280 \, \mathrm{K}$, 
$R \about 300 \, \mathrm{J \, kg^{-1} \, K^{-1}}$, the
pressure scale height $H_p \approx 8 \, \mathrm{km}$. The pressure
scale height is a decreasing function of planet mass (through $g$)
and an increasing function of planet temperature.

Because the saturation vapor pressure of condensable species 
depend strongly on temperature, it is possible for the scale height of
water---and other condensables---to differ greatly from the pressure
scale height.  The scale height of water vapor $H_w$ can be estimated
as follows
\begin{equation}
\frac{1}{H_w} =  \frac{\partial \ln e_s}{\partial z} =
\frac{\partial \ln e_s}{\partial T} \frac{\partial T}{\partial z}
=  \frac{L}{R_v T^2} \frac{\partial T}{\partial z},
\label{eqn-e_scale}
\end{equation}
where the chain rule and Clausius-Clapeyron relationship
\eqref{eqn-CC} have been used and vertical variations in relative
humidity have been neglected.  The scale height of water vapor is the
inverse of the product of the lapse rate $\Gamma = - \partial T/\partial z$ and
the fractional change of vapor pressure with temperature (i.e., the
Clausius-Clapeyron relationship). The scale height of water increases
with temperature [through \eqref{eqn-CC}] and with decreasing lapse
rate (i.e., less rapid vertical temperature decrease).  

Using
Earth-like values in \eqref{eqn-p_scale} and \eqref{eqn-e_scale} of $g
\about 10 \, \mathrm{m \, s^{-2}}$, $T \about 280 \, \mathrm{K}$, $R
\about 300 \, \mathrm{J \, kg^{-1} \, K^{-1}}$, $R_v \about 450 \,
\mathrm{J \, kg^{-1} \, K^{-1}}$, $L \about 2.5 \times 10^6 \,
\mathrm{J \, kg^{-1} }$, the ratio of the water vapor and pressure
scale heights $H_w / H_p$ is $\about 0.2$ for dry adiabatic lapse rate
($\Gamma = g/c_p$).  Thus, for Earth-like situations, the scale height
of water vapor is significantly less than the pressure scale height. This
implies that near-surface water vapor dominates the column water vapor (the
mass-weighted vertical integral of water vapor from the surface to the
top of atmosphere).  Moreover, the saturation vapor pressure at the
tropical tropopause is $\sim$$10^4$ times less than at the surface.
As a result, the middle and upper atmosphere (stratosphere and
above) are very dry.  Still, that the bulk of the atmospheric water vapor
resides near the surface does not mean that the upper tropospheric
humidity is unimportant: small concentrations there can be important
in affecting the radiation balance as water vapor is a greenhouse gas
\citep[e.g.,][]{held-soden-2000}; moreover, upper-tropospheric water 
controls the formation and distribution of cirrus cloud, which
is important in the atmospheric radiation balance.

The condensable gas will be
a larger fraction of the pressure scale height if the latent heat is
smaller or if the tropospheric stratification is greater.  Titan
provides an example in our solar system; the methane abundance at the
tropopause is only a factor of $\sim$3 smaller than near the surface
\citep{niemann-etal-2005}.  In such a situation, the condensable
vapor abundance in the stratosphere may be considerable, which can
have implications for planetary evolution because it will affect
the rate at which such species (whether water or methane) can be
irreversibly lost via photolytic breakup and escape of hydrogen to
space.

\subsection{Global Precipitation}
\label{global-precip}

A planet's precipitation is a fundamental part of its hydrological
cycle and has important implications for global climate feedbacks
such as the carbonate-silicate feedback cycle (Section~\ref{climate}).
At equilibrium, a planet's precipitation equals its evaporation in the
global mean.  Global precipitation is controlled by energy balance
requirements, either of the surface (to constrain evaporation) or of
the atmospheric column (to constrain the latent heating from
precipitation).  More complete discussions and reviews of the control
of global precipitation can be found in \citet{held-2000},
\citet{pierrehumbert-2002}, \citet{allen-ingram-2002},
\citet{schneider-etal-2010}, and \citet{ogorman-etal-2011}.

The dry static energy $s = c_p T + g z$ is the sum of the specific
enthalpy and potential energy. The dry static energy is materially conserved in
hydrostatic atmospheres in the absence of diabatic processes (e.g.,
radiation or latent heat release), if the kinetic energy of the
atmosphere is neglected \citep[cf.][]{betts-1974,
  peixoto-oort-1992}. The dry static energy budget is
\begin{equation}
\frac{\partial s}{\partial t} + {\bf v} \cdot \nabla s = Q_r + Q_c + SH,
\label{eqn-dse}
\end{equation}
where $Q_r$ is the radiative component of the diabatic tendency,
$Q_c$ is the latent-heat release component of the diabatic tendency,
and $SH$ is the sensible component of the surface enthalpy
flux. Taking the time- and global-mean of the dry static energy budget
eliminates the left-hand side of \eqref{eqn-dse}, so that when it is
integrated vertically over the atmospheric column, the global-mean
precipitation is a function of the radiation and sensible surface flux
\begin{equation}
L \langle P \rangle = \langle R_{atm} \rangle + \langle SH \rangle = \langle R_{TOA} \rangle - \langle R_{sfc} \rangle +  \langle SH \rangle,
\label{eqn-atm_en}
\end{equation}
with latent heat of vaporization $L$, precipitation $P$, net
atmospheric radiation $R_{atm}$, net top-of-atmosphere (TOA) radiation
$R_{TOA}$, and net surface radiation $R_{sfc}$. The operator $\langle
\cdot \rangle$ denotes the time- and global-mean. The radiation terms
are evaluated only at the surface and top-of-atmosphere because of the
vertical integration over the atmospheric column, and they can be
further decomposed into components associated with the shortwave and
longwave parts of the radiation spectrum.

The time- and global-mean of the surface energy budget
\eqref{eqn-surface_energy_budget} is
\begin{equation}
L \langle E \rangle = \langle R_{sfc} \rangle - \langle SH \rangle,
\end{equation}
with evaporation $E$ and other variables defined as in
\eqref{eqn-atm_en}.  In both budgets, the sensible turbulent surface
flux (the dry component of the surface enthalpy flux) enters, so the
global precipitation is not solely a function of the radiative
properties of the atmosphere.  The turbulent surface fluxes are
represented by bulk aerodynamic formula in GCMs \citep{garratt-1994}
[e.g., $SH = \rho c_d
\| {\bf v}_{\rm surf} \| (T_s - T_a)$], which depend on the
temperature difference between the surface and surface air $(T_s -
T_a)$, surface wind speed $\| {\bf v}_{\rm surf} \|$, and surface
roughness (through $c_d$).

From these budgets, one expects global precipitation to increase if
the solar constant increases (Fig.~\ref{fig-globalprecip}b), the
top-of-atmosphere or surface albedo decreases, or sensible surface
fluxes decrease.  Reductions in longwave radiation at the surface that
are greater than those at the top-of-atmosphere will increase the
global precipitation (Fig.~\ref{fig-globalprecip}a), as is expected
for increased greenhouse gas concentrations on Earth.  This illustrates
that the vertical structure of radiative changes is important. Another
example is that changes in TOA albedo and surface albedo differ in
their effect on global mean precipitation because they differ in their
effect on shortwave absorption by the atmosphere [a component of
$R_{atm}$ in (4)].  Thus, it is not possible to infer global-mean
precipitation from measurements of the radiances at the
top-of-atmosphere $R_{TOA}$ because the sensible surface flux $SH$ and
net surface radiation $R_{sfc}$ matter.

The surface energy budget suggests the existence of an approximate
limit to the magnitude increases in global-mean precipitation due to
increased greenhouse gas concentration. Under greenhouse gas warming,
the air--sea temperature difference at the surface generally
decreases. In the limit of no air--sea temperature difference, the net
surface longwave radiation and sensible surface fluxes vanish, which
leaves the cooling from latent surface fluxes to balance the warming
from net solar radiation at the surface. This limit is approximately
attained in idealized GCM simulations
\citep[Fig.~\ref{fig-globalprecip}a,][]{ogorman-schneider-2008}. However,
sensible surface fluxes may reverse sign (from cooling the surface to
warming it), so the warmest simulation in Fig.~\ref{fig-globalprecip}a
exceeds the precipitation expected from all of the surface solar
radiation being being converted into evaporation.  The sensitivity
of precipitation to greenhouse gas changes has implications for the
rate of silicate weathering (see section~\ref{climate})
\citep{pierrehumbert-2002,lehir-etal-2009}.

\begin{figure*}[!tb]
\begin{center}
  \noindent\includegraphics[width=30pc,angle=0]{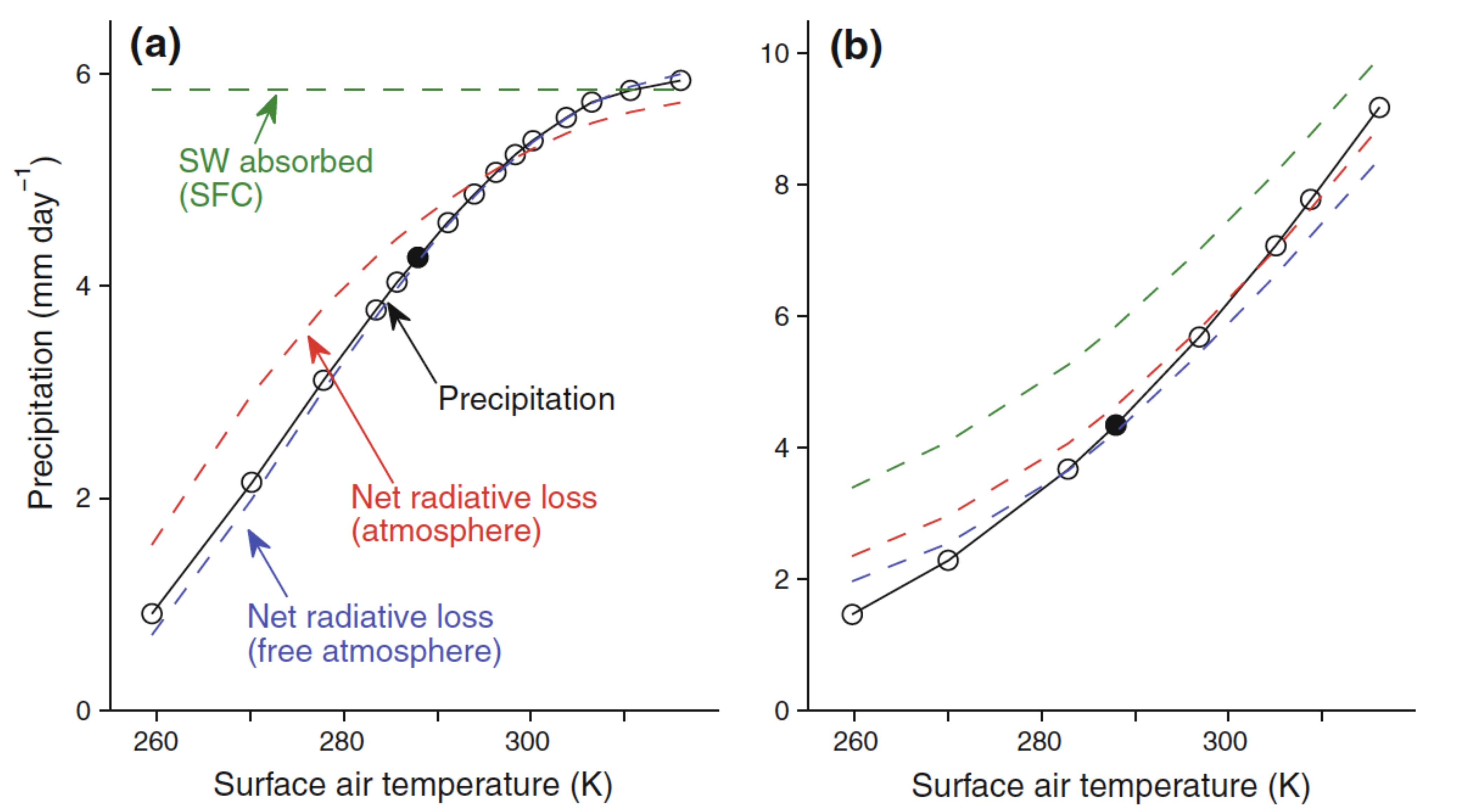}
\end{center}
\caption{ From \citet{ogorman-etal-2011}: Global-mean precipitation
  (solid line with circles) versus global-mean surface air temperature
  in two series of statistical-equilibrium simulations with an
  idealized GCM in which (a) the optical depth of the longwave absorber
  is varied and (b) the solar constant is varied (from about $800 \,
  \mathrm{W \, m^{-2}}$ to about $2300 \, \mathrm{W \, m^{-2}}$). The
  filled circles indicate the reference simulation (common to both
  series) which has the climate most similar to present-day
  Earth's. The red dashed lines show the net radiative loss of the
  atmosphere, the blue dashed lines show the net radiative loss of the
  free atmosphere (above $\sigma = p/p_s = 0.86$), and the green dashed lines
  show the net absorbed solar radiation at the surface (all in
  equivalent precipitation units of $\mathrm{mm \, day^{-1}}$).  }
\label{fig-globalprecip}
\end{figure*}

The energetic perspectives on global precipitation illustrates how the
rate of change of global precipitation and atmospheric humidity can
differ.  If the relative humidity does not change, the atmospheric
humidity increases with temperature at the rate given by the
Clausius-Clapeyron relation \eqref{eqn-CC}. In contrast, the global
precipitation depends on radiative fluxes such as the
top-of-atmosphere insolation and longwave radiation.  The radiative
fluxes depend in part on the atmospheric humidity through water
vapor's absorption of shortwave and longwave radiation (which gives
rise to the water vapor feedback), but this dependence does not constrain
the radiative changes to vary with the water vapor concentration at
the rate given by \eqref{eqn-CC}.  Precipitation and humidity have
different dimension, as well: precipitation and evaporation are fluxes
of water, whereas the atmospheric humidity is a concentration of
water. The suggestion that precipitation changes in proportion to
atmospheric humidity has been termed the ``saturation fallacy''
\citep{pierrehumbert-etal-2007}.

Another result that can be understood by considering the energy
budget is that there is not a unique relationship between
the global mean temperature and precipitation.
For example, in geoengineering schemes that offset warming due
to increased CO$_2$ by increasing albedo (reducing the solar radiation),
two climates with the same global mean temperature may 
have differing global mean precipitation because the sensitivity
of global precipitation per degree temperature change are not equal
for different types of radiation changes 
\citep{bala-etal-2008,ogorman-etal-2011}.
Figure~\ref{fig-globalprecip} shows that warming forced by solar
constant changes results in a larger increase in global-mean
precipitation than warming forced by changes in the longwave
optical depth in an idealized GCM \citep{ogorman-etal-2011}.

While the global mean precipitation is potentially a variable of
interest in characterizing a planet's habitability, regional
precipitation variations are substantial. Earth has regions of dry
deserts and rainy tropical regions, where the annual-mean
precipitation differs by a factor of $\about 20$. 
The regional precipitation is far from the global mean
because of water vapor transports by the atmospheric circulation.
Furthermore, the sensitivity of regional precipitation to external
perturbations can be of opposite sign as the global mean changes
\citep[for example, some regions may dry as the global mean precipitation
increases][]{chou-neelin-2004, held-soden-2006}.

\subsection{Regional Precipitation}
\label{regional-precip}


\begin{figure*}[!tb]
\begin{center}
  \noindent\includegraphics[width=30pc,angle=0]{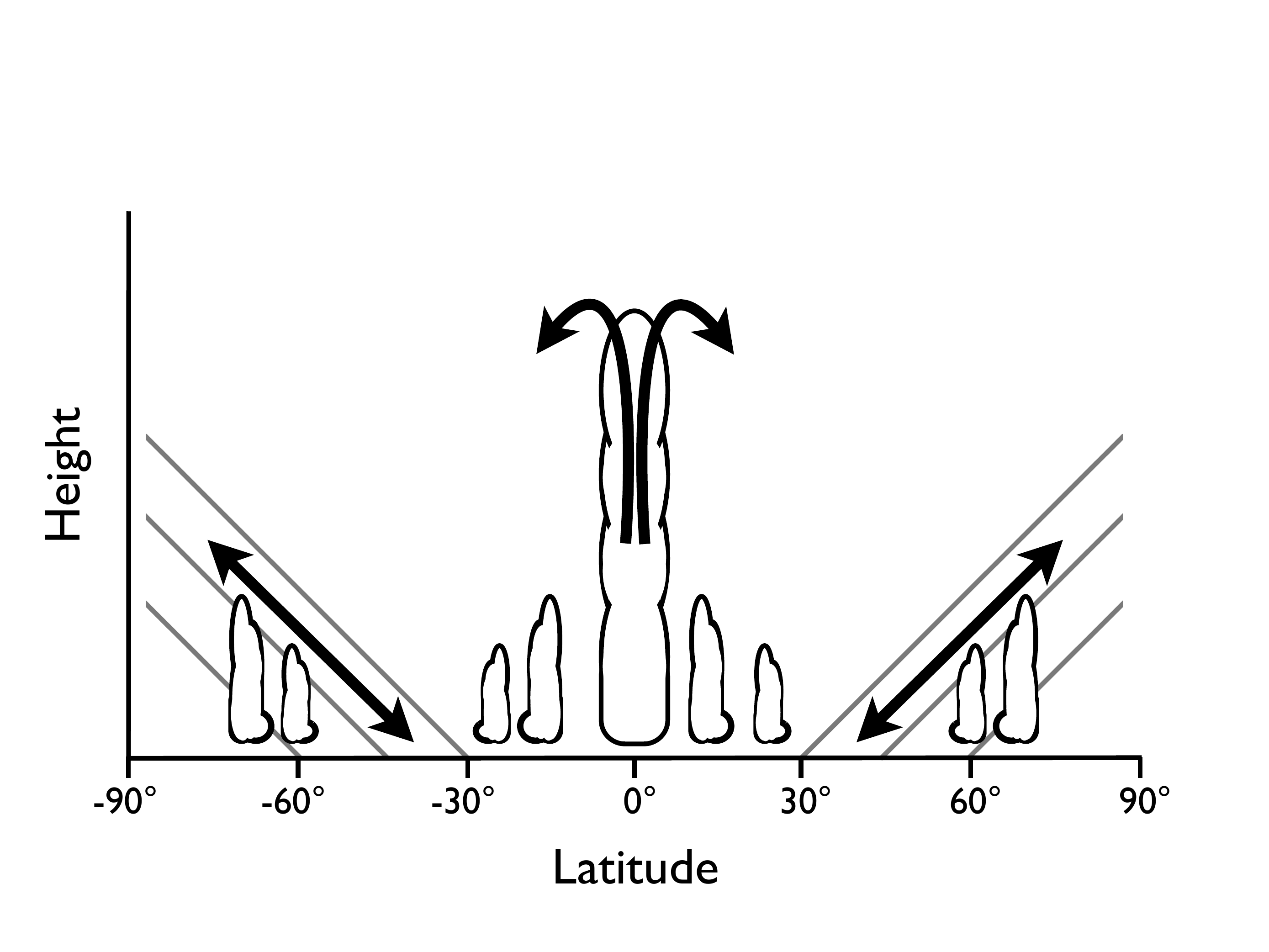}
\end{center}
\caption{
 Processes leading to condensation and affecting atmospheric humidity
 for Earth-like planets, based on Figure 8 of \citet{held-soden-2000}. 
 Gray lines indicate surfaces of constant entropy which air parcels
 approximately follow.  In the tropics, the ascending branch of
the Hadley cell leads to significant condensation and rainfall,
whereas the descending branches (at latitudes of $\sim$20--$30^{\circ}$
on Earth) exhibit less condensation and fewer clouds.  In the 
extratropics, the rising motion in baroclinic eddies transporting
energy poleward and upward (indicated schematically with sloping
lines) leads to significant condensation.
}
\label{fig-sketch}
\end{figure*}

To understand the variations in regional precipitation, we first
illustrate the processes that give rise to condensation
(Fig.~\ref{fig-sketch}).  In tropical latitudes of Earth-like planets,
there are time-mean regions of ascent where moist air is continually
brought from the surface to the colder upper troposphere which leads
to supersaturation and condensation (near the equator in
Fig.~\ref{fig-sketch}). There are time-mean regions of subsidence
where dry air from the upper troposphere descends toward the surface,
so the mean circulation leads to subsaturation and precipitation
occurs during transient events, when it does occur (near $30^\circ$ in
Fig.~\ref{fig-sketch}, corresponding to the descending branches of the
Hadley cells).  In the extratropical regions of Earth-like planets,
air parcels approximately follow surfaces of constant entropy which
slope upward and poleward. As parcels ascend in the warm, poleward
moving branches of extratropical cyclones (i.e., transient baroclinic
eddies like those described in Section~\ref{baroclinic}), they become
supersaturated and condense in the colder regions of the atmosphere.
This is illustrated in Fig.~\ref{fig-sketch}, where gray lines
indicate entropy surfaces and black arrows indicate the trajectories
of air masses.  This description is focused on mean meridional
circulations in the tropics and transient eddies in the
extratropics. Superimposed on these processes affecting the zonal-mean
precipitation are stationary eddies (that is, eddy structures that are
fixed rather than traveling in longitude) or zonally asymmetric mean
flows that give rise to important longitudinal variations in
precipitation on Earth.

While the zonal mean is an excellent starting point for Earth's
climate and atmospheric circulations, stationary eddies or zonally
asymmetric mean flows are of central importance for tidally locked
planets. The mean circulation aspect of the description will have an
east--west component: the near-surface branch of the circulation will
flow from the night side to the day side, ascend and precipitate
there, and return to the night side aloft
\citep{merlis-schneider-2010}.  One can rotate the time-mean
circulation in Fig.~\ref{fig-sketch} from a north-south orientation
(Hadley circulation) to an east-west orientation (Walker circulation)
to conceptualize the effect of the night-side to day-side circulations
on precipitation; however, the factors controlling the circulations
themselves may differ as their angular momentum balances differ (the
extent to which the circulation strength is slaved to the momentum
balance or responds directly to radiation).  In contrast, transient
eddies (that is, eddies that travel in longitude)
exist because temperature gradients are not aligned with the
planet's spin axis, which gives rise to baroclinic instability
(Section~\ref{baroclinic}), so the conceptual picture cannot simply be
rotated. The affect of zonal asymmetry of background state on
baroclinic eddies is not settled; in particular, the extent to which
stationary eddies and baroclinic eddies are separable (i.e., the
extent to which they are linear) is a research area
\citep{held-etal-2002,kaspi-schneider-2011}.

\begin{figure*}[!tb]
\begin{center}
  \noindent\includegraphics[width=30pc,angle=0]{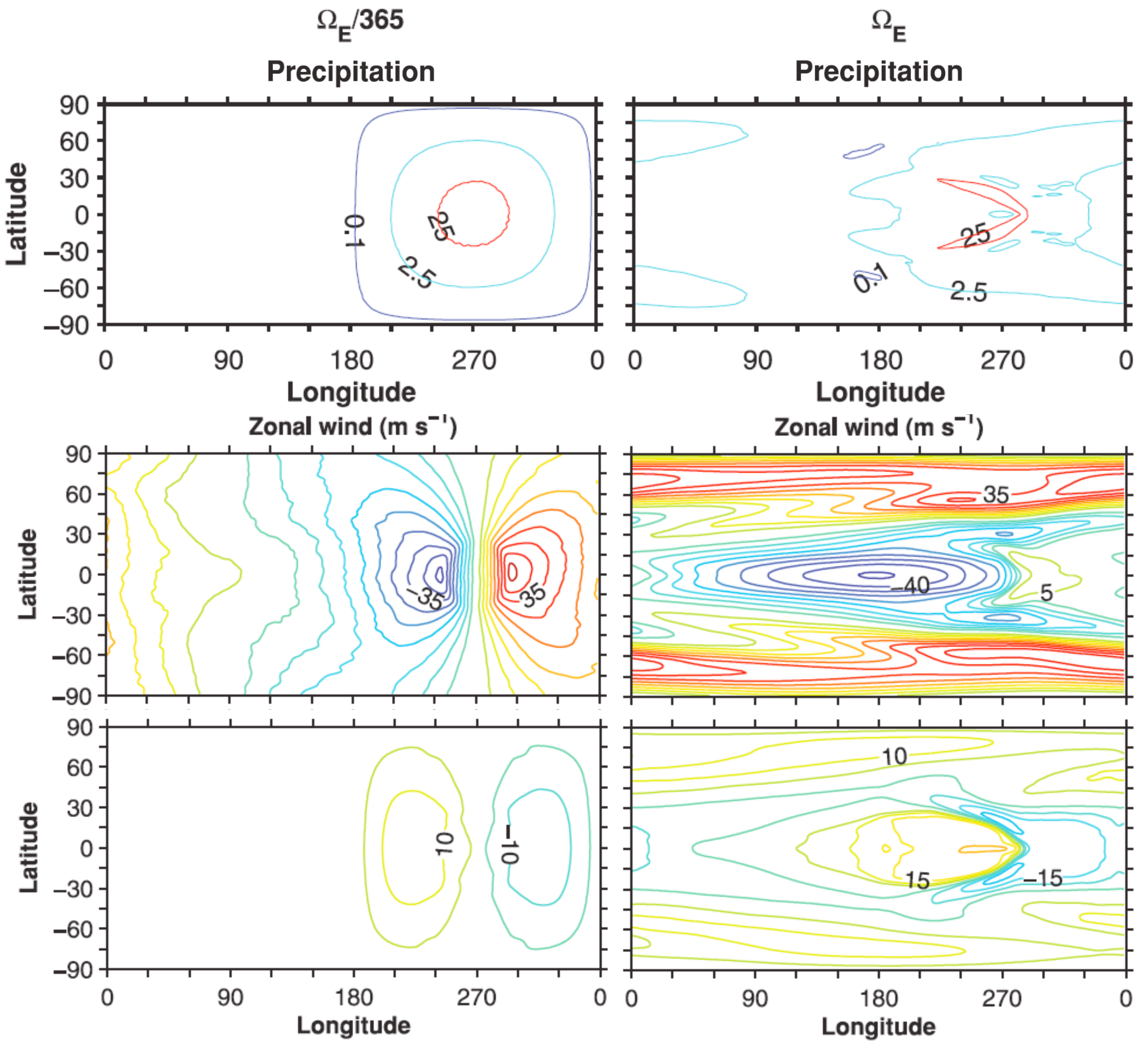}
\end{center}
\caption{ Time-mean circulation in two GCM simulations of Earth-like
  synchronously rotating exoplanets; models with rotation periods of
  one Earth year and one Earth day, respectively, are shown on the
  left and right.  Top row shows time-mean precipitation (contours of
  $0.1$, $2.5$, and $25.0 \, \mathrm{mm \, day^{-1}}$).  Middle and
  bottom rows show time-mean zonal wind on the $\sigma = p/p_s = 0.28$
  model level and at the surface, respectively (contour interval is $5
  \, \mathrm{m \, s^{-1}}$; the zero contour is not displayed).  The
  subsolar point is fixed at $270^\circ$ longitude in both models.
  Adapted from Figs.~2,4,11 of \citet{merlis-schneider-2010}.  }
\label{merlis_composite}
\end{figure*}

Simulations of Earth-like tidally locked exoplanets from
\citet{merlis-schneider-2010} illustrate the importance of zonally
asymmetric (east--west) mean atmospheric circulations in determining
precipitation (Fig.~\ref{merlis_composite}).  For both slowly and
rapidly rotating simulations, the surface zonal wind converges
[$\partial_x [u] < 0$, with the $[\cdot]$ indicating a
time-mean] on
the day-side near the subsolar point, ascends, and diverges
($\partial_x [u] > 0$) in the upper troposphere.  The night
side features convergent mean zonal winds in the upper troposphere,
subsidence, and divergent mean zonal winds near the surface.  These
are thermally direct circulations that transport energetic air from
the day side to the night side.  The water vapor is converging near
the surface on the day side where there is substantial precipitation.
There is a clear connection between the pattern of convergence in the
surface winds and precipitation---the rapidly rotating case, in
particular, has a crescent-shaped precipitation field.  The shape of
the convergence zone is similar to that of the Gill model
\citep{gill-1980, merlis-schneider-2010}.

To be quantitative about regional precipitation, consideration of the
water vapor budget, a conservation equation for the atmospheric
humidity, can be useful.  For timescales long enough that the humidity
tendency is small, the time-mean (denoted $[\cdot]$) net
precipitation, $[ P - E ]$, is balanced by the mass-weighted vertical
integral (denoted $\{ \cdot \}$) of the convergence of the water vapor
flux, $-\nabla \cdot [ {\bf u} \, q ] \,$:
\begin{equation}
[ P - E ] = - \int_0^{p_s} \! \nabla \cdot
[ {\bf u} \, q ] \; \frac{\mathrm{d}p}{g} \, =
- \nabla \cdot \{ [ {\bf u} \, q ] \},
\label{eqn-wv_budget}
\end{equation}
with horizontal wind vector ${\bf u}$ and specific humidity $q$.
The atmospheric circulation converges water vapor into
regions of net precipitation ($P>E$) from regions of net evaporation
($E>P$), where there is a divergence of water vapor.
The water vapor budget makes explicit that the atmospheric circulation
is fundamental in determining this basic aspect of climate.

The small scale height of water vapor in Earth-like
atmospheres emphasizes the near-surface water
vapor concentration, so a conceptually useful approximation to
\eqref{eqn-wv_budget} is to consider the near-surface region of the
atmosphere.  For example, the precipitation distribution follows the
pattern of the convergence of the surface wind in Fig.~\ref{merlis_composite}.

Note that the surface temperature does not directly appear in
\eqref{eqn-wv_budget}. In spite of this, there have been many attempts
to relate precipitation to surface temperature for Earth's climate.
The surface temperature may vary together with aspects of climate that
do directly determine the amount of precipitation (e.g., high
near-surface humidity or the largest magnitude near-surface
convergence of the horizontal winds may occur in the warmest regions
of the tropics).  There are conceptual models that relate the surface
temperature to pressure gradients that determine regions of
convergence through the momentum balance \citep{lindzen-nigam-1987,
  back-bretherton-2009a}.  Likewise, parcel stability considerations
(e.g., where is there more convective available potential energy) may
be approximately related to regions of anomalously high surface
temperature \citep{sobel-2007, back-bretherton-2009b}.

\subsection{Relative humidity}
\label{humidity}

Reviews by \citet{pierrehumbert-etal-2007} and
\citet{sherwood-etal-2010a} discuss the processes affecting Earth's
relative humidity and its sensitivity to climate changes.

The atmosphere of a planet is generally subsaturated---the relative
humidity is less than $100\%$.  The relative humidity of the
atmosphere depends on the combined effect of inhomogeneities in
temperature and the atmospheric circulation.

For Earth-like planets, the atmospheric circulation produces dry air
by advecting air upwards and/or polewards to cold temperatures where
the air becomes supersaturated and water vapor condenses
(Fig.~\ref{fig-sketch}). When this air is advected downward or
equatorward, the water vapor concentration will be subsaturated with
respect to the local temperature.

The nonlocal influence of the temperature field on the 
atmospheric humidity through the atmospheric circulation
can be formalized by the tracer of last saturation paradigm
\citep{galewsky-etal-2005, pierrehumbert-etal-2007}.
In this framework, the humidity of a given region in the atmosphere is 
decomposed into the saturation humidity at the non-local (colder)
region of last saturation, with weights given by the probability
that last saturation occurred there.

In addition to the adiabatic advection, convection affects the
relative humidity and can act to moisten or dry the atmosphere
\citep[e.g.,][]{emanuel-1994}.  In subsiding regions of Earth's
tropics, convection moistens the free troposphere to offset the
substantial drying associated with subsidence from the cold and dry
upper troposphere \citep[e.g.,][]{couhert-etal-2010}.  Therefore, both
the water vapor advection by the large-scale circulations and
small-scale convective processes, with the latter
parameterized in GCMs, are important in determining the relative
humidity.

These relative humidity dynamics are important for various aspects of
planetary atmospheres: they influence the radiation balance,
atmospheric circulations, and precipitation. The processes controlling
the relative humidity are tied directly to three-dimensional
atmospheric circulations, which is a difficulty of single column
models. Single column models do not explicitly simulate the
atmospheric eddies or mean circulations that advect water and set the
global-mean relative humidity. Furthermore, spatial variations in
relative humidity are radiatively important. For example, the same
global-mean relative humidity with a different spatial distribution
has a different global-mean radiative cooling
\citep{pierrehumbert-1995, pierrehumbert-etal-2007}.  Therefore, the
magnitude the day-side to night-side relative humidity contrasts of
tidally locked exoplanets may be important in determining the global
climate.

For greenhouse gas-forced climate changes on Earth, relative humidity
changes ($< 1\% \, \mathrm{K}^{-1}$) are typically smaller than
specific humidity changes ($\about 7 \% \, \mathrm{K}^{-1}$)
\citep{held-2000, sherwood-etal-2010b}. Therefore, it is a useful
state variable for climate feedback analysis that avoids strongly
canceling positive water vapor feedback and negative lapse rate
feedback found in the conventional feedback analysis that uses
specific humidity as a state variable \citep{held-shell-2012}. Climate
feedback analysis depends on the mean state, which clearly differs
between tidally locked exoplanets and Earth-like planets. A climate
feedback analysis has not been performed for a tidally locked planet
to diagnosing the relative contributions of the water vapor, lapse
rate, albedo, and cloud feedbacks to the climate sensitivity.

\subsection{Clouds}
\label{clouds}

The formation of clouds from the phase change of condensible
species---like water vapor---is often associated with ascending
motion. Ascent may be the result of convergent horizontal winds or
convective instability, so clouds typically have small space and time
scales. Therefore, the spatial distribution of clouds and their
sensitivity to external parameters such as the rotation rate or solar
constant depend on atmospheric turbulence of a variety of scales: from
the scales of the global circulations ($\about 1000 \, \mathrm{km}$)
down to those of moist convection and boundary layer turbulence
($\about 10 \, \mathrm{km}$ and smaller).

Clouds can affect the radiation balance substantially.  To illustrate,
we consider an Earth-like case, although other possibilities exist
\citep[e.g.,][]{wordsworth-etal-2010}.  First,
consider the longwave component of the radiation and assume that
clouds are completely opaque in that region of the spectrum.  If a
cloud is low in the atmosphere (near the surface), the temperature at
the top of the cloud, from where the radiation is re-emitted after
absorption, will be close to the temperature at which the radiation
would be emitted in the absence of any clouds (e.g., from the surface
or the clear air near the surface). In this case, the cloud will not
substantially affect the longwave radiation relative to cloud-free
conditions.  If a cloud is high in the atmosphere (near the
tropopause, for example), the temperature at the top of the cloud will
be substantially cooler than the emission temperature in the absence
of clouds.  So, high clouds enhance the atmospheric greenhouse by
absorbing longwave radiation emitted at the higher temperatures of
atmosphere and surface below and re-emitting longwave radiation at the
colder cloud temperature.  Simply characterizing the effect of clouds
on the shortwave radiation depends more subtly on the optical
characteristics. There can be a substantial reflection (e.g., cumulus
or stratus) or little reflection (e.g., cirrus) of shortwave
radiation. Additionally, the effect of clouds on the planetary albedo
depends on the surface albedo; for example, in a Snowball-Earth state
(Section~\ref{snowball-section}), a planet may be largely ice covered,
in which case clouds will not significantly modify the albedo.  Considering
all these effects, the net radiative effect of clouds (the
sum of shortwave and longwave) can be warming, cooling, or close to
neutral.  In the current Earth climate, these three possibilities are
manifest with cirrus, stratocumulus, and convecting cumulus clouds,
respectively.

The cloud radiative forcing (the difference in top-of-atmosphere net
radiation between cloudy and clear conditions) on Earth is negative: 
the Earth would be warmer without clouds.
This does not imply that clouds are expected to damp (negative
radiative feedback) rather than amplify (positive radiative feedback)
radiative perturbations.  In comprehensive general circulation model
projections of climate change, the cloud feedback is on average
positive (i.e., the change in cloud properties amplifies the radiative
perturbation from CO$_2$ changes) \citep{soden-held-2006}. However,
there is substantial uncertainty in the magnitude, and the physical
basis of this projection, if there is one, is still being elucidated
\citep[e.g.,][]{bony-etal-2006,zelinka-hartmann-2010}.

In addition to their important role in determining climate through the
radiation balance, clouds likewise affect remote (e.g., space-based)
observations of planets.  Interpreting cloud observations may help
reveal aspects of the atmospheric circulation, such as the
characteristic space and timescales of a planet's ``weather'' (e.g.,
cloud tracking on Jupiter), and climate (e.g., clouds on Titan
indicate regions of precipitation).  Preliminary efforts are being made
to explore the effects of clouds on lightcurves and spectra of terrestrial
exoplanets  \citep[e.g.][]{palle-etal-2008,
cowan-etal-2009, fujii-etal-2011, robinson-etal-2011, kawahara-fujii-2011,
fujii-kawahara-2012, sanroma-palle-2012, gomez-leal-etal-2012,
karalidi-etal-2012}.

%

\subsection{The hydrological cycle's effect on atmospheric temperature 
and circulations}
\label{moist-circulation}

While much of our discussion has focused on the role of atmospheric
circulations in shaping the distributions of precipitation, humidity,
and clouds, water vapor is not a passive tracer and can affect the
atmospheric temperature and winds\footnote{Condensible species also
  affect the atmospheric mass, which is a minor effect on Earth, but
  is important on Mars and is potentially important on some
  exoplanets.  See Section~\ref{sec:collapse}.}.  Here, we outline some ways in which the latent heat
release of condensation affects the atmosphere's thermal structure and
water vapor influences the energetics of atmospheric circulations.
Additional material may be found in reviews by \citet{sobel-2007} and
\citet{schneider-etal-2010}.

\subsubsection{Thermal structure}

The atmosphere's stratification in low latitudes (or slowly rotating
regions more generally) is largely determined by convection. The
tropical atmosphere is close to a moist adiabat on Earth
\citep[e.g.,][]{xu-emanuel-1989}, although entrainment and ice-phase
processes affect the upper troposphere \citep{romps-2010}. Simulations
of Earth-like tidally locked planets \citep{merlis-schneider-2010}
also have near moist adiabatic stratification over large regions of
the day side.  The influence of convecting regions is nonlocal:
gravity wave propagate away density variations (see
Section~\ref{wave-adjustment}), whereas gradients can be rotationally
balanced in the extratropics (more generally, in regions where the
Coriolis force is important in the horizontal momentum balance).  The
relative importance of moist convection in determining the tropical
stratification will be mediated by the geography of condensate
reservoirs (e.g., Titan or possible dry exoplanets) and the vertical
distribution of radiatively active agents (e.g., Venus and Titan).
The convective lapse rate is the starting point for theories of the
depth of the troposphere (i.e., the height of the tropopause) in low
latitudes \citep{held-1982, thuburn-craig-2000, schneider-2007}, which
may have implications for the interpretation of observed radiance.

Latent heat release also affects the stratification of the
extratropical atmosphere. The warm and moist sectors of extratropical
cyclones condense as they move poleward and upward
(Fig.~\ref{fig-sketch}).  The vertical component of these transports
is important in determining the lapse rate in the
extratropics. Accounting for this interaction between extratropical
cyclones and condensation or convection is an on-going area of
research for Earth-like planets
\citep{juckes-2000,frierson-etal-2006,ogorman-2011}.

In addition to the importance of the latent heat of condensation in
determining the atmosphere's vertical thermal structure, it helps
determine the horizontal temperature gradients.  On Earth, about half
of the meridional atmospheric energy transport in midlatitudes is in
the form of water vapor \citep[e.g.,][]{trenberth-caron-2001}.  This
is primarily accomplished by transient baroclinic eddies. The
transient eddy moisture flux can be parameterized in terms of mean
humidities and characteristic eddy velocities
\citep{pierrehumbert-2002, ogorman-schneider-2008,
  caballero-hanley-2012}. In the low latitudes of Earth, the mean
meridional circulation transports water vapor equatorward. This
thermally indirect latent energy flux is a consequence of the small
scale height of water vapor. The dry component of the energy flux is
larger and poleward, so the total energy transport by the Hadley
circulation is thermally direct.  For Earth-like exoplanets, the
energy transports by the day-side to night-side circulation will have
a similar decomposition---a thermally indirect latent component and a
thermally direct dry component \citep{merlis-schneider-2010}.

\subsubsection{Circulation energetics}

Latent heat release affects atmospheric circulations in a number of
subtle ways. On the one hand, latent heat release has the potential
to directly energize atmospheric circulations. On the other hand, the
effect of latent heat release on the thermal structure of the
atmosphere may weaken atmospheric circulations by increasing the
static stability or decreasing horizontal temperature gradients.

The quantification of the kinetic energy cycle of the atmosphere is
typically based on the dry thermodynamic budget, which is manipulated
to form the Lorenz energy cycle \citep{lorenz-1955,
  peixoto-oort-1992}. Latent heat release appears as an energy source
in these budgets.  For Earth's atmosphere, the diabatic generation of
eddy kinetic energy from latent heat release is up to $\about 1 \,
\mathrm{W \, m^{-2}}$ \citep{chang-2001,chang-etal-2002}, which is a
small fraction of the $\approx 80 \, \mathrm{W \, m^{-2}}$ global-mean
latent heat release \citep[e.g.,][]{trenberth-etal-2009}. This low
efficiency of converting latent heat into kinetic energy is consistent
with the overall low efficiency of the kinetic energy cycle
\citep{peixoto-oort-1992}.  There are also alternative formulations of
the kinetic energy cycle that incorporate moisture into the definition
of the available potential energy \citep{lorenz-1978, lorenz-1979,
  ogorman-2010}.

The energy transport requirements can play a role in determining the
strength of the tropical circulations. 
For example, in angular momentum-conserving Hadley cell theories,
the circulation energy transport is constrained to balance the
top-of-atmosphere energy budget \citep{held-hou-1980,held-2000}.
Such an energy transport requirement does not directly determine 
the circulation mass transport, however. The ratio of
the energy to mass transport is a variable known as
 the gross moist stability \citep[other related
  definitions exist,][]{raymond-etal-2009}.  The gross moist stability
expresses that the energy transport of tropical circulations results
from the residual between the largely offsetting thermally indirect
latent heat transport and the thermally direct dry static energy.
Therefore, the gross moist stability is much smaller than the dry
static stability for Earth-like atmospheres.  The gross moist
stability is an important, though uncertain, parameter in theories of
tropical circulations \citep{neelin-held-1987, held-2000, held-2001,
  sobel-2007, frierson-2007, merlis-etal-2013a}.

\section{ATMOSPHERIC CIRCULATION AND CLIMATE}
\label{climate}

On any terrestrial planet, the influence of the atmospheric
circulation on the climate is fundamental. On Earth, climate is
generally understood as the statistics (long-term averages and
variability) of surface pressure, temperature and wind, along with
variations in the atmospheric content of water vapor and
precipitation of liquid water or ice \citep{hartmann-1994-book}.
For terrestrial bodies in
general, this definition can readily be extended to include the
atmospheric transport of other condensable species, as in the cases of
the CO$_2$ cycle on Mars or the methane cycle on Titan.

The close interaction between atmospheric dynamics and climate is
readily apparent if we examine even the most basic features of the
inner planets in our solar system. For example, Venus has a mean
surface temperature of around 735~K, with almost no variation between
the equator and the poles. In contrast, Mars' mean surface temperature
is only 210~K, but in summer daytime temperatures can reach 290~K,
while in the polar winter conditions are cold enough for atmospheric
CO$_2$ to condense out as ice on the surface. Nonetheless, annual mean
temperatures near the top of Olympus Mons are barely different from
those in Mars' northern lowland plains. On Earth, surface temperature
variations lie between these two extremes---the variation in climate
with latitude is significant, but surface temperatures also rapidly
decrease with altitude due to thermal coupling with the
atmosphere. The differences in insolation, atmospheric pressure and
composition (92 bar CO$_2$ for Venus vs. around 1~bar N$_2$/O$_2$ for
Earth and 0.006 bar CO$_2$ for Mars) go a long way towards explaining
these gross climatic differences. In particular, from planetary
boundary layer theory it can be shown that the sensible heat exchange
between the surface and atmosphere is
\begin{eqnarray}
F_{\rm sens} &=& \overline{\rho_a w'(c_p T')} \nonumber \\
         &\simeq& c_p \rho_a C_D \|{\bf v}_{\rm surf}\| (T_{\rm surf} - T_a)\label{eq:PBL}
\end{eqnarray}
where the overbar denotes a time or spatial mean, $w'$ and $T'$ are
deviations from the mean vertical velocity and temperature, $T_{\rm
  surf}$ is surface temperature, $C_D$ is a drag coefficient, $c_p$ is
the specific atmospheric heat capacity, and $\rho_a$, $\|{\bf v}_{\rm
  surf}\|$ and $T_a$ are the density, mean wind speed and temperature
of the atmosphere near the surface \citep{pierrehumbert-2010}.
Because $F_{\rm sens}\propto
\rho_a$, the thermal coupling between the atmosphere and surface will
generally increase with the atmospheric pressure. Given the tendency
for dense, slowly rotating atmospheres to homogenize horizontal
temperature differences rapidly (see Section 2.2), it should be
intuitively clear why on planets like Venus, variations in annual mean
temperature as a function of latitude and longitude tend to be
extremely weak.

In detail, the picture can be more complex, because $\|{\bf v}_{\rm
  surf}\|$ is also a function of the atmospheric composition and
insolation pattern, and $C_D$ depends on both the surface roughness
and degree of boundary layer stratification. Beyond this, there are
many other effects that couple with atmospheric dynamics. As discussed
in Section \ref{hydrological}, a key additional process on Earth is
the hydrological cycle, which affects climate through latent heat
transport and the radiative properties of water vapour and water
ice/liquid clouds.  Examples from other Solar System planets include
the dust and CO$_2$ cycles on Mars, and the methane/hydrocarbon haze
cycle on Titan. These processes, which can be fascinatingly complex,
are a large part of the reason why detailed long-term predictions of
climate are challenging even for the relatively well-observed planets
of the Solar System.

Given these complexities, it is unsurprising that for terrestrial
exoplanets, study of the coupling between atmospheric circulation and
climate in generalized cases is still in its infancy. The present
state of knowledge is still heavily based on limited exploration of a
few Earth-like or Earth-similar cases. Rather than attempting an
overview of the entire subject, therefore, here we review a small
selection of the many situations where atmospheric dynamics is
expected to have a key influence on planetary climate.

\subsection{Influence of dynamics on the runaway H$_2$O greenhouse}
\label{sec:runaway}

\begin{figure*}
 \epsscale{1.2}
\plotone{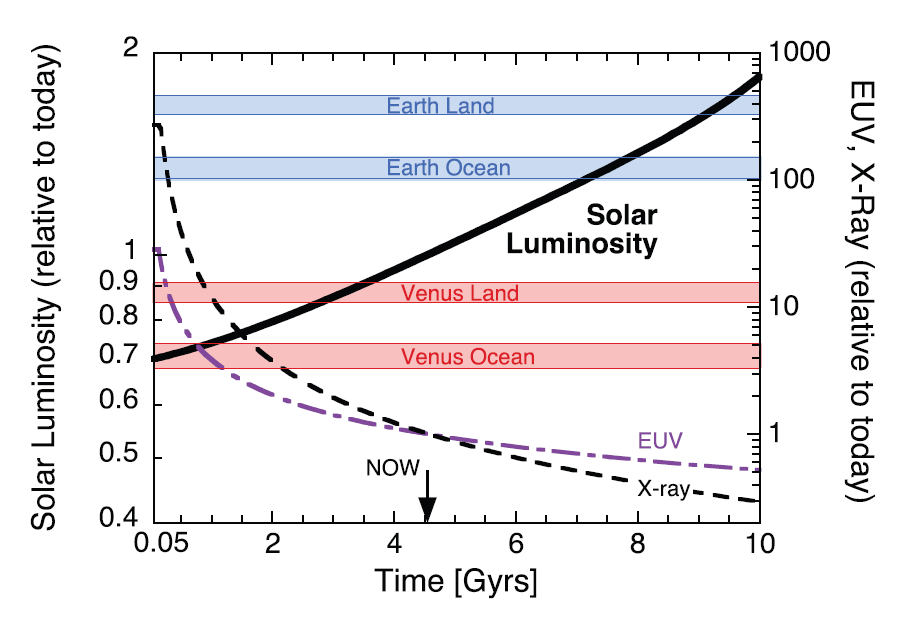}
 \caption{Solar luminosity as a function of time, alongside
 runaway H$_2$O greenhouse thresholds for land and aqua
   planets at Venus and Earth's orbits \citep[from][]{abe-etal-2011}. With a
   depleted water inventory, Venus could conceivably have remained a habitable
   planet until as little as 1~billion years ago.}
\label{runaway-greenhouse}
 \end{figure*}

The well-known runaway H$_2$O greenhouse effect occurs due to the
feedback between surface temperature and atmospheric infrared opacity
on a planet with surface liquid water. Above a given threshold for the
incoming stellar flux $F_0>F_{limit}$ (see Fig.\ref{runaway-greenhouse}),
the thermal radiation leaving the planet no
longer depends on the surface temperature, and water will continue to
evaporate until all surface sources have disappeared
\citep{kombayashi-1967,ingersoll-1969}.  Classical runaway greenhouse
calculations \citep[e.g.,][]{kasting-etal-1993} were performed in 1D
and assumed homogenous atmospheres with either 100\% relative humidity
or a fixed vertical profile.  In reality, however, the variations in
relative humidity with latitude due to dynamical processes should have
a fundamental effect on the planet's radiative budget and hence on the
critical stellar flux at which the transition to a runaway greenhouse
occurs.  The global variations of relative humidity in Earth's
atmosphere are strongly dependent on the ascending and descending
motion of the Hadley cells (Section~\ref{hadley}) and subtropical
mixing due to synoptic-scale eddies (Section~\ref{baroclinic}). Even
on Earth today, without heat transport to higher latitudes by the
atmosphere and ocean, the tropics would most likely be in a runaway greenhouse
state\footnote{Assuming, that is, that the negative feedback from
  cloud albedo increases is not so strong as to overwhelm the increase
  in solar forcing past the conventional runaway greenhouse
  threshold.}. \citet{pierrehumbert-1995} proposed that regulation of
mean tropical sea surface temperatures is in fact governed by a
balance between the radiative heating in saturated upwelling regions
and radiative cooling by ``radiator fins'' in larger regions of net
subsidence.

Very few researchers have yet taken on the challenge of modeling the
transition of an ocean planet to a runaway greenhouse state in a 3D
general circulation model (GCM).
\citet{ishiwatari-etal-2002} studied the
appearance of the runaway greenhouse state in an idealized 3D
circulation model with gray-gas radiative transfer. Their results
roughly corresponded to those found in 1D using a global mean relative
humidity of 60\%. However, the simplicity of their radiative scheme,
neglect of clouds and problems arising from the need for strong
vertical damping near the runaway state meant that they were unable to
constrain the greenhouse transition for cases such as early Venus
quantitatively. Later, \citet{ishiwatari-etal-2007} compared results
using a GCM with those from a 1D energy balance model (EBM). EBMs
are intermediate-complexity models that replace vertically resolved radiative
transfer with empirical functions for the fluxes at the top of the atmosphere
and represent all latitudinal heat transport processes by a simple 1D diffusion
equation. \citet{ishiwatari-etal-2007} found
that at intermediate values of the solar constant, multiple climate solutions
including runaway greenhouse, globally and partially glaciated states were
possible. Clearly, comparison of these results with simulations using a GCM
specifically designed to remain physically robust in the runaway limit would
be an interesting future exercise. As well as taking into account the non-gray
radiative transfer of water vapour, such a model would need to account for the
locally changing mass of water vapour in the large-scale dynamics and subgrid-scale
convection, and include some representation of the radiative effects of clouds.

The importance of clouds to the runaway greenhouse limit is central,
but our understanding of how they behave as the solar constant is
increased is still poor. \citet{kasting-1988} performed some
simulations in 1D with fixed water cloud layers and concluded that
because of their effect on the planetary albedo, Venus could easily
have once been able to maintain surface liquid water. However, they
did not include dynamical or microphysical cloud effects in their
model.  Going further, \citet{renno-1997} used a 1D model with a range
of empirical parameterizations of cumulus convection to study the
runaway transition, although with simplified cloud radiative transfer.
Perhaps unsurprisingly, they found that surface temperature depends
sensitively on cloud microphysical assumptions when solar forcing is
increased.  They also found that in general, mass flux schemes (i.e.,
convective parameterizations that capture the effect of environmental subsidence
associated with cumulus convection) caused a more rapid runaway
transition than adjustment schemes (i.e., convective parameterizations
that simply represent the effect of convection as a modification of
the local vertical temperature profile). Future 3D GCM studies of the
runaway greenhouse will be forced to confront these uncertainties in
sub-gridscale physical processes.

Cases where the initial planetary content of water is limited are also
of interest for the runaway greenhouse. \citet{abe-etal-2011} studied
inner habitable zone limits for ``land planets'', which they defined
as planets with limited total H$_2$O inventories (less than 1~m global
average liquid equivalent in their model) and no oceans. They showed than in these
cases, the tendency of the limited water inventory to become trapped
at the poles causes extreme drying of equatorial regions. This allows
the global mean outgoing longwave radiation to continue increasing
with temperature, delaying the transition to a runaway greenhouse
state. The critical average solar flux values they calculated in their
model were $415\W \m^{-2}$ and $330\W \m^{-2}$ for land and aqua
(global ocean) planets, respectively. Their results suggest planets
with limited water inventories could remain habitable much closer to
their host stars than the classical inner edge of the habitable zone
suggests. Because the Sun's luminosity has increased with time, their
results also suggest that a mainly dry early Venus could have stayed
cool enough to maintain regions of surface liquid water until as
recently as 1 billion years ago (see Figure \ref{runaway-greenhouse}).
However, scenarios where liquid water was never present on the surface
of Venus are also possible, and have been argued to be more consistent
with Ne/Ar isotopic ratios and the low oxygen content in the
present-day atmosphere \citep{gillmann-etal-2009,chassefiere-etal-2012}.

The insights of \citet{abe-etal-2011} on the importance of relative
humidity to the runaway greenhouse have recently been extended to the
case of tidally locked exoplanets around M-stars by
\citet{leconte-etal-2013}. On a tidally locked terrestrial planet, the
dark side may act as a cold trap for volatiles, which can lead to
total collapse of the atmosphere in extreme cases, as we discuss
later. For planets close to or inside the inner edge of the habitable
zone, however, the effects of this process on H$_2$O lead to an
interesting bistability in the climate. One state consists of a
classical runaway greenhouse where all the water is in the
atmosphere. In the other state, the vast majority of the H$_2$O is
present as ice on the planet's dark side, while the planet's day side
is relatively hot and extremely dry, allowing it to effectively
radiate the incoming solar radiation back to space (see Figure
\ref{runaway-bistable}).  In this scenario, depending on the thickness
of the ice sheet and the planet's thermal history, liquid water could
be present in some amount near the ice sheet's edges, allowing
marginal conditions for habitability well inside the classical inner
edge of the habitable zone. Further work may be required to assess
whether such a collapsed state would remain stable under the influence
of transient melting events due to e.g., meteorite impacts on the
planet's dark side.

\begin{figure*}
 \epsscale{1.}
\plotone{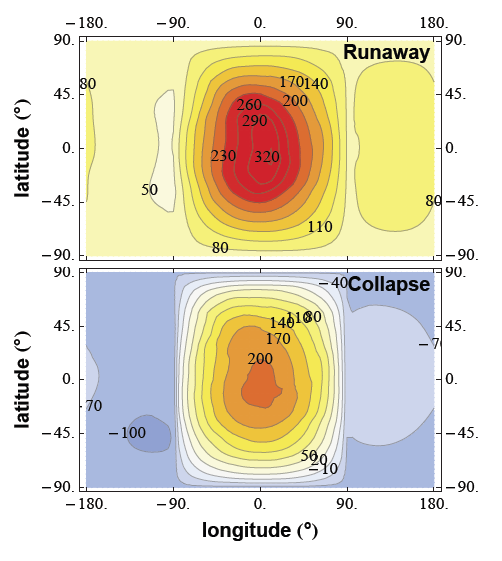}
 \caption{Surface temperature maps in $^\circ$~C from GCM simulations
   of a tidally locked, inner edge exoplanet with a 200-mbar
   equivalent incondensible atmospheric component and fixed total
   water inventory. In the top panel, the planet was initialized with
   a mean water vapor column of 250~kg~m$^{-2}$ and remains in a
   runaway greenhouse state. In the bottom panel, the smaller water
   vapor starting inventory (150~kg~m$^{-2}$) has led to a transition
   to a collapsed state, with lower surface temperatures and ice on
   the planet's dark side.}
\label{runaway-bistable}
 \end{figure*}

\subsection{Snowball Earth dynamics and climate}
\label{snowball-section}

Runaway greenhouse states occur because of the effectiveness of gaseous H$_2$O as an
infrared absorber. However, the ability of water to cause fundamental changes in
climate systems also extends to colder conditions.
Solid H$_2$O on the surface of a planet
can have an equally drastic effect on circulation and climate, in this case due to
its properties in the visible part of the spectrum.

Since the 1960s, it has been known that the Earth's climate has an
alternative equilibrium state to present-day conditions. If the ice sheets
that are currently confined to the poles instead extended all the way
to the equator, the elevation in surface albedo would cause so much
sunlight to be reflected back to space that mean global temperatures
would drop to around 230-250~K. As a result, surface and ocean ice would
be stable even in the tropics, and the Earth could remain in a stable
``snowball'' state unless the greenhouse effect of the atmosphere
increased dramatically. Snowball events are believed to have occurred
at least twice previously in Earth's history: once around 640~Ma and
again 710~Ma, in the so-called Marinoan and Sturtian ice ages. Geological
evidence (specifically, carbon isotope ratio excursions and
observations of cap carbonates above glacial deposits) suggests that
these events were ended by build-up of atmospheric CO$_2$ to extremely
high levels due to volcanic outgassing \citep{hoffman-schrag-2002,pierrehumbert-etal-2011}.
This scenario is particularly plausible because the land weathering of silicates
and associated drawdown of CO$_2$ into carbonates should be greatly reduced in a snowball climate
(see Section \ref{sec:carbonate-silicate}).

Snowball Earth transitions are interesting in an exoplanetary context
first because they represent a clear challenge to the concept of the
habitability zone. If a habitable planet falls victim
to snowball glaciation and cannot escape it, it may be irrelevant if
it continues to receive the same flux from its host star. Identifying
how these events can occur and how they end is hence of key importance
in understanding the uniqueness of Earth's current climate. Snowball
glaciation is also interesting from a purely physical viewpoint, as it
is another example of a situation where the interplay between
atmospheric dynamics and climate is fundamental to the problem. As we
have seen in Section~\ref{overview}, heat transport between a planet's regions of
high and low mean insolation is strongly dependent on the properties
of the atmosphere and planetary rotation rate.
As a result, the physics of the snowball transition
should vary considerably between different exoplanet cases.

Recently, it has been hypothesized that changes in surface albedo due
to atmospheric transport of dust may be critical to terminating
Snowball Earth episodes \citep{abbot-pierrehumbert-2010}. The reduced
surface temperature and stellar energy absorption at the equator and
absence of ocean heat transport on a globally frozen planet mean that
typically, the tropics become a region of net ablation (i.e., mean
evaporation exceeds precipitation). As a result, dust is slowly
transported to the surface at the equator by a combination of
horizontal glacial flow and vertical ice advection (see Figure
\ref{dust_accumul}). After a long time
period, the build-up of dust on the surface lowers the albedo
sufficiently to melt the surrounding ice and exit the planet from a
snowball state. The fingerprints of the ``mudball'' state that
presumably would follow this may have been imprinted on Earth's
geological record, in the form of variations in thickness with
(paleo)-latitude of clay drape deposits.

Other possibilities for snowball deglaciation come from water ice clouds.
\citet{pierrehumbert-2002} demonstrated using a one-dimensional radiative
model that clouds have a net warming effect on a snowball planet.
Because the surface albedo is high, clouds do not significantly increase the
planetary albedo by reflecting incoming radiation to space. Nonetheless, they still
effectively absorb outgoing IR radiation and hence cause greenhouse warming.
These conclusions were broadly confirmed by a later 3D GCM intercomparison
\citep{abbot-etal-2012}, although differences in cloud radiative forcing between
models was found to be significant. Just as for the runaway greenhouse, future
detailed studies of the physics of cloud formation and radiative transfer under snowball
conditions, possibly using cloud-resolving models, will be important for elucidating
their role.

\begin{figure*}
 \epsscale{1.0}
\plotone{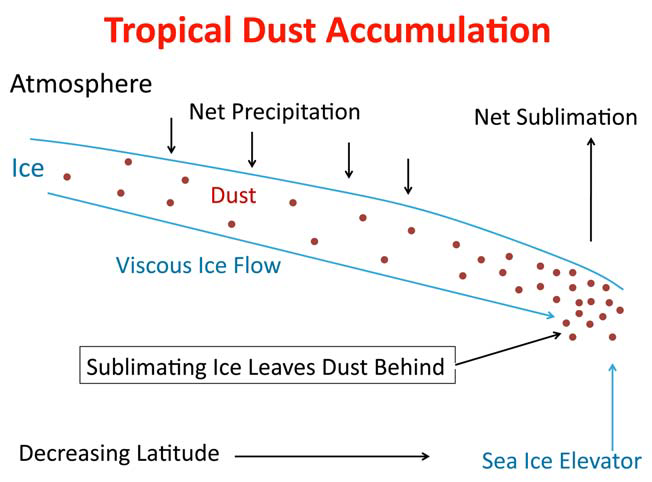}
 \caption{Schematic of the process by which tropical dust accumulates
   on a globally glaciated planet. The net sublimation at the tropics
   and precipitation in sub-tropical regions occurs primarily due to
   the reduced surface heat capacity of the surface when a thick ice
   layer is present \citep[from][]{abbot-pierrehumbert-2010}.}
\label{dust_accumul}
 \end{figure*}

Further interesting couplings between atmospheric dynamics and climate
arises if a planet near the snowball transition has a limited H$_2$O
inventory. \citet{abe-etal-2011} studied idealized planets where the total
amount of water (atmospheric and surface) only amounted to a 20--60 cm
globally averaged layer. They showed that for these so-called ``land''
planets, the drying of tropical regions by a dynamical process similar
to that responsible for dust accumulation would prevent the formation
of stable surface ice there for a solar flux as low as 80\% of that
received by Earth today. As a result, the planetary albedo could
remain low even if global mean surface temperatures dropped below the
freezing point of H$_2$O, and transient regions of liquid water could
persist longer than if the planet possessed as much water as Earth.

Finally, if the planet orbits a star of different spectral type to the
Sun, further complications emerge. Recent work
\citep{joshi-haberle-2011} has demonstrated that around red dwarf
M-stars, the mean albedos of snow and ice will be significantly
reduced due to the red-shifted incident spectrum. As a result, the
ice-albedo feedback is expected to be weaker, and snowball events may
not be as difficult to exit as on Earth. \citet{joshi-haberle-2011}
suggested that this effect may also help extend the outer edge of the
habitable zone around M-stars. However, it is likely to be of lesser
importance for planets with dense CO$_2$ atmospheres, because the increased
visible optical depth of the atmosphere and presence of CO$_2$ clouds
means the dependence of the planetary albedo on surface properties is
already weak in these cases \citep{wordsworth-etal-2011}.

\subsection{Carbonate-silicate weathering and atmospheric dynamics}\label{sec:carbonate-silicate}

So far, we have focused on the coupling between circulation and climate
in the context of H$_2$O. However, the role of other gases in climate may be
equally significant. Carbon dioxide is of particular interest, given that it
is the majority constituent of the atmospheres of both Venus and Mars.
It is also the key greenhouse gas in our own planet's atmosphere, because
of its special role in maintaining habitable surface conditions.

On Earth, the carbonate-silicate cycle is fundamental to the long-term
evolution of carbon dioxide in the atmosphere. In particular, it is
believed that the dependence of land silicate weathering on
temperature leads to a negative feedback between atmospheric CO$_2$
and mean surface temperature, rendering the climate stable to
relatively large variations in solar luminosity
\citep{walker-hayes-kasting-1981}. Indeed, the assumption of an
efficient carbonate-silicate cycle lies behind the classical
definition of the habitable zone \citep{kasting-etal-1993}.

The idea that climate could naturally self-regulate on rocky planets
due to an abiotic process is a highly attractive one, and probably a
major part of the explanation for why Earth has maintained a (mostly)
clement climate throughout its lifetime, despite significant increases
in the solar flux (around 30\% since 4.4~Ga).  Nonetheless, there are
still major uncertainties in the nature of Earth's carbon cycle,
including the importance of seafloor weathering
\citep{sleep-zahnle-2001,lehir-etal-2008}, the exact role of plants in land weathering,
and the dependence of the cycle on the details of plate tectonics.

Even given these uncertainties, it is already clear that substantial
differences from the standard carbon cycle arise when a planet becomes
synchronously locked. For example, \cite{edson-etal-2012}
investigated the relationship between weathering, atmospheric CO$_2$
concentrations, and the distribution of continents for tidally locked
planets around M-class stars. They coupled Earth GCM simulations with
a simple parametrization for the global CO$_2$ weathering rate, which
following \cite{walker-hayes-kasting-1981} was written as
\begin{eqnarray}
\frac{d \mbox{CO}_2 |_{tot} }{d t} &=& \frac 12 \int W_0 \left( \frac{f_{\mbox{CO}2}}{355\mbox{ ppmv}} \right)^{0.3}  \nonumber \\
                              &\quad& \times \left( \frac{\mathcal R}{0.665\mbox{ mm d}^{-1}}\right) \exp{\frac{T-T_0}{T_U}} \mbox{dA}
\label{eq:weatherCO2}
\end{eqnarray}
where $f_{\mbox{CO}2}$ is the atmospheric CO$_2$ volume mixing ratio, $W_0 = 8.4543\times10^{-10}$~C~$ s^{-1}~m^{-2}$ is the estimated present-day terrestrial weathering rate, $\mathcal R$ is the runoff rate, $T$ is temperature, $T_0 = 288$~K, $T_U = 17.7$~K, and the integral is over the planet's weatherable (i.e., land) surface. By iterating between the simulations and Eqn. (\ref{eq:weatherCO2}), \cite{edson-etal-2012} derived self-consistent $f_{\mbox{CO}2}$ values as a function of the (constant) substellar longitude, given Earth's present-day geography. They found that $f_{\mbox{CO}2}$ could vary by a factor of $\sim1\times10^4$ (7~ppmv vs. 60331~ppmv), depending on whether the substellar point was located over the Atlantic or Pacific oceans. As a result, the mean surface air
temperatures in their simulations varied between 247 and 282~K, with the climate in both cases similar to the ``eyeball'' state discussed in \cite{pierrehumbert-2011} (see next section).

The differences in temperature and equilibrium atmospheric CO$_2$
abundance between the two cases were caused by the differences in land
area in the illuminated hemisphere. The illuminated hemisphere of a
tidally locked planet is warmer, with higher precipitation rates (see
Fig. \ref{merlis_composite}), so a greater land area there implies
enhanced global weathering rates and hence more rapid drawdown of
CO$_2$.  Assuming that global CO$_2$ levels are controlled by an
expression like (\ref{eq:weatherCO2}) is clearly an oversimplification
given the complexity of Earth's real carbon cycle. In particular,
basalt carbonization on the seafloor, which has a much weaker
dependence on surface temperatures, is still poorly understood, but
could have played a vital role in CO$_2$ weathering throughout Earth's
history. Nonetheless, the study highlights the dramatic differences
that can be caused by 3D effects, and shows that a more detailed
understanding of dynamical couplings with the carbonate-silicate cycle
will be vital in future.

\cite{kite-etal-2011} also explored the possible nature of climate-weathering feedbacks on tidally locked planets using
an idealized energy balance climate model coupled to a simplified parametrization of silicate weathering. They noted that because
 heat transport efficiency depends on the atmospheric mass, in some cases reducing atmospheric pressure can increase the weathering rate
 and/or the liquid water volume on the day side of the planet. This can cause positive feedbacks that lead to a decrease in atmospheric mass.
The mechanism described by \cite{kite-etal-2011} is most likely to be important for planets like Mars, with thin, CO$_2$-dominated atmospheres.

\subsection{Collapse of condensable atmospheres}
\label{sec:collapse}

In Section \ref{sec:runaway}, we briefly discussed situations 
where water vapour can become trapped as ice on the dark sides of 
tidally locked planets. Here, we discuss the extension of this process 
to the majority constituent of a planet's atmosphere. As we will see, 
atmospheric collapse is a problem of major importance 
to planetary habitability that sensitively depends on the details  
of the coupling between circulation and climate.

The interest in the habitability of tidally locked planets stems from 
the fact that 
cool, faint red dwarf stars (M- and K- spectral class) are
significantly more common in the galaxy than stars like the Sun, and
they host some of the lowest mass and best characterized
exoplanets so far discovered \citep[e.g.,][]{udry-etal-2007,
  mayor-etal-2009, charbonneau-etal-2009,bean-etal-2010}.  
In the earliest detailed study of exoplanet habitability
\citep{kasting-etal-1993}, it was shown that planets around M-dwarf
stars could sustain surface liquid water if they were in sufficiently
close orbits. As discussed previously, the tidal interaction with their host stars
means that 
such planets will in most cases have resonant or synchronous rotation rates
and low obliquities, and hence permanent regions where little or no
starlight reaches the surface. 
Initially, this was thought to be a
potentially insurmountable obstacle to habitability, because the regions
receiving no light could become so cold that any gas (including
nitrogen) would freeze out on the surface there, depleting the
atmosphere until the planet eventually became completely airless.

The problem of atmospheric collapse was first investigated
quantitatively in a series of pioneering papers by Manoj Joshi and Bob
Haberle \citep{joshi-etal-1997,joshi-2003}. Using a
combination of basic scale analysis and 3D atmospheric modeling,
they showed that the collapse of the atmosphere was only
inevitable if it was inefficient at transporting heat. As we described
earlier, the efficiency
of atmospheric heat transport depends primarily on the composition
(via the specific heat capacity and the infrared opacity), the average
wind speed and (most critically) the total surface
pressure. Assuming a surface emissivity of unity,
the surface heat budget for a planet with condensible
atmospheric species can be written
\begin{equation}
\alpha \frac{\partial T_{surf}}{\partial t} = F^{dn}_{sw} +  (F^{dn}_{lw} - \sigma T_{surf}^4) - LE - SH 
\label{eq:surf_heat_budget}
\end{equation}
where $\sigma$ is the Stefan-Boltzmann constant, $\alpha$ is the
surface heat capacity, $F^{dn}_{sw}$ and $F^{dn}_{lw}$ are the net
downwelling fluxes of short- and longwave radiation from above,
$LE$ is the latent heat flux, and $SH$ is the sensible heat
flux, as described in (\ref{eq:PBL}). Equation
(\ref{eq:surf_heat_budget}) is essentially the unaveraged version of
(\ref{eqn-atm_en}), except that for a single-component atmosphere,
$LE$ only becomes important when the atmosphere begins to
condense on the surface (as on present-day Mars, for example). On the
dark side of a tidally locked planet $F^{dn}_{sw} = 0$, so radiative
cooling of the surface until the atmosphere begins to condense can be
prevented in two ways: tight thermal coupling with the atmosphere via
convection (the $SH$ term), or radiative heating of the surface
by an optically thick atmosphere ($F^{dn}_{lw}$). Both effects can be
important in principle, although the temperature inversion caused by
contact of warmer air from the dayside with the nightside surface may
shut down convection, reducing the magnitude of the sensible
heat flux or even reversing its sign.
The strength of $F^{dn}_{lw}$ will depend on
both the gas opacity of the atmosphere and the effects of clouds and
aerosols, if present, and hence can be expected to vary with
composition as well as atmospheric pressure.  In any case, the
calculations of \cite{joshi-etal-1997} showed that for CO$_2$
atmospheres with surface pressures of around 100~mBar or more on
Earth-like planets, both atmospheric transport and surface coupling
become efficient enough to avoid the possibility of collapse and the
climate can remain stable.

While Joshi and Haberle's simulations showed that synchronous rotation
does not rule out the possibility of habitable climates around M-class
stars, their work was focused on planets of Earth's mass and net
stellar flux, and hence could not be applied to more general
cases. Interest in this issue was reignited with the discovery of
potentially habitable planets in the Gliese 581 system in 2007
\citep{udry-etal-2007,mayor-etal-2009}.  Two planets, GJ581c and d,
orbit near the outer and inner edges of their system's habitable zone,
and have estimated masses that suggest terrestrial rather than gas
giant compositions. Simple one-dimensional models indicated that with
a CO$_2$-rich atmosphere, the ``d'' planet in particular could support
surface liquid water \citep{wordsworth-etal-2010, von-paris-etal-2010,
  hu-ding-2011, kaltenegger-etal-2011}.  Because M-stars have
red-shifted spectra compared to the Sun, CO$_2$ clouds typically cause
less warming for planets around them, because the increase in
planetary albedo they cause becomes nearly as significant as the
greenhouse effect caused by their IR scattering properties. However,
in CO$_2$-rich atmospheres this is more than compensated for by
increased near-IR absorption in the middle and lower atmosphere, which
lowers the planetary albedo and hence increases warming.

The results from 1D models for GJ581d were suggestive, but the increase in condensation
temperature with pressure means that atmospheric collapse becomes more
of a threat as the amount of CO$_2$ in the atmosphere increases. For example,
reference to vapor pressure
data for CO$_2$ shows $p_{sat}$=~10~bar at $\sim 235$~K, which is comparable to annual mean
surface temperatures at Earth's south pole.
It was therefore initially unclear if high-CO$_2$ scenarios were plausible for GJ581d.

\begin{figure*}
 \epsscale{1.0}
\plotone{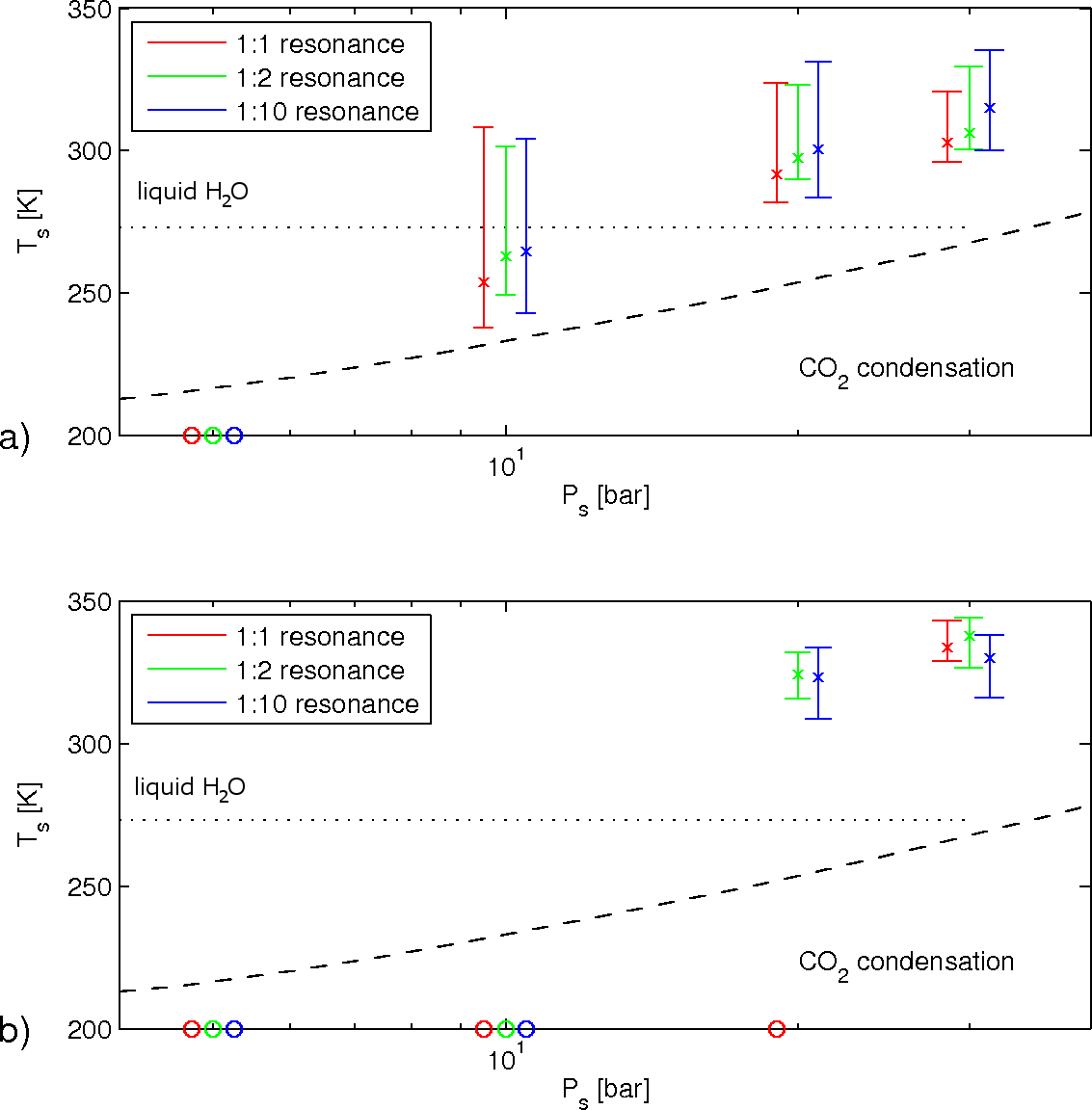}
 \caption{Simulated annual mean surface temperature (maximum, minimum
   and global average) as a function of atmospheric pressure and
   rotation rate for GJ581d assuming a) a pure CO$_2$ atmosphere and
   b) a mixed CO$_2$-H$_2$O atmosphere with infinite water source at
   the surface [from \citet{wordsworth-etal-2011}].  Data plotted with
   circles indicate where the atmosphere had begun to collapse on the
   surface in the simulations, and hence no steady-state temperature
   could be recorded. In the legend, 1:1 resonance refers to a
   synchronous rotation state, and 1:2 and 1:10 resonances refer to
   despun but asynchronous spin-orbit configurations.}
\label{GJ581d_temperatures}
 \end{figure*}

\citet{wordsworth-etal-2011} investigated the stability of CO$_2$-rich
atmospheres on GJ581d using a general circulation model with
band-resolved radiative transfer, simplified cloud physics, and
evaporation/condensation cycles for CO$_2$ and H$_2$O included. They
found that for this planet, which receives only around 30\% of Earth's
insolation, the atmosphere could indeed be unstable to collapse even
for pressures as high as 2--5 bar.  Nonetheless, at higher pressures
the simulated atmospheres stabilized due to a combination of the
increased greenhouse warming and homogenization of the surface
temperature by atmospheric heat transport.  For atmospheric CO$_2$
pressures of 10--30 bar (which are plausible given Venus and Earth's
total CO$_2$ inventories), \citet{wordsworth-etal-2011} modeled
surface temperatures for GJ581d in the 270--320~K range, with
horizontal temperature variations of order 10--50~K (see
Fig. \ref{GJ581d_temperatures}).  Sensitivity studies indicated that
while the exact transition pressure from unstable to stable
atmospheres was dependent on the assumed cloud microphysical
properties, the general conclusion of stability at high CO$_2$
pressure was not.  The authors concluded that despite the low stellar
flux it receives, GJ581d could therefore potentially support surface
liquid water.  As such, it is one of the most interesting currently
known targets for follow-up characterization.

Any atmospheric observations of planets around Gliese~581 will be
challenging, because the system does not appear to be aligned
correctly for transit spectroscopy, and the planet-star contrast ratio
constraints required for direct observations are
severe. Distinguishing between habitable scenarios for GJ581d and
other possibilities, such as a thin/collapsed atmosphere or a
hydrogen-helium envelope, may be best accomplished via a search for
absorption bands of CO$_2$/H$_2$O in the planet's emitted IR radiation
that do not significantly vary over the course of one orbit (see
Fig. \ref{GJ581d_spectra}). As described in \cite{selsis-etal-2011},
estimates of a non-transiting planet's atmospheric density are also
possible, in principle, via analysis of the IR ``variation spectrum''
derived from phase curves. Despite the challenges, at
$\sim$20~l.y. distance from the Sun, GJ581 is still a relatively close
neighbor, and a future dedicated mission such as NASA's proposed
Terrestrial Planet Finder (TPF) or ESA's Darwin would have the
necessary capability to perform such observations.

 \begin{figure*}
 \epsscale{1.0}
\plotone{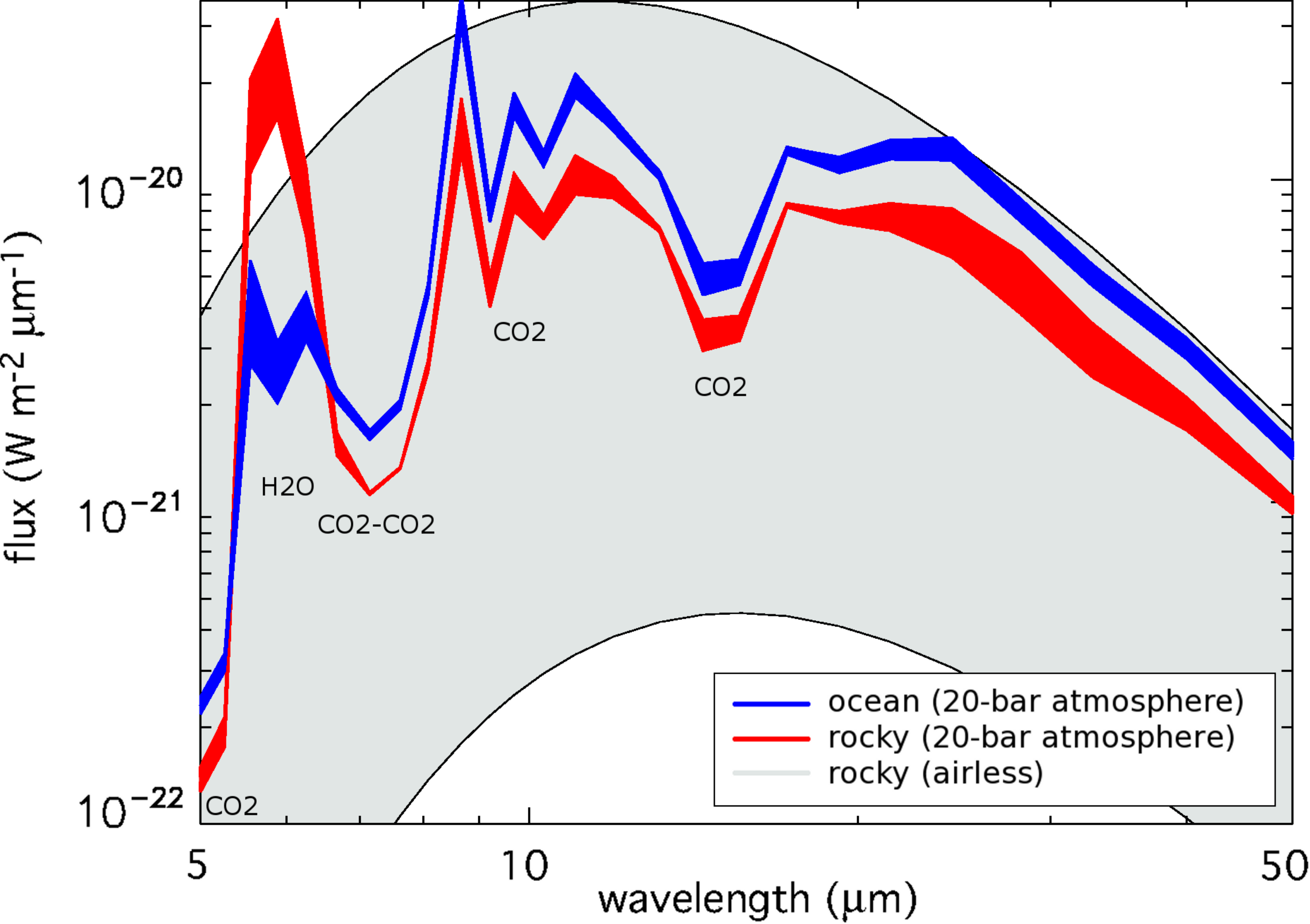}
 \caption{Simulated emission spectra for GJ581d given various
   atmospheric scenarios [from \citet{wordsworth-etal-2011}].  In all
   cases the thicknesses of the lines correspond to the range of
   variation predicted over the course of one planetary orbit. The
   relatively low variation in the red/blue cases is a characteristic
   signature of a planet with a stable, dense atmosphere with
   efficient horizontal heat redistribution.}
\label{GJ581d_spectra}
 \end{figure*}

Other recent studies have noted that even if global habitability is
ruled out by the freezing of surface water on a planet's night side,
it is still possible to maintain local regions of liquid
water. \citet{pierrehumbert-2011} investigated a series of possible
climates for a 3.1~$M_\oplus$ tidally locked exoplanet around an
M-star receiving 63\% of Earth's incident solar flux. He hypothesized
that if the planet's composition was dominated by H$_2$O, it would
most likely have an icy surface and hence a high albedo. However, if
the atmosphere were extremely thin and hence inefficient at
transporting heat, surface temperatures near the substellar point
could allow some H$_2$O to melt, even though the global mean
temperature would be around 192~K if heat transport were
efficient. Given a denser mixed N$_2$-CO$_2$-H$_2$O atmosphere, an
``eyeball'' climate could also be maintained where permanent ice is
present on the night side, but temperatures are high enough on the day
side to allow an ocean to form (see Fig. \ref{pierrehumbert-eyeball}).
Although such a scenario is admittedly hypothetical, it raises many
interesting questions on the general nature of exoplanet climate and
may help with the interpretation of observations in the future.

\begin{figure*}
 \epsscale{1.}
\plotone{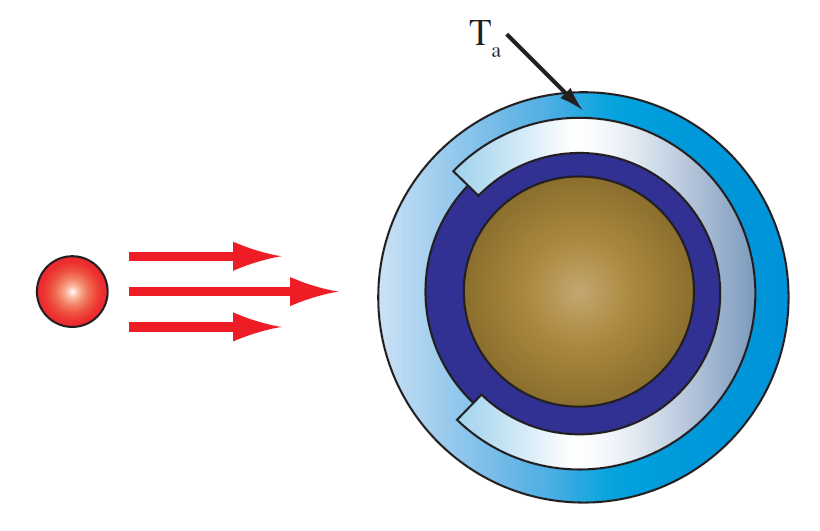}
 \caption{Schematic of the ``eyeball'' climate state for synchronously
   rotating terrestrial planet \citep{pierrehumbert-2011}. Sea ice is
   present across the permanent night side of the planet, but stops at
   a stellar zenith angle primarily determined by the balance between
   the local stellar flux and horizontal atmospheric and oceanic heat
   transport.}
\label{pierrehumbert-eyeball}
 \end{figure*}

Clearly, the likelihood of atmospheric collapse also depends strongly
on the condensation temperature of the majority gas in the
atmosphere. Despite this progress in modeling exotic exoplanet
atmospheres in specific cases, the general problem of atmospheric
condensation in a planet's cold trap(s) for arbitrary conditions
remains unsolved. Nonetheless, scale analysis has recently been used
to place some basic general constraints on atmospheric stability by
\cite{heng-kopparla-2012}. These authors assumed the criterion
$\tau_{\rm rad}>\tau_{\rm adv}$ as a condition for stability, where
$\tau_{\rm rad}$ and $\tau_{\rm adv}$ are representative timescales
for radiative cooling and advection, respectively. The radiative
timescale can be approximately defined as $\tau_{\rm rad}=c_p p \slash g
\sigma T_{\rm rad}^3$, with $p$ the total pressure, $g$ gravity and
$T_{\rm rad}$ the characteristic emission temperature, while $\tau_{\rm adv}\sim
L\slash U$, with $L$ and $U$ a characteristic horizontal length and
wind speed, respectively.  Evaluating $\tau_{\rm rad}$ is straightforward for
a known atmospheric composition, but $U$ and hence $\tau_{\rm adv}$ are
difficult to assess \emph{a priori} without a general theory for the
response of the atmospheric circulation to different forcing
scenarios.  \citet{heng-kopparla-2012} neatly circumvented this
problem by simply taking $U$ to be the speed of sound, allowing a
potentially strict criterion for atmospheric collapse based entirely
on known properties of the atmosphere.

Despite its elegance, their approach was limited by its neglect of the
role of thermodynamics in the problem.  In particular, a moving gas
parcel with a majority constituent that condenses at low temperatures
may continue to radiate until the local value of $\tau_{\rm rad}$ becomes
large, which tends to make an atmosphere much more resistant to
collapse.  There is still considerable scope for theoretical
development of the atmospheric collapse problem, and given its
fundamental importance to the understanding of both atmospheric
circulation and habitability around M-stars, it is likely that it will
continue to receive attention for some time.

The dependence of collapse on saturation vapour temperature is
particularly interesting when considered in the context of more exotic
scenarios for terrestrial planet atmospheric composition.  In
particular, the saturation vapour temperatures of hydrogen and helium
at 1~bar are under 25~K, rendering them essentially incondensible on
planets receiving even very little energy from their host stars.

Many rocky planets are believed to form with thin
primordial envelopes of these gases, which have large optical
depths in the infrared for pressures above a fraction of a bar. For
planets of Earth's mass or lower, these envelopes generally escape rapidly
under the elevated stellar wind and extreme ultraviolet (XUV) fluxes
from young stars. However, higher mass or more distant planets may
experience a range of situations where hydrogen and helium can remain
in their atmospheres for longer periods. \citet{stevenson-1999} suggested
that some planetary embryos ejected from their system during formation
could sustain liquid oceans if they kept their primordial hydrogen
envelopes. \citet{pierrehumbert-gaidos-2011} extended this idea to planets
in distant orbits from their stars. These planets could be detectable in
theory due to gravitational microlensing, although given their low equilibrium
temperatures, atmospheric characterization would be extremely challenging even
with the most ambitious planned missions. \citet{wordsworth-2012} noted that
in general, very few planets are left with just the right amount of
hydrogen in their atmospheres to maintain surface liquid water after
the initial period of intense XUV fluxes and rapid hydrogen
escape. Nonetheless, they showed that transient periods of surface
habitability (between $\sim$10,000~yr and $\sim$100~Myr)
still occur for essentially any planets
receiving less flux than Earth that form with a hydrogen envelope but
later lose it. In addition, XUV photolysis in hydrogen-rich atmospheres
can lead to the formation of a range of pre-biotic compounds
\citep{miller-1953, dewitt-etal-2009}.
Conditions on young terrestrial planets that at least
allow life to form could hence be much more common than previously
assumed.

\section{CONCLUSIONS}
\label{conclusion}

Given the diversity emerging among more massive exoplanets, it is
almost certain that terrestrial exoplanets will occupy an incredible
range of orbital and physical parameters, including orbital semi-major
axis, orbital eccentricity, incident stellar flux, incident stellar
spectrum, planetary rotation rate, obliquity, atmospheric mass,
composition, volatile inventory (including existence and mass of
oceans), and evolutionary history.  Since all of these parameters
affect the atmospheric dynamics, it is likewise probable that
terrestrial exoplanets will exhibit incredible diversity in the
specific details of their atmospheric circulation patterns and
climate.  Only a small fraction of this diversity has yet been
explored in GCMs and similar models.  As we have emphasized in this
review, existing theory suggests that a number of unifying themes will
emerge governing the atmospheric circulation on this broad class of
bodies.  This theory---summarized here---will provide a foundation for
understanding, and provide a broad context for, the results of GCM
investigations of particular objects.

Exoplanets that rotate sufficiently rapidly will exhibit extratropics
at high latitudes, where the dynamics are approximately geostrophic,
and where eddies resulting from baroclinic instabilities control the
meridional heat transport, equator-to-pole temperature differences,
meridional mixing rates, existence of jet streams, and thermal
stratification in the troposphere.  Regions where rotation is less
dominant---near the equator, and globally on slowly rotating
exoplanets---will exhibit tropical regimes where wave adjustment and
Hadley circulations typically act to minimize horizontal temperature
differences.  Significant interactions between the tropics and
extratropics can occur by a variety of mechanisms, perhaps the most
important of which is the propagation of Rossby waves between the two
regions.  The hydrological cycle on many terrestrial exoplanets will
exert significant effects both on the mean climate---through the
greenhouse effect and clouds---and on the circulation patterns.
Existing studies demonstrate that the circulation can influence
global-scale climate feedbacks, including the runaway greenhouse,
transitions to snowball-Earth states, and atmospheric collapse,
leading to a partial control of atmospheric circulation on the mean
climate and therefore planetary habitability.  

Exoplanets can exhibit greatly different regimes of thermal forcing
than occur on solar system planets, a topic which has yet to be fully
explored.  The day-night thermal forcing on sychronously rotating and
despun exoplanets will play a more important role than on the Earth,
leading in many cases to equatorial superrotation and other dynamical
effects.  When the heating rates are sufficiently great (in the limit
of thin and/or hot atmospheres), the heating and cooling gradients can
overwhelm the ability of the atmosphere to transport heat
horizontally, leading to freeze out of the more refractory components
(such as water or carbon dioxide) on the nightside or poles.  This
regime remains poorly understood, although it is important for
understanding the structure, and even existence, of atmospheres on hot
terrestrial planets and super Earths.

Observations will be needed to move the field forward.  These in many
cases will represent extensions to terrestrial planets of techniques
currently being applied to hot Jupiters.  Short-term, gains are most
likely for transiting planets, particularly those around small M
stars: a terrestrial planet orbiting such a star exhibits a
planet/star radius ratio similar to that of a gas giant orbiting a
Sun-like star, thereby making characterization (relatively) easier.
Transit spectroscopy of such systems will yield constraints on
atmospheric gas composition and existence of hazes
\citep[e.g.,][]{barman-2007, sing-etal-2008, pont-etal-2013,
  bean-etal-2010, desert-etal-2011, berta-etal-2012}. In the
longer-term future, such spectroscopy could also provide direct
measurements of atmospheric wind speeds at the terminator
\citep{snellen-etal-2010, hedelt-etal-2011}.  Secondary eclipse
detections and, eventually, full-orbit lightcurves will provide
information on the emission spectrum as a function of longitude,
allowing constraints on the vertical temperature profile and day-night
temperature difference to be inferred.  This will first be possible
for hot systems \citep{demory-etal-2012}, but with significant
investment of resources from the James Webb Space Telescope
will also be possible for planets in the habitable zones
\citep{seager-etal-2008}. Eclipse mapping may also eventually be
possible, as is now being performed for hot Jupiters
\citep{majeau-etal-2012, de-wit-etal-2012}.   Using direct
imaging to obtain spectroscopy of planets is another observational
avenue---not limited to transiting systems---that could be
performed from a platform like the Terrestrial Planet Finder.  
All of these observations will be a challenge, but the payoff will
be significant: the first characterization of terrestrial worlds
around stars in the solar neighborhood.

In the meantime, theory and models can help to address open questions
concerning the behavior of atmospheric circulation and climate across
a wider range of parameter space than encountered in the solar system.
Major questions emphasized in this review include the following: What
is the dependence of three-dimensional temperature structure,
humidity, and horizontal heat flux on planetary rotation rate, incident stellar
flux, atmospheric mass, atmospheric composition, planetary radius and
gravity, and other conditions?  What are the regimes of atmospheric
wind? What is the influence of an ocean in modulating the atmospheric
climate?  Under what conditions can clouds form, and what is their
three-dimensional distribution for planets of various types?  How do
seasonal cycles (due to non-zero obliquity and/or orbital
eccentricity) affect the atmospheric circulation and climate on
planets generally?  What is the role of the atmospheric
circulation---through its control of temperature, humidity,
cloudiness, and precipitation---in affecting longer-term climatic
processes including runaway greenhouses, ice-age cycles, transitions
to snowball-Earth states, the carbonate-silicate feedback, and
collapse of condensable atmospheric constituents onto the poles or
nightside?  And what is the continuum of atmospheric behaviors 
on exoplanets ranging from sub-Earth-sized terrestrial
planets to super Earths and mini-Neptunes?  While a detailed
understanding of particular exoplanets must await observations,
significant insights into the fundamental physical mechanisms
controlling atmospheric circulation and climate---of planets
generally---can be made
now with current theoretical and modeling tools.  This will not
only lay the groundwork for understanding future observations but
will place the atmospheric dynamics and climate of solar-system
worlds, including Earth, into its proper planetary context.

\bigskip
\textbf{ Acknowledgments.} We thank Jonathan Mitchell and an
anonymous referee for thorough reviews of the manuscript.
This paper was supported by NASA grants
NNX10AB91G and NNX12AI79G to APS.  TMM was supported by a 
Princeton Center for Theoretical Science fellowship.  \\

\bigskip
\parskip=0pt
{\small
\baselineskip=11pt

\bibliographystyle{ametsoc}
\bibliography{showman-bib}

\begin{thebibliography}{278}
\expandafter\ifx\csname natexlab\endcsname\relax\def\natexlab#1{#1}\fi

\bibitem[{{Abbot} and {Pierrehumbert}(2010){\it {Abbot} and
  {Pierrehumbert}\/}}]{abbot-pierrehumbert-2010}
{Abbot}, D.~S., and R.~T. {Pierrehumbert}, 2010:
\newblock {Mudball: Surface dust and Snowball Earth deglaciation}.
\newblock {\it Journal of Geophysical Research (Atmospheres)\/}, {\bf
  115}(D14), 3104.

\bibitem[{{Abbot} et~al.(2012){\it {Abbot}, {Voigt}, {Branson},
  {Pierrehumbert}, {Pollard}, {Le Hir}, and {Koll}\/}}]{abbot-etal-2012}
{Abbot}, D.~S., A.~{Voigt}, M.~{Branson}, R.~T. {Pierrehumbert}, D.~{Pollard},
  G.~{Le Hir}, and D.~D.~B. {Koll}, 2012:
\newblock {Clouds and Snowball Earth deglaciation}.
\newblock {\it \grl\/}, {\bf 39}, 20,711.

\bibitem[{{Abe} et~al.(2011){\it {Abe}, {Abe-Ouchi}, {Sleep}, and
  {Zahnle}\/}}]{abe-etal-2011}
{Abe}, Y., A.~{Abe-Ouchi}, N.~H. {Sleep}, and K.~J. {Zahnle}, 2011:
\newblock {Habitable Zone Limits for Dry Planets}.
\newblock {\it Astrobiology\/}, {\bf 11}, 443--460.

\bibitem[{{Adam} and {Paldor}(2009){\it {Adam} and
  {Paldor}\/}}]{adam-paldor-2009}
{Adam}, O., and N.~{Paldor}, 2009:
\newblock {Global Circulation in an Axially Symmetric Shallow Water Model
  Forced by Equinoctial Differential Heating}.
\newblock {\it Journal of Atmospheric Sciences\/}, {\bf 66}, 1418.

\bibitem[{{Adam} and {Paldor}(2010){\it {Adam} and
  {Paldor}\/}}]{adam-paldor-2010}
{Adam}, O., and N.~{Paldor}, 2010:
\newblock {Global Circulation in an Axially Symmetric Shallow-Water Model,
  Forced by Off-Equatorial Differential Heating}.
\newblock {\it Journal of Atmospheric Sciences\/}, {\bf 67}, 1275--1286.

\bibitem[{Allen and Ingram(2002){\it Allen and Ingram\/}}]{allen-ingram-2002}
Allen, M.~R., and W.~J. Ingram, 2002:
\newblock Constraints on future changes in climate and the hydrologic cycle.
\newblock {\it Nature\/}, {\bf 419}, 224--232.

\bibitem[{{Andrews} et~al.(1987){\it {Andrews}, {Holton}, and
  {Leovy}\/}}]{andrews-etal-1987}
{Andrews}, D.~G., J.~R. {Holton}, and C.~B. {Leovy}, 1987:
\newblock {\it Middle Atmosphere Dynamics\/}.
\newblock Academic Press, New York.

\bibitem[{{Arnold} et~al.(2012){\it {Arnold}, {Tziperman}, and
  {Farrell}\/}}]{arnold-etal-2012}
{Arnold}, N.~P., E.~{Tziperman}, and B.~{Farrell}, 2012:
\newblock {Abrupt Transition to Strong Superrotation Driven by Equatorial Wave
  Resonance in an Idealized GCM}.
\newblock {\it Journal of Atmospheric Sciences\/}, {\bf 69}, 626--640.

\bibitem[{Back and Bretherton(2009{\natexlab{a}}){\it Back and
  Bretherton\/}}]{back-bretherton-2009a}
Back, L.~E., and C.~S. Bretherton, 2009{\natexlab{a}}:
\newblock On the relationship between {SST} gradients, boundary layer winds,
  and convergence over the tropical oceans.
\newblock {\it J. Climate\/}, {\bf 22}, 4182--4196.

\bibitem[{Back and Bretherton(2009{\natexlab{b}}){\it Back and
  Bretherton\/}}]{back-bretherton-2009b}
Back, L.~E., and C.~S. Bretherton, 2009{\natexlab{b}}:
\newblock A simple model of climatological rainfall and vertical motion
  patterns over the tropical oceans.
\newblock {\it J. Climate\/}, {\bf 22}, 6477--6497.

\bibitem[{Bala et~al.(2008){\it Bala, Duffy, and Taylor\/}}]{bala-etal-2008}
Bala, G., P.~B. Duffy, and K.~E. Taylor, 2008:
\newblock Impact of geoengineering schemes on the global hydrological cycle.
\newblock {\it Proc. Nat. Acad. Sci.\/}, {\bf 105}, 7664--7669.

\bibitem[{{Barman}(2007){\it {Barman}\/}}]{barman-2007}
{Barman}, T., 2007:
\newblock {Identification of Absorption Features in an Extrasolar Planet
  Atmosphere}.
\newblock {\it \apjl\/}, {\bf 661}, L191--L194.

\bibitem[{{Barry} et~al.(2002){\it {Barry}, {Craig}, and
  {Thuburn}\/}}]{barry-etal-2002}
{Barry}, L., G.~C. {Craig}, and J.~{Thuburn}, 2002:
\newblock {Poleward heat transport by the atmospheric heat engine}.
\newblock {\it Nature\/}, {\bf 415}, 774--777.

\bibitem[{{Bean} et~al.(2010){\it {Bean}, {Kempton}, and
  {Homeier}\/}}]{bean-etal-2010}
{Bean}, J.~L., E.~{Kempton}, and D.~{Homeier}, 2010:
\newblock {A ground-based transmission spectrum of the super-Earth exoplanet GJ
  1214b}.
\newblock {\it \nat\/}, {\bf 468}, 669--672.

\bibitem[{{Beron-Vera} et~al.(2008){\it {Beron-Vera}, {Brown}, {Olascoaga},
  {Rypina}, {Ko{\c c}ak}, and {Udovydchenkov}\/}}]{beron-vera-etal-2008}
{Beron-Vera}, F.~J., M.~G. {Brown}, M.~J. {Olascoaga}, I.~I. {Rypina},
  H.~{Ko{\c c}ak}, and I.~A. {Udovydchenkov}, 2008:
\newblock {Zonal Jets as Transport Barriers in Planetary Atmospheres}.
\newblock {\it Journal of Atmospheric Sciences\/}, {\bf 65}, 3316.

\bibitem[{{Berta} et~al.(2012){\it {Berta}, et~al.\/}}]{berta-etal-2012}
{Berta}, Z.~K., et~al., 2012:
\newblock {The Flat Transmission Spectrum of the Super-Earth GJ1214b from Wide
  Field Camera 3 on the Hubble Space Telescope}.
\newblock {\it \apj\/}, {\bf 747}, 35.

\bibitem[{Betts(1974){\it Betts\/}}]{betts-1974}
Betts, A.~K., 1974:
\newblock Further comments on comparison of the equivalent potential
  temperature and the static energy.
\newblock {\it J. Atmos. Sci.\/}, {\bf 31}, 1713--1715.

\bibitem[{Bony et~al.(2006){\it Bony, et~al.\/}}]{bony-etal-2006}
Bony, S., et~al., 2006:
\newblock How well do we understand and evaluate climate change feedback
  processes?
\newblock {\it J. Climate\/}, {\bf 19}, 3445--3482.

\bibitem[{{Bordoni} and {Schneider}(2008){\it {Bordoni} and
  {Schneider}\/}}]{bordoni-schneider-2008}
{Bordoni}, S., and T.~{Schneider}, 2008:
\newblock {Monsoons as eddy-mediated regime transitions of the tropical
  overturning circulation}.
\newblock {\it Nature Geoscience\/}, {\bf 1}, 515--519.

\bibitem[{{Bordoni} and {Schneider}(2010){\it {Bordoni} and
  {Schneider}\/}}]{bordoni-schneider-2010}
{Bordoni}, S., and T.~{Schneider}, 2010:
\newblock {Regime Transitions of Steady and Time-Dependent Hadley Circulations:
  Comparison of Axisymmetric and Eddy-Permitting Simulations}.
\newblock {\it Journal of Atmospheric Sciences\/}, {\bf 67}, 1643--1654.

\bibitem[{{Borucki} et~al.(2011){\it {Borucki}, et~al.\/}}]{borucki-etal-2011}
{Borucki}, W.~J., et~al., 2011:
\newblock {Characteristics of Planetary Candidates Observed by Kepler. II.
  Analysis of the First Four Months of Data}.
\newblock {\it \apj\/}, {\bf 736}, 19.

\bibitem[{{Borucki} et~al.(2012){\it {Borucki}, et~al.\/}}]{borucki-etal-2012}
{Borucki}, W.~J., et~al., 2012:
\newblock {Kepler-22b: A 2.4 Earth-radius Planet in the Habitable Zone of a
  Sun-like Star}.
\newblock {\it \apj\/}, {\bf 745}, 120.

\bibitem[{{Borucki} et~al.(2013){\it {Borucki}, et~al.\/}}]{borucki-etal-2013}
{Borucki}, W.~J., et~al., 2013:
\newblock {Kepler-62: A five-planet system with planets of 1.4 and 1.6 Earth
  radii in the habitable zone}.
\newblock {\it Science (Sciencexpress)\/}.

\bibitem[{{Bretherton} and {Smolarkiewicz}(1989){\it {Bretherton} and
  {Smolarkiewicz}\/}}]{bretherton-smolarkiewicz-1989}
{Bretherton}, C.~S., and P.~K. {Smolarkiewicz}, 1989:
\newblock {Gravity Waves, Compensating Subsidence and Detrainment around
  Cumulus Clouds.}
\newblock {\it Journal of Atmospheric Sciences\/}, {\bf 46}, 740--759.

\bibitem[{Caballero and Hanley(2012){\it Caballero and
  Hanley\/}}]{caballero-hanley-2012}
Caballero, R., and J.~Hanley, 2012:
\newblock Midlatitude eddies, storm-track diffusivity and poleward moisture
  transport in warm climates.
\newblock {\it J. Atmos. Sci.\/}, p. in press.

\bibitem[{{Caballero} and {Huber}(2010){\it {Caballero} and
  {Huber}\/}}]{caballero-huber-2010}
{Caballero}, R., and M.~H. {Huber}, 2010:
\newblock {Spontaneous transition to superrotation in warm climates simulated
  by CAM3}.
\newblock {\it \grl\/}, p. in press.

\bibitem[{{Caballero} et~al.(2008){\it {Caballero}, {Pierrehumbert}, and
  {Mitchell}\/}}]{caballero-etal-2008}
{Caballero}, R., R.~T. {Pierrehumbert}, and J.~L. {Mitchell}, 2008:
\newblock {Axisymmetric, nearly inviscid circulations in non-condensing
  radiative-convective atmospheres}.
\newblock {\it Quarterly Journal of the Royal Meteorological Society\/}, {\bf
  134}, 1269--1285.

\bibitem[{{Castan} and {Menou}(2011){\it {Castan} and
  {Menou}\/}}]{castan-menou-2011}
{Castan}, T., and K.~{Menou}, 2011:
\newblock {Atmospheres of Hot Super-Earths}.
\newblock {\it \apjl\/}, {\bf 743}, L36.

\bibitem[{Chang(2001){\it Chang\/}}]{chang-2001}
Chang, E. K.~M., 2001:
\newblock {GCM} and observational diagnoses of the seasonal and interannual
  variations of the {P}acific storm track during the cool season.
\newblock {\it J. Atmos. Sci.\/}, {\bf 58}, 1784--1800.

\bibitem[{Chang et~al.(2002){\it Chang, Lee, and Swanson\/}}]{chang-etal-2002}
Chang, E. K.~M., S.~Lee, and K.~L. Swanson, 2002:
\newblock Storm track dynamics.
\newblock {\it J. Climate\/}, {\bf 15}, 2163--2183.

\bibitem[{{Charbonneau} et~al.(2009){\it {Charbonneau},
  et~al.\/}}]{charbonneau-etal-2009}
{Charbonneau}, D., et~al., 2009:
\newblock {A super-Earth transiting a nearby low-mass star}.
\newblock {\it \nat\/}, {\bf 462}, 891--894.

\bibitem[{{Charney}(1963){\it {Charney}\/}}]{charney-1963}
{Charney}, J.~G., 1963:
\newblock {A Note on Large-Scale Motions in the Tropics.}
\newblock {\it Journal of Atmospheric Sciences\/}, {\bf 20}, 607--608.

\bibitem[{{Chassefi{\`e}re} et~al.(2012){\it {Chassefi{\`e}re}, {Wieler},
  {Marty}, and {Leblanc}\/}}]{chassefiere-etal-2012}
{Chassefi{\`e}re}, E., R.~{Wieler}, B.~{Marty}, and F.~{Leblanc}, 2012:
\newblock {The evolution of Venus: Present state of knowledge and future
  exploration}.
\newblock {\it Planetary and Space Science\/}, {\bf 63}, 15--23.

\bibitem[{{Cho} and {Polvani}(1996){\it {Cho} and
  {Polvani}\/}}]{cho-polvani-1996a}
{Cho}, J.~Y.-K., and L.~M. {Polvani}, 1996:
\newblock The morphogenesis of bands and zonal winds in the atmospheres on the
  giant outer planets.
\newblock {\it Science\/}, {\bf 8}(1), 1--12.

\bibitem[{Chou and Neelin(2004){\it Chou and Neelin\/}}]{chou-neelin-2004}
Chou, C., and J.~D. Neelin, 2004:
\newblock Mechanisms of global warming impacts on regional tropical
  precipitation.
\newblock {\it J. Climate\/}, {\bf 17}, 2688--2701.

\bibitem[{{Cooper} and {Showman}(2005){\it {Cooper} and
  {Showman}\/}}]{cooper-showman-2005}
{Cooper}, C.~S., and A.~P. {Showman}, 2005:
\newblock {Dynamic Meteorology at the Photosphere of HD 209458b}.
\newblock {\it \apjl\/}, {\bf 629}, L45--L48.

\bibitem[{Couhert et~al.(2010){\it Couhert, Schneider, Li, Waliser, and
  Tompkins\/}}]{couhert-etal-2010}
Couhert, A., T.~Schneider, J.~Li, D.~E. Waliser, and A.~M. Tompkins, 2010:
\newblock The maintenance of the relative humidity of the subtropical free
  troposphere.
\newblock {\it J. Climate\/}, {\bf 23}, 390--403.

\bibitem[{{Cowan} and {Agol}(2011){\it {Cowan} and {Agol}\/}}]{cowan-agol-2011}
{Cowan}, N.~B., and E.~{Agol}, 2011:
\newblock {The Statistics of Albedo and Heat Recirculation on Hot Exoplanets}.
\newblock {\it \apj\/}, {\bf 729}, 54.

\bibitem[{{Cowan} et~al.(2009){\it {Cowan}, et~al.\/}}]{cowan-etal-2009}
{Cowan}, N.~B., et~al., 2009:
\newblock {Alien Maps of an Ocean-bearing World}.
\newblock {\it \apj\/}, {\bf 700}, 915--923.

\bibitem[{{de Wit} et~al.(2012){\it {de Wit}, {Gillon}, {Demory}, and
  {Seager}\/}}]{de-wit-etal-2012}
{de Wit}, J., M.~{Gillon}, B.-O. {Demory}, and S.~{Seager}, 2012:
\newblock {Towards consistent mapping of distant worlds: secondary-eclipse
  scanning of the exoplanet HD 189733b}.
\newblock {\it \aap\/}, {\bf 548}, A128.

\bibitem[{{Del Genio} and {Suozzo}(1987){\it {Del Genio} and
  {Suozzo}\/}}]{delgenio-suozzo-1987}
{Del Genio}, A.~D., and R.~J. {Suozzo}, 1987:
\newblock {A comparative study of rapidly and slowly rotating dynamical regimes
  in a terrestrial general circulation model}.
\newblock {\it Journal of Atmospheric Sciences\/}, {\bf 44}, 973--986.

\bibitem[{{Del Genio} and {Zhou}(1996){\it {Del Genio} and
  {Zhou}\/}}]{delgenio-zhou-1996}
{Del Genio}, A.~D., and W.~{Zhou}, 1996:
\newblock {Simulations of Superrotation on Slowly Rotating Planets: Sensitivity
  to Rotation and Initial Condition}.
\newblock {\it Icarus\/}, {\bf 120}, 332--343.

\bibitem[{{Del Genio} et~al.(1993){\it {Del Genio}, {Zhou}, and
  {Eichler}\/}}]{delgenio-etal-1993}
{Del Genio}, A.~D., W.~{Zhou}, and T.~P. {Eichler}, 1993:
\newblock {Equatorial superrotation in a slowly rotating GCM - Implications for
  Titan and Venus}.
\newblock {\it Icarus\/}, {\bf 101}, 1--17.

\bibitem[{{Del Genio} et~al.(2009){\it {Del Genio}, {Achterberg}, {Baines},
  {Flasar}, {Read}, {S{\'a}nchez-Lavega}, and
  {Showman}\/}}]{delgenio-etal-2009}
{Del Genio}, A.~D., R.~K. {Achterberg}, K.~H. {Baines}, F.~M. {Flasar}, P.~L.
  {Read}, A.~{S{\'a}nchez-Lavega}, and A.~P. {Showman}, 2009:
\newblock {Saturn Atmospheric Structure and Dynamics}.
\newblock  {\it Saturn from Cassini-Huygens (Dougherty, M.~K., Esposito, L.~W.
  and Krimigis, S.~M., Eds.)\/}, Springer, p. 113.

\bibitem[{{Deming} and {Seager}(2009){\it {Deming} and
  {Seager}\/}}]{deming-seager-2009}
{Deming}, D., and S.~{Seager}, 2009:
\newblock {Light and shadow from distant worlds}.
\newblock {\it \nat\/}, {\bf 462}, 301--306.

\bibitem[{{Demory} et~al.(2012){\it {Demory}, {Gillon}, {Seager}, {Benneke},
  {Deming}, and {Jackson}\/}}]{demory-etal-2012}
{Demory}, B.-O., M.~{Gillon}, S.~{Seager}, B.~{Benneke}, D.~{Deming}, and
  B.~{Jackson}, 2012:
\newblock {Detection of Thermal Emission from a Super-Earth}.
\newblock {\it \apjl\/}, {\bf 751}, L28.

\bibitem[{{D{\'e}sert} et~al.(2011){\it {D{\'e}sert}, {Bean}, {Miller-Ricci
  Kempton}, {Berta}, {Charbonneau}, {Irwin}, {Fortney}, {Burke}, and
  {Nutzman}\/}}]{desert-etal-2011}
{D{\'e}sert}, J.-M., J.~{Bean}, E.~{Miller-Ricci Kempton}, Z.~K. {Berta},
  D.~{Charbonneau}, J.~{Irwin}, J.~{Fortney}, C.~J. {Burke}, and P.~{Nutzman},
  2011:
\newblock {Observational Evidence for a Metal-rich Atmosphere on the
  Super-Earth GJ1214b}.
\newblock {\it \apjl\/}, {\bf 731}, L40.

\bibitem[{{DeWitt} et~al.(2009){\it {DeWitt}, {Trainer}, {Pavlov}, {Hasenkopf},
  {Aiken}, {Jimenez}, {McKay}, {Toon}, and {Tolbert}\/}}]{dewitt-etal-2009}
{DeWitt}, H.~L., M.~G. {Trainer}, A.~A. {Pavlov}, C.~A. {Hasenkopf}, A.~C.
  {Aiken}, J.~L. {Jimenez}, C.~P. {McKay}, O.~B. {Toon}, and M.~A. {Tolbert},
  2009:
\newblock {Reduction in Haze Formation Rate on Prebiotic Earth in the Presence
  of Hydrogen}.
\newblock {\it Astrobiology\/}, {\bf 9}, 447--453.

\bibitem[{{Dritschel} and {McIntyre}(2008){\it {Dritschel} and
  {McIntyre}\/}}]{dritschel-mcintyre-2008}
{Dritschel}, D.~G., and M.~E. {McIntyre}, 2008:
\newblock {Multiple Jets as PV Staircases: The Phillips Effect and the
  Resilience of Eddy-Transport Barriers}.
\newblock {\it Journal of Atmospheric Sciences\/}, {\bf 65}, 855.

\bibitem[{{Dritschel} and {Scott}(2011){\it {Dritschel} and
  {Scott}\/}}]{dritschel-scott-2011}
{Dritschel}, D.~G., and R.~K. {Scott}, 2011:
\newblock {Jet sharpening by turbulent mixing}.
\newblock {\it Phil. Trans. Roy. Soc. A\/}, {\bf 369}, 754--770.

\bibitem[{{Dunkerton} and {Scott}(2008){\it {Dunkerton} and
  {Scott}\/}}]{dunkerton-scott-2008}
{Dunkerton}, T.~J., and R.~K. {Scott}, 2008:
\newblock {A Barotropic Model of the Angular Momentum Conserving Potential
  Vorticity Staircase in Spherical Geometry}.
\newblock {\it Journal of Atmospheric Sciences\/}, {\bf 65}, 1105.

\bibitem[{{Edson} et~al.(2011){\it {Edson}, {Lee}, {Bannon}, {Kasting}, and
  {Pollard}\/}}]{edson-etal-2011}
{Edson}, A., S.~{Lee}, P.~{Bannon}, J.~F. {Kasting}, and D.~{Pollard}, 2011:
\newblock {Atmospheric circulations of terrestrial planets orbiting low-mass
  stars}.
\newblock {\it \icarus\/}, {\bf 212}, 1--13.

\bibitem[{{Edson} et~al.(2012){\it {Edson}, {Kasting}, {Pollard}, {Lee}, and
  {Bannon}\/}}]{edson-etal-2012}
{Edson}, A.~R., J.~F. {Kasting}, D.~{Pollard}, S.~{Lee}, and P.~R. {Bannon},
  2012:
\newblock {The Carbonate-Silicate Cycle and CO2/Climate Feedbacks on Tidally
  Locked Terrestrial Planets}.
\newblock {\it Astrobiology\/}, {\bf 12}, 562--571.

\bibitem[{Emanuel(1994){\it Emanuel\/}}]{emanuel-1994}
Emanuel, K.~A., 1994:
\newblock {\it Atmospheric Convection\/}.
\newblock Oxford University Press.

\bibitem[{Enderton and Marshall(2009){\it Enderton and
  Marshall\/}}]{enderton-marshall-09}
Enderton, D., and J.~Marshall, 2009:
\newblock Explorations of atmosphere-ocean-ice climates on an aquaplanet and
  their meridional energy transports.
\newblock {\it J. Atmos. Sci.\/}, {\bf 66}, 1593--1611.

\bibitem[{{Fang} and {Tung}(1996){\it {Fang} and {Tung}\/}}]{fang-tung-1996}
{Fang}, M., and K.~K. {Tung}, 1996:
\newblock {A simple study of nonlinear Hadley circulation with an ITCZ:
  analytic and numerical solutions}.
\newblock {\it J. Atmos. Sci.\/}, {\bf 53}, 1241--1261.

\bibitem[{{Farrell}(1990){\it {Farrell}\/}}]{farrell-1990}
{Farrell}, B.~F., 1990:
\newblock {Equable Climate Dynamics.}
\newblock {\it Journal of Atmospheric Sciences\/}, {\bf 47}, 2986--2995.

\bibitem[{Ferrari and Ferreira(2011){\it Ferrari and
  Ferreira\/}}]{ferrari-ferreira-2011}
Ferrari, R., and D.~Ferreira, 2011:
\newblock What processes drive the ocean heat transport?
\newblock {\it Ocean Modelling\/}, {\bf 38}, 171--186.

\bibitem[{{Flasar} et~al.(2009){\it {Flasar}, {Baines}, {Bird}, {Tokano}, and
  {West}\/}}]{flasar-etal-2009}
{Flasar}, F.~M., K.~H. {Baines}, M.~K. {Bird}, T.~{Tokano}, and R.~A. {West},
  2009:
\newblock {Atmospheric dynamics and meteorology}.
\newblock  {\it Titan from Cassini-Huygens (Brown, R.H., Lebreton, J-P., and
  Waite, J.H., Eds.)\/}, Springer, pp. 323--352.

\bibitem[{{Fressin} et~al.(2012){\it {Fressin}, et~al.\/}}]{fressin-etal-2012}
{Fressin}, F., et~al., 2012:
\newblock {Two Earth-sized planets orbiting Kepler-20}.
\newblock {\it \nat\/}, {\bf 482}, 195--198.

\bibitem[{Frierson(2007){\it Frierson\/}}]{frierson-2007}
Frierson, D. M.~W., 2007:
\newblock The dynamics of idealized convection schemes and their effect on the
  zonally averaged tropical circulation.
\newblock {\it J. Atmos. Sci.\/}, {\bf 64}, 1959--1976.

\bibitem[{{Frierson} et~al.(2006){\it {Frierson}, {Held}, and
  {Zurita-Gotor}\/}}]{frierson-etal-2006}
{Frierson}, D.~M.~W., I.~M. {Held}, and P.~{Zurita-Gotor}, 2006:
\newblock {A Gray-Radiation Aquaplanet Moist GCM. Part I: Static Stability and
  Eddy Scale}.
\newblock {\it Journal of Atmospheric Sciences\/}, {\bf 63}, 2548--2566.

\bibitem[{{Fujii} and {Kawahara}(2012){\it {Fujii} and
  {Kawahara}\/}}]{fujii-kawahara-2012}
{Fujii}, Y., and H.~{Kawahara}, 2012:
\newblock {Mapping Earth Analogs from Photometric Variability: Spin-Orbit
  Tomography for Planets in Inclined Orbits}.
\newblock {\it \apj\/}, {\bf 755}, 101.

\bibitem[{{Fujii} et~al.(2011){\it {Fujii}, {Kawahara}, {Suto}, {Fukuda},
  {Nakajima}, {Livengood}, and {Turner}\/}}]{fujii-etal-2011}
{Fujii}, Y., H.~{Kawahara}, Y.~{Suto}, S.~{Fukuda}, T.~{Nakajima}, T.~A.
  {Livengood}, and E.~L. {Turner}, 2011:
\newblock {Colors of a Second Earth. II. Effects of Clouds on Photometric
  Characterization of Earth-like Exoplanets}.
\newblock {\it \apj\/}, {\bf 738}, 184.

\bibitem[{Galewsky et~al.(2005){\it Galewsky, Sobel, and
  Held\/}}]{galewsky-etal-2005}
Galewsky, J., A.~Sobel, and I.~Held, 2005:
\newblock Diagnosis of subtropical humidity dynamics using tracers of last
  saturation.
\newblock {\it J. Atmos. Sci.\/}, {\bf 62}, 3353--3367.

\bibitem[{{Garcia} and {Salby}(1987){\it {Garcia} and
  {Salby}\/}}]{garcia-salby-1987}
{Garcia}, R.~R., and M.~L. {Salby}, 1987:
\newblock {Transient Response to Localized Episodic Heating in the Tropics.
  Part II: Far-Field Behavior.}
\newblock {\it Journal of Atmospheric Sciences\/}, {\bf 44}, 499--532.

\bibitem[{{Garratt}(1994){\it {Garratt}\/}}]{garratt-1994}
{Garratt}, J.~R., 1994:
\newblock {\it The Atmospheric Boundary Layer\/}.
\newblock Cambridge University Press.

\bibitem[{{Gierasch} et~al.(1997){\it {Gierasch},
  et~al.\/}}]{gierasch-etal-1997}
{Gierasch}, P.~J., et~al., 1997:
\newblock {The General Circulation of the Venus Atmosphere: an Assessment}.
\newblock  {\it Venus II: Geology, Geophysics, Atmosphere, and Solar Wind
  Environment\/}, S.~W. {Bougher}, D.~M. {Hunten}, and R.~J. {Philips}, Eds.,
  pp. 459--500.

\bibitem[{{Gill}(1980){\it {Gill}\/}}]{gill-1980}
{Gill}, A.~E., 1980:
\newblock {Some simple solutions for heat-induced tropical circulation}.
\newblock {\it Q. J. Roy. Meteor. Soc.\/}, {\bf 106}, 447--462.

\bibitem[{{Gillmann} et~al.(2009){\it {Gillmann}, {Chassefi{\`e}re}, and
  {Lognonn{\'e}}\/}}]{gillmann-etal-2009}
{Gillmann}, C., E.~{Chassefi{\`e}re}, and P.~{Lognonn{\'e}}, 2009:
\newblock {A consistent picture of early hydrodynamic escape of Venus
  atmosphere explaining present Ne and Ar isotopic ratios and low oxygen
  atmospheric content}.
\newblock {\it Earth and Planetary Science Letters\/}, {\bf 286}, 503--513.

\bibitem[{{G{\'o}mez-Leal} et~al.(2012){\it {G{\'o}mez-Leal}, {Pall{\'e}}, and
  {Selsis}\/}}]{gomez-leal-etal-2012}
{G{\'o}mez-Leal}, I., E.~{Pall{\'e}}, and F.~{Selsis}, 2012:
\newblock {Photometric Variability of the Disk-integrated Thermal Emission of
  the Earth}.
\newblock {\it \apj\/}, {\bf 752}, 28.

\bibitem[{{Green}(1970){\it {Green}\/}}]{green-1970}
{Green}, J.~S.~A., 1970:
\newblock {Transfer properties of the large-scale eddies and the general
  circulation of the atmosphere}.
\newblock {\it Quarterly Journal of the Royal Meteorological Society\/}, {\bf
  96}, 157--185.

\bibitem[{{Haine} and {Marshall}(1998){\it {Haine} and
  {Marshall}\/}}]{haine-marshall-1998}
{Haine}, T.~W.~N., and J.~{Marshall}, 1998:
\newblock {Gravitational, Symmetric, and Baroclinic Instability of the Ocean
  Mixed Layer}.
\newblock {\it Journal of Physical Oceanography\/}, {\bf 28}, 634--658.

\bibitem[{Hartmann(1994{\natexlab{a}}){\it Hartmann\/}}]{hartmann-1994}
Hartmann, D.~L., 1994{\natexlab{a}}:
\newblock A {PV} view of zonal mean flow vacillation, submitted to {\em J.
  Atmos. Sci.}

\bibitem[{Hartmann(1994{\natexlab{b}}){\it Hartmann\/}}]{hartmann-1994-book}
Hartmann, D.~L., 1994{\natexlab{b}}:
\newblock {\it Global physical climatology\/}.
\newblock vol.~56, Academic press.

\bibitem[{{Hayashi} et~al.(2000){\it {Hayashi}, {Ishioka}, {Yamada}, and
  {Yoden}\/}}]{hayashi-etal-2000}
{Hayashi}, Y.-Y., K.~{Ishioka}, M.~{Yamada}, and S.~{Yoden}, 2000:
\newblock {Emergence of circumpolar vortex in two dimensional turbulence on a
  rotating sphere}.
\newblock  {\it Proceedings of the IUTAM Symposium on Developments in
  Geophysical Turbulence (Fluid Mechanics and its Applications V. 58\/}, R.~M.
  Kerr and Y.~Kimura, Eds., Kluwer Academic Pub., pp. 179--192.

\bibitem[{{Hayashi} et~al.(2007){\it {Hayashi}, {Nishizawa}, {Takehiro},
  {Yamada}, {Ishioka}, and {Yoden}\/}}]{hayashi-etal-2007}
{Hayashi}, Y.-Y., S.~{Nishizawa}, S.-I. {Takehiro}, M.~{Yamada}, K.~{Ishioka},
  and S.~{Yoden}, 2007:
\newblock {Rossby Waves and Jets in Two-Dimensional Decaying Turbulence on a
  Rotating Sphere}.
\newblock {\it Journal of Atmospheric Sciences\/}, {\bf 64}, 4246--4269.

\bibitem[{{Hedelt} et~al.(2011){\it {Hedelt}, {Alonso}, {Brown}, {Collados
  Vera}, {Rauer}, {Schleicher}, {Schmidt}, {Schreier}, and
  {Titz}\/}}]{hedelt-etal-2011}
{Hedelt}, P., R.~{Alonso}, T.~{Brown}, M.~{Collados Vera}, H.~{Rauer},
  H.~{Schleicher}, W.~{Schmidt}, F.~{Schreier}, and R.~{Titz}, 2011:
\newblock {Venus transit 2004: Illustrating the capability of exoplanet
  transmission spectroscopy}.
\newblock {\it \aap\/}, {\bf 533}, A136.

\bibitem[{{Held}(1975){\it {Held}\/}}]{held-1975}
{Held}, I.~M., 1975:
\newblock {Momentum transport by quasi-geostrophic eddies}.
\newblock {\it Journal of the Atmospheric Sciences\/}, {\bf 32}, 1494--1497.

\bibitem[{Held(1982){\it Held\/}}]{held-1982}
Held, I.~M., 1982:
\newblock On the height of the tropopause and the static stability of the
  troposphere.
\newblock {\it J. Atmos. Sci.\/}, {\bf 39}, 412--417.

\bibitem[{{Held}(1999{\natexlab{a}}){\it {Held}\/}}]{held-1999}
{Held}, I.~M., 1999{\natexlab{a}}:
\newblock {The macroturbulence of the troposphere}.
\newblock {\it Tellus\/}, {\bf 51A-B}, 59--70.

\bibitem[{{Held}(1999{\natexlab{b}}){\it {Held}\/}}]{held-1999b}
{Held}, I.~M., 1999{\natexlab{b}}:
\newblock {Equatorial superrotation in Earth-like atmospheric models}.
\newblock {\it Bernhard Haurwitz Memorial Lecture\/}, {\bf American
  Meteorological Society}, available at
  www.gfdl.noaa.gov/isaac--held--homepage.

\bibitem[{{Held}(2000){\it {Held}\/}}]{held-2000}
{Held}, I.~M., 2000:
\newblock {The general circulation of the atmosphere}.
\newblock {\it Paper presented at 2000 Woods Hole Oceanographic Institute
  Geophysical Fluid Dynamics Program, Woods Hole Oceanographic Institute, Woods
  Hole, MA (available at http://www.whoi.edu/page.do?pid=13076)\/}.

\bibitem[{Held(2001){\it Held\/}}]{held-2001}
Held, I.~M., 2001:
\newblock The partitioning of the poleward energy transport between the
  tropical ocean and atmosphere.
\newblock {\it J. Atmos. Sci.\/}, {\bf 58}, 943--948.

\bibitem[{{Held} and {Hou}(1980){\it {Held} and {Hou}\/}}]{held-hou-1980}
{Held}, I.~M., and A.~Y. {Hou}, 1980:
\newblock {Nonlinear Axially Symmetric Circulations in a Nearly Inviscid
  Atmosphere.}
\newblock {\it Journal of Atmospheric Sciences\/}, {\bf 37}, 515--533.

\bibitem[{{Held} and {Larichev}(1996){\it {Held} and
  {Larichev}\/}}]{held-larichev-1996}
{Held}, I.~M., and V.~D. {Larichev}, 1996:
\newblock {A scaling theory for horizontally homogeneous, baroclinically
  unstable flow on a beta plane}.
\newblock {\it Journal of Atmospheric Sciences\/}, {\bf 53}, 946--952.

\bibitem[{Held and Shell(2012){\it Held and Shell\/}}]{held-shell-2012}
Held, I.~M., and K.~M. Shell, 2012:
\newblock Using relative humidity as a state variable in climate feedback
  analysis.
\newblock {\it J. Climate\/}, {\bf 25}, 2578--2582.

\bibitem[{{Held} and {Soden}(2000){\it {Held} and {Soden}\/}}]{held-soden-2000}
{Held}, I.~M., and B.~J. {Soden}, 2000:
\newblock {Water vapor feedback anad global warming}.
\newblock {\it Annu. Rev. Energy Environ.\/}, {\bf 25}, 441--475.

\bibitem[{{Held} and {Soden}(2006){\it {Held} and {Soden}\/}}]{held-soden-2006}
{Held}, I.~M., and B.~J. {Soden}, 2006:
\newblock {Robust Responses of the Hydrological Cycle to Global Warming}.
\newblock {\it Journal of Climate\/}, {\bf 19}, 5686.

\bibitem[{Held et~al.(2002){\it Held, Ting, and Wang\/}}]{held-etal-2002}
Held, I.~M., M.~Ting, and H.~Wang, 2002:
\newblock Northern winter stationary waves: theory and modeling.
\newblock {\it J. Climate\/}, {\bf 15}, 2125--2144.

\bibitem[{{Heng} and {Kopparla}(2012){\it {Heng} and
  {Kopparla}\/}}]{heng-kopparla-2012}
{Heng}, K., and P.~{Kopparla}, 2012:
\newblock {On the Stability of Super-Earth Atmospheres}.
\newblock {\it \apj\/}, {\bf 754}, 60.

\bibitem[{{Heng} and {Vogt}(2010){\it {Heng} and {Vogt}\/}}]{heng-vogt-2011}
{Heng}, K., and S.~S. {Vogt}, 2010:
\newblock {Gliese 581g as a scaled-up version of Earth: atmospheric circulation
  simulations}.
\newblock {\it ArXiv e-prints\/}.

\bibitem[{{Heng} et~al.(2011{\natexlab{a}}){\it {Heng}, {Frierson}, and
  {Phillipps}\/}}]{heng-etal-2011b}
{Heng}, K., D.~M.~W. {Frierson}, and P.~J. {Phillipps}, 2011{\natexlab{a}}:
\newblock {Atmospheric circulation of tidally locked exoplanets: II. Dual-band
  radiative transfer and convective adjustment}.
\newblock {\it \mnras\/}, {\bf 418}, 2669--2696.

\bibitem[{{Heng} et~al.(2011{\natexlab{b}}){\it {Heng}, {Menou}, and
  {Phillipps}\/}}]{heng-etal-2011}
{Heng}, K., K.~{Menou}, and P.~J. {Phillipps}, 2011{\natexlab{b}}:
\newblock {Atmospheric circulation of tidally locked exoplanets: a suite of
  benchmark tests for dynamical solvers}.
\newblock {\it \mnras\/}, {\bf 413}, 2380--2402.

\bibitem[{{Herrnstein} and {Dowling}(2007){\it {Herrnstein} and
  {Dowling}\/}}]{herrnstein-dowling-2007}
{Herrnstein}, A., and T.~E. {Dowling}, 2007:
\newblock {Effects of topography on the spin-up of a Venus atmospheric model}.
\newblock {\it Journal of Geophysical Research (Planets)\/}, {\bf 112}({E11}).

\bibitem[{{Hide}(1969){\it {Hide}\/}}]{hide-1969}
{Hide}, R., 1969:
\newblock {Dynamics of the Atmospheres of the Major Planets with an Appendix on
  the Viscous Boundary Layer at the Rigid Bounding Surface of an
  Electrically-Conducting Rotating Fluid in the Presence of a Magnetic Field.}
\newblock {\it J. Atmos. Sci.\/}, {\bf 26}, 841--853.

\bibitem[{{Hoffman} and {Schrag}(2002){\it {Hoffman} and
  {Schrag}\/}}]{hoffman-schrag-2002}
{Hoffman}, P.~F., and D.~{Schrag}, 2002:
\newblock {Review article: The snowball Earth hypothesis: testing the limits of
  global change}.
\newblock {\it Terra Nova\/}, {\bf 14}, 129--155.

\bibitem[{{Hollingsworth} et~al.(2007){\it {Hollingsworth}, {Young},
  {Schubert}, {Covey}, and {Grossman}\/}}]{hollingsworth-etal-2007}
{Hollingsworth}, J.~L., R.~E. {Young}, G.~{Schubert}, C.~{Covey}, and A.~S.
  {Grossman}, 2007:
\newblock {A simple-physics global circulation model for Venus: Sensitivity
  assessments of atmospheric superrotation}.
\newblock {\it \grl\/}, {\bf 34}, 5202.

\bibitem[{{Holton}(2004){\it {Holton}\/}}]{holton-2004}
{Holton}, J.~R., 2004:
\newblock {\it An Introduction to Dynamic Meteorology, 4th Ed.\/}.
\newblock Academic Press, San Diego.

\bibitem[{{Hoskins} et~al.(1999){\it {Hoskins}, {Neale}, {Rodwell}, and
  {Yang}\/}}]{hoskins-etal-1999}
{Hoskins}, B., R.~{Neale}, M.~{Rodwell}, and G.~{Yang}, 1999:
\newblock {Aspects of the large-scale tropical atmospheric circulation}.
\newblock {\it Tellus Series B Chemical and Physical Meteorology B\/}, {\bf
  51}, 33--44.

\bibitem[{{Hoskins} and {Karoly}(1981){\it {Hoskins} and
  {Karoly}\/}}]{hoskins-karoly-1981}
{Hoskins}, B.~J., and D.~J. {Karoly}, 1981:
\newblock {The Steady Linear Response of a Spherical Atmosphere to Thermal and
  Orographic Forcing.}
\newblock {\it Journal of Atmospheric Sciences\/}, {\bf 38}, 1179--1196.

\bibitem[{{Hu} and {Ding}(2011){\it {Hu} and {Ding}\/}}]{hu-ding-2011}
{Hu}, Y., and F.~{Ding}, 2011:
\newblock {Radiative constraints on the habitability of exoplanets Gliese 581c
  and Gliese 581d}.
\newblock {\it \aap\/}, {\bf 526}, A135.

\bibitem[{{Huang} and {Robinson}(1998){\it {Huang} and
  {Robinson}\/}}]{huang-robinson-1998}
{Huang}, H.-P., and W.~A. {Robinson}, 1998:
\newblock {Two-Dimensional Turbulence and Persistent Zonal Jets in a Global
  Barotropic Model.}
\newblock {\it Journal of Atmospheric Sciences\/}, {\bf 55}, 611--632.

\bibitem[{{Hunt}(1979){\it {Hunt}\/}}]{hunt-1979}
{Hunt}, B.~G., 1979:
\newblock {The Influence of the Earth's Rotation Rate on the General
  Circulation of the Atmosphere.}
\newblock {\it Journal of Atmospheric Sciences\/}, {\bf 36}, 1392--1408.

\bibitem[{{Ingersoll}(1969){\it {Ingersoll}\/}}]{ingersoll-1969}
{Ingersoll}, A.~P., 1969:
\newblock {The Runaway Greenhouse: A History of Water on Venus.}
\newblock {\it Journal of Atmospheric Sciences\/}, {\bf 26}, 1191--1198.

\bibitem[{{Ingersoll}(1989){\it {Ingersoll}\/}}]{ingersoll-1989}
{Ingersoll}, A.~P., 1989:
\newblock {Io meteorology - How atmospheric pressure is controlled locally by
  volcanos and surface frosts}.
\newblock {\it \icarus\/}, {\bf 81}, 298--313.

\bibitem[{{Ingersoll} et~al.(1985){\it {Ingersoll}, {Summers}, and
  {Schlipf}\/}}]{ingersoll-etal-1985}
{Ingersoll}, A.~P., M.~E. {Summers}, and S.~G. {Schlipf}, 1985:
\newblock {Supersonic meteorology of Io - Sublimation-driven flow of SO2}.
\newblock {\it \icarus\/}, {\bf 64}, 375--390.

\bibitem[{{Ingersoll} et~al.(2004){\it {Ingersoll}, {Dowling}, {Gierasch},
  {Orton}, {Read}, {S{\'a}nchez-Lavega}, {Showman}, {Simon-Miller}, and
  {Vasavada}\/}}]{ingersoll-etal-2004}
{Ingersoll}, A.~P., T.~E. {Dowling}, P.~J. {Gierasch}, G.~S. {Orton}, P.~L.
  {Read}, A.~{S{\'a}nchez-Lavega}, A.~P. {Showman}, A.~A. {Simon-Miller}, and
  A.~R. {Vasavada}, 2004:
\newblock {\it {Dynamics of Jupiter's atmosphere}\/}, pp. 105--128,
  Jupiter.~The Planet, Satellites and Magnetosphere.

\bibitem[{{Ishioka} et~al.(1999){\it {Ishioka}, {Yamada}, {Hayashi}, and
  {Yoden}\/}}]{ishioka-etal-1999}
{Ishioka}, K., M.~{Yamada}, Y.-Y. {Hayashi}, and S.~{Yoden}, 1999:
\newblock {Pattern formation from two-dimensional decaying turbulence on a
  rotating sphere}.
\newblock {\it Nagare Multimedia, The Japan Society of Fluid Mechanics\/}, {\bf
  [available online at http://www.nagare.or.jp/mm/99/ishioka/]}.

\bibitem[{{Ishiwatari} et~al.(2002){\it {Ishiwatari}, {Takehiro}, {Nakajima},
  and {Hayashi}\/}}]{ishiwatari-etal-2002}
{Ishiwatari}, M., S.-I. {Takehiro}, K.~{Nakajima}, and Y.-Y. {Hayashi}, 2002:
\newblock {A Numerical Study on Appearance of the Runaway Greenhouse State of a
  Three-Dimensional Gray Atmosphere.}
\newblock {\it Journal of Atmospheric Sciences\/}, {\bf 59}, 3223--3238.

\bibitem[{{Ishiwatari} et~al.(2007){\it {Ishiwatari}, {Nakajima}, {Takehiro},
  and {Hayashi}\/}}]{ishiwatari-etal-2007}
{Ishiwatari}, M., K.~{Nakajima}, S.~{Takehiro}, and Y.-Y. {Hayashi}, 2007:
\newblock {Dependence of climate states of gray atmosphere on solar constant:
  From the runaway greenhouse to the snowball states}.
\newblock {\it Journal of Geophysical Research (Atmospheres)\/}, {\bf
  112}(D11), 13,120.

\bibitem[{{Jansen} and {Ferrari}(2012){\it {Jansen} and
  {Ferrari}\/}}]{jansen-ferrari-2012}
{Jansen}, M., and R.~{Ferrari}, 2012:
\newblock {Macroturbulent Equilibration in a Thermally Forced Primitive
  Equation System}.
\newblock {\it Journal of Atmospheric Sciences\/}, {\bf 69}, 695--713.

\bibitem[{{Joshi}(2003){\it {Joshi}\/}}]{joshi-2003}
{Joshi}, M., 2003:
\newblock {Climate Model Studies of Synchronously Rotating Planets}.
\newblock {\it Astrobiology\/}, {\bf 3}, 415--427.

\bibitem[{{Joshi} and {Haberle}(2012){\it {Joshi} and
  {Haberle}\/}}]{joshi-haberle-2011}
{Joshi}, M.~M., and R.~M. {Haberle}, 2012:
\newblock {Suppression of the Water Ice and Snow Albedo Feedback on Planets
  Orbiting Red Dwarf Stars and the Subsequent Widening of the Habitable Zone}.
\newblock {\it Astrobiology\/}, {\bf 12}, 3--8.

\bibitem[{{Joshi} et~al.(1997){\it {Joshi}, {Haberle}, and
  {Reynolds}\/}}]{joshi-etal-1997}
{Joshi}, M.~M., R.~M. {Haberle}, and R.~T. {Reynolds}, 1997:
\newblock {Simulations of the Atmospheres of Synchronously Rotating Terrestrial
  Planets Orbiting M Dwarfs: Conditions for Atmospheric Collapse and the
  Implications for Habitability}.
\newblock {\it Icarus\/}, {\bf 129}, 450--465.

\bibitem[{Juckes(2000){\it Juckes\/}}]{juckes-2000}
Juckes, M.~N., 2000:
\newblock The static stability of the midlatitude troposphere: The relevance of
  moisture.
\newblock {\it J. Atmos. Sci.\/}, {\bf 57}, 3050--3057.

\bibitem[{{Kaltenegger} et~al.(2011){\it {Kaltenegger}, {Segura}, and
  {Mohanty}\/}}]{kaltenegger-etal-2011}
{Kaltenegger}, L., A.~{Segura}, and S.~{Mohanty}, 2011:
\newblock {Model Spectra of the First Potentially Habitable Super-Earth -
  Gl581d}.
\newblock {\it \apj\/}, {\bf 733}, 35.

\bibitem[{{Karalidi} et~al.(2012){\it {Karalidi}, {Stam}, and
  {Hovenier}\/}}]{karalidi-etal-2012}
{Karalidi}, T., D.~M. {Stam}, and J.~W. {Hovenier}, 2012:
\newblock {Looking for the rainbow on exoplanets covered by liquid and icy
  water clouds}.
\newblock {\it \aap\/}, {\bf 548}, A90.

\bibitem[{{Kaspi} and {Schneider}(2011){\it {Kaspi} and
  {Schneider}\/}}]{kaspi-schneider-2011}
{Kaspi}, Y., and T.~{Schneider}, 2011:
\newblock {Downstream Self-Destruction of Storm Tracks}.
\newblock {\it Journal of Atmospheric Sciences\/}, {\bf 68}, 2459--2464.

\bibitem[{{Kaspi} and {Showman}(2012){\it {Kaspi} and
  {Showman}\/}}]{kaspi-showman-2012}
{Kaspi}, Y., and A.~P. {Showman}, 2012:
\newblock {Three-dimensional Atmospheric Circulation and Climate of Terrestrial
  Exoplanets and Super Earths}.
\newblock  {\it AAS/Division for Planetary Sciences Meeting Abstracts\/},
  vol.~44 of {\it AAS/Division for Planetary Sciences Meeting Abstracts\/}, p.
  \#208.04.

\bibitem[{{Kasting}(1988){\it {Kasting}\/}}]{kasting-1988}
{Kasting}, J.~F., 1988:
\newblock {Runaway and moist greenhouse atmospheres and the evolution of earth
  and Venus}.
\newblock {\it Icarus\/}, {\bf 74}, 472--494.

\bibitem[{{Kasting} et~al.(1993){\it {Kasting}, {Whitmire}, and
  {Reynolds}\/}}]{kasting-etal-1993}
{Kasting}, J.~F., D.~P. {Whitmire}, and R.~T. {Reynolds}, 1993:
\newblock {Habitable Zones around Main Sequence Stars}.
\newblock {\it Icarus\/}, {\bf 101}, 108--128.

\bibitem[{{Kawahara} and {Fujii}(2011){\it {Kawahara} and
  {Fujii}\/}}]{kawahara-fujii-2011}
{Kawahara}, H., and Y.~{Fujii}, 2011:
\newblock {Mapping Clouds and Terrain of Earth-like Planets from Photometric
  Variability: Demonstration with Planets in Face-on Orbits}.
\newblock {\it \apjl\/}, {\bf 739}, L62.

\bibitem[{{Kim} and {Lee}(2001){\it {Kim} and {Lee}\/}}]{kim-lee-2001b}
{Kim}, H.-K., and S.~{Lee}, 2001:
\newblock {Hadley Cell Dynamics in a Primitive Equation Model. Part II:
  Nonaxisymmetric Flow.}
\newblock {\it Journal of Atmospheric Sciences\/}, {\bf 58}, 2859--2871.

\bibitem[{{Kite} et~al.(2011){\it {Kite}, {Gaidos}, and
  {Manga}\/}}]{kite-etal-2011}
{Kite}, E.~S., E.~{Gaidos}, and M.~{Manga}, 2011:
\newblock {Climate Instability on Tidally Locked Exoplanets}.
\newblock {\it \apj\/}, {\bf 743}, 41.

\bibitem[{{Knutson} et~al.(2007){\it {Knutson}, {Charbonneau}, {Allen},
  {Fortney}, {Agol}, {Cowan}, {Showman}, {Cooper}, and
  {Megeath}\/}}]{knutson-etal-2007b}
{Knutson}, H.~A., D.~{Charbonneau}, L.~E. {Allen}, J.~J. {Fortney}, E.~{Agol},
  N.~B. {Cowan}, A.~P. {Showman}, C.~S. {Cooper}, and S.~T. {Megeath}, 2007:
\newblock {A map of the day-night contrast of the extrasolar planet HD
  189733b}.
\newblock {\it \nat\/}, {\bf 447}, 183--186.

\bibitem[{{Kombayashi}(1967){\it {Kombayashi}\/}}]{kombayashi-1967}
{Kombayashi}, M., 1967:
\newblock {Discrete equilibrium temperatures of a hypothetical planet with the
  atmosphere and the hydrosphere of one component-two phase system under
  constant solar radiation.}
\newblock {\it J. Meteor. Soc. Japan\/}, {\bf 45}, 137--138.

\bibitem[{{Kraucunas} and {Hartmann}(2005){\it {Kraucunas} and
  {Hartmann}\/}}]{kraucunas-hartmann-2005}
{Kraucunas}, I., and D.~L. {Hartmann}, 2005:
\newblock {Equatorial Superrotation and the Factors Controlling the Zonal-Mean
  Zonal Winds in the Tropical Upper Troposphere.}
\newblock {\it J. Atmos. Sci.\/}, {\bf 62}, 371--389.

\bibitem[{{Kuo} and {Polvani}(1997){\it {Kuo} and
  {Polvani}\/}}]{kuo-polvani-1997}
{Kuo}, A.~C., and L.~M. {Polvani}, 1997:
\newblock {Time-Dependent Fully Nonlinear Geostrophic Adjustment}.
\newblock {\it Journal of Physical Oceanography\/}, {\bf 27}, 1614--1634.

\bibitem[{{Larichev} and {Held}(1995){\it {Larichev} and
  {Held}\/}}]{larichev-held-1995}
{Larichev}, V.~D., and I.~M. {Held}, 1995:
\newblock {Eddy amplitudes and fluxes in a homogenous model of fully developed
  baroclinic instability}.
\newblock {\it Journal of Physical Oceanography\/}, {\bf 25}, 2285--2297.

\bibitem[{Le~Hir et~al.(2008){\it Le~Hir, Ramstein, Donnadieu, and
  Godd{\'e}ris\/}}]{lehir-etal-2008}
Le~Hir, G., G.~Ramstein, Y.~Donnadieu, and Y.~Godd{\'e}ris, 2008:
\newblock Scenario for the evolution of atmospheric pco2 during a snowball
  earth.
\newblock {\it Geology\/}, {\bf 36}(1), 47--50.

\bibitem[{Le~Hir et~al.(2009){\it Le~Hir, Donnadieu, Godd{\'e}ris,
  Pierrehumbert, Halverson, Macouin, N{\'e}d{\'e}lec, and
  Ramstein\/}}]{lehir-etal-2009}
Le~Hir, G., Y.~Donnadieu, Y.~Godd{\'e}ris, R.~T. Pierrehumbert, G.~P.
  Halverson, M.~Macouin, A.~N{\'e}d{\'e}lec, and G.~Ramstein, 2009:
\newblock The snowball earth aftermath: Exploring the limits of continental
  weathering processes.
\newblock {\it Earth and Planetary Science Letters\/}, {\bf 277}(3), 453--463.

\bibitem[{{Lebonnois} et~al.(2012){\it {Lebonnois}, {Covey}, {Grossman},
  {Parish}, {Schubert}, {Walterscheid}, {Lauritzen}, and
  {Jablonowski}\/}}]{lebonnois-etal-2012}
{Lebonnois}, S., C.~{Covey}, A.~{Grossman}, H.~{Parish}, G.~{Schubert},
  R.~{Walterscheid}, P.~{Lauritzen}, and C.~{Jablonowski}, 2012:
\newblock {Angular momentum budget in General Circulation Models of
  superrotating atmospheres: A critical diagnostic}.
\newblock {\it Journal of Geophysical Research (Planets)\/}, {\bf 117}(E16),
  12,004.

\bibitem[{{Leconte} et~al.(2013){\it {Leconte}, {Forget}, {Charnay},
  {Wordsworth}, {Selsis}, {Millour}, and {Spiga}\/}}]{leconte-etal-2013}
{Leconte}, J., F.~{Forget}, B.~{Charnay}, R.~{Wordsworth}, F.~{Selsis},
  E.~{Millour}, and A.~{Spiga}, 2013:
\newblock {3D climate modeling of close-in land planets: Circulation patterns,
  climate moist bistability, and habitability}.
\newblock {\it Astronomy and Astrophysics\/}, {\bf in press}.

\bibitem[{{Lee} et~al.(2007){\it {Lee}, {Lewis}, and {Read}\/}}]{lee-etal-2007}
{Lee}, C., S.~R. {Lewis}, and P.~L. {Read}, 2007:
\newblock {Superrotation in a Venus general circulation model}.
\newblock {\it Journal of Geophysical Research (Planets)\/}, {\bf 112}({E11}).

\bibitem[{{L{\'e}ger} et~al.(2004){\it {L{\'e}ger},
  et~al.\/}}]{leger-etal-2004}
{L{\'e}ger}, A., et~al., 2004:
\newblock {A new family of planets? ``Ocean-Planets''}.
\newblock {\it \icarus\/}, {\bf 169}, 499--504.

\bibitem[{{L{\'e}ger} et~al.(2011){\it {L{\'e}ger},
  et~al.\/}}]{leger-etal-2011}
{L{\'e}ger}, A., et~al., 2011:
\newblock {The extreme physical properties of the CoRoT-7b super-Earth}.
\newblock {\it \icarus\/}, {\bf 213}, 1--11.

\bibitem[{{Leovy}(2001){\it {Leovy}\/}}]{leovy-2001}
{Leovy}, C., 2001:
\newblock {Weather and climate on Mars}.
\newblock {\it \nat\/}, {\bf 412}, 245--249.

\bibitem[{{Lindzen} and {Hou}(1988){\it {Lindzen} and
  {Hou}\/}}]{lindzen-hou-1988}
{Lindzen}, R.~S., and A.~V. {Hou}, 1988:
\newblock {Hadley Circulations for Zonally Averaged Heating Centered off the
  Equator.}
\newblock {\it Journal of Atmospheric Sciences\/}, {\bf 45}, 2416--2427.

\bibitem[{Lindzen and Nigam(1987){\it Lindzen and
  Nigam\/}}]{lindzen-nigam-1987}
Lindzen, R.~S., and S.~Nigam, 1987:
\newblock On the role of sea surface temperature gradients in forcing low-level
  winds and convergence in the {T}ropics.
\newblock {\it J. Atmos. Sci.\/}, {\bf 44}, 2418--2436.

\bibitem[{{Lissauer} et~al.(2011){\it {Lissauer},
  et~al.\/}}]{lissauer-etal-2011}
{Lissauer}, J.~J., et~al., 2011:
\newblock {A closely packed system of low-mass, low-density planets transiting
  Kepler-11}.
\newblock {\it \nat\/}, {\bf 470}, 53--58.

\bibitem[{{Lopez} et~al.(2012){\it {Lopez}, {Fortney}, and
  {Miller}\/}}]{lopez-etal-2012}
{Lopez}, E.~D., J.~J. {Fortney}, and N.~{Miller}, 2012:
\newblock {How Thermal Evolution and Mass-loss Sculpt Populations of
  Super-Earths and Sub-Neptunes: Application to the Kepler-11 System and
  Beyond}.
\newblock {\it \apj\/}, {\bf 761}, 59.

\bibitem[{Lorenz(1955){\it Lorenz\/}}]{lorenz-1955}
Lorenz, E.~N., 1955:
\newblock Available potential energy and the maintenance of the general
  circulation.
\newblock {\it Tellus\/}, {\bf 7}, 157--167.

\bibitem[{Lorenz(1978){\it Lorenz\/}}]{lorenz-1978}
Lorenz, E.~N., 1978:
\newblock Available energy and the maintenance of a moist circulation.
\newblock {\it Tellus\/}, {\bf 30}, 15--31.

\bibitem[{Lorenz(1979){\it Lorenz\/}}]{lorenz-1979}
Lorenz, E.~N., 1979:
\newblock Numerical evaluation of moist available energy.
\newblock {\it Tellus\/}, {\bf 31}, 230--235.

\bibitem[{{Majeau} et~al.(2012){\it {Majeau}, {Agol}, and
  {Cowan}\/}}]{majeau-etal-2012}
{Majeau}, C., E.~{Agol}, and N.~B. {Cowan}, 2012:
\newblock {A Two-dimensional Infrared Map of the Extrasolar Planet HD 189733b}.
\newblock {\it \apjl\/}, {\bf 747}, L20.

\bibitem[{{Maltrud} and {Vallis}(1993){\it {Maltrud} and
  {Vallis}\/}}]{maltrud-vallis-1993}
{Maltrud}, M.~E., and G.~K. {Vallis}, 1993:
\newblock {Energy and enstrophy transfer in numerical simulations of
  two-dimensional turbulence}.
\newblock {\it Physics of Fluids\/}, {\bf 5}, 1760--1775.

\bibitem[{{Marcus} and {Lee}(1998){\it {Marcus} and {Lee}\/}}]{marcus-lee-1998}
{Marcus}, P.~S., and C.~{Lee}, 1998:
\newblock {A model for eastward and westward jets in laboratory experiments and
  planetary atmospheres}.
\newblock {\it Physics of Fluids\/}, {\bf 10}, 1474--1489.

\bibitem[{{Marshall} and {Plumb}(2007){\it {Marshall} and
  {Plumb}\/}}]{marshall-plumb-2007}
{Marshall}, J., and R.~A. {Plumb}, 2007:
\newblock {\it Atmosphere, Ocean, and Climate Dynamics\/}.
\newblock Academic Press.

\bibitem[{{Matsuno}(1966){\it {Matsuno}\/}}]{matsuno-1966}
{Matsuno}, T., 1966:
\newblock {Quasi-geostrophic motions in the equatorial area}.
\newblock {\it J. Meteorol. Soc. Japan\/}, {\bf 44}, 25--43.

\bibitem[{{Mayor} et~al.(2009){\it {Mayor}, et~al.\/}}]{mayor-etal-2009}
{Mayor}, M., et~al., 2009:
\newblock {The HARPS search for southern extra-solar planets. XVIII. An
  Earth-mass planet in the GJ 581 planetary system}.
\newblock {\it \aap\/}, {\bf 507}, 487--494.

\bibitem[{{McWilliams}(2006){\it {McWilliams}\/}}]{mcwilliams-2006}
{McWilliams}, J.~C., 2006:
\newblock {\it Fundamentals of Geophysical Fluid Dynamics\/}.
\newblock Cambridge Univ. Press, Cambridge, UK.

\bibitem[{{Menou} and {Rauscher}(2009){\it {Menou} and
  {Rauscher}\/}}]{menou-rauscher-2009}
{Menou}, K., and E.~{Rauscher}, 2009:
\newblock {Atmospheric Circulation of Hot Jupiters: A Shallow Three-Dimensional
  Model}.
\newblock {\it \apj\/}, {\bf 700}, 887--897.

\bibitem[{{Merlis} and {Schneider}(2010){\it {Merlis} and
  {Schneider}\/}}]{merlis-schneider-2010}
{Merlis}, T.~M., and T.~{Schneider}, 2010:
\newblock {Atmospheric dynamics of Earth-like tidally locked aquaplanets}.
\newblock {\it ArXiv e-prints\/}.

\bibitem[{Merlis et~al.(2012){\it Merlis, Schneider, Bordoni, and
  Eisenman\/}}]{merlis-etal-2013a}
Merlis, T.~M., T.~Schneider, S.~Bordoni, and I.~Eisenman, 2012:
\newblock Hadley circulation response to orbital precession. {P}art~{I}:
  Aquaplanets.
\newblock {\it J. Climate\/}, p. in press.

\bibitem[{{Miller}(1953){\it {Miller}\/}}]{miller-1953}
{Miller}, S.~L., 1953:
\newblock {A Production of Amino Acids under Possible Primitive Earth
  Conditions}.
\newblock {\it Science\/}, {\bf 117}, 528--529.

\bibitem[{{Mitchell} and {Vallis}(2010){\it {Mitchell} and
  {Vallis}\/}}]{mitchell-vallis-2010}
{Mitchell}, J.~L., and G.~K. {Vallis}, 2010:
\newblock {The transition to superrotation in terrestrial atmospheres}.
\newblock {\it Journal of Geophysical Research (Planets)\/}, {\bf 115}(E14),
  12,008.

\bibitem[{{Mitchell} et~al.(2006){\it {Mitchell}, {Pierrehumbert}, {Frierson},
  and {Caballero}\/}}]{mitchell-etal-2006}
{Mitchell}, J.~L., R.~T. {Pierrehumbert}, D.~M.~W. {Frierson}, and
  R.~{Caballero}, 2006:
\newblock {The dynamics behind Titan's methane clouds}.
\newblock {\it Proceedings of the National Academy of Science\/}, {\bf 103},
  18,421--18,426.

\bibitem[{{Mitchell} et~al.(2009){\it {Mitchell}, {Pierrehumbert}, {Frierson},
  and {Caballero}\/}}]{mitchell-etal-2009}
{Mitchell}, J.~L., R.~T. {Pierrehumbert}, D.~M.~W. {Frierson}, and
  R.~{Caballero}, 2009:
\newblock {The impact of methane thermodynamics on seasonal convection and
  circulation in a model Titan atmosphere}.
\newblock {\it \icarus\/}, {\bf 203}, 250--264.

\bibitem[{{Navarra} and {Boccaletti}(2002){\it {Navarra} and
  {Boccaletti}\/}}]{navarra-boccaletti-2002}
{Navarra}, A., and G.~{Boccaletti}, 2002:
\newblock {Numerical general circulation experiments of sensitivity to Earth
  rotation rate}.
\newblock {\it Climate Dynamics\/}, {\bf 19}, 467--483.

\bibitem[{Neelin and Held(1987){\it Neelin and Held\/}}]{neelin-held-1987}
Neelin, J.~D., and I.~M. Held, 1987:
\newblock Modeling tropical convergence based on the moist static energy
  budget.
\newblock {\it Mon. Wea. Rev.\/}, {\bf 115}, 3--12.

\bibitem[{{Nettelmann} et~al.(2011){\it {Nettelmann}, {Fortney}, {Kramm}, and
  {Redmer}\/}}]{nettelmann-etal-2011}
{Nettelmann}, N., J.~J. {Fortney}, U.~{Kramm}, and R.~{Redmer}, 2011:
\newblock {Thermal Evolution and Structure Models of the Transiting Super-Earth
  GJ 1214b}.
\newblock {\it \apj\/}, {\bf 733}, 2.

\bibitem[{{Niemann} et~al.(2005){\it {Niemann}, et~al.\/}}]{niemann-etal-2005}
{Niemann}, H.~B., et~al., 2005:
\newblock {The abundances of constituents of Titan's atmosphere from the GCMS
  instrument on the Huygens probe}.
\newblock {\it \nat\/}, {\bf 438}, 779--784.

\bibitem[{{Norton}(2006){\it {Norton}\/}}]{norton-2006}
{Norton}, W.~A., 2006:
\newblock {Tropical Wave Driving of the Annual Cycle in Tropical Tropopause
  Temperatures. Part II: Model Results.}
\newblock {\it J. Atmos. Sci.\/}, {\bf 63}, 1420--1431.

\bibitem[{{Nozawa} and {Yoden}(1997){\it {Nozawa} and
  {Yoden}\/}}]{nozawa-yoden-1997a}
{Nozawa}, T., and S.~{Yoden}, 1997:
\newblock {Formation of zonal band structure in forced two-dimensional
  turbulence on a rotating sphere}.
\newblock {\it Physics of Fluids\/}, {\bf 9}, 2081--2093.

\bibitem[{O'Gorman(2010){\it O'Gorman\/}}]{ogorman-2010}
O'Gorman, P.~A., 2010:
\newblock Understanding the varied response of the extratropical storm tracks
  to climate change.
\newblock {\it Proc. Nat. Aacad. Sci.\/}, {\bf 107}, 19,176--19,180.

\bibitem[{O'Gorman(2011){\it O'Gorman\/}}]{ogorman-2011}
O'Gorman, P.~A., 2011:
\newblock The effective static stability experienced by eddies in a moist
  atmosphere.
\newblock {\it J. Atmos. Sci.\/}, {\bf 68}, 75--90.

\bibitem[{O'Gorman and Schneider(2008){\it O'Gorman and
  Schneider\/}}]{ogorman-schneider-2008}
O'Gorman, P.~A., and T.~Schneider, 2008:
\newblock The hydrological cycle over a wide range of climates simulated with
  an idealized {GCM}.
\newblock {\it J. Climate\/}, {\bf 21}, 3815--3832.

\bibitem[{O'Gorman et~al.(2011){\it O'Gorman, Allan, Byrne, and
  Previdi\/}}]{ogorman-etal-2011}
O'Gorman, P.~A., R.~P. Allan, M.~P. Byrne, and M.~Previdi, 2011:
\newblock Energetic constraints on precipitation under climate change.
\newblock {\it Surv. Geophys.\/}, pp. 1--24.

\bibitem[{{Olbers} et~al.(2011){\it {Olbers}, {Willebrand}, and
  {Eden}\/}}]{olbers-etal-2012}
{Olbers}, R.~G., J.~{Willebrand}, and C.~{Eden}, 2011:
\newblock {\it Ocean Dynamics\/}.
\newblock Springer-Verlag, New York.

\bibitem[{{Pall{\'e}} et~al.(2008){\it {Pall{\'e}}, {Ford}, {Seager},
  {Monta{\~n}{\'e}s-Rodr{\'{\i}}guez}, and {Vazquez}\/}}]{palle-etal-2008}
{Pall{\'e}}, E., E.~B. {Ford}, S.~{Seager},
  P.~{Monta{\~n}{\'e}s-Rodr{\'{\i}}guez}, and M.~{Vazquez}, 2008:
\newblock {Identifying the Rotation Rate and the Presence of Dynamic Weather on
  Extrasolar Earth-like Planets from Photometric Observations}.
\newblock {\it \apj\/}, {\bf 676}, 1319--1329.

\bibitem[{{Parish} et~al.(2011){\it {Parish}, {Schubert}, {Covey},
  {Walterscheid}, {Grossman}, and {Lebonnois}\/}}]{parrish-etal-2011}
{Parish}, H.~F., G.~{Schubert}, C.~{Covey}, R.~L. {Walterscheid},
  A.~{Grossman}, and S.~{Lebonnois}, 2011:
\newblock {Decadal variations in a Venus general circulation model}.
\newblock {\it \icarus\/}, {\bf 212}, 42--65.

\bibitem[{{Pavan} and {Held}(1996){\it {Pavan} and {Held}\/}}]{pavan-held-1996}
{Pavan}, V., and I.~M. {Held}, 1996:
\newblock {The diffusive approximation for eddy fluxes in baroclinically
  unstable flows}.
\newblock {\it Journal of the Atmospheric Sciences\/}, {\bf 53}, 1262--1272.

\bibitem[{{Pedlosky}(1987){\it {Pedlosky}\/}}]{pedlosky-1987}
{Pedlosky}, J., 1987:
\newblock {\it Geophysical Fluid Dynamics, 2nd Ed.\/}.
\newblock Springer-Verlag, New York.

\bibitem[{{Pedlosky}(2003){\it {Pedlosky}\/}}]{pedlosky-2003}
{Pedlosky}, J., 2003:
\newblock {\it Waves in the Ocean and Atmosphere\/}.
\newblock Springer-Verlag, New York.

\bibitem[{{Pedlosky}(2004){\it {Pedlosky}\/}}]{pedlosky-2004}
{Pedlosky}, J., 2004:
\newblock {\it Ocean Circulation Theory\/}.
\newblock Springer-Verlag, New York.

\bibitem[{{Peixoto} and {Oort}(1992){\it {Peixoto} and
  {Oort}\/}}]{peixoto-oort-1992}
{Peixoto}, J.~P., and A.~H. {Oort}, 1992:
\newblock {\it Physics of Climate\/}.
\newblock American Institute of Physics, New York.

\bibitem[{{Perna} et~al.(2010){\it {Perna}, {Menou}, and
  {Rauscher}\/}}]{perna-etal-2010}
{Perna}, R., K.~{Menou}, and E.~{Rauscher}, 2010:
\newblock {Magnetic Drag on Hot Jupiter Atmospheric Winds}.
\newblock {\it \apj\/}, {\bf 719}, 1421--1426.

\bibitem[{{Perna} et~al.(2012){\it {Perna}, {Heng}, and
  {Pont}\/}}]{perna-etal-2012}
{Perna}, R., K.~{Heng}, and F.~{Pont}, 2012:
\newblock {The Effects of Irradiation on Hot Jovian Atmospheres: Heat
  Redistribution and Energy Dissipation}.
\newblock {\it \apj\/}, {\bf 751}, 59.

\bibitem[{{Pierrehumbert} and {Gaidos}(2011){\it {Pierrehumbert} and
  {Gaidos}\/}}]{pierrehumbert-gaidos-2011}
{Pierrehumbert}, R., and E.~{Gaidos}, 2011:
\newblock {Hydrogen Greenhouse Planets Beyond the Habitable Zone}.
\newblock {\it \apjl\/}, {\bf 734}, L13.

\bibitem[{Pierrehumbert(1995){\it Pierrehumbert\/}}]{pierrehumbert-1995}
Pierrehumbert, R.~T., 1995:
\newblock Thermostats, radiator fins, and the local runaway greenhouse.
\newblock {\it J. Atmos. Sci.\/}, {\bf 52}, 1784--1806.

\bibitem[{Pierrehumbert(2002){\it Pierrehumbert\/}}]{pierrehumbert-2002}
Pierrehumbert, R.~T., 2002:
\newblock The hydrologic cycle in deep-time climate problems.
\newblock {\it Nature\/}, {\bf 419}, 191--198.

\bibitem[{{Pierrehumbert}(2010){\it {Pierrehumbert}\/}}]{pierrehumbert-2010}
{Pierrehumbert}, R.~T., 2010:
\newblock {\it {Principles of Planetary Climate}\/}.
\newblock Cambridge University Press.

\bibitem[{{Pierrehumbert}(2011){\it {Pierrehumbert}\/}}]{pierrehumbert-2011}
{Pierrehumbert}, R.~T., 2011:
\newblock {A Palette of Climates for Gliese 581g}.
\newblock {\it \apjl\/}, {\bf 726}, L8.

\bibitem[{Pierrehumbert and Swanson(1995){\it Pierrehumbert and
  Swanson\/}}]{pierrehumbert-swanson-1995}
Pierrehumbert, R.~T., and K.~L. Swanson, 1995:
\newblock {Baroclinic Instability}.
\newblock {\it Annu. Rev. Fluid Mech.\/}, {\bf 27}, 419--467.

\bibitem[{Pierrehumbert et~al.(2007){\it Pierrehumbert, Brogniez, and
  Roca\/}}]{pierrehumbert-etal-2007}
Pierrehumbert, R.~T., H.~Brogniez, and R.~Roca, 2007:
\newblock On the relative humidity of the atmosphere.
\newblock  {\it The Global Circulation of the Atmosphere\/}, T.~Schneider and
  A.~H. Sobel, Eds., Princeton University Press, pp. 143--185.

\bibitem[{{Pierrehumbert} et~al.(2011){\it {Pierrehumbert}, {Abbot}, {Voigt},
  and {Koll}\/}}]{pierrehumbert-etal-2011}
{Pierrehumbert}, R.~T., D.~S. {Abbot}, A.~{Voigt}, and D.~{Koll}, 2011:
\newblock {Climate of the Neoproterozoic}.
\newblock {\it Annual Review of Earth and Planetary Sciences\/}, {\bf 39},
  417--460.

\bibitem[{{Polvani} and {Sobel}(2002){\it {Polvani} and
  {Sobel}\/}}]{polvani-sobel-2002}
{Polvani}, L.~M., and A.~H. {Sobel}, 2002:
\newblock {The Hadley Circulation and the Weak Temperature Gradient
  Approximation.}
\newblock {\it Journal of Atmospheric Sciences\/}, {\bf 59}, 1744--1752.

\bibitem[{{Polvani} et~al.(1995){\it {Polvani}, {Waugh}, and
  {Plumb}\/}}]{polvani-etal-1995}
{Polvani}, L.~M., D.~W. {Waugh}, and R.~A. {Plumb}, 1995:
\newblock {On the Subtropical Edge of the Stratospheric Surf Zone.}
\newblock {\it Journal of Atmospheric Sciences\/}, {\bf 52}, 1288--1309.

\bibitem[{{Pont} et~al.(2012){\it {Pont}, {Sing}, {Gibson}, {Aigrain}, {Henry},
  and {Husnoo}\/}}]{pont-etal-2013}
{Pont}, F., D.~K. {Sing}, N.~P. {Gibson}, S.~{Aigrain}, G.~{Henry}, and
  N.~{Husnoo}, 2012:
\newblock {The prevalence of dust on the exoplanet HD 189733b from Hubble and
  Spitzer observations}.
\newblock {\it ArXiv e-prints\/}.

\bibitem[{{Randel} and {Held}(1991){\it {Randel} and
  {Held}\/}}]{randel-held-1991}
{Randel}, W.~J., and I.~M. {Held}, 1991:
\newblock {Phase Speed Spectra of Transient Eddy Fluxes and Critical Layer
  Absorption.}
\newblock {\it Journal of Atmospheric Sciences\/}, {\bf 48}, 688--697.

\bibitem[{{Rauscher} and {Menou}(2010){\it {Rauscher} and
  {Menou}\/}}]{rauscher-menou-2010}
{Rauscher}, E., and K.~{Menou}, 2010:
\newblock {Three-dimensional Modeling of Hot Jupiter Atmospheric Flows}.
\newblock {\it \apj\/}, {\bf 714}, 1334--1342.

\bibitem[{{Rauscher} and {Menou}(2012){\it {Rauscher} and
  {Menou}\/}}]{rauscher-menou-2012}
{Rauscher}, E., and K.~{Menou}, 2012:
\newblock {The Role of Drag in the Energetics of Strongly Forced Exoplanet
  Atmospheres}.
\newblock {\it \apj\/}, {\bf 745}, 78.

\bibitem[{Raymond et~al.(2009){\it Raymond, Sessions, Sobel, and
  Fuchs\/}}]{raymond-etal-2009}
Raymond, D.~J., S.~Sessions, A.~H. Sobel, and Z.~Fuchs, 2009:
\newblock The mechanics of gross moist stability.
\newblock {\it J. Adv. Model. Earth Syst.\/}, {\bf 1}, Art. \#9, 20 pp.

\bibitem[{{Read}(2011){\it {Read}\/}}]{read-2011}
{Read}, P.~L., 2011:
\newblock {Dynamics and circulation regimes of terrestrial planets}.
\newblock {\it \planss\/}, {\bf 59}, 900--914.

\bibitem[{{Read} and {Lewis}(2004){\it {Read} and {Lewis}\/}}]{read-lewis-2004}
{Read}, P.~L., and S.~R. {Lewis}, 2004:
\newblock {\it The Martian Climate Revisited\/}.
\newblock Springer/Praxis, New York.

\bibitem[{{Read} et~al.(2007){\it {Read}, {Yamazaki}, {Lewis}, {Williams},
  {Wordsworth}, {Miki-Yamazaki}, {Sommeria}, and {Didelle}\/}}]{read-etal-2007}
{Read}, P.~L., Y.~H. {Yamazaki}, S.~R. {Lewis}, P.~D. {Williams},
  R.~{Wordsworth}, K.~{Miki-Yamazaki}, J.~{Sommeria}, and H.~{Didelle}, 2007:
\newblock {Dynamics of Convectively Driven Banded Jets in the Laboratory}.
\newblock {\it Journal of Atmospheric Sciences\/}, {\bf 64}, 4031.

\bibitem[{{Renn{\'o}}(1997){\it {Renn{\'o}}\/}}]{renno-1997}
{Renn{\'o}}, N.~O., 1997:
\newblock {Multiple equilibria in radiative-convective atmospheres}.
\newblock {\it Tellus Series A\/}, {\bf 49}, 423.

\bibitem[{{Rhines}(1975){\it {Rhines}\/}}]{rhines-1975}
{Rhines}, P.~B., 1975:
\newblock {Waves and turbulence on a beta-plane}.
\newblock {\it Journal of Fluid Mechanics\/}, {\bf 69}, 417--443.

\bibitem[{{Richardson} et~al.(2007){\it {Richardson}, {Deming}, {Horning},
  {Seager}, and {Harrington}\/}}]{richardson-etal-2007}
{Richardson}, L.~J., D.~{Deming}, K.~{Horning}, S.~{Seager}, and
  J.~{Harrington}, 2007:
\newblock {A spectrum of an extrasolar planet}.
\newblock {\it \nat\/}, {\bf 445}, 892--895.

\bibitem[{{Robinson} et~al.(2011){\it {Robinson},
  et~al.\/}}]{robinson-etal-2011}
{Robinson}, T.~D., et~al., 2011:
\newblock {Earth as an Extrasolar Planet: Earth Model Validation Using EPOXI
  Earth Observations}.
\newblock {\it Astrobiology\/}, {\bf 11}, 393--408.

\bibitem[{{Rogers} and {Seager}(2010){\it {Rogers} and
  {Seager}\/}}]{rogers-seager-2010b}
{Rogers}, L.~A., and S.~{Seager}, 2010:
\newblock {Three Possible Origins for the Gas Layer on GJ 1214b}.
\newblock {\it \apj\/}, {\bf 716}, 1208--1216.

\bibitem[{{Rogers} et~al.(2011){\it {Rogers}, {Bodenheimer}, {Lissauer}, and
  {Seager}\/}}]{rogers-etal-2011}
{Rogers}, L.~A., P.~{Bodenheimer}, J.~J. {Lissauer}, and S.~{Seager}, 2011:
\newblock {Formation and Structure of Low-density exo-Neptunes}.
\newblock {\it \apj\/}, {\bf 738}, 59.

\bibitem[{Romps and Kuang(2010){\it Romps and Kuang\/}}]{romps-2010}
Romps, D.~M., and Z.~Kuang, 2010:
\newblock Do undiluted convective plumes exist in the upper tropical
  troposphere?
\newblock {\it J. Atmos. Sci.\/}, {\bf 67}, 468--484.

\bibitem[{{Salby} and {Garcia}(1987){\it {Salby} and
  {Garcia}\/}}]{salby-garcia-1987}
{Salby}, M.~L., and R.~R. {Garcia}, 1987:
\newblock {Transient Response to Localized Episodic Heating in the Tropics.
  Part I: Excitation and Short-Time Near-Field Behavior.}
\newblock {\it Journal of Atmospheric Sciences\/}, {\bf 44}, 458--498.

\bibitem[{{Sanrom{\'a}} and {Pall{\'e}}(2012){\it {Sanrom{\'a}} and
  {Pall{\'e}}\/}}]{sanroma-palle-2012}
{Sanrom{\'a}}, E., and E.~{Pall{\'e}}, 2012:
\newblock {Reconstructing the Photometric Light Curves of Earth as a Planet
  along Its History}.
\newblock {\it \apj\/}, {\bf 744}, 188.

\bibitem[{{Saravanan}(1993){\it {Saravanan}\/}}]{saravanan-1993}
{Saravanan}, R., 1993:
\newblock {Equatorial Superrotation and Maintenance of the General Circulation
  in Two-Level Models.}
\newblock {\it J. Atmos. Sci.\/}, {\bf 50}, 1211--1227.

\bibitem[{Sarmiento and Gruber(2006){\it Sarmiento and
  Gruber\/}}]{sarmiento-gruber-2006}
Sarmiento, J.~L., and N.~Gruber, 2006:
\newblock {\it Ocean biogeochemical dynamics\/}.
\newblock Princeton University Press.

\bibitem[{{Schneider}(2006){\it {Schneider}\/}}]{schneider-2006}
{Schneider}, T., 2006:
\newblock {The General Circulation of the Atmosphere}.
\newblock {\it Annual Review of Earth and Planetary Sciences\/}, {\bf 34},
  655--688.

\bibitem[{Schneider(2007){\it Schneider\/}}]{schneider-2007}
Schneider, T., 2007:
\newblock The thermal stratification of the extratropical troposphere.
\newblock  {\it The Global Circulation of the Atmosphere\/}, T.~Schneider and
  A.~H. Sobel, Eds., Princeton University Press, pp. 47--77.

\bibitem[{{Schneider} and {Bordoni}(2008){\it {Schneider} and
  {Bordoni}\/}}]{schneider-bordoni-2008}
{Schneider}, T., and S.~{Bordoni}, 2008:
\newblock {Eddy-Mediated Regime Transitions in the Seasonal Cycle of a Hadley
  Circulation and Implications for Monsoon Dynamics}.
\newblock {\it Journal of Atmospheric Sciences\/}, {\bf 65}, 915.

\bibitem[{{Schneider} and {Walker}(2006){\it {Schneider} and
  {Walker}\/}}]{schneider-walker-2006}
{Schneider}, T., and C.~C. {Walker}, 2006:
\newblock {Self-Organization of Atmospheric Macroturbulence into Critical
  States of Weak Nonlinear Eddy Eddy Interactions.}
\newblock {\it Journal of Atmospheric Sciences\/}, {\bf 63}, 1569--1586.

\bibitem[{{Schneider} and {Walker}(2008){\it {Schneider} and
  {Walker}\/}}]{schneider-walker-2008}
{Schneider}, T., and C.~C. {Walker}, 2008:
\newblock {Scaling laws and regime transitions of macroturbulence in dry
  atmospheres}.
\newblock {\it J. Atmos. Sci.\/}, {\bf 65}, 2153--2173.

\bibitem[{{Schneider} et~al.(2010){\it {Schneider}, {O'Gorman}, and
  {Levine}\/}}]{schneider-etal-2010}
{Schneider}, T., P.~A. {O'Gorman}, and X.~J. {Levine}, 2010:
\newblock {Water vapor and the dynamics of climate changes}.
\newblock {\it Reviews of Geophysics\/}, {\bf 48}, 3001.

\bibitem[{{Schumacher} et~al.(2004){\it {Schumacher}, {Houze}, and
  {Kraucunas}\/}}]{schumacher-etal-2004}
{Schumacher}, C., R.~A. {Houze}, Jr., and I.~{Kraucunas}, 2004:
\newblock {The Tropical Dynamical Response to Latent Heating Estimates Derived
  from the TRMM Precipitation Radar.}
\newblock {\it Journal of Atmospheric Sciences\/}, {\bf 61}, 1341--1358.

\bibitem[{{Scott}(2010){\it {Scott}\/}}]{scott-2010}
{Scott}, R.~K., 2010:
\newblock {\it {{\rm The structure of zonal jets in shallow water turbulence on
  the sphere}. {\rm In} {\it IUTAM Symposium on Turbulence in the Atmosphere
  and Oceans} (D.G. Dritschel, Ed.)}\/}, pp. 243--252, Springer.

\bibitem[{{Scott} and {Dritschel}(2012){\it {Scott} and
  {Dritschel}\/}}]{scott-dritschel-2012}
{Scott}, R.~K., and D.~G. {Dritschel}, 2012:
\newblock {The structure of zonal jets in geostrophic turbulence}.
\newblock {\it J. Fluid Mech.\/}, {\bf 711}, 576--598.

\bibitem[{{Scott} and {Tissier}(2012){\it {Scott} and
  {Tissier}\/}}]{scott-tissier-2012}
{Scott}, R.~K., and A.-S. {Tissier}, 2012:
\newblock {The generation of zonal jets by large-scale mixing}.
\newblock {\it Physics of Fluids\/}, {\bf 24}(12), 126,601.

\bibitem[{{Seager} and {Deming}(2010){\it {Seager} and
  {Deming}\/}}]{seager-deming-2010}
{Seager}, S., and D.~{Deming}, 2010:
\newblock {Exoplanet Atmospheres}.
\newblock {\it \araa\/}, {\bf 48}, 631--672.

\bibitem[{{Seager} et~al.(2008){\it {Seager}, {Deming}, and
  {Valenti}\/}}]{seager-etal-2008}
{Seager}, S., D.~{Deming}, and J.~A. {Valenti}, 2008:
\newblock {Transiting Exoplanets with JWST}.
\newblock {\it ArXiv e-prints\/}.

\bibitem[{{Selsis} et~al.(2011){\it {Selsis}, {Wordsworth}, and
  {Forget}\/}}]{selsis-etal-2011}
{Selsis}, F., R.~D. {Wordsworth}, and F.~{Forget}, 2011:
\newblock {Thermal phase curves of nontransiting terrestrial exoplanets. I.
  Characterizing atmospheres}.
\newblock {\it \aap\/}, {\bf 532}, A1.

\bibitem[{{Shell} and {Held}(2004){\it {Shell} and {Held}\/}}]{shell-held-2004}
{Shell}, K.~M., and I.~M. {Held}, 2004:
\newblock {Abrupt Transition to Strong Superrotation in an Axisymmetric Model
  of the Upper Troposphere.}
\newblock {\it J. Atmos. Sci.\/}, {\bf 61}, 2928--2935.

\bibitem[{Sherwood et~al.(2010{\natexlab{a}}){\it Sherwood, Ingram, Tsushima,
  Satoh, Roberts, Vidale, and O'Gorman\/}}]{sherwood-etal-2010b}
Sherwood, S.~C., W.~Ingram, Y.~Tsushima, M.~Satoh, M.~Roberts, P.~L. Vidale,
  and P.~A. O'Gorman, 2010{\natexlab{a}}:
\newblock Relative humidity changes in a warmer climate.
\newblock {\it J. Geophys. Res.\/}, {\bf 115}, 09,104.

\bibitem[{Sherwood et~al.(2010{\natexlab{b}}){\it Sherwood, Roca, Weckwerth,
  and Andronova\/}}]{sherwood-etal-2010a}
Sherwood, S.~C., R.~Roca, T.~M. Weckwerth, and N.~G. Andronova,
  2010{\natexlab{b}}:
\newblock Tropospheric water vapor, convection, and climate.
\newblock {\it Reviews of Geophysics\/}, {\bf 48}, RG2001.

\bibitem[{{Showman} and {Guillot}(2002){\it {Showman} and
  {Guillot}\/}}]{showman-guillot-2002}
{Showman}, A.~P., and T.~{Guillot}, 2002:
\newblock {Atmospheric circulation and tides of ``51 Pegasus b-like'' planets}.
\newblock {\it \aap\/}, {\bf 385}, 166--180.

\bibitem[{{Showman} and {Polvani}(2010){\it {Showman} and
  {Polvani}\/}}]{showman-polvani-2010}
{Showman}, A.~P., and L.~M. {Polvani}, 2010:
\newblock {The Matsuno-Gill model and equatorial superrotation}.
\newblock {\it \grl\/}, {\bf 37}, 18,811.

\bibitem[{{Showman} and {Polvani}(2011){\it {Showman} and
  {Polvani}\/}}]{showman-polvani-2011}
{Showman}, A.~P., and L.~M. {Polvani}, 2011:
\newblock {Equatorial Superrotation on Tidally Locked Exoplanets}.
\newblock {\it \apj\/}, {\bf 738}, 71.

\bibitem[{{Showman} et~al.(2008){\it {Showman}, {Cooper}, {Fortney}, and
  {Marley}\/}}]{showman-etal-2008a}
{Showman}, A.~P., C.~S. {Cooper}, J.~J. {Fortney}, and M.~S. {Marley}, 2008:
\newblock {Atmospheric Circulation of Hot Jupiters: Three-dimensional
  Circulation Models of HD 209458b and HD 189733b with Simplified Forcing}.
\newblock {\it \apj\/}, {\bf 682}, 559--576.

\bibitem[{{Showman} et~al.(2009){\it {Showman}, {Fortney}, {Lian}, {Marley},
  {Freedman}, {Knutson}, and {Charbonneau}\/}}]{showman-etal-2009}
{Showman}, A.~P., J.~J. {Fortney}, Y.~{Lian}, M.~S. {Marley}, R.~S. {Freedman},
  H.~A. {Knutson}, and D.~{Charbonneau}, 2009:
\newblock {Atmospheric Circulation of Hot Jupiters: Coupled Radiative-Dynamical
  General Circulation Model Simulations of HD 189733b and HD 209458b}.
\newblock {\it \apj\/}, {\bf 699}, 564--584.

\bibitem[{{Showman} et~al.(2010){\it {Showman}, {Cho}, and
  {Menou}\/}}]{showman-etal-2010}
{Showman}, A.~P., J.~Y.-K. {Cho}, and K.~{Menou}, 2010:
\newblock {\it {\rm Atmospheric circulation of Exoplanets}. {\rm In} {\it
  Exoplanets} (S. Seager, Ed.)\/}, pp. 471--516, Univ. Arizona Press.

\bibitem[{{Showman} et~al.(2013){\it {Showman}, {Fortney}, {Lewis}, and
  {Shabram}\/}}]{showman-etal-2013}
{Showman}, A.~P., J.~J. {Fortney}, N.~K. {Lewis}, and M.~{Shabram}, 2013:
\newblock {Doppler Signatures of the Atmospheric Circulation on Hot Jupiters}.
\newblock {\it \apj\/}, {\bf 762}, 24.

\bibitem[{{Siedler} et~al.(2001){\it {Siedler}, {Church}, and
  {Gould}\/}}]{siedler-etal-2001}
{Siedler}, G., J.~{Church}, and J.~{Gould}, 2001:
\newblock {\it Ocean Circulation and Climate: Observing and Modeling the Global
  Ocean\/}.
\newblock Academic Press, International Geophysics Series, Vol. 77.

\bibitem[{{Sing} et~al.(2008){\it {Sing}, {Vidal-Madjar}, {Lecavelier des
  Etangs}, {Desert}, {Ballester}, and {Ehrenreich}\/}}]{sing-etal-2008}
{Sing}, D.~K., A.~{Vidal-Madjar}, A.~{Lecavelier des Etangs}, J.~M. {Desert},
  G.~{Ballester}, and D.~{Ehrenreich}, 2008:
\newblock {Determining atmospheric conditions at the terminator of the
  hot-Jupiter HD209458b}.
\newblock {\it ArXiv e-prints\/}, {\bf 803}.

\bibitem[{Sleep and Zahnle(2001){\it Sleep and Zahnle\/}}]{sleep-zahnle-2001}
Sleep, N.~H., and K.~Zahnle, 2001:
\newblock Carbon dioxide cycling and implications for climate on ancient earth.
\newblock {\it Journal of Geophysical Research: Planets (1991--2012)\/}, {\bf
  106}(E1), 1373--1399.

\bibitem[{{Smith}(2008){\it {Smith}\/}}]{smith-2008}
{Smith}, M.~D., 2008:
\newblock {Spacecraft Observations of the Martian Atmosphere}.
\newblock {\it Annual Review of Earth and Planetary Sciences\/}, {\bf 36},
  191--219.

\bibitem[{{Snellen} et~al.(2010){\it {Snellen}, {de Kok}, {de Mooij}, and
  {Albrecht}\/}}]{snellen-etal-2010}
{Snellen}, I.~A.~G., R.~J. {de Kok}, E.~J.~W. {de Mooij}, and S.~{Albrecht},
  2010:
\newblock {The orbital motion, absolute mass and high-altitude winds of
  exoplanet HD209458b}.
\newblock {\it \nat\/}, {\bf 465}, 1049--1051.

\bibitem[{{Sobel}(2002){\it {Sobel}\/}}]{sobel-2002}
{Sobel}, A.~H., 2002:
\newblock {Water vapor as an active scalar in tropical atmospheric dynamics}.
\newblock {\it Chaos\/}, {\bf 12}, 451--459.

\bibitem[{Sobel(2007){\it Sobel\/}}]{sobel-2007}
Sobel, A.~H., 2007:
\newblock Simple models of ensemble-averaged tropical precipitation and surface
  wind, given the sea surface temperature.
\newblock  {\it The Global Circulation of the Atmosphere\/}, T.~Schneider and
  A.~H. Sobel, Eds., Princeton University Press, pp. 219--251.

\bibitem[{{Sobel} et~al.(2001){\it {Sobel}, {Nilsson}, and
  {Polvani}\/}}]{sobel-etal-2001}
{Sobel}, A.~H., J.~{Nilsson}, and L.~M. {Polvani}, 2001:
\newblock {The Weak Temperature Gradient Approximation and Balanced Tropical
  Moisture Waves$^{*}$.}
\newblock {\it Journal of Atmospheric Sciences\/}, {\bf 58}, 3650--3665.

\bibitem[{Soden and Held(2006){\it Soden and Held\/}}]{soden-held-2006}
Soden, B.~J., and I.~M. Held, 2006:
\newblock An assessment of climate feedbacks in coupled ocean-atmosphere
  models.
\newblock {\it J. Climate\/}, {\bf 19}, 3354--3360.

\bibitem[{{Stevenson}(1999){\it {Stevenson}\/}}]{stevenson-1999}
{Stevenson}, D.~J., 1999:
\newblock {Life-sustaining planets in interstellar space?}
\newblock {\it \nat\/}, {\bf 400}, 32.

\bibitem[{{Stone}(1972){\it {Stone}\/}}]{stone-1972}
{Stone}, P.~H., 1972:
\newblock {A simplified radiative-dynamical model for the static stability of
  rotating atmospheres}.
\newblock {\it Journal of the Atmospheric Sciences\/}, {\bf 29}, 406--418.

\bibitem[{{Suarez} and {Duffy}(1992){\it {Suarez} and
  {Duffy}\/}}]{suarez-duffy-1992}
{Suarez}, M.~J., and D.~G. {Duffy}, 1992:
\newblock {Terrestrial Superrotation: A Bifurcation of the General
  Circulation.}
\newblock {\it J. Atmos. Sci.\/}, {\bf 49}, 1541--1556.

\bibitem[{{Sukoriansky} et~al.(2007){\it {Sukoriansky}, {Dikovskaya}, and
  {Galperin}\/}}]{sukoriansky-etal-2007}
{Sukoriansky}, S., N.~{Dikovskaya}, and B.~{Galperin}, 2007:
\newblock {On the "arrest" of inverse energy cascade and the Rhines scale}.
\newblock {\it J. Atmos. Sci.\/}, {\bf 64}, 3312--3327.

\bibitem[{{Thompson} and {Young}(2006){\it {Thompson} and
  {Young}\/}}]{thompson-young-2006}
{Thompson}, A.~F., and W.~R. {Young}, 2006:
\newblock {Scaling Baroclinic Eddy Fluxes: Vortices and Energy Balance}.
\newblock {\it Journal of Physical Oceanography\/}, {\bf 36}, 720.

\bibitem[{{Thompson} and {Young}(2007){\it {Thompson} and
  {Young}\/}}]{thompson-young-2007}
{Thompson}, A.~F., and W.~R. {Young}, 2007:
\newblock {Two-layer baroclinic eddy heat fluxes: zonal flows and energy
  balance}.
\newblock {\it J. Atmos. Sci.\/}, {\bf 64}, 3214--3231.

\bibitem[{{Thompson}(1971){\it {Thompson}\/}}]{thompson-1971}
{Thompson}, R.~O.~R.~Y., 1971:
\newblock {Why There is an Intense Eastward Current in the North Atlantic but
  not in the South Atlantic}.
\newblock {\it Journal of Physical Oceanography\/}, {\bf 1}, 235--238.

\bibitem[{Thuburn and Craig(2000){\it Thuburn and
  Craig\/}}]{thuburn-craig-2000}
Thuburn, J., and G.~C. Craig, 2000:
\newblock Stratospheric influence on tropopause height: The radiative
  constraint.
\newblock {\it J. Atmos. Sci.\/}, {\bf 57}, 17--28.

\bibitem[{Trenberth and Caron(2001){\it Trenberth and
  Caron\/}}]{trenberth-caron-2001}
Trenberth, K., and J.~Caron, 2001:
\newblock Estimates of meridional atmosphere and ocean heat transports.
\newblock {\it J. Climate\/}, {\bf 14}, 3433--3443.

\bibitem[{Trenberth et~al.(2009){\it Trenberth, Fasullo, and
  Kiehl\/}}]{trenberth-etal-2009}
Trenberth, K.~E., J.~T. Fasullo, and J.~Kiehl, 2009:
\newblock Earth's global energy budget.
\newblock {\it Bull. Amer. Meteor. Soc.\/}, {\bf 90}, 311--323.

\bibitem[{{Tziperman} and {Farrell}(2009){\it {Tziperman} and
  {Farrell}\/}}]{tziperman-farrell-2009}
{Tziperman}, E., and B.~{Farrell}, 2009:
\newblock {Pliocene equatorial temperature: Lessons from atmospheric
  superrotation}.
\newblock {\it Paleoceanography\/}, {\bf 24}(26), A261,101+.

\bibitem[{{Udry} et~al.(2007){\it {Udry}, et~al.\/}}]{udry-etal-2007}
{Udry}, S., et~al., 2007:
\newblock {The HARPS search for southern extra-solar planets. XI. Super-Earths
  (5 and 8 M$_\oplus$) in a 3-planet system}.
\newblock {\it \aap\/}, {\bf 469}, L43--L47.

\bibitem[{{Vallis}(2006){\it {Vallis}\/}}]{vallis-2006}
{Vallis}, G.~K., 2006:
\newblock {\it Atmospheric and Oceanic Fluid Dynamics: Fundamentals and
  Large-Scale Circulation\/}.
\newblock Cambridge Univ. Press, Cambridge, UK.

\bibitem[{{Vallis}(2011){\it {Vallis}\/}}]{vallis-2011}
{Vallis}, G.~K., 2011:
\newblock {\it Climate and the Oceans\/}.
\newblock Princeton Primers in Climate series; Princeton Univ. Press.

\bibitem[{Vallis and Farneti(2009){\it Vallis and
  Farneti\/}}]{vallis-farneti-2009}
Vallis, G.~K., and R.~Farneti, 2009:
\newblock Meridional energy transport in the coupled atmosphere--ocean system:
  {S}caling and numerical experiments.
\newblock {\it Quart. J. Roy. Meteor. Soc.\/}, {\bf 135}, 1643--1660.

\bibitem[{{Vallis} and {Maltrud}(1993){\it {Vallis} and
  {Maltrud}\/}}]{vallis-maltrud-1993}
{Vallis}, G.~K., and M.~E. {Maltrud}, 1993:
\newblock {Generation of mean flows and jets on a beta plane and over
  topography}.
\newblock {\it J. Phys. Oceanography\/}, {\bf 23}, 1346--1362.

\bibitem[{{Vasavada} and {Showman}(2005){\it {Vasavada} and
  {Showman}\/}}]{vasavada-showman-2005}
{Vasavada}, A.~R., and A.~P. {Showman}, 2005:
\newblock {Jovian atmospheric dynamics: an update after Galileo and Cassini}.
\newblock {\it Reports of Progress in Physics\/}, {\bf 68}, 1935--1996.

\bibitem[{{von Paris} et~al.(2010){\it {von Paris}, {Gebauer}, {Godolt},
  {Grenfell}, {Hedelt}, {Kitzmann}, {Patzer}, {Rauer}, and
  {Stracke}\/}}]{von-paris-etal-2010}
{von Paris}, P., S.~{Gebauer}, M.~{Godolt}, J.~L. {Grenfell}, P.~{Hedelt},
  D.~{Kitzmann}, A.~B.~C. {Patzer}, H.~{Rauer}, and B.~{Stracke}, 2010:
\newblock {The extrasolar planet Gliese 581d: a potentially habitable planet?}
\newblock {\it \aap\/}, {\bf 522}, A23.

\bibitem[{{Walker} and {Schneider}(2005){\it {Walker} and
  {Schneider}\/}}]{walker-schneider-2005}
{Walker}, C.~C., and T.~{Schneider}, 2005:
\newblock {Response of idealized Hadley circulations to seasonally varying
  heating}.
\newblock {\it \grl\/}, {\bf 32}, L06,813.

\bibitem[{{Walker} and {Schneider}(2006){\it {Walker} and
  {Schneider}\/}}]{walker-schneider-2006}
{Walker}, C.~C., and T.~{Schneider}, 2006:
\newblock {Eddy Influences on Hadley Circulations: Simulations with an
  Idealized GCM}.
\newblock {\it Journal of Atmospheric Sciences\/}, {\bf 63}, 3333--3350.

\bibitem[{{Walker} et~al.(1981){\it {Walker}, {Hays}, and
  {Kasting}\/}}]{walker-hayes-kasting-1981}
{Walker}, J.~C.~G., P.~B. {Hays}, and J.~F. {Kasting}, 1981:
\newblock {A negative feedback mechanism for the long-term stabilization of the
  earth's surface temperature}.
\newblock {\it \jgr\/}, {\bf 86}, 9776--9782.

\bibitem[{{Wiktorowicz} and {Ingersoll}(2007){\it {Wiktorowicz} and
  {Ingersoll}\/}}]{wiktorowicz-ingersoll-2007}
{Wiktorowicz}, S.~J., and A.~P. {Ingersoll}, 2007:
\newblock {Liquid water oceans in ice giants}.
\newblock {\it \icarus\/}, {\bf 186}, 436--447.

\bibitem[{{Williams} and {Pollard}(2003){\it {Williams} and
  {Pollard}\/}}]{williams-pollard-2003}
{Williams}, D.~M., and D.~{Pollard}, 2003:
\newblock {Extraordinary climates of Earth-like planets: three-dimensional
  climate simulations at extreme obliquity}.
\newblock {\it International Journal of Astrobiology\/}, {\bf 2}, 1--19.

\bibitem[{{Williams}(1978){\it {Williams}\/}}]{williams-1978}
{Williams}, G.~P., 1978:
\newblock {Planetary circulations. I - Barotropic representation of Jovian and
  terrestrial turbulence}.
\newblock {\it Journal of Atmospheric Sciences\/}, {\bf 35}, 1399--1426.

\bibitem[{{Williams}(1988{\natexlab{a}}){\it {Williams}\/}}]{williams-1988a}
{Williams}, G.~P., 1988{\natexlab{a}}:
\newblock {The dynamical range of global circulations -- I}.
\newblock {\it Climate Dynamics\/}, {\bf 2}, 205--260.

\bibitem[{{Williams}(1988{\natexlab{b}}){\it {Williams}\/}}]{williams-1988b}
{Williams}, G.~P., 1988{\natexlab{b}}:
\newblock {The dynamical range of global circulations -- II}.
\newblock {\it Climate Dynamics\/}, {\bf 3}, 45--84.

\bibitem[{{Williams} and {Holloway}(1982){\it {Williams} and
  {Holloway}\/}}]{williams-holloway-1982}
{Williams}, G.~P., and J.~L. {Holloway}, 1982:
\newblock {The range and unity of planetary circulations}.
\newblock {\it \nat\/}, {\bf 297}, 295--299.

\bibitem[{{Williams} and {Follows}(2011){\it {Williams} and
  {Follows}\/}}]{williams-follows-2011}
{Williams}, R.~G., and M.~J. {Follows}, 2011:
\newblock {\it Ocean Dynamics and the Carbon Cycle\/}.
\newblock Cambridge Univ. Press, Cambridge, UK.

\bibitem[{{Wood} and {McIntyre}(2010){\it {Wood} and
  {McIntyre}\/}}]{wood-mcintyre-2010}
{Wood}, R.~B., and M.~E. {McIntyre}, 2010:
\newblock {A General Theorem on Angular-Momentum Changes due to Potential
  Vorticity Mixing and on Potential-Energy Changes due to Buoyancy Mixing}.
\newblock {\it Journal of Atmospheric Sciences\/}, {\bf 67}, 1261--1274.

\bibitem[{{Wordsworth}(2012){\it {Wordsworth}\/}}]{wordsworth-2012}
{Wordsworth}, R., 2012:
\newblock {Transient conditions for biogenesis on low-mass exoplanets with
  escaping hydrogen atmospheres}.
\newblock {\it \icarus\/}, {\bf 219}, 267--273.

\bibitem[{{Wordsworth} et~al.(2008){\it {Wordsworth}, {Read}, and
  {Yamazaki}\/}}]{wordsworth-etal-2008}
{Wordsworth}, R.~D., P.~L. {Read}, and Y.~H. {Yamazaki}, 2008:
\newblock {Turbulence, waves, and jets in a differentially heated rotating
  annulus experiment}.
\newblock {\it Physics of Fluids\/}, {\bf 20}(12), 126,602.

\bibitem[{{Wordsworth} et~al.(2010){\it {Wordsworth}, {Forget}, {Selsis},
  {Madeleine}, {Millour}, and {Eymet}\/}}]{wordsworth-etal-2010}
{Wordsworth}, R.~D., F.~{Forget}, F.~{Selsis}, J.-B. {Madeleine}, E.~{Millour},
  and V.~{Eymet}, 2010:
\newblock {Is Gliese 581d habitable? Some constraints from radiative-convective
  climate modeling}.
\newblock {\it \aap\/}, {\bf 522}, A22.

\bibitem[{{Wordsworth} et~al.(2011){\it {Wordsworth}, {Forget}, {Selsis},
  {Millour}, {Charnay}, and {Madeleine}\/}}]{wordsworth-etal-2011}
{Wordsworth}, R.~D., F.~{Forget}, F.~{Selsis}, E.~{Millour}, B.~{Charnay}, and
  J.-B. {Madeleine}, 2011:
\newblock {Gliese 581d is the First Discovered Terrestrial-mass Exoplanet in
  the Habitable Zone}.
\newblock {\it \apjl\/}, {\bf 733}, L48.

\bibitem[{Xu and Emanuel(1989){\it Xu and Emanuel\/}}]{xu-emanuel-1989}
Xu, K.-M., and K.~A. Emanuel, 1989:
\newblock Is the tropical atmosphere conditionally unstable?
\newblock {\it Mon. Wea. Rev.\/}, {\bf 117}, 1471--1479.

\bibitem[{{Yamamoto} and {Takahashi}(2003){\it {Yamamoto} and
  {Takahashi}\/}}]{yamamoto-takahashi-2003}
{Yamamoto}, M., and M.~{Takahashi}, 2003:
\newblock {The Fully Developed Superrotation Simulated by a General Circulation
  Model of a Venus-like Atmosphere.}
\newblock {\it Journal of Atmospheric Sciences\/}, {\bf 60}, 561--574.

\bibitem[{{Yoden} et~al.(1999){\it {Yoden}, {Ishioka}, {Hayashi}, and
  {Yamada}\/}}]{yoden-etal-1999}
{Yoden}, S., K.~{Ishioka}, Y.-Y. {Hayashi}, and M.~{Yamada}, 1999:
\newblock {A further experiment on two-dimensional decaying turbulence on a
  rotating sphere}.
\newblock {\it Nuovo Cimento C Geophysics Space Physics C\/}, {\bf 22},
  803--812.

\bibitem[{Zelinka and Hartmann(2010){\it Zelinka and
  Hartmann\/}}]{zelinka-hartmann-2010}
Zelinka, M.~D., and D.~L. Hartmann, 2010:
\newblock Why is longwave cloud feedback positive?
\newblock {\it J. Geophys. Res.\/}, {\bf 115}, D16,117.

\bibitem[{{Zurita-Gotor} and {Vallis}(2009){\it {Zurita-Gotor} and
  {Vallis}\/}}]{zurita-gotor-vallis-2009}
{Zurita-Gotor}, P., and G.~K. {Vallis}, 2009:
\newblock {Equilibration of baroclinic turbulence in primitive equations and
  quasigeostrophic models}.
\newblock {\it J. Atmos. Sci.\/}, {\bf 66}, 837--863.

\end{thebibliography}


\end{document}